\documentclass[3p]{elsarticle}
\journal{Physics Reports}
\usepackage{epsfig}
\usepackage{amsmath}
\usepackage{graphicx}
\usepackage{rotating}
\usepackage{color}
\usepackage{amssymb,amsmath}
\usepackage{calc}
\usepackage[numbers]{natbib}
\usepackage{url}

\usepackage[labelfont=bf,textfont={sl}]{caption}
\usepackage[hang, raggedright, nooneline]{subfigure}
\usepackage{pdflscape}
\usepackage{threeparttable}
\usepackage{booktabs}
\usepackage{changes}

\newcommand{\cora}{CR }

\def\deg{^{\circ}}
\def\be{\begin{equation}}
\def\ee{\end{equation}}

\definecolor{rappblau}{rgb}{0.086, 0.254, 0.574}
\definecolor{rapprot}{rgb}{0.527, 0.043, 0.238}

\usepackage{amssymb}  

\begin{document}

\begin{frontmatter}
\title{Closing in on the origin of Galactic cosmic rays using multimessenger information}
\author[ad1]{Julia Becker Tjus},
\author[ad1, ad2]{Lukas Merten}
\address[ad1]{RAPP Center and Theoretische Physik IV, Fakult\"at f\"ur Physik \& Astronomie, Ruhr-Universit\"at Bochum, 44780 Bochum, Germany}
\address[ad2]{Institute for Astro- and Particle Physics, University of Innsbruck, 6020 Innsbruck, Austria}
\begin{abstract}
In cosmic ray physics extensive progress has been made in recent years, both concerning theory and observation. The vast details in direct, indirect and secondary detections on the one hand provide the basis for a detailed modeling of the signatures via cosmic-ray transport and interaction, paving the way for the identification of Galactic cosmic-ray sources. On the other hand, the large number of constraints from different channels of cosmic-ray observables challenges these models frequently.   

In this review, we will summarize the state-of-the art of the detection of cosmic rays and their secondaries, followed by a discussion what we can learn from coupling our knowledge of the cosmic-ray observables to the theory of cosmic-ray transport in the Galactic magnetic field. Finally, information from neutral secondaries will be added to draw a multimessenger-picture of the non-thermal sky, in which the hypothesis of supernova remnants as the dominant sources survives best. While this has been known since the 1930s, evidence for this scenario is steadily growing, with the first possible detection of hadronic signatures at GeV energies detected for three SNRs with Fermi. The existence of SNRs as PeVatrons, however, is not validated yet. The discussion of this and other open questions concerning the level of anisotropy, composition and spectral shape of the cosmic-ray energy spectrum is reviewed. Future perspectives of how to find the smoking cosmic-ray source gun concludes this review.
  
\end{abstract}

\end{frontmatter}
\tableofcontents
\newpage
\section{Introduction}
\begin{figure}[htbp]
\centering{
\includegraphics[trim = 0mm 0mm 0mm 0mm, clip, width=0.8\textwidth]{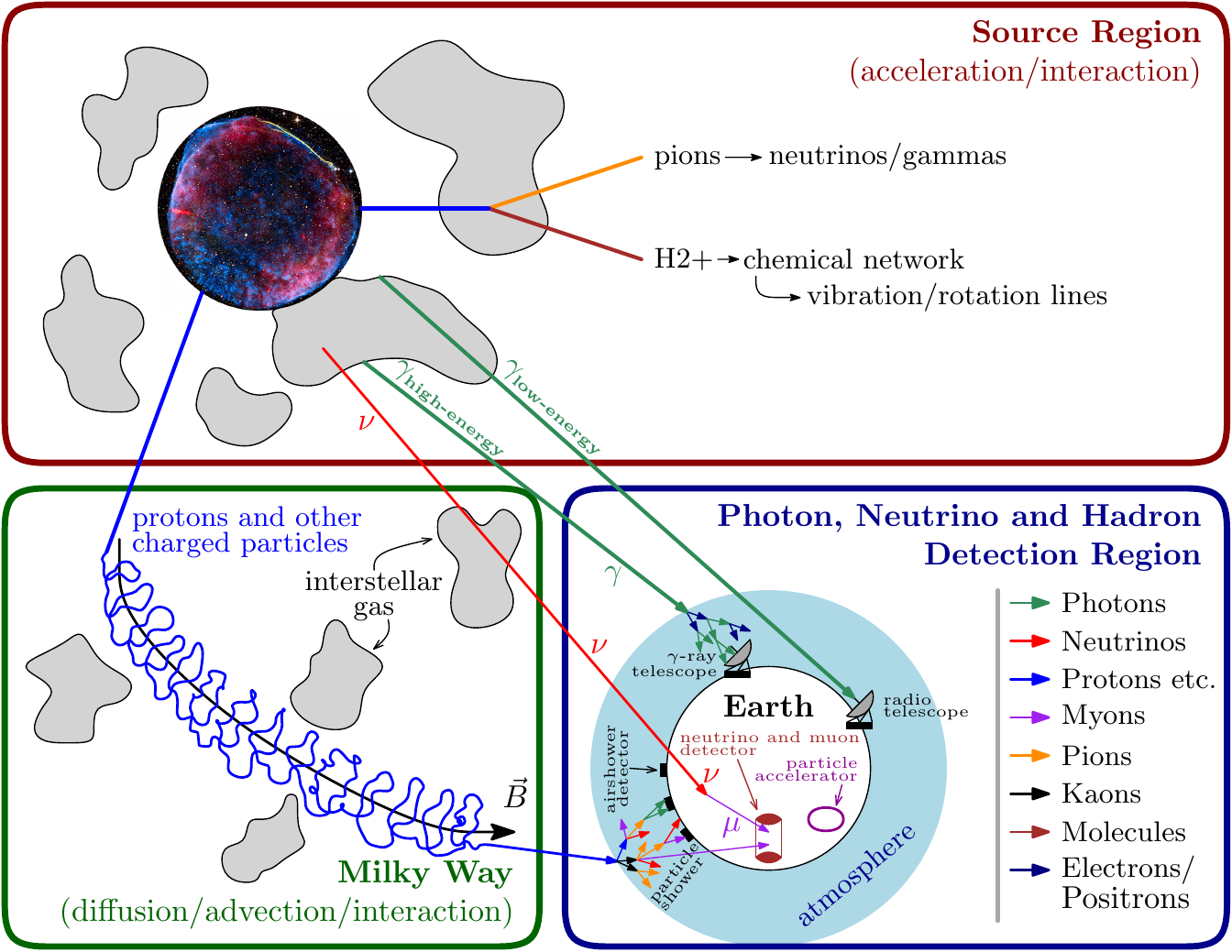}
\caption{Schematic picture of the non-thermal multimessenger emission from hadronic cosmic-ray sources in the Galaxy. Figure inspired by \citet{wagner2004}.
\label{multimessenger:fig}}
}
\end{figure}
  In 1934, 22 years after the first detection of cosmic rays, \citet{baade_zwicky1934} suggested supernova remnants as the primary sources of this high-energy radiation. With 
increasing knowledge on the cosmic-ray energy spectrum over the years, this hypothesis was strengthened, in particular as the class of supernova remnants in the Milky Way appears 
to be the only reasonable candidate to be able to explain the high flux of cosmic rays: it can deliver an intensity of around 1 particle per second and square-meter on Earth at 
$\sim $~50 GeV particle energies, which can only be obtained when 10\% of the energy released in supernova explosions of massive stars in the Galaxy goes into cosmic-ray 
production. Still, more than 100 years after the first detection of cosmic rays, the mystery of their origin is subject to state-of-the-art research: how can supernova remnants 
reach energies up to or above $10^{15}$~eV? How can the observed large- and small-scale anisotropy level of $0.1\% - 0.01\%$ be reproduced in the TeV -- PeV range? 

  Enormous progress has been made in encompassing details of cosmic rays and their tracers during the past $10-15$ years by putting a large variety of top observatories in 
operation. 
As for direct cosmic-ray measurements, composition information about the cosmic-ray spectrum is known from GeV to $>$PeV energies, several breaks in addition to the long-known 
\emph{knee} and \emph{ankle} features have been detected and the observation of the anisotropy level in the TeV -- PeV range has been established on small and large scales. 
Concerning the direct view on the sources, more than 80 Galactic TeV emitters have been identified and a diffuse high-energy neutrino flux was confirmed for the first time in 2013 \citep{icecube2014}, although not dominantly arising in the Galaxy. 

  These many details of multimessenger information as summarized in Fig.\ \ref{multimessenger:fig} are both a blessing and a curse: as more and more details are revealed,
  the simplified models that have been used for many years in order to make the case for supernova remnants as primary source candidates for the cosmic-ray flux are not able to reproduce all these details. At the same time, these insights have triggered the development of complex multimessenger methods and models. 
  
  In this review article, the current state of the art of our knowledge and ignorance of Galactic cosmic-ray sources will be presented. 
  In particular, primary cosmic-ray observables from MeV to EeV will be presented as well as information from secondary cosmic rays, produced in cosmic-ray interactions at 
the production site or in the interstellar medium on their way to Earth. This includes a (potentially) direct view of the sources via e.g.\ ionization signatures, neutrinos and 
gamma-rays as well as those pieces of information provided by cosmic rays themselves, i.e.\ spectrum, chemical composition and arrival directions. 

Here, astrophysical parameters of the source candidates and of the interstellar medium through which the propagation of cosmic rays needs to be 
modeled will be presented, in particular concerning the description of magnetic field configurations, gas densities, etc. The modeling of hadronic cosmic-ray interactions with 
cross-sections that match recent measurements at LHC is another aspect that needs to be considered in order to properly include hadronic secondary production. Finally, radiation 
from cosmic-ray electrons, in particular synchrotron radiation, Inverse Compton (IC) and bremsstrahlung (brems), on the one hand are used to investigate the acceleration 
environment. The high-energy signatures  also need to be disentangled from hadronically produced gamma-rays.

  It is the aim of this article to show how recent advances in experimental setups for the detection of cosmic-ray signatures in combination with theoretical modeling start to shed 
light on the answer to one of the central questions in physics and astrophysics: \textit{what is the origin of cosmic rays?}

  This review is organized as follows:  a short review on the theoretical modeling of cosmic-ray electron and hadron interaction and radiation signatures in the context of non-thermal astrophysics is given in Section \ref{multimessenger_sources:sec}. Section \ref{data:sec} presents the state-of-the-art of cosmic-ray data, including the cosmic ray all-particle spectrum from GeV to EeV energies together with what is known about composition, (an-)isostropy and the detection of cosmic-ray interaction products, gamma-rays and neutrinos. Section \ref{candidates:sec} presents the knowledge gained from the connection of theoretical modeling of cosmic-ray transport and data. In particular, basic arguments are presented why cosmic rays below the knee should be of Galactic origin and why the population above the ankle should be of extragalactic origin (galaxy clusters, active galaxies and/or gamma-ray bursts). These considerations serve as a starting point concerning the detailed multimessenger modeling of local signatures of neutral secondaries that are detected directly from (potential) cosmic-ray sources and diffuse signatures of the photon spectrum that are discussed in connection with cosmic rays (Section \ref{multimessenger_modeling:sec}).  
  Section \ref{summary:sec} presents summary and conclusions together with an outlook.   
\section{Theory on signatures from cosmic-ray secondaries \label{multimessenger_sources:sec}}
Cosmic-ray interactions with gas or photon fields are useful diagnostics in order to identify the origin of these charged particles. What makes the situation complex is the co-existence of non-thermal nuclei, protons \textit{and} electrons. All three show similar signatures of interactions, in particular concerning their high-energy gamma-ray spectra. Anomalous signatures of positrons and anti-particles, as well as neutrino emission are clear signs of a hadronic origin, if not dark matter signals.

Here we review continuum signatures (Section \ref{continuum:sec}) as well as line emission (Section \ref{line:sec}). For the continuum radiation, we focus on nuclei-nuclei interactions and mention photohadronic processes, which are typically sub-dominant in the Galactic context. In the same sub-section, we further discuss leptonic emission (synchrotron, IC and bremsstrahlung) in order to have a full picture. Line emission
is discussed in the context of the excitation or ionization of hydrogen through cosmic rays.

There are entire reviews and books dedicated to these interaction processes. 
In this section, we will review the different interaction processes that lead to interesting signatures concerning their potential of identifying the origin of Galactic cosmic rays. For the details of the leptonic continuum processes, we refer to the books of \citet{rybicki_lightman1979,schlickeiser2002} and the review by \citet{blumenthal_gould1970}. Inelastic hadronic interactions are reviewed in the textbooks of \citet{schlickeiser2002,sigl2017} and the research paper of \citet{schlickeiser_mannheim1994}. Line emission is reviewed in \citet{grenier2015}.

        \subsection{Continuum signatures \label{continuum:sec}}
\subsubsection{Hadronic secondaries}

Hadron-hadron interactions have been studied in the Galactic context on different levels, in particular concerning the diffuse signature from cosmic-ray interactions with the Galactic gas, see e.g.\ \citet{strong_propagation_1998} for early, groundbreaking work. Here, the production of pions via these interactions is of high importance, as these in turn deliver signatures of high-energy gamma-rays and neutrinos, as well as positrons:
\begin{eqnarray}
  p\,p&\rightarrow& \sum\pi^{\pm,0}\\
  p\,\gamma&\rightarrow&\left\{\begin{array}{lll}
      \Delta^{+}\rightarrow p\pi^{0}/n\pi^{+}&&\rm Delta-resonance\\
      \sum\pi^{\pm,0}&&\rm Multi-pion production\\
      \end{array}\right.\\
  \pi^{+}&\rightarrow& \mu^{+}+\nu_{\mu}\rightarrow (e^{+}\,\nu_{e}\,\overline{\nu}_{\mu})+\nu_{\mu}\,\\
  \pi^{-}&\rightarrow& \mu^{-}+\overline{\nu}_{\mu}\rightarrow (e^{-}\,\overline{\nu}_{e}\,{\nu}_{\mu})+\overline{\nu}_{\mu}\,\\
  \pi^{0}&\rightarrow& \gamma\gamma\,.
\end{eqnarray}
The neutral secondaries ($\nu$, $\gamma$) can help to identify the sources of cosmic-ray as --- in contrast to charged hadrons --- they are pointing back to the origin of their production.\footnote{It should be noted that the contribution from Kaons can be important as well, in particular at higher energies. The production of hadrons with heavier quark contents that decay quicker via the electromagnetic interaction can become important in very dense environments, in which the pions, decaying via the weak interaction, interact before they can decay. This effect is in particular important in the Earth's atmosphere, but can also play a role in cores of active or starburst galaxies. Kaons and charm-quark containing hadrons like $\Lambda_c,\,D^{\pm},\,D^0$ and $D_s$ can contribute significantly \citep{sibyll2019}.}

Interactions of higher nuclei have usually been treated with a \textit{mass scaling approach}. Nuclei are described as a superposition of the individual nucleons. In this approach, the nuclei cross section is scaled as $\sigma_{\mathrm{Nuc}\gamma}=A^\alpha\,\sigma_{p\gamma}$, where $A$ is the mass number of the nucleon. This scaling is used in many semi-analytical studies and different propagation codes, e.g., \citet{anchordoqui_2008, murase2008, crpropa30, SimProp2017, boncioli_2017}. More recent studies, e.g. \citet{morejon_2019}, have assumed that the scaling index $\alpha$ is not constant but changes with energy from $\alpha\approx 1$ to $\alpha=2/3$ at the highest interaction energies. 

With a scaling at hand, a detailed description of proton-proton interactions is sufficient which then can  be extended to heavier nuclei as described above. From proton-proton interactions, the number of pions $dN_{\pi}$ in the energy interval $(E_{\pi},\,E_{\pi}+dE_{\pi})$ is determined by the spectrum resulting from multi-pion production via $p\,p\rightarrow \pi^{\pm/0}$:
\begin{equation}
  q_{\pi^{\pm,0}}(E_{\pi}) = \int_{E_{\rm th}}^{\infty} dE_{p} F_{\pi^{\pm,0}}(E_{\pi}(E_{p}))\, \int_{0}^{\tau}d\tau'\,q_p(\tau')\,.
\end{equation}
Here, $F_{\pi}$ is the distribution function of  pions that results from the interaction of a proton of energy $E_p$ and $q_p(E_p)$ is the original proton spectrum at the point of interaction $q_p$. Assuming a proton spectrum at the source of acceleration $j_p(E_p)$, for a homogeneous target density $n$, the proton spectrum at the point of interaction is damped due to previous interactions, i.e.\ $q_p=j_p\cdot \exp(-\tau)$. The integration is performed starting at the threshold for inelastic proton-proton interactions to happen, $E_{\rm th}$. Further, $\tau=l\cdot n\cdot \sigma_{pp}$ is the optical depth with $l$ as the propagation length after acceleration, $n$ is the target density and $\sigma_{pp}\approx 3\times 10^{-26}$~cm$^{2}$ is the inelastic proton-proton cross-section at a center of mass energy of $\sqrt{s}\approx10$~GeV. It should be noted that this cross-section increases logarithmically in energy, thus leading to a slight flattening of the secondary spectra with respect to the primaries \citep[Fig.\ 51.6]{pdg2018}.  Furthermore, depending on the specific source scenario, the optical depth effect could also result in a softening of the spectrum.

A full description of the pion distribution function can be achieved via event generators like SIBYLL \citep{sibyll2019}, QGSJET \citep{qgsjet2011}, EPOS \citep{epos2008} or DPMJET \citep{dpmjet2008}. An analytical solution by fitting the SIBYLL results is presented by \citet{kelner_pgamma2008}. These results are accurate above 100~GeV and represent a simple way for a precise treatment of the distribution functions. 

To receive a first order-of-magnitude result, the so-called \textit{Delta-Approximation} presented in e.g.\ \citet{schlickeiser_mannheim1994,becker2014} can be used. This approach assumes that all pions in an interaction are produced at the average energy, i.e.\
\begin{equation}
  F_{\pi}=\xi_{\pi^{\pm,0}}\cdot \delta(E_{\pi}-\langle E_{\pi}\rangle)\,.
\end{equation}
Here, the pion multiplicity of an interaction at low energies can roughly be approximated as
\begin{eqnarray}
  \xi_{\pi^{\pm}}&=&2\cdot \left(\frac{E_p-E_{\rm th}}{\rm GeV}\right)^{1/4}\,,\\
  \xi_{\pi^{0}}&=&\xi_{\pi^{\pm}}/2\,.
  \label{xi_delta:equ}
\end{eqnarray}
For a power-law injection spectrum in kinetic energy, $j_p=A_p\, ([E_p-m_p\,c^2]/GeV)^{-\alpha}$, and for an approximate scaling of the average pion energy $\langle E_{\pi}\rangle \sim 1/6\,([E_p-m_p\,c^2]/GeV)^{3/4}$, the pion spectrum can be written as 

\begin{equation}
q_{\pi^{\pm}}\approx 20\cdot N_H\,A_p\,\sigma_{pp}\,\left(\frac{6\cdot E_{\pi}}{\rm GeV}\right)^{-4/3\,(\alpha-1/2)}\,.
\end{equation}
Here, the column density $N_H:=l\,n$ was introduced and it was assumed that the threshold energy for pion production is $E_{\rm th}\approx m_p\,c^2$. The $\pi^{0}$-spectrum is half of that value according to Equ.\ (\ref{xi_delta:equ}). The neutrino spectrum of a certain flavor $\nu_i=\overline{\nu}_e,\nu_e,\overline{\nu}_{\mu},\nu_{\mu}$ can be determined by assuming that $1/4$th of the energy of the original pion goes into each of the flavors:
\begin{equation}
q_{\nu_i}(E_{\nu_i})=q_{\pi^{\pm}}(E_{\pi})dE_{\pi}/dE_{\nu_i}=4\cdot q_{\pi^{\pm}}(4E_{\nu_i})\,.
\end{equation}
For gamma-rays, the $\pi^0$ decays into particles of the same energy,
\begin{equation}
q_{\gamma}(E_{\gamma})=q_{\pi^{0}}(E_{\pi})dE_{\pi}/dE_{\gamma}=2\, q_{\pi^{0}}(2E_{\gamma})\,.
\end{equation}
Thus, in rough approximation, the resulting total neutrino and gamma spectra become\footnote{For the neutrinos, we do not distinguish between anti-particles and particles in this representation.} \citep{becker2014}

\begin{eqnarray}
  q_{\nu,\,tot}=q_{\nu_{\mu}}^{(1)}+q_{\nu_{\mu}}^{(2)}+q_{\nu_{e}}&\approx& 300\cdot N_H\cdot A_p\cdot \sigma_{pp}\cdot \left(\frac{24\cdot E_{\nu}}{\rm GeV}\right)^{-4\alpha/3+2/3}\\
  q_{\gamma,\,tot}=2\cdot q_{\gamma}&\approx & 50\cdot N_H\cdot A_p\cdot \sigma_{pp}\cdot \left( \frac{12\cdot E_{\gamma}}{\rm GeV}\right)^{-4\alpha/3+2/3}\,.
\end{eqnarray}
Here, an additional factor of 1.6 was multiplied to the fluxes $q_i$ taking into account an additional contribution from Helium following \citep{schlickeiser2002}. Here, it is assumed that all other elements are present at a negligible fraction at the sub-percent level.

This simplified result still has its approximate validity --- in particular, today's event generators are typically optimized toward the highest energies and results are usually valid above $\sim 80-100$~GeV. For lower-energy interactions, low-energy particle interaction models like Fluka \citep{fluka2014} or URQMD \citep{urqmd2008} need to be applied instead. An interaction model that is valid over a broad range of energies does not exist as of today.

When applying the delta-approximation, it still needs to be kept in mind that (a) the reproduced power-law behavior of the secondary spectra is not correct, mostly because the cross-section is energy dependent and rather leads to a further flattening of the total spectrum \citep{kelner_pp2006}; (b) the deviation becomes particularly large (factor $2$ and more) toward the highest energies, i.e.\ in the TeV--PeV region and above \citep{kelner_pp2006}; the three  neutrinos produced in one interaction have slightly different spectra, in particular concerning their low-energy cutoff and their normalization. While the second muon neutrino and the electron neutrino are produced in coincidence and have the same spectrum, the neutrino from the first interaction receives a slightly different spectrum, about 20\% reduced in flux intensity as compared to the other two neutrino spectra \citep{kelner_pp2006}.

For neutrinos produced in interactions at the acceleration site of Galactic cosmic-ray sources or during cosmic-ray propagation through the ISM, the following aspects are relevant:
\begin{itemize}
    \item Neutrino oscillations do not change the total neutrino flux. However, when considering only muon neutrinos or another flavor, the ratios have to be adjusted from $(\nu_\mu:\nu_e:\nu_\tau)\left|_{\rm source}\right.=(1:2:0)$ at the source to $(\nu_\mu:\nu_e:\nu_\tau)\left|_{\rm \oplus}\right.\sim (1:1:1)$ at Earth.
     \item For target column densities exceeding $N_{H}\gtrsim 10^{25}$~cm$^{-2}$, charged pions interact at optical depths $\tau_{p\pi}>1$, their flux is reduced and heavy quark contributions become important instead. For the ISM in the Milky Way, with a typical density of $n_{H}\sim 1$~cm$^{-3}$ and scales of $1-10$~kpc, these effects can typically be neglected. For cores of starburst galaxies, in which the density can become as large as $1000$~cm$^{3}$ with propagation scales of $>$~kpc, these effects could start to become important.
\end{itemize}

\subsubsection{Leptonic secondaries}
Accelerated electrons (and positrons) cause radiation via synchrotron emission, bremsstrahlung and IC scattering. Here, we shortly review the main properties of these signatures. 

\paragraph{Synchrotron radiation}
A single electron emits photons in a frequency spectrum $dI/d\omega(x)$ with $x:=\omega/\omega_c$ as the radial frequency of the emission normalized to the critical frequency $\omega_c:=3/2\,\gamma^3\,\omega_B\,(1-\mu^2)^{1/2}$ at which the spectrum is expected to cut off,  with $\gamma=E_e/(mc^2)$ as the Lorentz factor and $\omega_B$ as the gyro-frequency of the particle. The exact description for a single electron is derived within the concept of electrodynamics (in cgs-units, see e.g.\ \citet{jackson1962}) as
\begin{equation}
\frac{dI}{d\omega}=2\cdot\sqrt{3}\,\frac{e^2}{c}\,\gamma\,\frac{\omega}{\omega_c}\,\int_{2\omega/\omega_c}K_{5/3}(x)dx\,.
\end{equation}
Here, $K_{5/3}$ is the modified Bessel function. The spectral behavior can be approximated in the limits $x\ll 1$ and $x\gg 1$ as
\begin{equation}
  \frac{\mathrm{d}I}{\mathrm{d}\omega}\propto\left\{\begin{array}{lll}x^{1/3}&\rm for&x\ll 1\\
  x^{1/2}\,\exp(-2\,x)&\rm for&x\gg 1\end{array}\right.\,.
\end{equation}
For a population of electrons, the total frequency spectrum becomes
\begin{equation}
\left.\frac{\mathrm{d}I}{\mathrm{d}\omega}\right|_{tot}=\int \frac{\mathrm{d}P}{\mathrm{d}\omega}(x(E_e))\,\frac{\mathrm{d}N_e}{\mathrm{d}E_e}\,\mathrm{d}E_e\,.
\end{equation}
If we assume a power-law spectrum, $\mathrm{d}N_e/\mathrm{d}E_e\propto E_{e}^{-\alpha}$, the total frequency spectrum can be described as
\begin{equation}
\left.\frac{\mathrm{d}I}{\mathrm{d}\omega}\right|_{tot}\propto \omega^{-a}
\end{equation}
with an exponent that depends on the electron spectral index
\begin{equation}
  a:=\frac{\alpha-1}{2}\,.
  \label{indices:equ}
\end{equation}
This result has two implications:
\begin{enumerate}
\item The measurement of a power-law spectral behavior in the radio to X-ray range can be interpreted as non-thermal emission from electrons that also follow a power-law. A process that accelerates electrons should even be able to accelerate hadrons, as the acceleration process only\footnote{At the lowest (injection) energies, the spectrum is actually expected to be influenced by the gyroradius of the particle, thus resulting in different expectations for protons and electrons. At energies above the rest mass of the particle, however, these effects are negligible, see e.g.\ \citep{merten2017a} and references therein for a discussion.} depends on the amount of charge $|Z|$.
\item The spectral index of the primary distribution $\alpha$ can be deduced from such measurements according to Equ.\ (\ref{indices:equ}). One pitfall with this interpretation is that the question of a synchrotron (or Inverse Compton) loss-dominated region has to be answered first: if the region is dominated by the cooling term, the spectrum of electrons itself is steepened by one power as the cooling time scale behaves as $\tau_{\rm synch/IC}\propto E^{-1}$, i.e.\ $\alpha_{\rm cool} =\alpha+1$ (for $\alpha>1$) \citep{rybicki_lightman1979}. Thus, the spectral index received from Equ.\ (\ref{indices:equ}) is the cooled one $\alpha_{\rm cool}$ rather than the spectral index of acceleration. As an example, FR galaxies have a spectral index of $a\sim 0.8\pm 0.2$ \citep{bbr2005}. Using the non-cooled interpretation of Equ.\ (\ref{indices:equ}), the spectrum of the electrons is $E^{-2.6}$. The cooling scenario is rather unlikely as the index is not steep enough with $a<1$ for most sources.

  In a cooling-dominated environment, the electron index cannot directly be identified with the proton index, as protons do not cool as quickly as electrons: the pitch-angle averaged energy loss rate for a particle of charge $q$ in cgs units is given as
  \begin{equation}
    P_{\rm synch} :=\frac{dE}{dt}=\frac{4}{9}\,\frac{q^{4}}{m^{4}\,c^7}E^{2}\,B^2\,.
\label{synchr_loss:equ}
  \end{equation}
Making use of Equ.\ (\ref{synchr_loss:equ}), the corresponding loss time scale is given by
\begin{equation}
\tau_{\rm loss}\approx |E\cdot \left(\frac{dE}{dt}\right)^{-1}|=\frac{9}{4}c^7\,\frac{m^4}{q^4}\,E^{-1}\,B^{-2}\,.
  \end{equation}  
  Thus, at equal energies $E_e=E_p=E$ and in the same magnetic field, the radiation of a proton is a factor of $(m_e/m_p)^{4}\approx (2000)^{-4}\sim 10^{-13}$ weaker as compared to electrons. The comparison at the same Lorentz factor $d\gamma/dt$ results in a scaling  $\propto m^{-3}$, but still results in a reduced emission from protons at a level of $10^{-10}$. Thus, sources of synchrotron radiation need to be investigated with respect to their loss limit concerning electrons, while protons in \textit{most}\footnote{but not all (see e.g.\ \citet{muecke2003,biermann_agn_synch2009})} environments will not show significant synchrotron emission.
\end{enumerate}

\paragraph{Relativistic bremsstrahlung}
The radiation that is emitted spontaneously when an electron passes close to a nucleus is called bremsstrahlung. This effect exists both for the interaction of electrons with nuclei and with neutral atoms. Details on the differences for a neutral and fully ionized gas are discussed below.

As described by \citet{blumenthal_gould1970}, the differential cross-section for an electron with initial and final energies $E_i$ and $E_f$ is given as
\begin{equation}
 \frac{\mathrm d\sigma}{\mathrm d\epsilon} = \frac{\alpha r_0^2}{\epsilon E_i^2} \left[(E_i^2+E_f^2)\phi_1-\frac{2}{3}E_iE_f\phi_2\right]\,,
\end{equation}
In this process, a photon with energy $\epsilon =E_i-E_f$ is produced. Here, $\alpha$ is the fine-structure constant and $r_0$ the Bohr radius. The functions $\phi_i$ depend on the nature of the Coulomb field with which the particles scatter. These functions can therefore naturally include the changes in the field due to the electrons in the shell of an atom. For an unshielded particle, 
the functions become $\phi_1=\phi_2=Z^2\phi_u$ with 
\begin{equation}
  \phi_u = 4\left(\ln\left[(\Delta\,2\,\alpha)^{-1}\right]
  -\frac{1}{2}\right).
\end{equation}
with $\Delta = \epsilon\,mc^2/(2\,E_i\,E_f)$. This approximation holds for $\Delta >2$.
For a nucleus with one electron in the shell or more complex systems, the functions $\phi_u$ become more complicated as described in detail in \citet{blumenthal_gould1970}.

Molecular hydrogen typically has the highest abundance in galactic molecular clouds. Their cross section can be approximated by the one for atomic hydrogen with an error below 3\% \citep{Gould1969,schlickeiser2002}, hence $d\sigma_{H_2}/d\epsilon\approx 2\,d\sigma_{H-I}/d\epsilon$.

The total energy spectrum resulting from bremsstrahlung can be expressed as

\begin{eqnarray}
  \frac{\mathrm dN}{\mathrm dt  d\epsilon} &=& c\int\limits_{\epsilon+mc^2}^\infty \frac{dN_e}{dE_{e}}(E_{e})\sum\limits_s N_s\frac{\mathrm d\sigma_s}{\mathrm d\epsilon}\,\mathrm dE_e\\
  &\approx& 1.3\,c\int\limits_{\epsilon+mc^2}^\infty \frac{\mathrm{d}N_e}{\mathrm{d}E_{e}}(E_{e})\left( n_{H-{I}}+2n_{H_{2}}\right)\frac{\mathrm d\sigma_{H-{I}}}{\mathrm d\epsilon}\,\mathrm dE_{e}\,.
\end{eqnarray}
Here, we sum over the dominant species $s={\rm H-{I}},\,{\rm H_2},\,{\rm He}$ and assume a threshold energy of $E_{e,\min}=\epsilon+m\,c^2$ in the integral. The latter approximation is taking into account Helium by applying a global factor $1.3$ \citep{schlickeiser2002}.  Solving this integral for a primary power-law distribution, $dN_e/dE_e=N_{e,0}\, E_{e}^{-\alpha}$ in the limit of strongly shielded relativistic electrons results in a spectral distribution
\begin{equation}
\frac{\mathrm{d}N}{\mathrm{d}t \mathrm{d}\epsilon}\approx 2\times 10^{3}\,N_{e,0}(\vec{r})\frac{\alpha\,\sigma_T}{8\,\pi}\left[n_{H-{I}}+2n_{H_{2}}\right]\,\frac{1-2(\alpha-2)/(3\alpha(\alpha+1))}{\alpha-1}\,\epsilon^{-\alpha}\,.
\end{equation}
Here, $\sigma_T=e^2/(6\pi\,\epsilon_0\,m_e\,c^2)\approx 6.65\times 10^{-29}$~m$^2$ is the Thomson cross section.
Thus, in the limit of strong shielding, i.e.\ in a gas that is mostly neutral, the spectral energy distribution from bremsstrahlung follows the one of the primary particles. The integral cannot be evaluated analytically for the case of moderate shielding and needs to be treated numerically.

As for the energy loss rate, different cases have to be considered (strong and weak shielding as well as the general case). All three regimes have in common that the energy loss rate increases linearly with $E_e$. The loss rate is largest in the case of strong shielding \citep{blumenthal_gould1970}:
\begin{equation}
P_{\rm brems} \approx 45 \frac{\alpha\,c\,\sigma_T}{2\pi}\left[n_{H-I}+2\,n_{H_2}\right] E_{e}\,.
\end{equation}
Here, we use the approximation $\phi_{1,H-I}(\Delta=0)\approx \phi_{2,H-I}=:\phi_{H-I}\approx 45$.

The linear behavior with the electron energy is in contrast to synchrotron radiation and Inverse Compton scattering, for which the loss rates increase with $E_{e}^{2}$ (see detailed discussion in Section \ref{radiation_discussion:sec}).

\subsubsection{Inverse Compton Radiation}
In an environment with relativistic electrons and a low-energy photon field, energy is transferred to the photons, following from four-momentum conservation of the quantum-mechanical interaction:
\begin{equation}
\epsilon_{\rm IC}=\frac{2E_e\,\epsilon_{\rm low-E}}{\epsilon_{\rm low-E}\,(1-\cos\theta)+E_e(1+\cos\theta)-\frac{m_{e}^{2}}{2E_e}\cos\theta}\,.
\end{equation}
Here, $\epsilon_{\rm IC}$ is the energy of the photon after the interaction, $E_e$ and $\epsilon_{\rm low-E}$ the electron and photon energy before interaction, respectively. The angle $\theta$ is defined between the incoming electron and photon in the observer's frame.
In the limit of $E_e\rightarrow m_e\,c^2$, this general equation for the \textit{Inverse Compton Scattering} turns into the well-known energy transfer from a high-energy photon to the electron at rest, the classical Compton scattering. In the non-thermal environments, we expect a population of accelerated electrons, $dN_e/dE_e\propto E_{e}^{-\alpha}$. Low-energy photons are then bound to be present via synchrotron radiation of the same electrons. For interactions of electrons with their own synchrotron photons, the process is called \textit{Synchrotron Self Compton (SSC)}. Other possible sources of scattering are radiation fields from stars, peaking at UV energies, or the CMB, peaking at microwave-lengths.

The energy rate that goes into Inverse Compton (IC) in the classical Thomson limit ($ \Gamma := \frac{4\epsilon\,E_{e}}{m_{e}^{2}c^4}\ll 1$) is given as
\begin{equation}
P_{\rm IC}^{\rm Thomson}=\frac{4}{3}\,\sigma_T\,c\beta^2\,\gamma^2\,U_{\rm ph}
\end{equation}
with $U_{\rm ph}$ as the energy density in the low-energy photon field. In astrophysical environments, this is often a blackbody target field or a non-thermal radiation component, for instance the synchrotron field that is produced by the relativistic electrons themselves.

This result can be compared to the energy loss rate in synchrotron radiation. Expressing the latter in terms of the energy density of the magnetic field, $U_B=B^2/(2\mu_0)$, we receive
\begin{equation}
P_{\rm synch}=\frac{4}{3}\,\sigma_T\,c\gamma^2\,\beta^2\,U_B = \frac{U_B}{U_{\rm ph}}\,P_{\rm IC}\,.
\end{equation}
Thus, for stationary photon and magnetic fields, the loss rates of synchrotron and IC radiation scale as a constant ratio of the two energy densities. Both terms increase with $\gamma^2$. They are therefore typically combined into one single loss term in the transport equation, see Equ.\ (\ref{transport:equ}). A typical approach concerning the fitting of non-thermal spectral energy distributions (SEDs), in particular concerning the ones of supernova remnants in the Galaxy, is to first fit the synchrotron bump, as its interpretation is rather unambiguous, and then subsequently fit the high-energy part by trying different combinations of hadronic and leptonic contributions. The scaling for the IC component can easily be predicted from the considerations above: First, the fitting of the synchrotron will result in the constraint of the quantity $n_e\,B^2$, which needs to be fixed in order to reach the proper normalization of the synchrotron bump. As the B-field is typically not known, it can then be varied, but keeping the product $n_e\,B^2$ constant. This in turn means that an increase of the B-field results in the decrease of the electron density. Therefore, for higher B-fields, at a fixed, observationally known synchrotron field, the IC component would be reduced. The normalization of the IC energy spectrum therefore scales as $B^{-2}$ (again, if the synchrotron field is observationally fixed). The knowledge of a multiwavelength-energy spectrum from radio wavelength to TeV energies therefore provides the possibility to fit the spectrum using different B-fields and identify the best-fit scenario. This in principle can narrow down the real-existing B-fields in the source. The complication with this approach is that other radiation fields, in particular bremsstrahlung and photons from hadronic interactions, also play a role, independent of the magnetic field.  Thus, the determination of the B-field via multiwavelength observation of the two-bump structure of non-thermal emission is not straight forward.

The spectral energy behavior of a Compton spectrum for a single electron is given as \citep{Jones1968}
\begin{equation}
  \frac{\mathrm dN_\gamma}{\mathrm d\epsilon\mathrm d\epsilon_{\rm IC}} = c\,n(\epsilon)\frac{2\pi r_0^2 m_{e}^{2}c^4}{E_{e}^{2}}\frac{1}{\epsilon}\,F(\Gamma,\,q)\,.
  \end{equation}
Here, $r_0$ is the Bohr radius, $\epsilon$ is the energy of the photons in the original photon field with a density $n(\epsilon)$ the differential photon density, $\epsilon_{\rm IC}$ is the energy of the up-scattered IC-photons. The function $F(\Gamma,\,q)$ is defined as
\begin{equation}
F(\Gamma,\,q):=\left[2q\ln q+(1+2q)(1-q)+\frac{1}{2}\frac{(\Gamma q)^2}{1+\Gamma q}(1-q)\right]\,.
  \end{equation}
The two dimension-less parameters $\Gamma$ and $q$ are defined as 
\begin{equation}
 \Gamma = \frac{4\epsilon\,E_{e}}{m_{e}^{2}c^4}, \quad q = \frac{\epsilon_{\rm IC}}{\Gamma\left(E_e-\epsilon_{\rm IC}\right)}\,.
\end{equation}
The limit $\Gamma\ll 1 $ corresponds to the classical Thomson limit, while $\Gamma \gg 1$ is the extreme Klein-Nishima limit.

The total energy spectrum from IC is then again given by the convolution of the single electron radiation field with the energy distribution of the relativistic electron population:
\begin{equation}
 \frac{\mathrm dN}{\mathrm d\epsilon_{\rm IC}} = \int\limits_0^\infty\int\limits_{E_\mathrm{min}}^{\infty} \frac{\mathrm dN_\mathrm{e}}{\mathrm dE_{e}}(E_{e})\frac{\mathrm dN_\gamma}{\mathrm d\epsilon\mathrm d\epsilon_{\rm IC}}(E,\epsilon,\epsilon_{\rm IC})\,\mathrm dE_{e}\mathrm d\epsilon\,.
\end{equation}
The lower integration limit $E_\mathrm{min}$ results from the kinematics of the problem \citep{blumenthal_gould1970}, leading to the inequality $\epsilon_{\rm IC}(1+\Gamma)\leq \Gamma E_e$. It can be expressed as
\begin{equation}
  E_\mathrm{min} = \frac{\epsilon}{2}\left[1+\sqrt{1+\frac{m_{e}^{2}c^4}{\epsilon_{\rm IC}\epsilon}}\right]\,.
\end{equation}
For a power-law distribution at injection, $dN_e/dE_e\propto E^{-\alpha}$, the spectral behavior of the Inverse Compton gamma-ray energy spectrum in the Thomson limit becomes
\begin{equation}
\frac{\mathrm dN}{\mathrm d\epsilon_{\rm IC}} \propto \epsilon^{-\frac{\alpha-1}{2}}\,.
\end{equation}
Thus, the spectral behavior of the IC process resembles the one of synchrotron radiation, but with its peak shifted to higher energies by a factor  $E_{\rm IC,\,\max}^{\rm Thomson}=4\,(E_e/(m_e\,c^2))^{2}\,\epsilon_{\max}$ in the Thomson limit.

The quantummechanical, extreme Klein-Nishina limit is given by  $\Gamma \gg 1$.  The loss rate in this case for an isotropic photon field can be approximated as
\begin{equation}
  P_{\rm IC}^{\rm KN} =  \frac{4}{3}\,\sigma_T\,c\,m_{e}^{2}\,c^4\,\int_{0}^{\infty}d\epsilon\,\frac{n_{\rm ph}}{\epsilon}\left(\ln(\Gamma)-\frac{11}{6} \right)\,.
  \label{IC_KN:equ}
\end{equation}
The loss rate is proportional to $P_{\rm IC,\,KN}\propto \ln(\Gamma)$ and thus $\propto \ln(E_e)$ \citep{schlickeiser2002}. This is opposed to the Thomson limit, in which the loss rate goes faster, as $E_{e}^{2}$.  However, the energy loss per interaction is stronger in the Klein-Nishina limit \citep{blumenthal_gould1970}.  The spectral behavior in the Klein-Nishina limit becomes significantly steeper than the Thomson-limit, 
\begin{equation}
\frac{\mathrm dN}{\mathrm d\epsilon_{\rm IC,\,KN}} \propto \epsilon_{\rm IC}^{-\alpha-1}\,.
\end{equation}
The maximum energy in the Klein Nishina limit is given by $E_{\rm IC,\,\max}^{\rm KN}=E_{e,\,max}$. It therefore depends linearly on the boost factor rather than the boost factor squared, as is the case in the Thomson regime.

Thus, depending on the energy range in which the target photon spectrum peaks, the distribution of the Inverse Compton spectrum will turn from an $\epsilon_{\rm IC}^{-(\alpha-1)/2}$-spectrum to an $\epsilon_{\rm IC}^{-\alpha-1}$-spectrum, e.g.\ $\epsilon_{\rm IC}^{-0.5}$ to $\epsilon_{\rm IC}^{-3}$ for an $E_{e}^{-2}$ primary spectrum.

\subsection{Line signatures from cosmic-ray impact\label{line:sec}} 
              There are three diagnostical methods connected to cosmic-ray induced line spectra --- Balmer-line signatures reveal features of cosmic-ray acceleration (see \citet{blasi2013} for a review) and the rovibrational lines of cosmic-ray spallation products help to identify cosmic-ray spallation yields and diffusion properties with it (see \citet{eastrogam2017,eastrogam2018} for reviews). One of the most prominent features  comes from cosmic-ray ionization, which can be made visible via lines of the chemical network that follows. 
              These chemical elements are produced in a rotational- and/or vibrational state are usually referred to as \textit{rovibrational lines} (see \citet{grenier2015} for a review).
              This method of observing cosmic-rays at their origin has become more prominent during the past decade as instruments like ALMA and Herschel can  identify places of enhanced ionization. Here, we will shortly summarize the physics and state-of-the-art on cosmic-ray ionization.

\begin{figure}[htbp]
\centering{
\includegraphics[trim = 0mm 0mm 0mm 0mm, clip, width=0.6\textwidth]{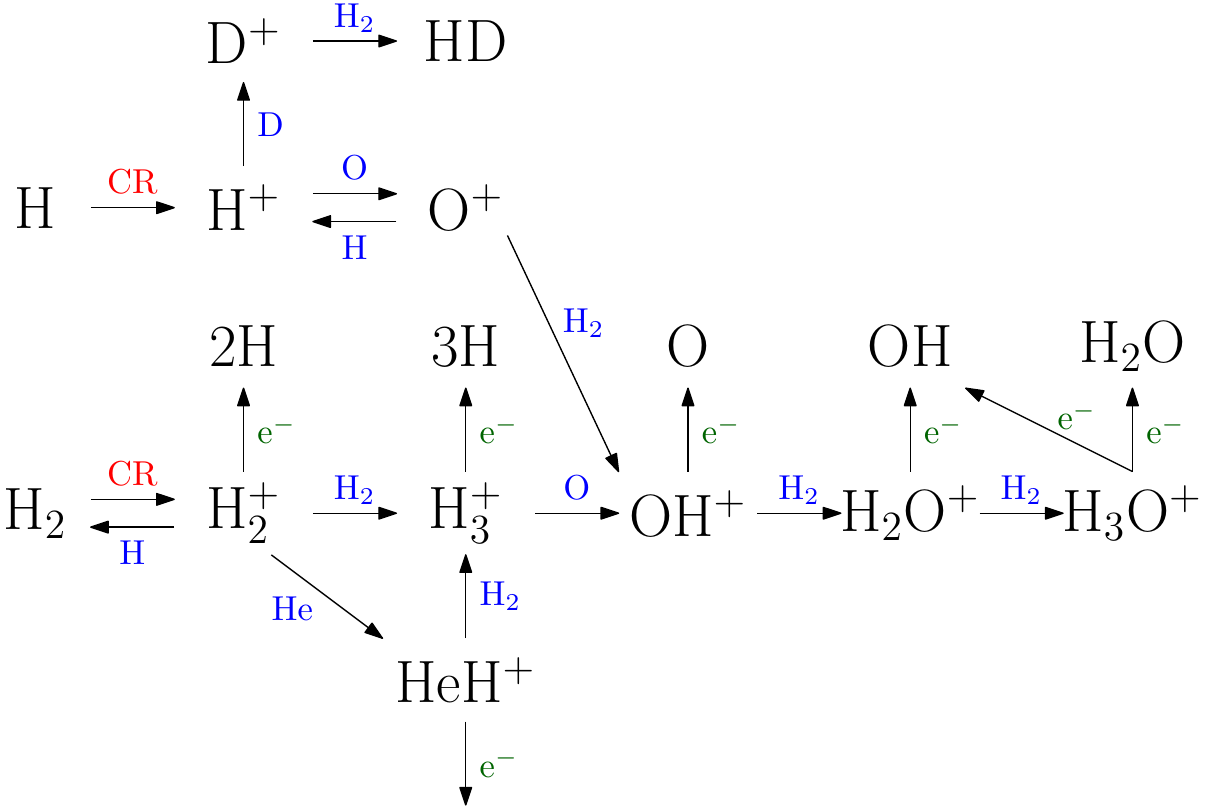}
\caption{Chemical network induced by cosmic-ray ionization. Figure changed and reproduced after \citep{grenier2015}, with kind permission from John Black. \label{chemical_network:fig}}
}
\end{figure}

Ionization of atomic or molecular hydrogen leads to the production of charged nuclei, i.e.\
\begin{eqnarray}
  p_{\rm CR} + H &\rightarrow& H^+ + e^- + p_{\rm CR}^{'}\\
  p_{\rm CR} + H_2 &\rightarrow& H_{2}^{+} + e^- + p_{\rm CR}^{'}\,.
  \label{ion_process:equ}
  \end{eqnarray}
In the chemical chain, $H_{2}^{+}$ reacts on short time scales with ambient electrons to form two hydrogen atoms, with Helium to form HeH$^+$ or with H$_2$ in order to produce the more stable element H$_{3}^{+}$. This chemical network is continued particularly via the H$_{3}^{+}$ production as schematically shown in Fig.\ \ref{chemical_network:fig}.

Conclusions about the ionization rate can be drawn from these detections with the best tracer H$_{3}^{+}$ and more indirect tracers farther along the chemical network chain \citep{black1998,black2012,grenier2015}. The launch of the Herschel Space Observatory made detailed measurements of H$_{3}^{+}$ regions in the Milky Way possible during the past decades, see e.g.\ \citep{grenier2015} and references therein.  While the Galactic average ionization rate is at a level of $\overline{\zeta}_{\rm Gal}\sim 2\times 10^{-16}$~s$^{-1}$ \citep{indriolo2013,indriolo2015}, the ionization rate near SNRs appears to be enhanced: \citet{indriolo2010} find a value close to $10^{-15}$~s$^{-1}$ near IC443. \citet{indriolo2015}  report regions close to a large sample of SNRs with $\zeta>5\times 10^{-16}$~s$^{-1}$ as strict lower limits. The true values are discussed to possibly be as large as $10^{-14}$~s$^{-1}$, but uncertainties in the parameters that are necessary to determine the original ionization rate do not permit to quantify these values further. \citet{ceccarelli2011} derive an ionization rate of $\zeta^{H_2}\sim 10^{-15}$~s$^{-1}$ from the measurement of the DCO$^+$ to HCO$^+$ ratio. Similarly high values are derived for the SNR W28 \citep{vaupre2014} using the same chemical ratio. The latter method for the determination of the ionization rate via DCO$^+/$HCO$^+$ should be taken with care as some possible ambiguities from poor knowledge on the distribution of temperature and density that enter the calculation of ionization rate from the measured isotope ratio as discussed in \citet{grenier2015}. Measurements of NH$_3$ at the SNR W28 infer an ionization rate at the level of $10^{-14}$~s$^{-1}$ as well as discussed in \citet{schuppan2012} and strengthen the argument of enhanced ionization in the region of W28 independent of the $HCO+/DCO+$ approach. Thus, there are independent pieces of evidence from a variety of measurements from different SNR-MC systems that they show an enhanced ionization rate.

\begin{figure}[htbp]
\centering{
\includegraphics[trim = 0mm 0mm 0mm 0mm, clip, width=0.8\textwidth]{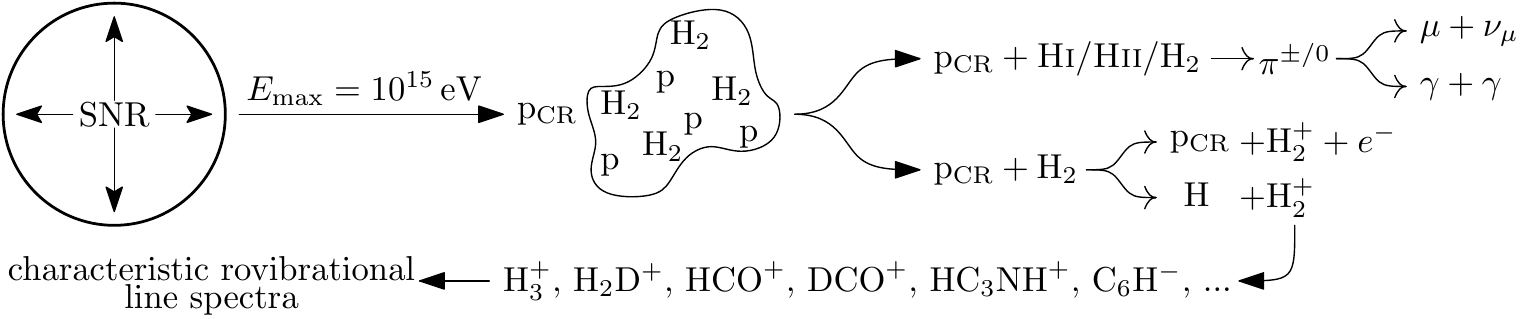}
\caption{Connection between the production of neutrinos, gammas and rotational-vibrational lines in hadron-hadron interactions in MCs adjacent to SNRs. Figure reproduced after the original idea from \citep{schuppan2014}. \label{scheme_ionization:fig}}
}
\end{figure}

What makes these detections so intriguing in the context of cosmic-ray diagnostics is the expected correlation to gamma-ray emission. Figure \ref{scheme_ionization:fig} shows the concept of the coincident emission: while cosmic-rays interact via inelastic hadronic processes above the kinematic threshold of $E_p>280$~MeV, the ionization cross-section peaks in the keV-MeV range and ionization can be relevant up to around $1$~GeV proton energy, see e.g.\ \citep{becker2011,schuppan2012}. Thus, gamma-ray spectra and ionization signatures probe different parts of the same cosmic-ray spectrum. While gamma-ray emission itself is ambiguous due to leptonic processes (IC, brems), ionization has a similar problem due to photo-ionization processes. A full multimessenger-modeling of the region of interest is important in order to understand the correlations in detail. What can be used in particular is to study the spatial distribution of ionization and gamma-rays. These details are discussed in Section \ref{multimessenger_modeling:sec}.

The ionization rate of molecular hydrogen by cosmic rays can be calculated as \citep{padovani2009}
\begin{equation}
\zeta^{H_2}=4\pi\,\left[\sum_k\int_{E_{\min}}^{E_{\max}}\frac{dN_k}{dE_k}\,[1+\phi_k(E_k)]\,\sigma_{k}^{ion}\,dE_k+\int_{0}^{E_{\rm max}}\frac{dN_p}{dE_p}(E_p)\,\sigma_{p}^{e.c.}(E_p)\right]\,.
\end{equation}
The first terms correspond to ionizations by cosmic-ray electrons ($k=e$) and protons ($k=p$), the last term is due to electron capture and often negligible.
The term $\phi_k$ is  added in order to account for ionization by secondary electrons, produced in the ionization process described in Equ.\ (\ref{ion_process:equ}). The integration limits are in principle $E_{\min}=0$ and $E_{\max}=E_{\max}^{\rm acc}$, but can be constrained further by assuming a certain range of validity for the energy spectrum of cosmic rays. The ionization cross-section by electrons and protons, $\sigma_{k}^{\rm ion}$ and by electron capture, $\sigma_{p}^{e.c.}$ are comprehensively described in e.g.\ \citep{padovani2009}.  The total ionization rate is typically dominated by proton ionization: the electron capture cross-section is significantly smaller than the proton ionization one. Ionization through primary electrons is negligible as low-energy electrons are not expected to be able to penetrate the diffuse clouds in which ionization is expected to be significant \citep{schuppan2012}. Secondary electrons through proton ionization, however, are produced inside the clouds and are of higher energies. They are therefore included via the term $\phi_k$ with $k=p$.

The loss time scale for cosmic-rays due to ionization is expressed as \citep{schuppan2012}
\begin{equation}
\tau_{loss}^{\rm ion}\approx \frac{p}{|\frac{dp}{dt}|}= 5.6\times 10^{15}\,Z^{-2}\,\left(\frac{n}{1\,{\rm cm}^{-3}}\right)^{-1}\,\left(\frac{p}{m\,c}\right)^{3}\quad \rm s\,
\end{equation}
above $\sim 10$~MeV/c. 

A competing process is photoionization --- it is calculated as \citep{schuppan2014}
\begin{equation}
\zeta_{\rm photo}^{H_{2}^{+}}=\frac{f_i}{I_{H_2}}\,\int_{E_{\min}}^{E_{\max}}F_{\rm photo}\cdot E\cdot \sigma_{\rm pa}(E)\,dE\,.
\end{equation}
Here, the factor $f_i$ is the fraction of photons that is absorbed through the ionization process, $I_{H_{2}}=15.4$~eV is the ionization potential of molecular hydrogen and $\sigma_{\rm pa}$ is the cross-section for photo-absorption. The function $F_{\rm photo}$ is the flux of photons that is available for ionization. The latter can be taken from observations. Here, the most relevant part of the photons spectrum is the X-ray flux, as the high energies are able to penetrate deep into the clouds due to the production of a larger number of secondary high-energy electrons that can maintain a significant flux even deep into the cloud and therefore can resemble the distribution of ionization signatures from cosmic-rays.

\citet{schuppan2014} have calculated the ionization profiles for four SNR-MC systems and point out that it depends on the individual conditions of the SNR if photo-ionization or cosmic-ray ionization dominates. Detailed simulations of the region of interest can help to identify those systems which are likely to be dominated by cosmic-ray ionization in order to include these line signatures in multimessenger studies even more systematically in the future.

              \subsection{The global picture of radiation signatures\label{radiation_discussion:sec}}

              Table \ref{radiation:tab} summarizes the processes discussed above. The most important features of the distributions are summarized here. In the Galactic context, photohadronic interactions are usually negligible. An exception could be the Galactic Center region or other regions of high star formation like the Cygnus complex, as the large number of high-mass stars can lead to a high overall UV field, which might be necessary to take into account, see more detailed discussion in Section \ref{multimessenger_modeling:sec}. For the interaction of cosmic-ray protons with target hydrogen (or heavier nuclei), the kinematic threshold of the problem can be calculated to $280$~MeV \citep[][e.g.]{schlickeiser_mannheim1994}. The neutral pion in the lowest-energy interaction receives $140$~MeV rest energy and decays into two photons, $\pi^{0}\rightarrow \gamma\,\gamma$. Therefore, the threshold energy for gamma-ray production via hadronic interactions is $70$~MeV. This \textit{pion peak} is a distinct signature that is expected only from hadronic interactions. It has been discussed in \citep{fermi_ic443,fermi_w44,tobias_sugar2015} that it has been observed in three SNRs, i.e.\ IC443, W44 and W51 \citep{fermi_ic443,tobias_sugar2015}. However, the turn-over might be at energies larger than $70$~MeV and thus rather be due to a change in the proton spectrum as discussed in \citet{strong_icrc2015,strong2018}. Even if the hadronic nature of these sources is indicated by the measurements, due to their very steep energy spectra $(\sim E_{p}^{-3}-E_{p}^{-4}$ in the GeV-TeV range), these objects are no candidates for the contribution to the total cosmic-ray energy spectrum, which is significantly flatter (roughly $E^{-2.7}$ in the GeV-TeV region as discussed in detail in Sections \ref{data:sec} and \ref{candidates:sec}).

              \begin{figure}[htbp]
\centering{
\includegraphics[trim = 0mm 0mm 0mm 0mm, clip, width=0.8\textwidth]{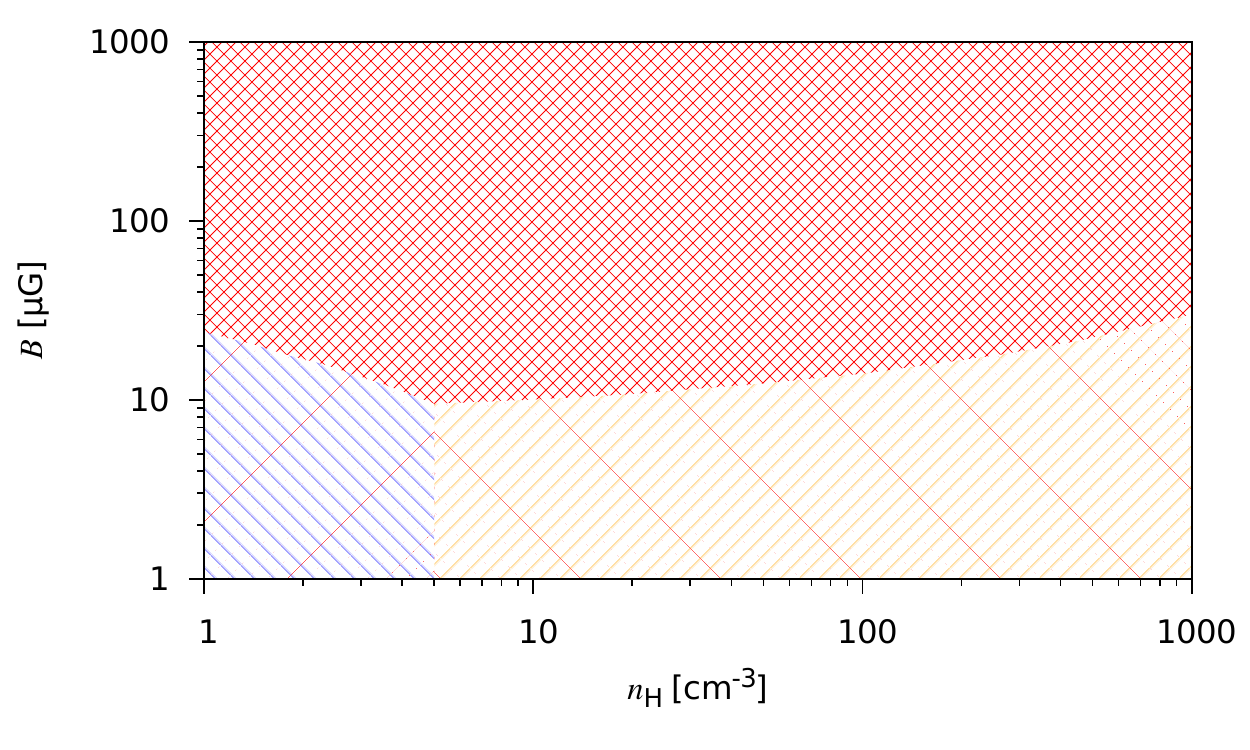}
\caption{Investigation of the dominance of different gamma-ray emission processes at $100\,$GeV depending on ambient proton density and magnetic field strength. The spectral index of the primary particle distributions, protons and electrons, has been chosen as $E^{-2}$. The synchrotron emission has been kept constant --- this corresponds to a scenario where the synchrotron emission is measured and the high-energy gamma-ray emission can be predicted under the constraints of the synchrotron emission. The proton-to-electron ratio is chosen to be $f_{ep}=0.02$. Figure following \citet{mandelartz2015}.  \label{interaction_processes:fig}}
}
\end{figure}

The pion bump at $\sim 70$~MeV is difficult to access with current instruments, as it is in an energy range in which no instrument has optimal settings and sufficient effective area for the detection of a larger number of sources. Thus, there are only three additional ways to distinguish leptonic and hadronic signatures in the high-energy spectra of Galactic sources:
\begin{enumerate}
\item \textit{Broad-band fits to the spectral energy distribution (SED) from objects/regions:} These are usually performed in a way that the synchrotron bump of the SED is fitted in a first step. This procedure fixes the combination of magnetic field strength and non-thermal electron density $n_e\cdot B^2=const$  together with the spectral index of the non-thermal electron population, $\alpha_e$. In a second step, the high-energy part of the spectrum can be fit for different combinations of the parameter set $(B,\,n_{\rm tot},\,\alpha_p)$. In particular, by changing $B$, the electron density is also changed in order to keep the synchrotron spectrum at the same level. The increase of $B$ implies the necessity to decrease $n_e$. This has direct consequences for the Inverse Compton spectrum. Due to the reduced number of available electrons, the IC emission  is decreased in intensity at the same time as $B$ is increased. Furthermore, by increasing the hadronic target density $n_{\rm tot}\approx n_{H-I}+2\,n_{H_2}$ the time scale for bremsstrahlung is decreased (see Table \ref{radiation:tab}) and it becomes dominant at some point. \citet{mandelartz2015} investigated the broadband SED from high-energy emitting SNRs in the Milky Way with such a procedure. Figure \ref{interaction_processes:fig} shows a general feature of such fits that is independent on the choice of the individual remnant: in the parameter space $(B,\,n_{\rm tot})$, it can be predicted which of the processes dominates at a certain energy. For low B-fields and low hadronic target densities, the spectrum at $100$~GeV will be dominated by Inverse Compton scattering. With increasing $B-$field, the electron population will be reduced in intensity in order to maintain the synchrotron spectrum at the same level and proton-proton interactions will start to dominate the spectrum. Finally, when increasing the total target density, bremsstrahlung will take over the spectrum, independent of the magnetic field. The details of this graph, in particular the exact critical values $(B_{\rm crit},\,n_{\rm tot,\,crit})$  will change somewhat with the chosen energy and other parameters like the proton-to-electron ratio, but the general message stays the same: extremely high densities will lead to the dominance of bremsstrahlung, high B-fields at moderate target densities will lead to the dominance of proton-proton interactions.
\item \textit{The detection of neutrinos from the region of interest:} The first detection of astrophysical high-energy neutrinos has been published not too long ago in 2013 \citep{icecube2013}. While Galactic emission as the dominant source of the signal can be excluded (see Section \ref{multimessenger_modeling:sec} for a detailed discussion), future measurement will be sensitive to Galactic emission via the explicit search for point sources and extended sources in the Galaxy, with relevant limits to the emission already published now \citep{icecube_galactic2017,galacticnus_icecube_antares2018}. Neutrino signatures are unambigous signs of hadronic emission and the only alternative explanation would be dark matter signatures.
  \item \textit{The coincident detection of ionization signatures and gamma-ray emission:} New instruments that measure line spectra in the submm range, in particular ALMA and Herschel, make it possible to quantify the local level of ionization in the Galaxy quite well. The enhancements around SNRs and SNR-MC systems that have been confirmed during the past decade indicate that it could be possible to correlate high-energy signatures (gamma-rays and neutrinos) with ionization signatures. First studies of theoretical predictions \citep{padovani2009,becker2011,schuppan2012,schuppan2014}  and observational correlations \citep{vaupre2014} have been performed and lead the way to systematic studies of hadronic emission from Galactic sources. 
\end{enumerate}
Connecting these three diagnostic tools and adding information from the charged particles is the only way to achieve a full understanding of the entire energy spectrum of (Galactic) cosmic rays.

In general, these processes are quite well understood. In particular,  the leptonic processes are well quantified. There is one crucial exception: the improvement of the description of the hadronic cross sections is a major ongoing effort in current research. Astroparticle physics has the unique setup of a particle with extreme energies hitting a fixed target. The energies of ultra-high energy cosmic rays cannot be reached at man-made accelerators. Therefore, measurements of the inclusive cross sections at these energies can only be made via astroparticle physics itself, as performed for nucleus-air interactions by the Pierre Auger Observatory (PAO) \citep{auger_cross_section2012} and for neutrino-nucleon interactions by IceCube \citep{icecube_cross_section2017}. These interactions in turn are most sensitive to the forward-scattering region due to the highly boosted interaction. The highest energies that are accessible from man-made accelerators are currently reached at the Large Hadron Collider (LHC) at CERN, with a center-of-mass energy in the collision of two protons up to $14$~TeV \citep{lhc2008}. Translated into a fixed-target collision, this would correspond to a proton with $10^{17}$~eV. Measurements at the LHC, however, are most sensitive to the transverse momentum of the cross section, as the beam pipe is located in forward direction. LHC-b is best-suited to perform measurements that are at large pseudo-rapidities\footnote{The pseudo-rapidity is a logarithmic measure for the scattering angle, defined as $\eta=\frac{1}{2}\ln\left(\frac{p+p_T}{p-p_T}\right)$ with $p_T$ as the transversal part of the momentum $\vec{p}$.} $2<\eta\lesssim 4.5$, so in quasi-forward direction and at the same time at high energies. The future accelerator FAIR in Darmstadt will be able to further improve knowledge in direct forward direction \citep{fair}. While these experiments have made and will make a more complete picture of the inclusive hadronic cross section possible, it is currently not possible to access the entire momentum space. Extrapolations are necessary when investigating interactions in the astrophysical context. And as these happen predominantly in the not well-constrained forward direction, the improvement of the knowledge on cross sections is one of the central issues in particle- and astroparticle physics as of today.

In addition to the uncertainties in the cross-section, the modeling of cosmic-ray propagation in combination with radiation signatures is complex. In order to understand the interplay, effects of energy-dependent diffusion, with a consideration of perpendicular transport, in combination with the necessity to disentangle hadronic and leptonic signatures at the same time. What makes these multimessenger studies even more challenging is that often  target distributions (gas and photons) and magnetic field structures are both complex and not well-known in detail. The details of the multimessenger modeling of individual objects in the Milky Way and more diffuse regions are reviewed in Section \ref{multimessenger_modeling:sec}.

\begin{threeparttable}
\caption{Summary of continuum radiation processes. Table adjusted after \citet{schlickeiser2002}.\label{radiation:tab}}
          
\begin{tabular}{l|lll}
\toprule
Radiation &Photon Spectrum&$\gamma$ energy range&loss time scale [$10^{7}$~yr]\\
\midrule
           
Synchrotron& $\epsilon^{-(\alpha-1)/2}$&radio - X-ray&$10^{3}\cdot(\frac{B}{\mu\,\rm G})^{-2}\,(\frac{E_e}{\rm GeV})^{-1}$\\
           
IC (Thomson) & $\epsilon^{-(\alpha-1)/2}$ &MeV - TeV&$50\cdot(\frac{U_{\rm ph}}{10^{-12}~{\rm erg/cm}^{-3}})^{-1}\,(\frac{E_e}{\rm GeV})^{-1}$\\
           
IC (KN) & $\epsilon^{-(\alpha+1)}$&MeV - TeV&see Equ.\ (\ref{IC_KN:equ})\\
           
brems (strong sh.) & $\epsilon^{-\alpha}$&MeV - TeV&$3.3\cdot(\frac{n_{H-I}+2\,n_{H_2}}{{\rm cm}^{-3}})^{-1}$\\
           
$pp\rightarrow \pi^{0}/X$& $\epsilon^{-\alpha}$&$70$~MeV - PeV&$500\cdot \left(\frac{n_H}{1\,{\rm cm}^{-3}}\right)^{-1}$\\
           
Ionization&Line emission&submm&$18\cdot(\frac{E_p}{\rm m_p\,c^2})^{3}(\frac{n_{H_I}+2n_{H_{2}}}{1\,{\rm cm}^3})^{-1}$\\
           \bottomrule
\end{tabular}
\end{threeparttable}
\section{Cosmic rays and their secondaries: observations \label{data:sec}}
The observation of cosmic rays and their secondaries can be divided into three categories: Energy spectrum, chemical composition and arrival directions. All of these properties are 
connected to each other. A complete theory of cosmic rays and their secondaries has to explain all of the features and observables simultaneously. Despite the theoretical 
challenges of describing cosmic-ray transport and interaction with proper astrophysical input, the observational results can be separated quite clearly. In this section, we present a 
summary of recent measurements by various observatories that in the past decades have revealed different features of the spectrum, composition, anisotropy. In addition, a short 
overview on the state-of-the-art concerning the detection of Galactic gamma-rays and neutrinos is presented.


\subsection{Cosmic ray energy spectrum \label{allparticle:sec}}

The cosmic ray all particle energy spectrum, typically given in units of an energy-weighted particle flux
$[\Phi\cdot E^{b}] = \mathrm{particles}\cdot\mathrm{length}^{-2}\cdot\mathrm{time}^{-1}\cdot$solid-angle$^{-1}\cdot\mathrm{energy}^{b-1}$, has been measured with increasing precision for 
nearly a century, covering an enormous energy range from above $10$~GeV up to $10^{11}$~GeV. In rough approximation, it can be described by a power law. When we look at the huge 
amount of different astrophysical objects, from protostars to super-massive black holes in active galaxies, with different properties and located in all kinds of environments, 
it is remarkable that cosmic rays show such a smooth energy distribution. Over more than 14 orders of magnitude in energy and 34 orders of magnitude in flux the cosmic-ray energy 
spectrum shows only small --- but nevertheless relevant --- features. Already at this stage one may conclude that: (1) Either only a limited number of astrophysical objects with 
similar properties accelerates cosmic rays or (2) cosmic rays are accelerated by the same physical mechanism that is to be found in different sources or even source classes. 
Which of these two scenarios is realized by nature is not fully understood yet. Probably a mixture of both is true: The number of possible sources can be reduced drastically by 
different energetic arguments and the theory of diffusive shock acceleration suggest a similar energy transfer mechanism for all source classes. We will discuss the details of 
these subtleties in the following sections of this review.

The huge range in energy and flux level of cosmic rays makes it impossible to observe the complete spectrum with one single instrument. The flux at the lowest energies as it is 
measured e.g.\ by the Voyager spacecrafts or the Alpha Magnetic Spectrometer (AMS-02) experiment yields hundreds of particles per second for a square-meter sized detector area 
which is challenging for the 
data processing but decreases statistical uncertainties fast. At the other edge of the spectrum observatories like the Telescope Array (TA) and the Pierre Auger Observatory (PAO) 
cover several hundred square-kilometer of surface 
area to collect a flux that amounts to only a few ultra-high energy cosmic rays (UHECRs) per square kilometer and year. Also the detection techniques differ very much for low and high 
energy cosmic rays: Where cosmic rays up to energies of $E=10^{4}-10^{5}$~GeV can be detected directly cosmic rays above this energy can only be made visible via their imprint on Earth via their produced air 
showers. This can lead to a systematic shift in the measured energy spectrum between different observatories and telescopes as it is discussed e.g.\ in 
\citet{Ivanov:2017juh,serap2013, polygonato}.

\begin{figure}[htbp]
\centering{
\includegraphics[trim = 2mm 5mm 5mm 3mm, clip, width=\textwidth]{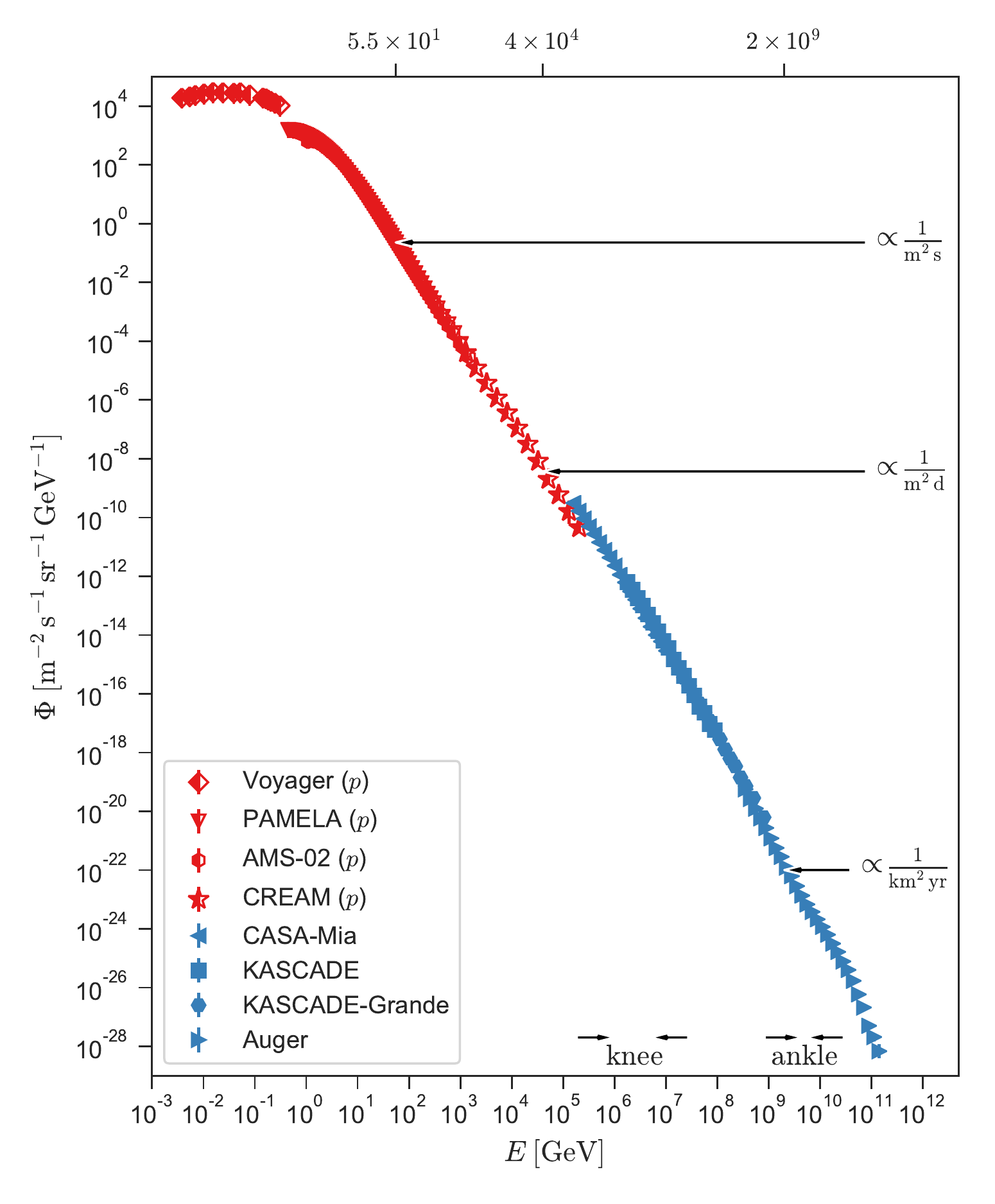}
\caption{All-particle spectrum of cosmic rays from MeV to ZeV energies. For energies below $E=10^4$~GeV often a total flux of all elements is not available, so here only the proton flux (left filled), which dominates the total spectrum, is given. For higher energies is the reported total flux (filled) shown. See references of the original publication for details. --- Voyager \cite{Stone150}, PAMELA \cite{PamelaFlux2011}, AMS-02 \cite{ams2015a, ams2015b}, CREAM \cite{0004-637X-728-2-122}, CASA-Mia \cite{CRDB_CasaMIA}, KASCADE \cite{ANTONI20051}, KASCADE-Grande \cite{2013PhRvD..87h1101A}, Auger \cite{Fenu:2017hlc}}
\label{all_particle:fig}
}
\end{figure}

Figure \ref{all_particle:fig} shows the unweighted cosmic-ray flux from the lowest to highest energies. This representation shows good agreement between all 
observatories, with some subtleties concerning the energy calibration as discussed already in \citet{berezinsky2006}.
In the figure it is also clearly visible that above $\sim 10$~GeV, the energy spectrum can be described by a power law in energy in a first-order approximation:
\begin{align}
\frac{\mathrm{d}N}{\mathrm{d}E} 
= \Phi(E) \propto E^{-\gamma}\quad .
\end{align}
Toward lower energies, i.e.\ below $E\sim 10$~GeV, a flattening of the spectrum, most likely resulting from cosmic-ray ionization processes of the interstellar medium (ISM) is clearly 
visible. The significant difference at the lowest energies between measurements by Voyager and AMS-02 is due to different locations of the experiments. The most recent Voyager data have been taken 
after leaving the heliosphere and therefore no longer influenced by the Sun. Data taken closer to Earth, measured by e.g.\ the PAMELA\footnote{Payload for Antimatter 
Matter Exploration and Light-nuclei Astrophysics} satellite, show a variation of the spectrum with the 22-year cycle of the solar magnetic field, revealing that cosmic 
rays below $\sim 10$~GeV are influenced by the solar magnetic field (see e.g.\ \citet{gleeson1968,strauss2012,potgieter2013,0004-637X-829-1-8} for a discussion of solar modulation of the cosmic-ray energy spectrum and 
composition). Above $\sim 10$~GeV cosmic rays are too energetic to be  influenced by the Sun significantly.

The second prominent feature is the significant steepening of the spectrum for energies above $\approx 3\times 10^{6}$~GeV, the break energy typically referred to as the 
\textit{cosmic-ray knee}. The third high-profile feature is the so-called \textit{cosmic-ray ankle}, at an energy of $\approx 10^{9.7}$~GeV, at which the spectrum flattens again. 
In the 1990s, these were the known features and the spectral index was determined to (e.g.\ \citet{wiebel_crs})\footnote{See Tab.\ \ref{tab:ankle} and Tab.\ \ref{tab:cutoff} for 
details}
\begin{equation}
\gamma\approx \left\{\begin{array}{lll}2.67\pm 0.02&&10<E/{\rm GeV}<10^{6.4}\\
3.07\pm 0.07&&10^{6.4}<E/{\rm GeV}<10^{9.7}\\
2.33\pm 0.33&&E/{\rm GeV} >10^{9.7}\end{array} \right. \quad .
\end{equation}
A cut-off in the spectrum at the highest energies is present at $\sim 5 \times 10^{10}$~GeV \citep{Fenu:2017hlc} with minor differences between 
the Northern and Southern hemisphere (see section \ref{ssec:CutOff} for details).

Today, the well-trained eye might also find some more features even without zooming in to the different energy ranges: As mentioned above there are several features in the energy spectrum of cosmic rays named after the joints of a human leg from lower to higher rigidities as we will discuss in detail in the following paragraphs, starting at low energies.
\subsubsection{The cosmic-ray hip --- the change in slope at $R=300$~GV}
The most recently confirmed deviation from the uniform power-law is also the lowest energetic one. In 2009 and 2010 a hardening of the cosmic-ray spectrum around a rigidity $R$ of a 
few hundred GeV was reported by the Advanced Thin Ionization Calorimeter (ATIC) \citep{panov2006} and the Cosmic Ray Energetics and Mass balloon (CREAM)  flights 
\citep{0004-637X-728-2-122,0004-637X-707-1-593, 0004-637X-707-1-593,maestro09}, respectively, for different 
elements. This kink in the observed flux was later confirmed with very high statistics by PAMELA \citep{PamelaFlux2011,pamela_B-C} and 
AMS-02 \citep{ams2015b,ams2015a,ams2017,ams2018}.

\begin{figure}[htbp]
\centering{
\includegraphics[trim = 0mm 0mm 5mm 0mm, clip, height=.95\textheight]{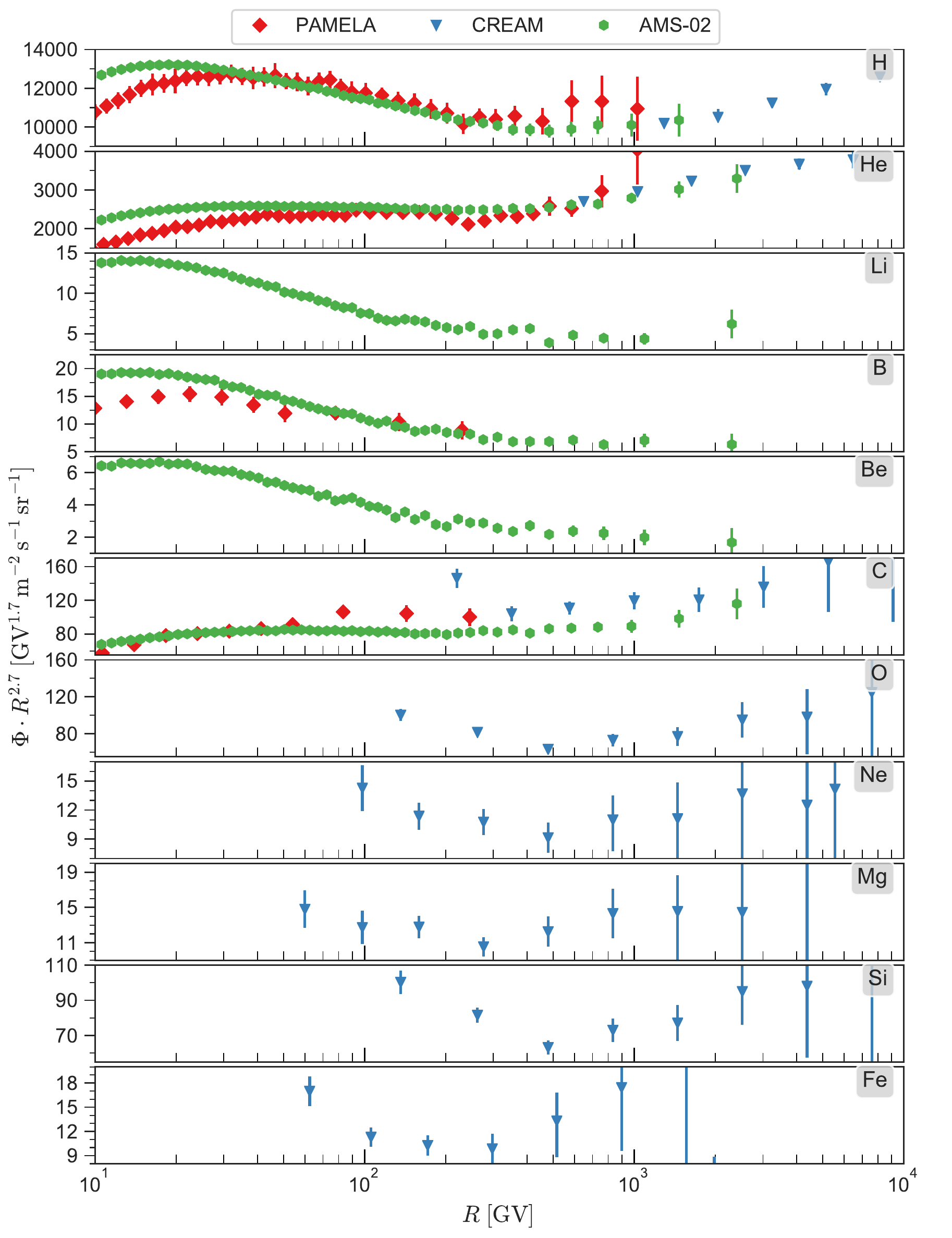}
\caption{Spectral break at a rigidity around $R=300$~GV for is observed for many different nuclei by AMS, CREAM and PAMELA. --- AMS \cite{ams2015a, ams2015b, ams2017, ams2018}, PAMELA \cite{PamelaFlux2011, pamela_B-C}, CREAM \cite{0004-637X-728-2-122, 0004-637X-707-1-593}}
\label{lowenergybreak:fig}
}
\end{figure}

Figure \ref{lowenergybreak:fig} shows the cosmic-ray spectrum for eleven elements from proton to iron. Data taken by AMS-02, CREAM and PAMELA are shown. The change in the slope 
is clearly visible for most of the elements. The different observatories see subtle differences between each other. Still, there is overall agreement on the change in the spectral index appearing around $R=300$~GV. Already by eye it becomes visible that the exact position of the break might slightly change for the different elements. A broken power-law 
can be fit to the data to determine the break rigidity. However, at least for heavier elements only a very limited number of data points are available, making it hard to get robust fitting results. The most recent fitting results can be 
found in Table \ref{tab:ams}. Here, the flattening of the spectral index is determined to be larger than $0.1$ in all available data. Furthermore, a significant difference in the 
spectral behavior of protons and heavier elements is clearly visible.

The physics of such a break will be discussed in Section \ref{candidates:sec}. Qualitatively, there are three possibilities for the origins of this change in slope: (1) A change 
in the energy behavior in the transport of the particles; (2) a change of the source(s) contributing to the spectrum; (3) a feature in the local properties of the source(s). As 
discussed in more detail in Section \ref{candidates:sec}, all three cases can be able to produce a feature at a single rigidity and to be independent of the particle energy since 
transport as well as acceleration of the cosmic rays dependent  on the ratio $R\approx E/q$ between energy and charge  and the energy alone.

\begin{sidewaystable}[hbp]
\centering
\begin{threeparttable}

\caption{Parameters of the hip. AMS-02 and PAMELA do report a break rigidity $R_0$ and the spectral index before ($\gamma_1$) and after ($\gamma_2$) it or the 
difference in the spectral index $\Delta \alpha$ (AMS-02). For AMS-02 the spectral second spectral index is calculated $\gamma_2=\gamma_1-\Delta\gamma$.}\label{tab:ams}

\begin{small}
\begin{tabular}{l|cccl}
\toprule
Exp.\ (Element) & $R_0$ [GeV] & $\gamma_1$ & $\gamma_2$  & Ref.\ \\
\midrule

AMS-02 (p)\tnote{1} & $336_{-44}^{+68}(\mathrm{fit})_{-28}^{+66}(\mathrm{sys})\pm 1(\mathrm{sol})$ & $2.849\pm 
0.002(\mathrm{fit})_{-0.003}^{+0.0037}(\mathrm{sys})_{-0.003}^{+0.004}(\mathrm{sol})$ & $2.716_{-0.021}^{+0.032}(\mathrm{fit})^{+0.046}_{-0.030}(\mathrm{sys})\pm 0.006(\mathrm{sol})$ \tnote{6} & [1] \\

AMS-02 (He)\tnote{2} & $245_{-31}^{+35}(\mathrm{fit})_{-30}^{+33}(\mathrm{sys})\pm 3(\mathrm{sol})$ & $2.780\pm 
0.005(\mathrm{fit})\pm 0.001(\mathrm{sys})\pm 0.004(\mathrm{sol})$ & $2.661^{+0.014}_{-0.011}(\mathrm{fit})_{-0.028}^{+0.033}(\mathrm{sys})\pm 0.006(\mathrm{sol})$ \tnote{7} & [2] \\

PAMELA (p)\tnote{3} & $232_{-30}^{+35}$ & $2.85\pm 
0.015(\mathrm{fit})\pm 0.004(\mathrm{sys})$ & $2.67\pm 0.03(\mathrm{fit})\pm 0.05(\mathrm{sys})$ & [3] \\

PAMELA (He)\tnote{3} & $243_{-31}^{+27}$ & $2.77\pm 
0.01(\mathrm{fit})\pm 0.027(\mathrm{sys})$ & $2.48\pm 0.06(\mathrm{fit})\pm 0.03(\mathrm{sys})$ & [3] \\

PAMELA (B)\tnote{4} & -- & $ 3.01\pm 0.13$ & -- & [4] \\

PAMELA (C)\tnote{4} & -- & $ 2.72\pm 0.06$ & -- & [4] \\

CREAM-III (p)\tnote{5} & -- & -- & $ 2.61\pm 0.01$ & [5] \\

CREAM-III (He)\tnote{5} & -- & -- & $ 2.55\pm 0.01$ & [5] \\

CREAM-II (C)\tnote{5} & -- & -- & $ 2.61\pm 0.07$ & [6] \\

CREAM-II (O)\tnote{5} & -- & -- & $ 2.67\pm 0.07$ & [6] \\

CREAM-II (Ne)\tnote{5} & -- & -- & $ 2.72\pm 0.10$ & [6] \\

CREAM-II (Mg)\tnote{5} & -- & -- & $ 2.66\pm 0.08$ & [6] \\

CREAM-II (Si)\tnote{5} & -- & -- & $ 2.67\pm 0.08$ & [6] \\

CREAM-II (Fe)\tnote{5} & -- & -- & $ 2.63\pm 0.07$ & [6] \\

\bottomrule
\end{tabular}

\begin{tablenotes}
\begin{minipage}{.49\linewidth}

      \item[1] Fitting range $45-1800$~GV 
      \item[2] Fitting range $45-3000$~GV 
      \item[3] Fitting range $80-1200$~GV 
      \item[4] Fitting only below $R\lesssim 200$~GV, therefore no break is reported 
      \item[5] Fitting only above $R\gtrsim 200$~GV, therefore no break is reported
      \item[6] Calculated from $\Delta\gamma=0.133_{-0.021}^{+0.032}(\mathrm{fit})_{-0.030}^{+0.046}(\mathrm{sys})\pm 
0.005(\mathrm{sol})$
	  \item[7] Calculated from $\Delta\gamma=0.119_{-0.010}^{+0.013}(\mathrm{fit})_{-0.028}^{+0.033}(\mathrm{sys})\pm 
0.004(\mathrm{sol})$ 
\end{minipage}%
\begin{minipage}{.49\linewidth}
 {[1]} \citet{ams2015a} \\
 {[2]} \citet{ams2015b} \\
 {[3]} \citet{PamelaFlux2011} \\
 {[4]}\citet{pamela_B-C} \\
 {[5]} \citet{0004-637X-728-2-122} \\
 {[6]} \citet{0004-637X-707-1-593} \\
 {}
\end{minipage}
\end{tablenotes}
\end{small}

\end{threeparttable}
\end{sidewaystable}
\subsubsection{The knee and the iron knee}
Moving upward in energy, the next feature is the so called \textit{cosmic-ray knee}. This feature is washed out in the total energy spectrum over more than one decade in energy
due to its rigidity dependence. The strongest change 
in slope is around an energy $E\approx (1-5)\times 10^6$~GeV (see Tab.\ \ref{tab:knee} for a summary). Here, the spectral index $\gamma$ increases by $0.4-0.5$ which corresponds to a 15\% larger index  \citep{aartsen_icetop2013,prosin2014, knurenko2020}. After the discovery of the first break of the knee the KASCADE-Grande\footnote{KArlsruhe Shower Core and Array 
DEtector-Grande} experiment \citep{kascade_grande2013} confirmed another knee-like structure in the energy spectrum around $E\approx 6\times 10^7$~GeV which is exactly at 26 times 
higher energy than the first knee. 
This feature is more pronounced when only the heavy component of is taken into account. This has driven the interpretation that the first and 
the second structure are produced by the same underlying charge dependent process, thus leading to a light ($E\approx (3-5)\times 10^6$~GeV) and a heavy (iron) ($E\approx 80\times 
10^6$~GeV) knee. The Tunka-133 and IceTop observatories do detect a break in the all particle energy spectrum at somewhat higher energies  around $E\approx 100\times 10^6$~GeV. The Telescope Array Low Energy extension (TALE) detects the first knee at $10^{6.6}$~GeV and a second knee at $100\times 10^{6.1}$~eV, compatible with Tunka and IceTop \citep{abbasi2018}. TALE also sees indications for an ankle-like structure in the all-particle spectrum at $10^{7.2}$~GeV.  
Unfortunately, composition resolved data to confirm the KASCADE-Grande result is not yet available. A more detailed summary 
can be found in \citet{haungs2015}.

\begin{figure}[htbp]
\centering{
\includegraphics[trim = 0mm 0mm 7mm 0mm, clip, width=0.8\textwidth]{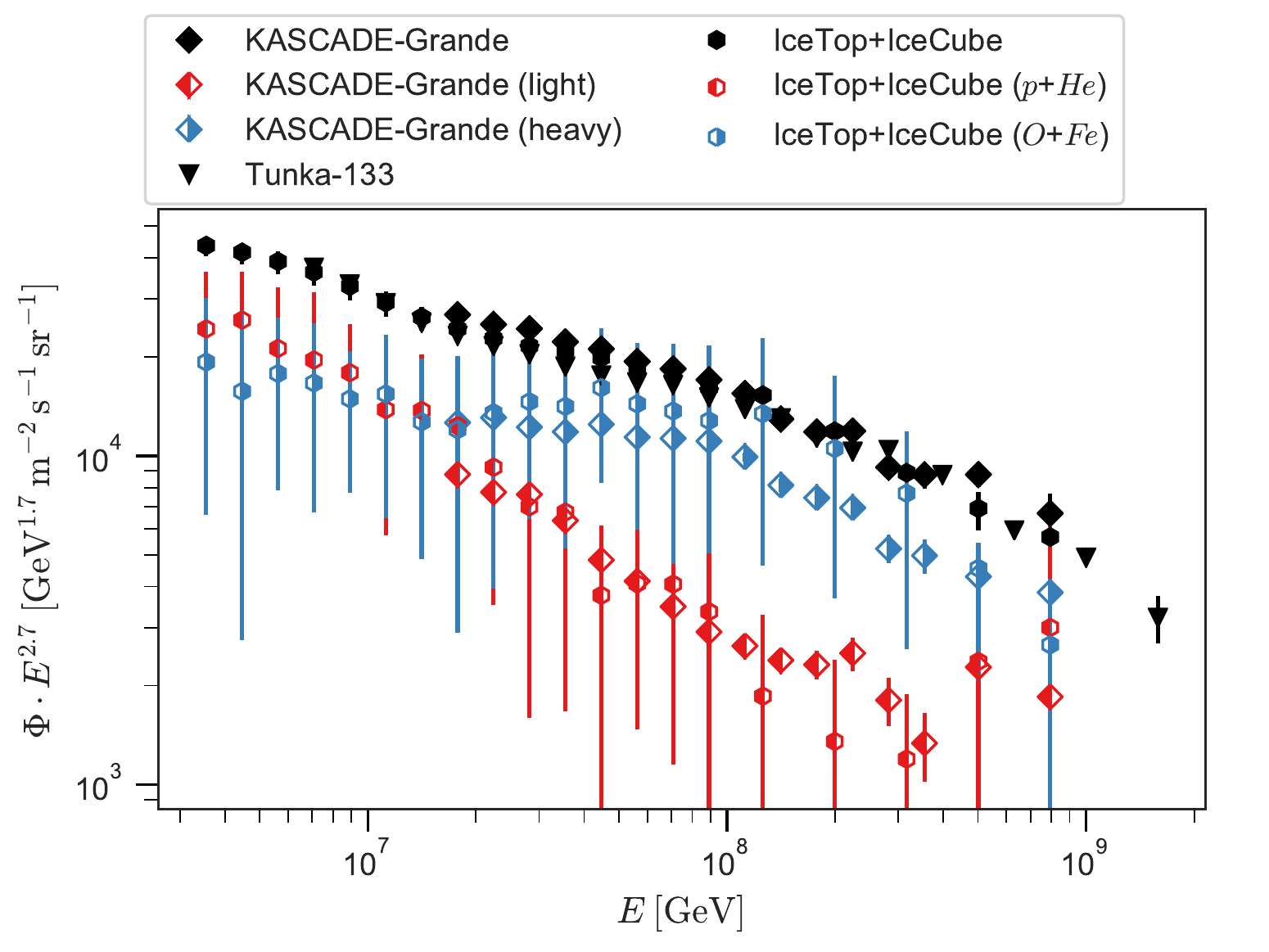}
\caption{The knee to ankle region measured by the three distinct observatories. The KASCADE-Grande data clearly show the differences in the light and heavy component of the 
cosmic-ray flux. --- KASCADE-Grande \citep{2013PhRvD..87h1101A}, Tunka-133 \citep{prosin2016}, IceCube+IceTop \cite{AndeenPlum:2019}.}
\label{fig:knee2ankle}
}
\end{figure}

Figure \ref{fig:knee2ankle} 
shows the total flux measured by KASCADE-Grande, Tunka-133 and IceTop where most of the above described features are visible. In addition, the distinction in light and heavy 
elements (associated with electron rich and poor air shower detection) for the KASCADE-Grande experiment are shown. All three experiments that are dedicated to this energy range 
(KASCADE-Grande, Tunka-133 and IceTop) agree relatively well, apart from a systematic energy shift on the order of some percent. Furthermore, all three groups do report another 
small structure within the energy range of the first knee. It is not as pronounced as the two others and shows a small hardening of the spectrum around $E\approx 6\times 
10^{7}$~GeV. This feature is probably dominated by light elements \citep{aartsen_icetop2013,prosin2014}.

For energies above the KASCADE/KASCADE-Grande measurements an element-wise discrimination of the cosmic rays is no longer possible and rather, an average value of the logarithm on the mass number of the cosmic-ray showers at a certain energy, derived from 
the $X_{\max}$ values of the air showers, is given,i.e.\ $\langle\ln(A)\rangle$. $X_{\max}$ significantly depends on the chosen 
interaction model used in the Monte-Carlo air shower simulation, but gives at least hints on the composition of cosmic rays at the highest energies. 

\begin{threeparttable}

\caption{Parameters of the knee} \label{tab:knee}

\begin{tabular}{l|cccl}
\toprule
Exp.\ (Software) & $E_\mathrm{knee}$ [PeV] & $\gamma_1$ & $\gamma_2$  & Ref.\ \\
\midrule

KASCADE (QGS-Jet) \tnote{1} & $4.0\pm 0.8$ & $2.70\pm 0.01$ & $3.10\pm 0.07$ & [1] \\

KASCADE (SIBYLL) \tnote{1} & $5.7\pm 1.6$ & $2.70\pm 0.06$ & $3.14\pm 0.06$ & [1] \\

CASA-Mia\tnote{2} & $(1.0-2.5)$ & $2.66\pm 0.02$ & $3.07\pm 0.05$ & [2] \\

CASA-Blanca\tnote{1,3} & $2_{-0.2}^{+0.4}$ & $2.72\pm0.02$ & $2.95\pm 0.02$ & [3] \\

EAS\tnote{2} & $(2.7-4.1)$ & $2.76\pm 0.03\;$\tnote{4} & $3.19\pm 0.06\;$\tnote{4} & [4] \\

AKENO & $4.68$ & $2.62\pm 0.12$ & $3.02\pm 0.05$ & [5] \\

Yakutsk EAS & $3$ & $2.70\pm 0.03$ & $3.12\pm 0.03$ & [6] \\
\bottomrule
\end{tabular}

\begin{small}
\begin{tablenotes}
\begin{minipage}{.49\linewidth}
      \item[1] Fitted spectrum: \\$\Phi\propto E^{-\gamma_1}\left(1+\left(\frac{E}{E_\mathrm{knee}}\right)^4\right)^{\frac{\gamma_1-\gamma_2}{4}}$
      \item[2] A smooth transition is observed
      \item[3] Others transition widths have been tested giving similar results.
      \item[4] Fitting range $(0.9-2.3)$~PeV
      \item[5] Fitting range $>5$~PeV
\end{minipage}
\begin{minipage}{.39\linewidth}
\item {[1]} \citet{ANTONI20051}
\item {[2]} \citet{CRDB_CasaMIA}
\item {[3]} \citet{FOWLER200149}
\item {[4]} \citet{AGLIETTA19991}
\item {[5]} \citet{AKENO_Knee}
\item {[6]} \citet{knurenko2020}
\end{minipage}
    \end{tablenotes}
\end{small}
\end{threeparttable}

\subsubsection{The second cosmic-ray knee}

A second knee has been identified around $E\approx 3.2\times 10^{8}$~GeV and marks another softening of the spectrum 
\citep{knurenko2020,prosin2014,abuzayyad2001}. Table \ref{tab:2knee} summarizes some of measured parameters and additional information may be found in \citet{bergman2007, haungs2015}. This structure is more subtle compared to the structure at $E_\mathrm{knee}=(3-100)\times 10^{6}$~GeV. Sometimes the iron 
knee described in the previous paragraph is also referred to as the \emph{second knee} \citep[e.g.][]{haungs2015}; this is not surprising since all features are relatively close to each other and high precision data is 
not available. Different measuring techniques and differences in the location of the observatories make a cross calibration, which might help to resolve the open questions, very 
hard. 

Since element-resolved flux data are not available at these energies, a full analysis of the feature is complicated. It is not yet clear if the knee and the second knee are differerent features of one population. A general rigidity-dependent process as suggested e.g.\ by 
\citet{unger2015} might be the cause. Such an explanation via a scaling of the maximum energy of cosmic rays of the same source class with the particle charge is problematic. A process like that 
does require a significant amount of trans-uranium elements in the cosmic composition around the second knee; simply due to the huge energy gap between the knees of a factor around 
100. A similar need for very heavy elements ($Z>50$) was also shown in fits to the total cosmic ray spectrum in \citet{serap2013}.

Another explanation is that cosmic rays are 
accelerated by two Galactic source classes which have different maximum rigidities leading to two distinct knees in the spectrum \citep[e.g.][]{biermann_apj2010, Biermann_WR}. 
Detailed composition analysis of the interesting energy range might help to solve this question. For further details on the theoretical background, we refer to Sections \ref{composition:sec} and \ref{multimessenger_modeling:sec}.

At higher energies, the lack of a large-scale anisotropy in the data suggests an extragalactic origin of the detected cosmic rays, see e.g.\ \citep{moncada2017,anchordoqui2019}.

\begin{threeparttable}

\caption{Parameters of the second knee}\label{tab:2knee}

\begin{tabular}{l|cccl}
\toprule
Exp.\ & $E_\mathrm{knee}$ [PeV] & $\gamma_1$ & $\gamma_2$  & Ref.\ \\
\midrule

Yakutsk EAS & $100$ & $2.92\pm 0.03$ & $3.24\pm 0.04$ & [1] \\

Tibet-III & $300$ & $3.00\pm 0.01\mathrm{(stat)}$ & $3.33\pm 0.15\mathrm{(stat)}$ & [2]\\ 
&& $\pm 0.05\mathrm{(sys)}$ & $\pm 0.05\mathrm{(sys)}$ & \\

Fly's eye & $400$ & $3.01\pm 0.06$ & $3.27\pm 0.02$ & [3]\\

Auger\tnote{1} & 141-400 & $3.07\pm0.01$ & $3.28\pm0.04$ & [4]\\
\bottomrule
\end{tabular}

\begin{small}
\begin{tablenotes}
\item[1] Auger does observe a smooth transition between $E=(141-446)$~PeV with a gradient of $\mathrm{d}\gamma/\mathrm{d}\log E\approx0.5$.
 \item {[1]} \citet{knurenko2020}
 \item {[2]} \citet{prosin2014} 
 \item {[3]} \citet{FlysEye}
 \item {[4]} \citet{AugerSecondKnee}
\end{tablenotes}
\end{small}

\end{threeparttable}

\subsubsection{The cosmic-ray ankle}
The last significant feature in the energy spectrum before the cut-off is the so called \textit{cosmic-ray ankle} around an energy $E_\mathrm{ankle}\approx 10^{9.7}$~GeV. The ankle 
is measured with high precision by HiRes \citep{abbasi2008,FlysEye}, TA \citep{Tinyakov:2017bju} and Auger \citep{abraham2010}. Around this energy the spectrum is hardening again with different spectral indices measured by the experiments (see Table \ref{tab:ankle}) in the range $\gamma\sim 2.5 - 2.8$. These differences are under investigation by the experiments and are probably associated with systematic uncertainties in the energy scale, but could also be an effect of the observation of different hemispheres.
Cosmic rays with energies above the ankle are believed to be of extragalactic origin for different reasons that are reviewed in detail in the following sections.

\begin{threeparttable}
\caption{Parameters of the ankle. All groups basically performed broken power-law fits to the total cosmic-ray spectrum, where $-\gamma_1$ and $-\gamma_2$ are the spectral indices 
before and after the ankle at $E_\mathrm{ankle}$, respectively.}\label{tab:ankle}

\begin{tabular}{l|cccl}
\toprule
Exp.\ & $\gamma_1$ & $\gamma_2$ & $E_\mathrm{ankle}$ [EeV] & Ref.\ \\
\midrule
Auger & $3.293\pm 0.002\pm 0.05$ & $2.53\pm 0.02 \pm 0.1$ & $5.08\pm 0.06\pm 0.8$ & [1] \\
TA & $3.226\pm 0.007$ & $2.66\pm 0.02$ & $5.25\pm 0.24$ & [2] \\
HiRes & $3.25\pm 0.01$ & $2.81\pm 0.03$ & $4.5\pm 0.5$ & [3]\\
\bottomrule
\end{tabular}

\begin{small}
\begin{tablenotes}
 \item {[1]} \citet{Fenu:2017hlc}
 \item {[2]} \citet{Tinyakov:2017bju} 
 \item {[3]} \citet{abbasi2008}
\end{tablenotes}
\end{small}

\end{threeparttable}

\subsubsection{The high-energy cut-off}
\label{ssec:CutOff} 
The cut-off at the highest energies is characterized by a strong and fast decrease of the total cosmic ray spectrum around  $E_\mathrm{cut-off}\approx 4\times 10^{10}$~GeV 
\citep{abraham2008b,abbasi2008,Tinyakov:2017bju}. Here, on the one hand the results cannot be compared straight away with each other because Auger and TA and High Resolution Fly's 
Eye (HiRes) use two different functional ansatzes for the UHECRs fitting procedure. TA and HiRes use simple broken power-laws, where the last break point is called $E_\mathrm{cut}$ Auger applies a more complicated functional form where the relevant energy scaling is given by $E_\mathrm{s}$. Details can 
be found in the references given in Table \ref{tab:cutoff}.  In the limit of $E\rightarrow\infty$ the spectral index at the cut off $\gamma_\mathrm{cut}$ given by TA and HiRes can be compared with the sum of the parameters
$\gamma_2+\Delta\alpha$ given by Auger. Here, $\gamma_2$ is the spectral index between the ankle and the cut-off (see Tab.\ \ref{tab:ankle}). Furthermore, the energy $E_{1/2}$, defined as the energy where the integrated flux drops by one half compared to the case without a cut-off, is given by all three experiments and can be used for comparison.

It has been discussed that this high-energy cut-off might be due to the GZK effect \citep{Greisen, ZatsepinKuzmin}. The GZK effect 
is caused by the interaction of cosmic rays with the cosmic microwave background leading to a very fast fragmentation of cosmic rays with energies above this energy regime. 
Generally the cut-off can also be explained by an absolute maximum energy of the acceleration mechanism. Current measurements show a value that is compatible with both a GZK 
cut-off and a maximum energy of the sources. 

\begin{threeparttable}
\caption{Parameters of the cut-off. Note: Due to different fitting functions Auger results cannot be compared directly to TA and HiRes.}\label{tab:cutoff}

\begin{tabular}{l|cccl}
\toprule
Exp.\ & $\gamma_\mathrm{cut}$ & $E_{1/2}$~[EeV] & $E_\mathrm{cut} / E_\mathrm{s}$~[EeV] & Ref.\ \\
\midrule
Auger & $5.8 \pm 0.1 \pm 0.4$\tnote{4} & $22.6\pm 0.8\pm 4$ & $39\pm 2\pm 8$ & [1] \\
TA & $4.7\pm 0.6$ & $60.3\pm 6.9$ & $63.1\pm 7.3$ & [2] \\
HiRes & $5.1\pm 0.7$ & $57.5\pm 9.3$ & $56.2\pm 5.2$ & [3]\\
\bottomrule
\end{tabular}

\begin{small}
\begin{tablenotes}
 \item {[1]} \citet{Fenu:2017hlc}
 \item {[2]} \citet{Tinyakov:2017bju} 
 \item {[3]} \citet{abbasi2008}
 \item {[4]} Calculated from $\gamma_2+\Delta \gamma = (2.5\pm 0.1\pm 0.4) + (3.293\pm 0.002\pm 0.05)$
\end{tablenotes}
\end{small}

\end{threeparttable}

\subsubsection{Summary}

In summary, the all particle cosmic-ray energy spectrum is dominated by two significant changes in slope, namely the knee and the ankle, and a cut-off at the highest energies. In 
addition, the region between the knee and the ankle where the transition between galactic and extragalactic sources of the cosmic rays is assumed, several more subtle features are 
still puzzling and in need of interpretation. The origin of the change in slope at lower energies as observed with AMS, CREAM and PAMELA is also still under debate as discussed below

\subsection{Cosmic-ray composition}
\label{sec:abundance}
Cosmic rays are naturally a mixture of all charged particle species which are existing at the acceleration site. In addition, interactions that occur during the propagation of these primary 
cosmic rays lead to the production of so called secondary cosmic rays; enriching the composition for example in light elements like Lithium, Beryllium and Boron. As we have already indicated in Section \ref{allparticle:sec}, there is nothing like \emph{the} cosmic-ray composition as it changes significantly with energy. One might integrate over 
the whole energy range to derive a single composition but this is only possible when a specific cosmic-ray model is chosen (see Section \ref{composition:sec}). In this section we will shortly discuss  the evolution of the individual cosmic-ray  energy spectra qualitatively and concentrate on the ratio of secondary to primary particles. Section \ref{composition:sec} will then discuss the energy-dependent cosmic-ray abundance quantitatively.

\begin{figure}[htbp]
\centering{
\includegraphics[trim = 0mm 5mm 7mm 9mm, clip, width=0.9\textwidth]{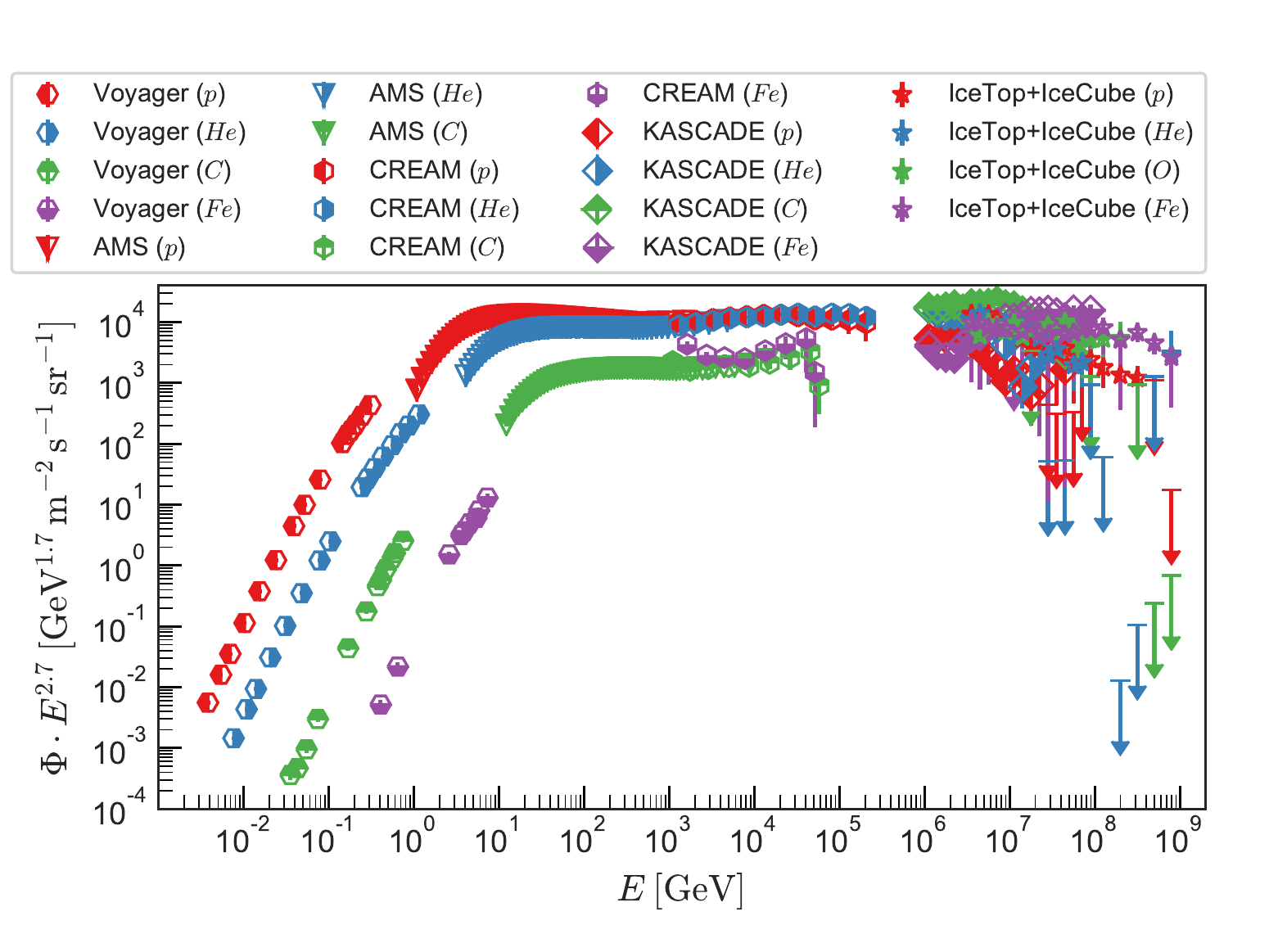}
\caption{The elemental resolved cosmic-ray energy spectrum shows a clear proton dominance up to several GeV. At higher energies the composition of cosmic ray is more complicated 
and shows several turnovers from light to heavy composition and vice-versa. --- Data: Voyager \cite{Stone150}, AMS-02 \cite{ams2015a, ams2015b, ams2017}, CREAM 
\cite{0004-637X-728-2-122, 0004-637X-707-1-593}, KASCADE \cite{ANTONI20051}, IceCube+IceTop \cite{AndeenPlum:2019}}
\label{fig:compositionVoyager}
}
\end{figure}

Figure \ref{fig:compositionVoyager} shows the element-resolved energy spectrum of cosmic rays measured with different experiments. In the energy range from 
$\sim(10^{-3}-10^{2})$~GeV the most abundant element is Hydrogen followed by Helium and heavier elements like Carbon, Nitrogen and Oxygen up to Iron. At around $E\approx 10^3$~GeV 
CREAM data reveal a turnover from a very light composition to heavier one with Helium becoming the dominating species. KASCADE data confirm this increase, where first the 
CNO-group is dominating at $\sim 10^6$~GeV and at even higher energies Iron becomes the most abundant element. KASCADE-Grande does not report individual elements (or groups of 
elements) but gives only a separation in a light and heavy component. Nevertheless, a trend for another change in the medium weight of the composition is visible: The composition 
becomes lighter between $(10^8 - 10^9)$~GeV. Not shown is the observed (e.g.\ Auger and TA) increase in the mean logarithm of the mass number $\langle \ln(A) \rangle$ at the very 
highest energies. 

\subsubsection{Primary-to-secondary ratio}
Per definition, we can divide cosmic rays into two groups: (1) Cosmic rays that are directly originating at the sources of acceleration --- so called primary cosmic 
rays\footnote{Often, primarily accelerated electrons are also included in the term cosmic rays, which makes it necessary to distinguish between hadronic cosmic rays and cosmic-ray 
electrons. In this review, as we focus on the hadronic component, we refer to the hadronic component of cosmic rays as \textit{cosmic rays} and specify cosmic-ray electrons 
explicitly when we discuss these.} --- and (2) cosmic rays that are produced by interactions or spallation processes during the propagation of primary cosmic rays. These are 
secondary cosmic rays, belonging to the class of cosmic rays that are created via interaction\footnote{Often, the term secondary cosmic rays is used for all interaction products 
that arise from primary cosmic rays, like the neutral end products neutrinos and gamma-rays that originate from (photo)hadronic interactions, spallation or even the particle 
showers that are created in the Earth's atmosphere.}.

As discussed above, the abundance of primary cosmic rays follows the one of elements that are generated in large numbers in fusion processes of stars. Especially the 
light elements, lithium, beryllium and boron, are (almost) not produced in stellar fusion. This means when a significant amount of e.g.\ boron is measured at Earth it has to be 
created during propagation. In the spallation process, a heavy nucleus (in this case from the cosmic-ray flux) is cloven by the interaction with a proton or a neutron of the 
ISM. Thus, it follows directly that the amount of secondary particles should be related to the total accumulated column density of primary cosmic rays. From this column density one may calculate  a cosmic ray propagation time assuming a constant target density in the Galaxy. When this value is compared with the propagation time derived from observations of radioactive cosmic-ray isotopes (so called \textit{cosmic clocks}), which is much larger, one may conclude that cosmic rays mainly propagate in the very dilute Galactic halo, see e.g.\ \citet{sigl2017}. An energy-resolved ratio allows for an 
analysis of the energy dependence of the transport process; this is also true if no additional information on the target density is used.

\begin{figure}[htbp]
\centering{
\includegraphics[trim = 5mm 6mm 2mm 2mm, clip, width=0.9\textwidth]{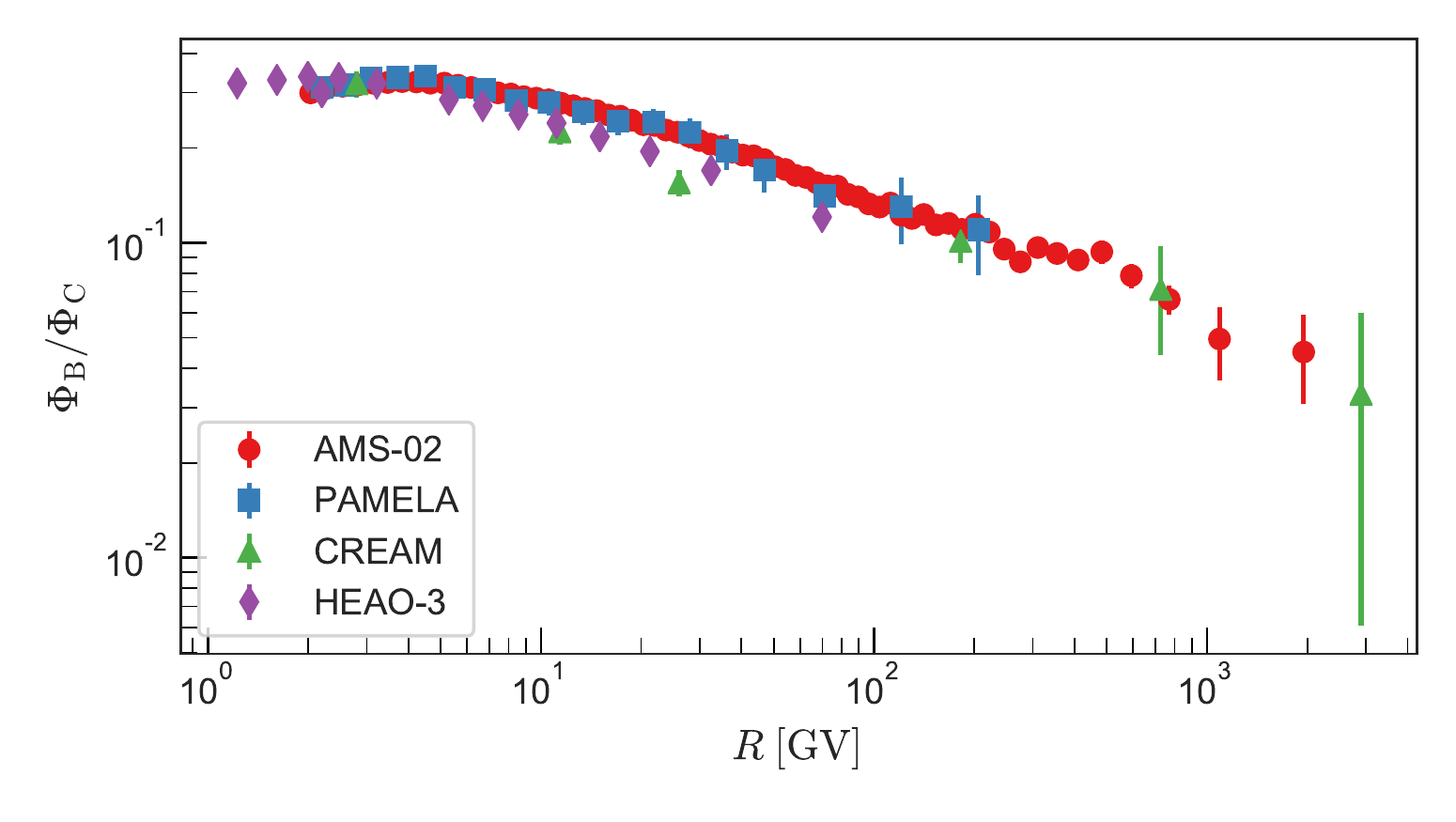}
\caption{Boron (B) over Carbon (C) ratio measured by the AMS-02 \citep{ams2016} and PAMELA \citep{pamela_B-C}, CREAM \citep{CREAM_BoC} and HEAO-3 \citep{HEAO_BoC} observatories. Note: HEAO measurements are shown without errorbars. Both measurements agree very well with each other in the whole rigidity range. A single power-law fit to the AMS-02 data above $R=65$~GV gives a spectral index of $\gamma=-0.333\pm 0.014(\mathrm{fit}) \pm 0.005(\mathrm {syst})$ \cite{ams2016}.}
\label{BoverC:fig}
}
\end{figure}

Since we can learn much from the secondary to primary ratios, observatories have been designed to measure it with very high precision. Figure \ref{BoverC:fig} shows rigidity 
resolved 
boron over carbon ratio $\Phi_\mathrm{B}/\Phi_\mathrm{C}$ measured with different telescopes. All groups see a decreasing ratio with increasing rigidity; so higher energetic 
primary cosmic rays are traversing less target material during there propagation. Ballistic propagation of cosmic rays would have led to a constant primary to secondary ratio. 
Opposed to this, a diffusive particle transport will lead to a rigidity-dependent transport process. The AMS-02 experiment on board of the International Space Station (ISS) 
detects a rigidity dependence $\propto R^{-0.33}$ above $R=65$~GV \citep{ams2016} which is expected for a Kolmogorov turbulence spectrum. The theoretical background is discussed 
in Section \ref{spallation:sec}.

\subsection{(An)isotropy}
\label{anisotropy_data:sec}
At all energies, the cosmic-ray flux is showing a very high level of isotropy. We want to emphasize this huge homogeneity in the arrival direction of cosmic rays as it is often no 
longer visible in figures of the detected anisotropies: these are given as the relative intensity difference with respect to the average flux and thus do not include the zeroth 
multipole order of the arrival directions. Even the largest deviation from the nearly perfectly isotropic sky are on the order of $10^{-3}-10^{-4}$. Only at 
EeV-energies, the anisotropy reaches the percent-level. Nonetheless, these small irregularities serve as important probes for theories of both the transport and accelerations of 
cosmic rays.

The relative intensity of cosmic rays, $I_{\rm rel}$, with respect to the solid-angle-averaged, isotropic, cosmic-ray distribution function $f_\mathrm{iso}$ 
arriving at 
Earth can be expressed as
\begin{equation}
I_{\rm rel}(\vec{r},\,p\vec{e}_n,\,t)=\frac{f(\vec{r},\,p\vec{e}_n,\,t)}{f_{\rm iso}}\label{rel_int:equ}\,.
\end{equation}
Here, $f(\vec{r},\,\vec{p},\,t)$ is the distribution function of cosmic rays\footnote{This quantity is related to the differential cosmic-ray intensity as $n_{\rm 
CR}(\vec{r},\,\vec{p},\,t)=p^2\cdot f(\vec{r},\,\vec{p},\,t)$.  
These 
quantities are discussed in more detail in Section \ref{candidates:sec}.}. Further, $\vec{e}_n=(\cos(\alpha)\sin(\delta), \sin(\alpha)\sin(\delta), \cos(\delta))$ is a unit vector 
in 
the direction of consideration. The natural system to describe cosmic ray anisotropies is the equatorial coordinate system parametrized by right ascension $\alpha$ and declination 
$\delta$.

In order to determine the relative intensity, it can be expressed in terms of a series of spherical harmonics, see e.g.\ \citet{berezinskii1990,sigl2017,AHLERS2017184}:
\begin{eqnarray}
I_{\rm rel}&=& 1 + \sum_{l\geq 1}\sum_{-l}^l a_{lm}Y_{lm}(\pi/2-\delta, \alpha)\,; \quad a^*_{lm}=(-1)^m a_{l-m}\label{irel:equ}\\
&=& 1+ \vec{\delta}\cdot\vec{e}_n(\alpha, \delta)+\mathcal{O}(\left\{a_{lm}\right\}_{l\geq 2}) \quad \,.
\end{eqnarray}
Here, $Y_{lm}$ are the spherical harmonics which form an orthogonal basis on the full sky and $a_{lm}$ are the multipole moments. The first term unity, $1$, is the isotropic contribution 
and the second term $\vec{\delta}\cdot\vec{e}_n$ described the dipole contribution with amplitude $|\vec{\delta}|$.

\begin{figure}[htbp]
\centering{
\includegraphics[trim = 0mm 40mm 35mm 40mm, clip, width=0.8\textwidth]{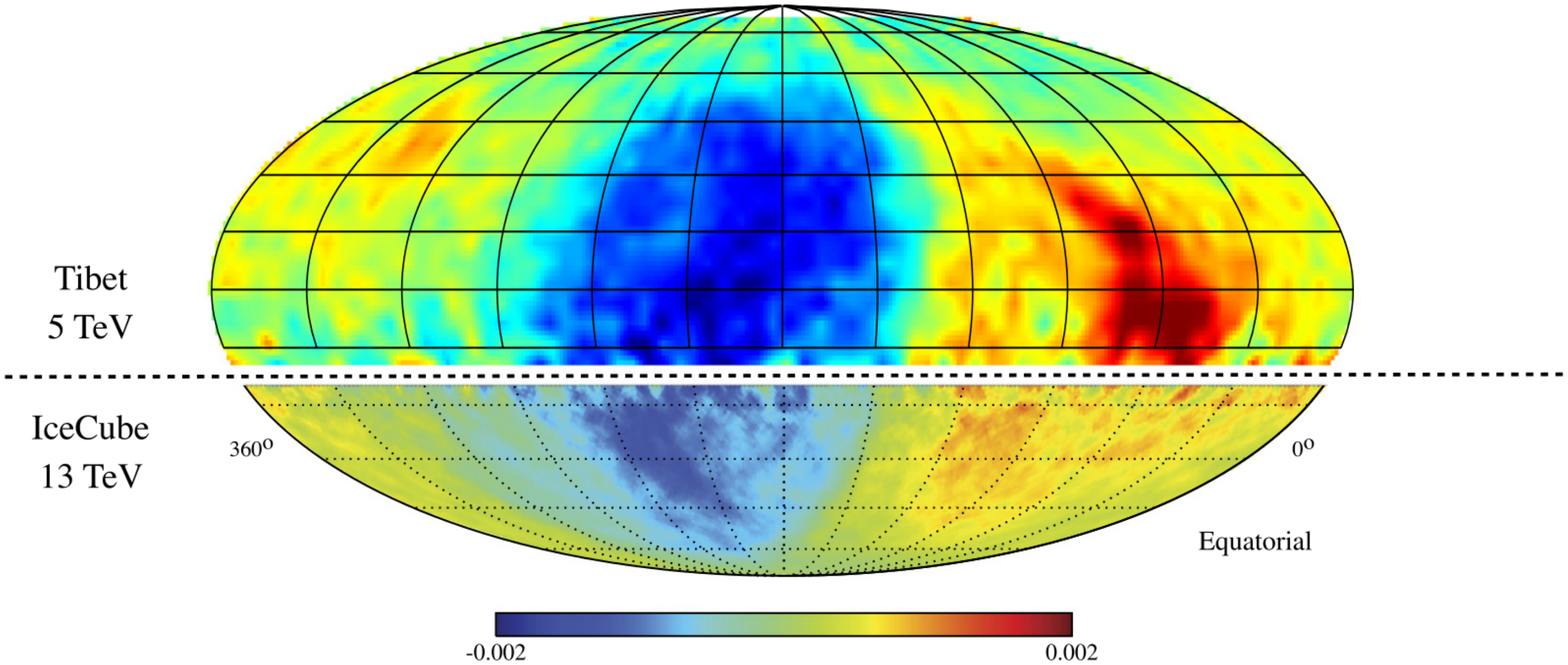}
\caption{Skymap of the observed arrival direction in equatorial coordinates measured by Tibet (upper panel) and IceCube (lower panel). Maps are smoothed by a $5^\circ$-kernel. 
Figure by the courtesy of the authors taken from \citet{AHLERS2017184}.}
\label{anisotropy_sky:fig}
}
\end{figure}

Figure \ref{anisotropy_sky:fig} shows the nearly all-sky anisotropy in the TeV energy range. The median energy $\bar{E}=5$~TeV of the $4.9\times 10^{10}$ Tibet-AS$\gamma$ events 
is slightly lower than for the $3.2\times10^{11}$ IceCube events ($\bar{E}=13$~TeV). The dipole is clearly visible when the isotropic component is subtracted. 
A recent combined analysis of HAWC and IceCube data can be found in \citep{HAWCIceCube}.

\subsubsection{Dipole anisotropy}\label{dipole:sec}
In the equatorial coordinate system the dipole amplitude $\vec{\delta}$ can be calculated as:
\begin{align}
\vec{\delta}(\delta_{0h}, \delta_{6h}, \delta_\mathrm{N}) = \sqrt{\frac{3}{2\pi}} \left(-\Re(a_{11}), \Im(a_{11}), a_{10}\right)\quad , \label{eq:dipole}
\end{align}
where, following the notation in \citet{ahlers2016} $\delta_{0h}$ is pointing in the local hour angle ($\alpha=0$) and $\delta_{6h}$ pointing in the direction of the vernal equinox ($\alpha=90^{\circ}$).  Further, $\Re$ and $\Im$ denote the real and imaginary parts of the coefficienct of the spherical harmonics, i.e.\ $a_{11}$ in Equ.\ (\ref{irel:equ}).

The observation of these anisotropies is a experimental challenge because of two things: (1) The signal has to be separated from the isotropic background that is at least a 
thousand times larger. Thus, very good knowledge of the systematics of the detector response is crucial and high event statistics is required. (2) The fact that --- especially at higher energies --- 
most experiments are ground-based makes them blind for anisotropies parallel to the rotation axis of the Earth. This eventually leads to the fact that $a_{10}$ in Equ.\ 
(\ref{eq:dipole}) can in general not be restricted by high energy telescopes leading to a two-dimensional representation of the dipole projection onto the equatorial plane with 
phase 
$\tilde{\alpha}$ and amplitude $A$:
\begin{align}
(\delta_{0h}, \delta_{6h}) = \left(A\cos(\tilde{\alpha}), A\sin(\tilde{\alpha})\right)\,.
\end{align}

\begin{figure}[htbp]
\centering{
\includegraphics[trim = 0mm 0mm 0mm 0mm, clip, width=\textwidth]{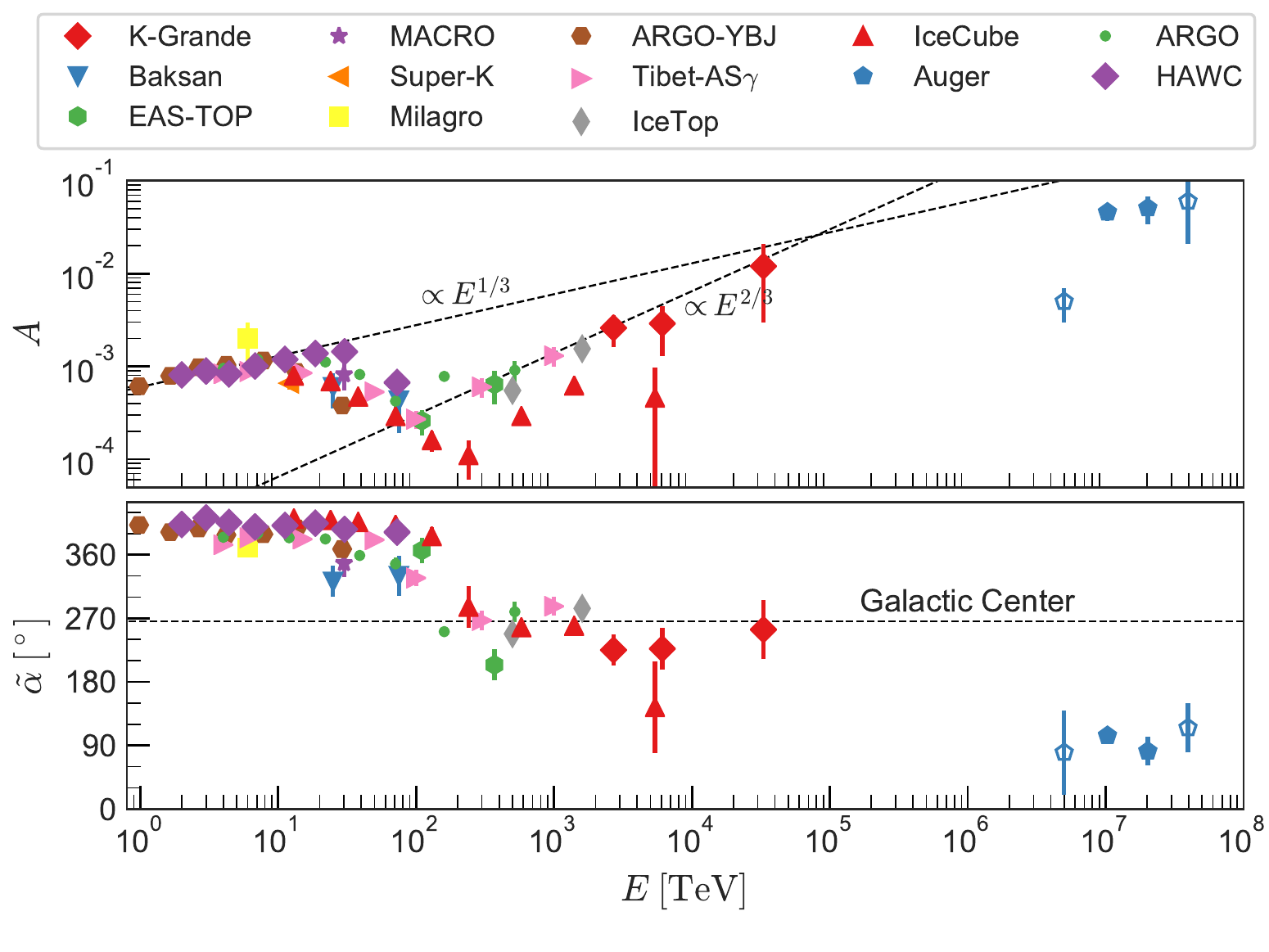}
\caption{Energy dependence of the amplitude (upper panel) and phase (lower panel) of the measured cosmic ray dipole anisotropy. Non filled markers do only have a small statistical significance. Figure after \citet{AHLERS2017184}. --- Data: Auger data is from \citep{augerDipole2018}, others taken from \citet{AHLERS2017184} and references therein.}
\label{anisotropy:fig}
}
\end{figure}

Figure \ref{anisotropy:fig} shows this amplitude\footnote{Note: The amplitude $A$ shown in the plot is always smaller or equal to the true dipole amplitude, $|\delta|\geqq A$} $A$ and 
phase $\tilde{\alpha}$ measured by different observatories from TeV up to EeV energies. The top panel shows the amplitude 
$A$ which is slowly increasing with energy. In the last years high statistics of the current generation of observatories could reduce the uncertainties significantly. But still, especially above $\approx 
100$~TeV, the different measurements do not line up exactly. This is most likely due to different locations of the observatories leading to different parts of the sky that are 
accessible for the Earth-bound telescopes. Furthermore, other systematic differences cannot generally be excluded. To guide the eye, we include two lines  $\propto 
E^{1/3,\; 2/3}$ (following the figure presented in \citet{AHLERS2017184}) which is 
expected for different diffusion scenarios\footnote{Note that these lines are not fitted to data, but are simply meant to guide the eye..}). The details of the physics and different models for this dipole-anisotropy will be 
discussed in Section \ref{anisotropy:sec}. Although the overall trend might agree with a simple Kolmogorov-type diffusion theory ($\propto E^{1/3}$) several deviations are visible 
in the data.

The lower panel shows the direction of the dipole projection. Here, the dipole direction is constant from $1-100$~TeV and at around $100$~TeV toward higher energies, an abrupt change of the dipole direction 
is visible. After this phase change the 
dipole overlaps with the direction of the Galactic Center (GC) at medium energies ($100-10^4$~TeV), in agreement with an increasing source density in this direction. The origin of 
the Auger dipole at EeV-energies is still 
under debate but is clearly not in the direction of the GC, which strengthens the case of cosmic rays beyond the ankle being of extragalactic origin 
\citep{augerDipole2017, augerDipole2018}.

For an extensive review --- including a study on the projection problem and other technical issues, as well as the modeling of large- and small-scale an-isotropies --- we refer 
to the review by \citet{AHLERS2017184} and references therein. A summary of the state-of-the-art concerning the interpretation of the dipole and multipole structures in the 
context of other experimental cosmic ray results and the connected multimessenger-modeling is presented in Section \ref{anisotropy:sec}.

\subsubsection{Multipole anisotropy}\label{multipole:sec}
In addition to the dipole anisotropy, especially in the TeV -- PeV energy range, small-scale anisotropies at multipoles $l>1$ are observed. Observatories like the High-Altitude Water Cherenkov 
Observatory (HAWC) \citep{2014ApJ...796..108A}, IceCube and IceTop \citep{2016ApJ...826..220A}, Tibet \citep{2010ApJ...711..119A} and others have reported anisotropies at least 
down to the level of $\approx 10^\circ$. At smaller scales intensity maps are often dominated by shot noise. To avoid an influence of these stochastic errors the intensity maps 
are usually smoothed with a kernel of about same width as the structures of interest.
The multipole directions and amplitudes are rigidity dependent making the comparison between different measurements complicated since the detectors have different energy 
thresholds and the analyses do not necessarily use the same composition models.

All of the observatories do agree with each other on the existence of small scale anisotropies. In recent years combined efforts in the observation of these 
anisotropies are for example made by the HAWC and IceCube observatories. The combination of data from the two observations has lead to nearly complete sky coverage reducing 
systematic uncertainties significantly.
We refer to \citet{desiati_lazarian2013, AHLERS2017184} for a detailed discussion of the status of the observations. The theoretical background and approaches to 
explain the small scale anisotropies is given in Section \ref{anisotropy:sec}.

\begin{figure}[htbp]
\centering
\includegraphics[trim = 0mm 0mm 0mm 0mm, clip, width=0.8\textwidth]{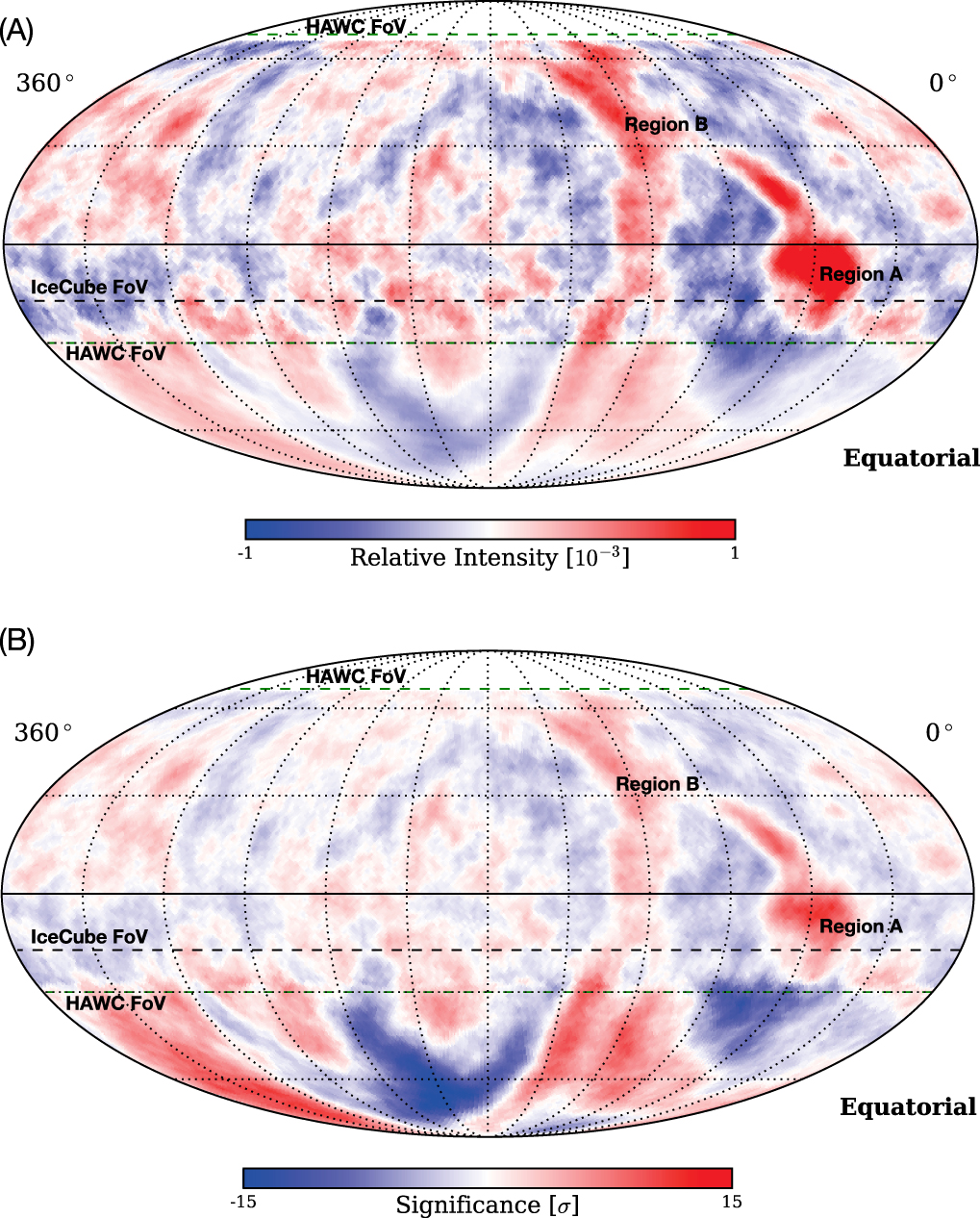}
\caption{Combined IceCube and HAWC intensity map after subtracting multipoles with $l\leq 3$ with median energy $E\approx 2$~TeV. The  regions A and B are observed now with very large significance for all hotspots. Figure\textsuperscript{\textcopyright} AAS --- reproduced with permission --- kindly provided by J.\ C. Díaz-Vélez originally published in \citet{HAWC_IceCube_2019}.
\label{fig:SmallScaleAnisotropy}
}
\end{figure}

Figure \ref{fig:SmallScaleAnisotropy} shows the small scale anisotropies measured by the HAWC and IceCube observatories with a median energy of $E=10$~TeV based on 5 years of IceCube data (05/2011-05/2016) and two years of HAWC data (05/2015-05/2017) \cite{HAWC_IceCube_2019}. The dipole, quadrupole and octopole have been removed to emphasize the small scale features. Here, two distinct regions A and B that deviate with very high statistical significance from isotropy are observed. The regions A and B have first been reported as hotspots A and B by the MILAGRO observatory \citep{milagro_hotspots}.

\begin{figure}[htbp]
\centering{
\includegraphics[trim = 0mm 0mm 0mm 0mm, clip, width=0.9\textwidth]{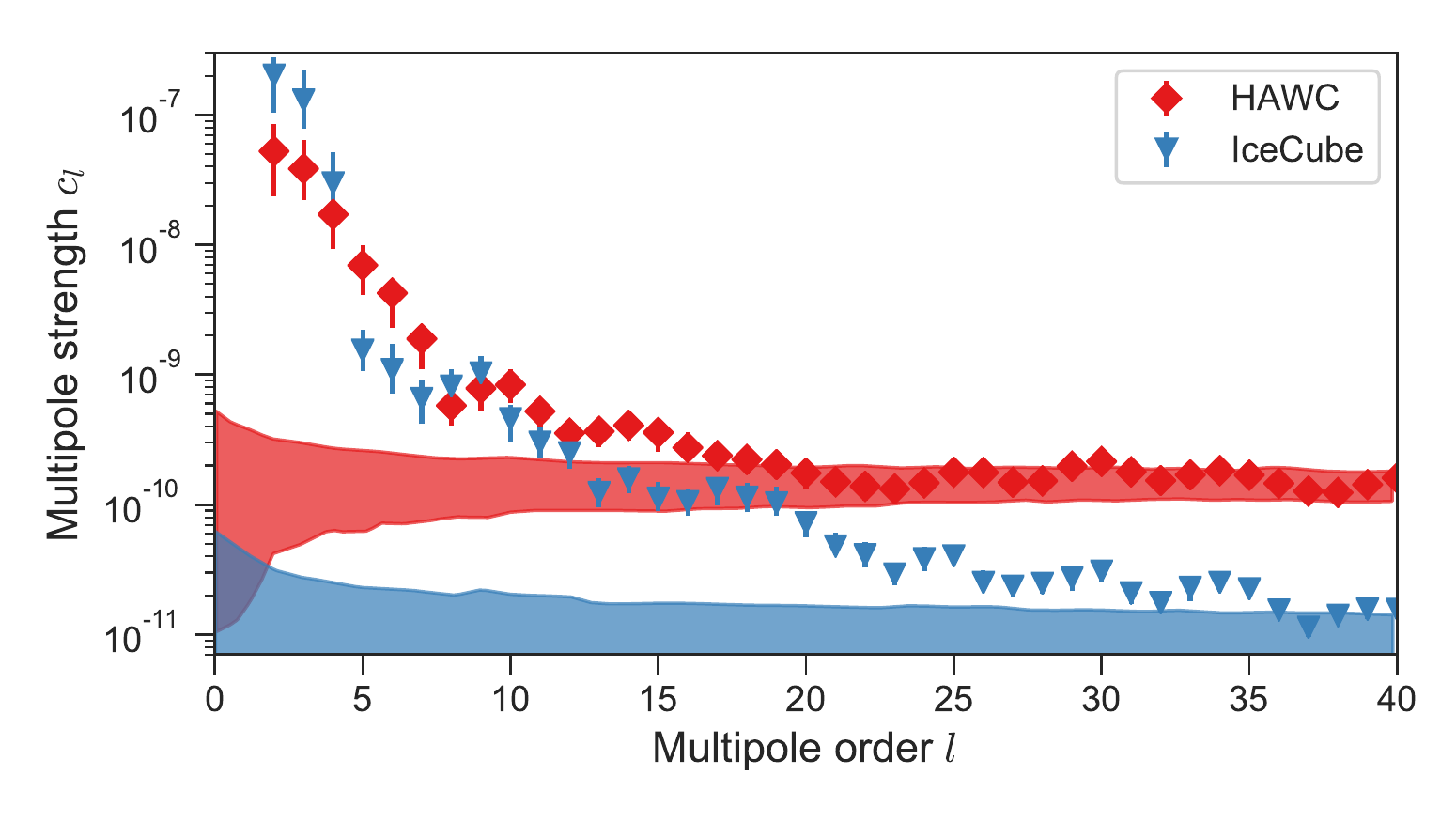}
\caption{Angular power spectra measured by the HAWC detector ($\bar{E}=2$~TeV, red squares, June 2013 - February 2014) and IceCube observatory ($\bar{E}=20$~TeV, blue triangles, May 2009 - May 2015). The difference between the two experiments can be explained by different fitting techniques and foremost by different noise levels; HAWC's noise level is about one order magnitude larger. In shaded colors the 95 percent isotropic expectation is shown for IceCube (blue) and HAWC (red) data. A clear deviation from an isotropic expectation can be found for all scales larger than $10^\circ \mathrel{\widehat{=}} l\approx 20$. Data by the courtesy of Markus Ahlers originally published in \citet{AHLERS2017184}. }
\label{fig:SmallScaleMultipoles}
}
\end{figure}

Figure \ref{fig:SmallScaleMultipoles} shows the multipole analysis by IceCube and HAWC \citep{AHLERS2017184}. Here, a significant deviation from isotropy is clearly observed until $l\approx 20$ which corresponds to scales of $10\deg$. 
    
\clearpage

\subsection{Leptons and antimatter}
\label{sec:LeptonsAntimatter}
From basic arguments on charge neutrality at the sources it is plausible that besides charged hadrons also leptons are accelerated, although the subtleties of the exact ratio are complicated (see a short discussion in \citet{merten2017a}). Some amount of anti-particles should be present in the total cosmic ray flux simply through the production via hadronic interactions. However, it is not yet clear if all of the anti-particles are produced via hadronic interactions or by pair-production in strong magnetic fields of pulsars.

\begin{figure}[htbp]
\centering{
\includegraphics[trim = 3mm 4mm 15mm 0mm, clip, width=0.8\textwidth]{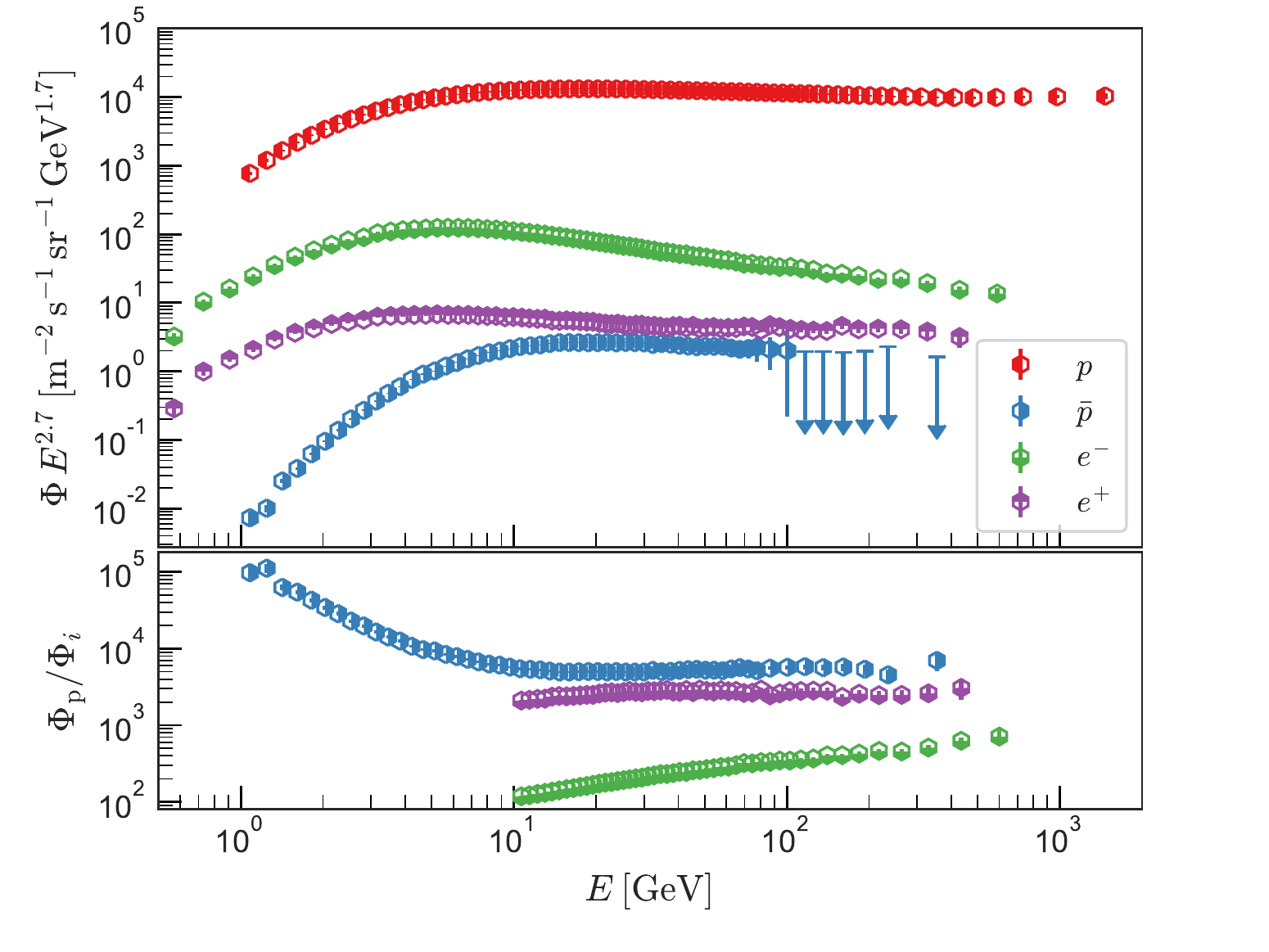}
\caption{Upper panel shows the proton $p$, anti-proton $\bar{p}$, electron $e^-$ and positron $e^+$ fluxes weighted with $E^{2.7}$ to scale. The lower panel shows the proton/(anti-proton/electron/positron) flux ratios $\Phi_p/\Phi_i$.  --- Data from \citet{ams2014,aguilar2016}.}
\label{fig:LeptonFluxes}
}
\end{figure}

\begin{figure}[htbp]
\centering
\includegraphics[trim = 2mm 2mm 15mm 11mm, clip, width=0.9\textwidth]{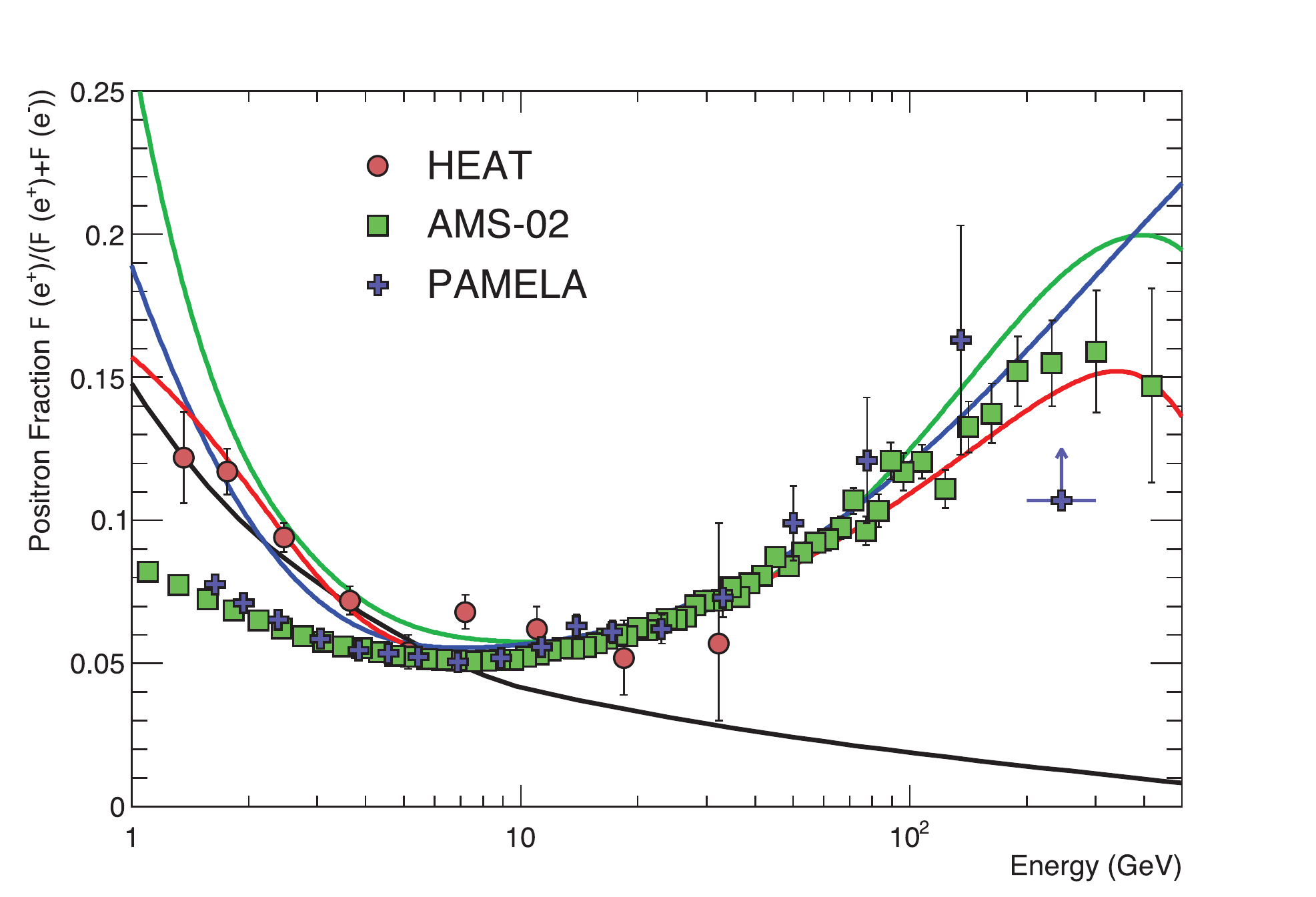}
\caption{The observed positron fraction compared with different models. Only a constant scalar diffusion model (black line) is ruled out. All other models --- space dependent anisotropic diffusion (blue line), production in pulsars (red line) and dark matter decay (green line) --- agree reasonably well with the data. Figure from \citet[Fig.\ 29.3]{pdg2018} --- Data: \citet{accardo2014,pamela2009,beatty2004}  --- Models: \citet[][black line]{strong_propagation_1998}, \citet[][blue line]{gaggero2013}, \citet[][red line]{yin2013}, \citet[][green line]{ibarra2013}.
}
\label{fig:ePlusFraction}
\end{figure}

Today, cosmic-ray observatories have measured electron, positron and anti-proton fluxes with high precision. Figure \ref{fig:LeptonFluxes} shows the observed $e^-$, $e^+$ and $\bar{p}$ fluxes in comparison with the proton flux and their corresponding ratios $\Phi_p/\Phi_i$ (lower panel) measured with the AMS-02 spectrometer. Three things are clearly visible:
\begin{enumerate}
\item the proton flux is dominant over the whole energy range;
\item above $E\approx 10$~GeV the ratios of protons to anti-protons $\Phi_p/\Phi_{\bar{p}}$ and protons to positrons $\Phi_{p}/\Phi_{e^+}$ are rather constant whereas the proton electron ratio $\Phi_{p}/\Phi_{e^-}$ is increasing;
\item the so-called positron fraction $f_{e^{+}}:=\Phi_{e^+}/(\Phi_{e^+}+\Phi_{e^-})$ is varying with energy, it decreases as expected in the standard model up to energies of $\sim 10$~GeV and starts to increase toward higher energies (see Fig.\ \ref{fig:ePlusFraction}).
  \end{enumerate}
Heavier anti-matter than anti-protons have not yet been detected with high confidence. AMS-02, however, has reported first evidence for the detection of anti-helium ($\overline{\mathrm{He}}^3$ and $\overline{\mathrm{He}}^4$, see \citep{sciencemag17}).

Figure \ref{fig:ePlusFraction} shows the fraction of positrons with respect to the total positron and electron flux $f_{e^+}$ as measured by HEAT, PAMELA and AMS-02. All measurements show a decrease of $f_{e^+}$ with increasing energy at energies between $1$~GeV and $10$~GeV. AMS02-data indicate that a maximum could be reached at several hundred GeV, but more data at the highest energies are necessary to confirm this trend.

Different ansatzes for the origin of this excess of positrons are discussed, also displayed in Fig.\ \ref{fig:ePlusFraction}. The standard model prediction of pure secondary production in the Galaxy \citep{strong_propagation_1998} is not able to explain the increase beyond $\sim 10$~GeV (black line). The three general possibilities for enhanced positron production are discussed in more detail in Section \ref{multimessenger_modeling:sec}. In summary, they are:
\begin{enumerate}
\item taking more sophisticated propagation effects into account (\citet{gaggero2013}, blue line in Fig.\ \ref{fig:ePlusFraction});
\item dark matter decay \citep[][green line]{ibarra2013};
  \item positron production in pulsars \citep[][red line]{yin2013} explains the current data equally good. 
    \end{enumerate}
It has to be noted that more recent models for the dark matter decay ansatz exist \citep{cirelli2009, cholis2013b, kopp2013_DM}. Especially the models by \citeauthor{cholis2013b} suggest that these possible dark matter particles are probably decaying first into some unstable intermediate states before eventually decaying into leptons.

\subsection{Hadronic interaction signatures}
\label{sec:HadronicInteraction}
The neutral cosmic-ray interaction products neutrinos and gamma-rays are crucial pieces in the development of a multimessenger picture of Galactic cosmic-ray sources and propagation. Most relevant in the Galactic environment is the interaction of cosmic rays with the ambient gas. At GeV-energies and above, it leads to the production of pions, which in turn decay into gamma-rays (neutral pions) and leptons (charged pions, with three neutrinos per charged pion and one electron). The details of this process are described in Section \ref{multimessenger_sources:sec}. 

By now, a large number of gamma-ray sources has been detected in the GeV-TeV range. However, there are competing processes that arise when electrons 
interact with their environment. Most relevant are the processes bremsstrahlung and Inverse Compton (also discussed in more detail in Section \ref{multimessenger_sources:sec}). It 
is difficult to disentangle these leptonic signatures from the hadronic parts, but as will be reviewed below and in Section \ref{multimessenger_sources:sec}, new data from instruments like Fermi, 
HAWC, H.E.S.S.\footnote{High Energy Stereoscopic System}, MAGIC-Telescopes\footnote{Major Atmospheric Gamma-Ray Imaging Cherenkov} and VERITAS\footnote{Very Energetic Radiation 
Imaging Telescope Array System} shed more light on the nature of Galactic gamma-ray emission, both concerning the diffuse emission and local emission signatures. 

In contrast to gamma-rays, neutrinos are difficult to detect due to their extremely small interaction cross-section: the neutrino-nucleon cross section at $1$~TeV is $\sigma_{\nu   N}\sim 6\times 10^{-36}$~cm$^{2}$ (see \citep{icecube_cross_section2017} and references therein). The integrated column depth of Earth (from one side through the center to the other side), is $N_{\oplus}\approx 
4\times 10^{33}$~cm$^{-2}$. Thus, the optical depth for a TeV neutrino crossing the Earth's inner core is $\tau \sim \sigma_{\nu N}\cdot N_{\oplus}\sim 2\times 10^{-2} \ll 1$.  
 What these numbers show is that (a) the 
detection of an astrophysical neutrino flux at Earth always happens for only a very small fraction of the total flux at Earth; (b) large detection volumes are needed in order to 
actually detect a significant flux. A further challenge in the detection of astrophysical neutrinos is the existence of atmospheric muons and neutrinos in large numbers, 
produced in cosmic-ray interactions in 
the Earth's atmosphere. These dominate the fluxes seen by super-GeV neutrino observatories: The atmospheric fluxes produce a background that is in general significantly 
larger than 
the signal itself. Only sophisticated analysis methods can reduce the strong background in order to detect the signal that we know is present since the first evidence and 
detection 
was published in 2013 \citep{icecube2013,icecube2014}. The IceCube detector as the largest operating neutrino observatory at the moment, detects $10^{11}$ atmospheric muons and 
$\sim 10^{6}$ atmospheric neutrinos per year \citet{halzen_klein2010}. These numbers compare to some $10-100$ detected astrophysical neutrinos per year. The fact that a high-energy neutrino flux actually has been detected by IceCube recently is owing to a combination of an elaborate detector configuration and innovative background reduction algorithms. The establishment of an alert system that triggers telescopes all over the world when a high-energy neutrino with a large probability of being 
of astrophysical nature is detected, a first hint for a correlation with a point-source, the blazar TXS 0506+056, was observed \citep{icecube_txs_fermi2018,icecube_txs2018}. This 
is a first indication of the diffuse signal being of extragalactic nature. However, there exists a strong limit on the contribution of gamma-ray emitting blazars to the diffuse 
flux, which makes modeling of this phenomenon more complex. Thus, there is no necessity for the entire signal to come from the same population of sources. While it is difficult to 
produce the entire diffuse neutrino flux in the environment of our Galaxy, part of this flux still might come from the Milky Way. The discussion of the neutrino origin is part of 
this review (see Section \ref{multimessenger_modeling:sec}).

A summary of the state of the art of gamma-ray and neutrino detection  in the context of Galactic emission are presented in this section.  Details on the physical processes leading to gamma and neutrino production and a discussion of the spectral behavior of the fluxes are presented in Section \ref{multimessenger_sources:sec}. Finally, the interpretation of the gamma and neutrino data within a multimessenger framework is done in Section \ref{multimessenger_modeling:sec}.

\subsubsection{Photons}
The past decades have been incredibly successful concerning the detection of gamma-ray sources. The first generation of Imaging Air Cherenkov Telescopes (IACT), 
HEGRA\footnote{High Energy Gamma-Ray Astronomy} and Whipple, could confirm a total of 13 gamma-ray sources, with four of these being of Galactic origin (see \citet{tevcat} for a 
summary of data). During that time, the EGRET\footnote{Energetic Gamma-Ray Energetic Telescope} instrument on board of the Compton Gamma-Ray Observatory (CGRO)  mapped the sky 
at GeV energies. These instruments provided us with diffuse gamma-ray maps for both extragalactic \citep{egret,egret_new} and Galactic emission 
\citep{hunter1997,egret_new}, as well as with detections of extended and point-like sources. The third EGRET source catalog 
\citep{egret_cat1999,egret_cat2008} reports around 200 detections with exact numbers depending on the specific analysis. Most of these sources remained unidentified 
at that point, i.e.\ they could not be matched to an astrophysical counterpart. Only a number of blazars ($\sim 90$) could be identified through correlations with other 
wavelengths. Thus, little was known about the gamma-ray emission of Galactic sources.

\begin{figure}[htbp]
\centering
\includegraphics[trim = 0mm 0mm 0mm 0mm, clip, width=0.8\textwidth]{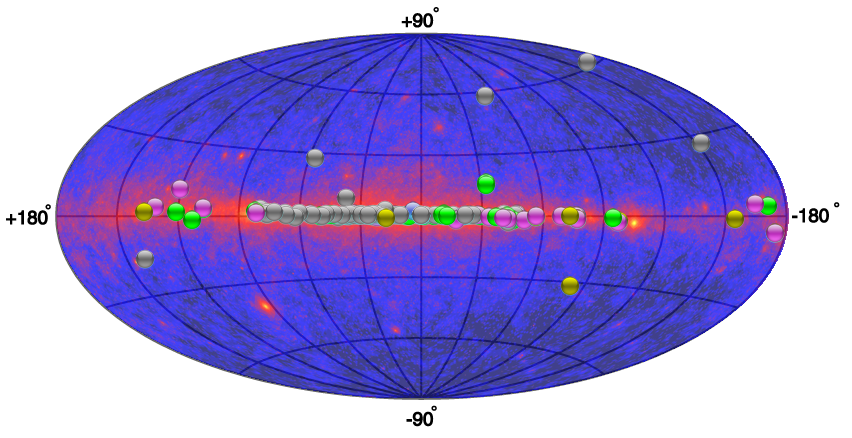}
\caption{Skymap of Galactic high-energy gamma-rays --- dots show results from Earth-bound observatories: green represents shell-type SNRs, SNR/MC-interactions, composite SNRs and superbubbles,  yellow are gamma-ray binaries, pink are PWNs, grey are unidentified sources. The color-map is from Fermi. Produced from \citet{tevcat}. 
\label{skymap_tevcat:fig}}
\end{figure}

\begin{figure}[htbp]
\centering{
\includegraphics[trim = 0mm 0mm 0mm 0mm, clip, width=0.8\textwidth]{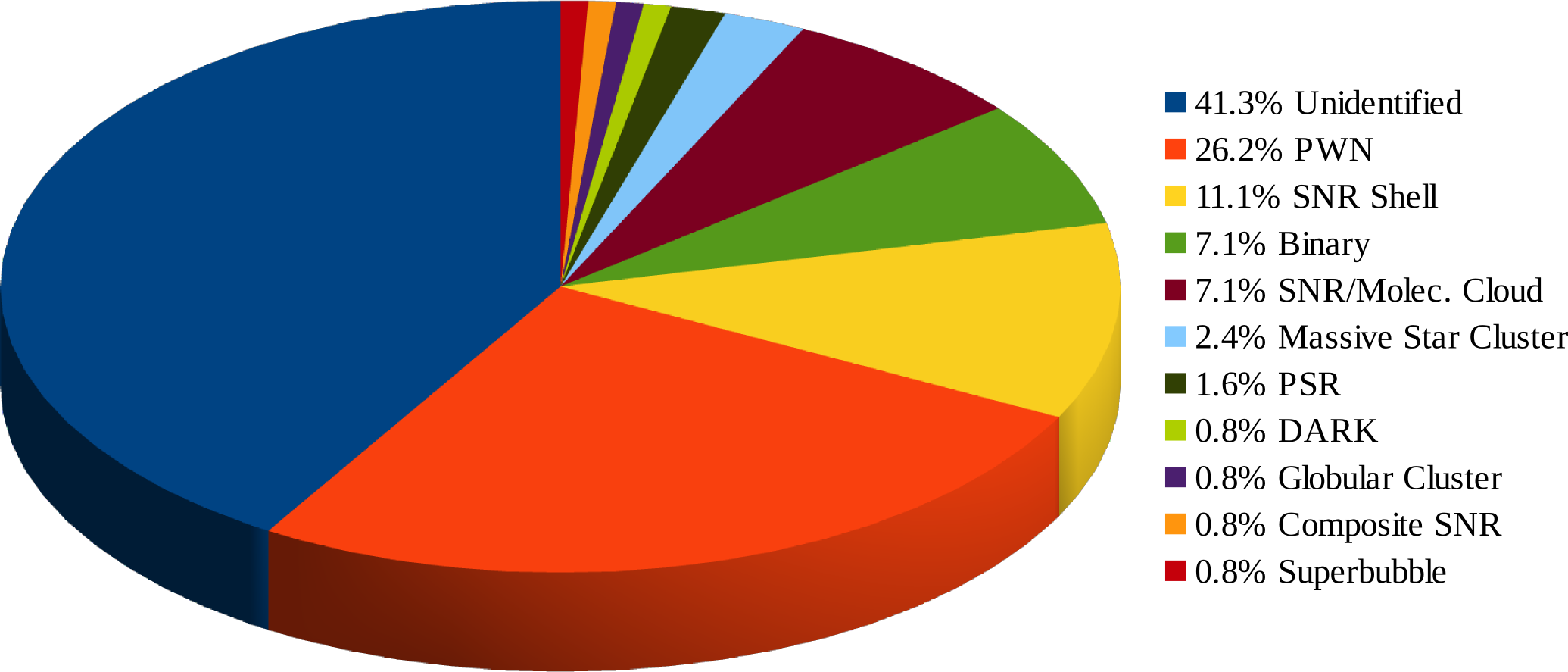}
\caption{Percentage of different source class contributions to the TeV gamma-ray sky. Dark-blue is unidentified sources, orange and yellow represent pulsar wind nebulae and supernova remnant shells, respectively. Dark-green and purple are binaries and SNR-MC associations, while light-blue and brown show a small percentage of massive star clusters and pulsars. One source each are seen in gamma-rays from globular clusters (dark purple), composite SNRs and superbubbles. Numbers are taken from \citet{tevcat}.
}
\label{photon_pie:fig}
}
\end{figure}

The number of sources increased drastically since the beginning of operation of second generation gamma-ray telescopes:
\begin{itemize}
\item In 8 years of data taking, the \textbf{Fermi-LAT} instrument on board of the Fermi Gamma-Ray Space Telescope could identify more than 5000 localized sources in the range $100$~MeV to $10$~GeV \citep{fermi_4fgl_2020}. Out of these, more than $3000$ are associated to or identified as blazars, more than $1.000$ have no clearly identified counterpart,  232 are pulsars, 24 are SNRs, 11 are PWN, 3 are star-forming regions, 5 are high-mass X-ray binaries and one low-mass X-ray binary, one binary system and one nova could be detected in the Galaxy. These are only the detections at a level with $\geq 4\sigma$ significance. For the case of SNRs, there are 16 more detections that are unresolved and the association cannot be made at this point.
\item \textbf{MILAGRO} \citep{milagro2004,abdo_milagro2007} was the first ground-based Cherenkov detection array, consisting of a water pool equipped with photomultiplier tubes 
(PMTs) plus an outer array of water tanks with PMTs, that could monitor the visible sky continuously. The advantage of this concept compared to the IAC technique is that they can 
detect diffuse or extended emission in the TeV-range. MILAGRO's major success was the detection of 6 sources in the Cygnus region that reveal flat gamma-ray spectra up to $\sim 
10$~TeV and are interesting candidates for hadronic emission \citep{milagro2004,abdo_milagro2007}. \textbf{HAWC}, in operation as a successor of MILAGRO since 2015, reports a 
detection of 39 sources in the TeV-range,  out of which 2 are associated with Pulsar Wind Nebulae (PWN), 2 are identified as SNRs, 2 as blazars. 23 are still unidentified 
\citep{hawc2017_sourcecat}. These measurements are very valuable, as they extend the energy range of Imaging Air Cherenkov Telescopes (IACTs) and as they represent the continuous 
monitoring of the sources via a partial-sky monitoring system, rather than a pointing technique that is used by the IACTs. In the future, the \textbf{Southern Wide-field Gamma-ray Observatory (SWGO)} is planned to further increase the sensitivity of this technique \citep{swgo2019}.
\item The currently operating second generation of IACTs are  the telescopes \textbf{H.E.S.S.}, \textbf{MAGIC} and \textbf{VERITAS}. Since the first light of H.E.S.S.\ in 2002, followed by MAGIC and VERITAS first light (2003 and 2007), more than 150 gamma-ray sources at TeV energies have been identified.  The H.E.S.S.\ Galactic Plane Survey \citep{hess_gpsurvey2018} is one recent success that highlights the importance of the IACTs - 78 sources were found at TeV energies in this scan, 31 of these have firm identifications with PWN, SNRs, composite SNRs and gamma-ray binaries. 47 are still unidentified, but have possible associations with objects seen at other wavelengths.
\end{itemize}

An up-to-date summary of the TeV-sky is presented at the webpage \url{http://tevcat.uchicago.edu/} \citep{tevcatICRC}. Created from the same web-page, Fig.\ 
\ref{skymap_tevcat:fig} shows the localized Galactic sources detected so far, as well as the Fermi color-code sky map in the background. More than 40\% of the of the detected TeV 
sources are still classified as \textit{unidentified}, often because the spatial resolution of the instrument does not allow for a proper correlation with one counterpart at other 
wavelengths, as there is usually a larger number of candidates. Of those sources with identified counterparts, emission from pulsar wind nebulae  is the leading source class 
($\sim 26\%$), followed by shell SNRs ($\sim 11\%$), binary systems and molecular clouds in the vicinity of shell SNRs (each $\sim 7\%$). Gamma-ray emission is a clear sign of 
particle acceleration. However, as there are different processes that can contribute to the high-energy part of the electromagnetic continuum, i.e.\ $\pi^{0}$-decay (hadronic interactions) as well as Inverse Compton scattering and bremsstrahlung (leptonic signatures). Thus, the sources cannot simply be assumed to be hadronic accelerators. A detailed discussion 
of the multimessenger physics of these sources is presented in Section \ref{multimessenger_modeling:sec}. 

The most important observational result in recent years concerning gamma-ray emission in the Milky Way are the following:

\begin{itemize}
\item The three SNR-MC systems IC443, W44 and W51 have been discussed to show emission of hadronic origin
\citep{fermi_ic443,tobias_sugar2015}. With the Fermi satellite, it was possible to identify the characteristic cut-off of the pion spectrum at $70$~MeV. It should be noted, however, that the sensitivity of Fermi at these low energies is small and that it has been discussed that the turn-over is rather due to the kinematic break in the primary proton energy spectrum that arises when the population follows a power-law in momentum space \citep{strong_icrc2015,strong2018}. Even this scenario is of hadronic nature, so that these three cases show strong indications of a hadronic acceleration scenario. Still, the primary spectrum of these sources is extremely steep. At higher energies, their spectral index is in the range $\sim 3$ to $\sim 4$  and they might cut off below TeV energies. Thus, 
while these  indications give further hints toward SNRs being hadronic accelerators, it is not clear yet if this source class in general is capable of accelerating particles up to the knee.
\item The detection of a gamma-ray signal up to $\sim 100$~TeV in the Galactic Center region has been announced by the H.E.S.S.\ collaboration \citep{hess_gc_2016}. To 
create a signal up to 100 TeV in gamma-ray energy with purely leptonic processes is difficult. Thus, this is a serious hint for a hadronic population in the Galactic Center, 
reaching up to PeV energies in cosmic rays, providing first evidence for a so-called \textit{Galactic PeVatron}, i.e.\ a source/region that emits cosmic rays up to the knee. 
\item In the past decade, outflow signatures have been identified from the Galactic Center toward the halo, so in the direction perpendicular to the Galactic disk. The so-called WMAP haze was discovered first \citep{finkbeiner2004}, followed by the Fermi-bubbles \citep{finkbeiner_galactic_wind2010,dobler2010}. These two features, the first at microwave and the second at GeV energies, respectively, have been shown to have coinciding edges \citep{dobler2012}. A similar feature has been identified at radio wavelengths as well \citep{carretti2013,Heywood2019}. Different interpretations are the possibility of dark matter decays, star formation signatures or a jet-like structure emerging from the Galactic Center.
    \item The inner Galaxy has long shown an excess in the gamma-ray flux that has been discussed to possibly arise from a dark matter signature \citep[e.g.][]{Calore2015}. Today, thanks to Fermi data, it is known that the excess is smaller than originally indicated by EGRET and that there is the need for a gradient in both cosmic-ray source density and spectral index in order to explain the \textit{cosmic-ray gradient problem}.
    \item  Six prominent MILAGRO sources show emission up to 10s of TeV in the Cygnus region, also indicating a hadronic component
\citep{abdo_milagro2007,abdo_milagro2008}. As Cygnus is a star-forming region, there is a lot of non-thermal activity. It thus represents one of the most interesting 
astrophysical laboratories in the Galaxy where signatures of cosmic rays can be studied. This region is particularly interesting for IceCube, which can detect neutrinos from the 
northern hemisphere, where Cygnus is located, at relatively low (TeV) energies with good pointing.
\end{itemize}

All these recent findings are milestones in gamma-ray astronomy that can lead the way to (a) understanding the evolution of the Milky Way with all its components (gas, cosmic-rays, magnetic field, dark matter) in a better way and (b) identifying the 
sources of cosmic rays at different energies in the next decades. Successor observatories and satellites motivate their performance with today's knowledge of these sources and 
extended regions in the Galaxy. At the MeV energies, projects like \textit{e-ASTROGAM} \citep{eastrogam2017} and \textit{AMEGO}\footnote{All-sky Medium Energy 
Gamma-ray Observatory} \citep{amego2017} will make it possible to study the so-called \textit{pion bump}, revealing the rising flank of the hadronic gamma-ray spectrum 
from $\pi^{0}$-decays in more detail. This would be a crucial step for the identification of hadronic emitters, as it is the only truly unique signature for an unambiguous identification of 
a hadronic source using only gamma-rays.
At PeV energies, CTA\footnote{Cherenkov Telescope Array} will be able to enhance the performance of the current generation of IACTs in sensitivity and pointing accuracy, see 
e.g.\ \citep{rieger2013,funk2017}. This way, Galactic PeVatrons can be identified in large(r) numbers and with the possibility to spatially resolve the regions of emission, thus 
helping to produce a map of potential cosmic-ray sources that accelerate up to the knee and beyond.

\subsubsection{Neutrinos}
The idea of detecting high-energy neutrinos by deploying a detector in large natural water or ice reservoirs was brought up by \citet{markov1960}, see e.g.\ \citet{spiering2012} 
for an elaborate review on the history of high-energy neutrino astronomy. This first idea was to install detectors in water, like huge lakes or the sea, in order to be able to 
deploy large-scale detection arrays. It was clear early-on that the large background of atmospheric neutrinos would be one of the largest challenges for the endeavor. The first attempt of a detector in the pacific was realized by the DUMAND\footnote{Deep Underwater Muon and Neutrino Detection} experiment off 
the shore of Hawaii, which unfortunately only worked for a short time. 
\paragraph*{First generation: Proof of concept} Several detectors were build as proof of concept and in order to understand the atmospheric neutrino flux at TeV energies better. Those detectors, first showing that the method of high-energy neutrino detection in deep water and ice works, were the \textit{Baikal Deep Underwater Neutrino 
Telescope} in lake Baikal and the \textit{Antarctic Muon And Neutrino Detection Array-II (AMANDA-II)} in the deep ice 
at the South Pole \citep{amanda2001}. It was realized quickly that in particular the Antarctic ice could be instrumented with a kilometer-scale detector which would enable scientists to observe the astrophysical neutrino flux. These arrays were followed by detectors in the Mediterranean, 
in particular the ANTARES\footnote{Astronomy with a Neutrino Telescope and Abyss environmental RESearch} detector, currently representing the most sensitive 
instrument in the Northern Hemisphere \citep{antares2011}. Above all, this generation proved the feasibility of the detection concept in lake and sea water as well as in 
the Antarctic ice.

 Apart from measuring the atmospheric neutrino spectrum up to super-TeV energies, the results were used to exclude several source models that were predicted at the time 
\citep{stacking_interpret, becker_review2008}. But also, the detector arrays were extremely useful to quantify the optical properties of ice and water. It was proven that the 
measurement of high-energy neutrinos is possible with these arrays, with reasonable pointing information ($\lesssim 1^{\circ}$ depending on the energy and the medium) and energy 
resolution ($\Delta \log E \lesssim$ 0.3). The proof-of-concept of AMANDA in particular showed that the Antarctic ice was bubble-free below $\sim 1.500$~m below the surface and 
that the ice posseses absorption lengths long enough to have the strings separated by more than $100$~m \citep{optical_properties1995}.
 
\begin{figure}[htbp]
\centering{
\includegraphics[trim = 0mm 0mm 0mm 0mm, clip, width=0.8\textwidth]{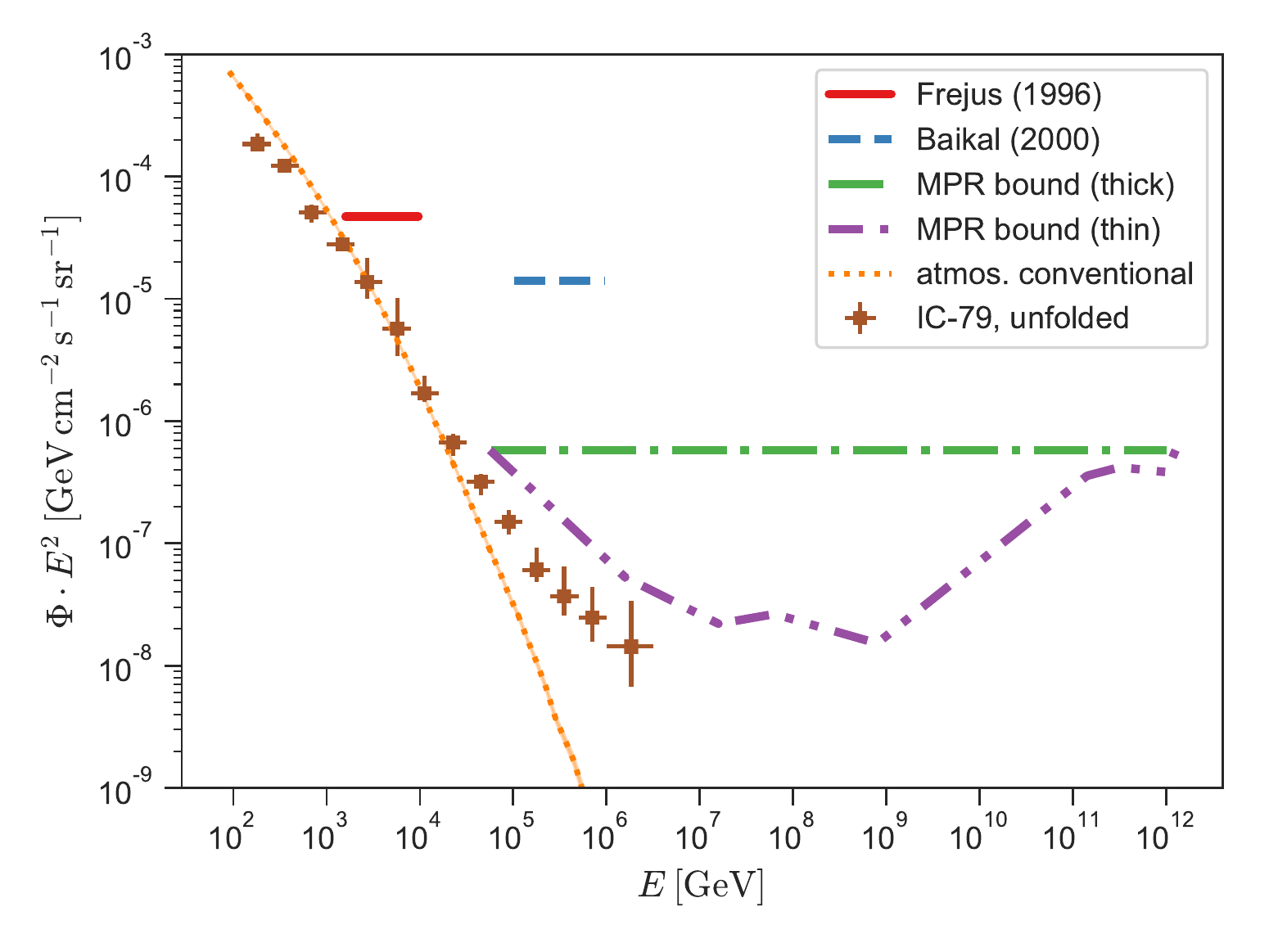}
\caption{Neutrino flux versus energy with limits from early observatories Fr{\'e}jus (red solid line) and Baikal (blue dashed line) together with the unfolded IceCube spectrum of the detector configuration IC-79 (filled quares), in which the deviation from the conventional atmospheric neutrino flux (\citep{fedynitch2012}, yellow dotted line) is visible for the first time. Astrophysical upper bounds for optically thin (green dot-dashed line) and optically thick (pink dot-dot dashed line) AGN as presented by \citet{mpr2001} are shown for comparison.
\label{spectrum_neutrinos:fig}}}
\end{figure}

 The great success of high-energy neutrino astronomy in the past two decades is demonstrated in Figure \ref{spectrum_neutrinos:fig}: first limits on the high-energy neutrino flux were 
published by Fr{\'e}jus  in 1996 \citep{frejus_limit} and a few years later by the Baikal collaboration \citep{baikal2000}. The sensitivity at that time was at a flux level of  
$\left.E_{\nu}^{2}\cdot dN_{\nu}/dE_{\nu}\right|_{\rm sens}\sim 10^{-4}$~GeV/(s sr cm$^2$). Today, the sensitivity is more than 4 orders of magnitude better and the energy range 
reaches PeV energies.  Enlarging the detection arrays significantly in combination with technical optimizations and elaborate analysis strategies in the recent years has led to 
the first detection of astrophysical high-energy neutrinos with the next generation telescope IceCube \cite{icecube2013,icecube2014}.

\paragraph*{Second generation: First detection}
 The proof-of-concept measurements led to the construction of IceCube in the years 2005 to 2010 --- the first cubic kilometer-sized array to detect high-energy neutrinos from 
outer space \citep{icecube2013,icecube2014}. In its final configuration, IceCube consists of  86 strings deployed in the Antarctic ice, on which 5,160 photomultiplier 
tubes (PMTs) are attached. These PMTs measure the Cherenkov light in the ice that is produced by the particles created in the interaction of a neutrino with a nucleon in the ice: 
via neutral current interactions, a hadronic cascade is released,  while charged current interactions lead to a hadronic cascade plus the signature from the charged lepton. In particular, 
muon-neutrino induced muons produce long track-like signatures in the ice that allow to reconstruct the original direction of the neutrino to better than $1^{\circ}$ resolution, dependent on the 
energy of the particle. Cascades can have resolutions up to $10^{\circ}$, derived from the temporal distribution of the spherical signal. The energy range covered by IceCube goes 
down to $\sim 10$~GeV  by using the dense DeepCore subarray. At these energies, searches for Dark Matter as well as measurements of neutrino oscillations are being performed. The 
highest energy events measured by IceCube are in the PeV range. At these highest energies, IceCube succeeded with the first measurement of the diffuse astrophysical neutrino flux 
and could show that it is at a level of $E_{\nu}^{2}\cdot dN_{\nu}/dE_{\nu}\sim 10^{-8}$~GeV/(s sr cm$^2$) \citep{icecube2013}. The first signal was detected in a high-energy 
starting track analysis, where all events that start in the detector were taken into account, unless they were coincident with an atmospheric signal entering the detector or 
the surface array IceTop simultaneously. 
This very restrictive selection leads to a small data sample. During 2 years of operation, 
 28 events could be identified this way, with an expected background of atmospheric muons and neutrinos of around $10$ \citep{icecube2013}. Particularly interesting were the two PeV events that 
became known as \textit{Ernie and Bert}. At these extreme energies, the expected atmospheric signal was close to zero. An astrophysical flux, on the other hand, is expected to be flatter than 
the atmospheric flux, thus more likely to produce PeV events. In addition, both arrival directions and flavor composition are inconsistent with the atmospheric background. All these arguments led 
to the first evidence of the existence of astrophysical neutrinos. Today, more than 8 years of data are available, bringing the detection to a level of  $\gg 5\sigma$.
 
 The energy spectrum deduced from this analysis was close to $dN_{\nu}/dE_{\nu}\sim E_{\nu}^{-2}$, compatible with what is expected from Fermi acceleration (see Section 
\ref{candidates:sec} for a detailed discussion of the expected spectral behavior).  The fact that astrophysical neutrinos with $>$PeV energies have been detected reveals a 
component of the cosmic-ray energy spectrum up to $\sim 100$~PeV. The detections by IceCube therefore probe the \textit{shin region} of the cosmic-ray energy spectrum, i.e.\ 
the region between the knee and the ankle.
 
As of today, 7.5 years of starting tracks have been analyzed. In total, 102 events have been detected, 60 of which lie above $60$~TeV energy \citep{kopper2018}. The signal has even 
been detected in several other data sets \citep{icecube_globalfit2015}. In particular, different analyses focussing on through-going muons (track-like events) could confirm a signal 
\citep{numu_signal2016,ic79_spectrum2017}. Figure \ref{spectrum_neutrinos:fig} shows the unfolded neutrino spectrum 
from atmospheric muon-neutrinos. While the results of the different analyses are compatible with each other, there are differences in the measured spectral index. As the 
sensitivity of the different analyses covers different energy ranges, this could point to a spectrum that is not a simple power-law, but has a more complex structure. There can be 
other reasons for this, like different astrophysical components or a contribution of the decay of heavy-quark-containing hadrons in the atmosphere, see e.g.\ \citet{halzen_klein2010} for a discussion. 
 
\begin{figure}[htbp]
\centering{
\includegraphics[trim = 0mm 0mm 0mm 0mm, clip, width=0.8\textwidth]{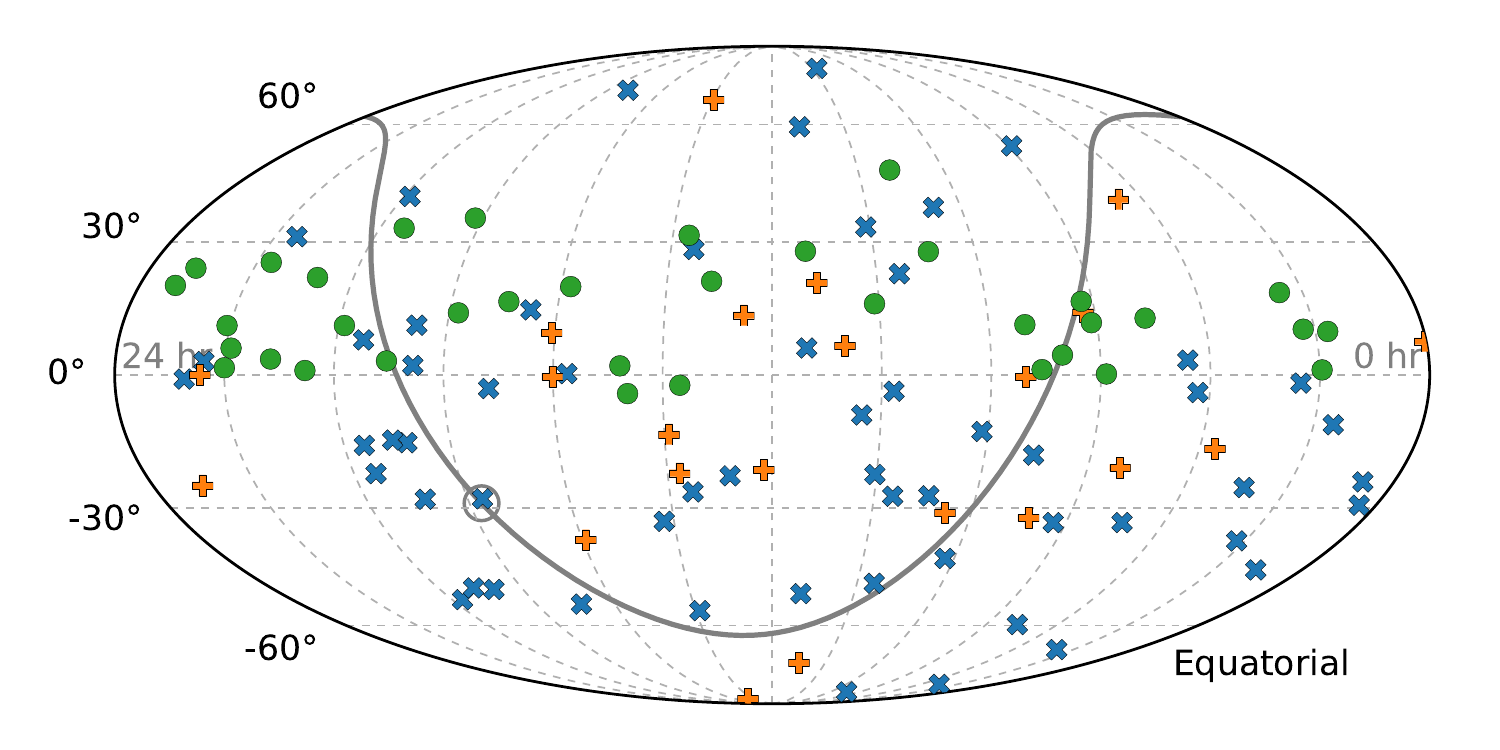}
\caption{
Skymap of high-energy neutrinos, with a large fraction being of astrophysical origin. Green circles represent throughgoing tracks with a deposited energy $>200$~TeV, blue crosses are cascades and orange pluses are starting tracks. The typical resolution for the track-like events is $\sim 1^{\circ}$, while cascade-like events have $\sim 10^{\circ} - 20^{\circ}$ \citep{icecube_cascades2017} resolution. Credits: Data come from the IceCube Collaboration. Figure kindly produced and provided by Jakob van Santen.
\label{skymap_neutrinos_jakob:fig}}
}
\end{figure}

For several years now, IceCube has established an alert strategy in which an event (or a temporal and spatial clustering of events) in the real-time analysis receives a probability of being of 
astrophysical nature. If certain criteria are fulfilled, an alert is sent out to the multiwavelength community. Most recently, an IceCube event 
was followed up this way by different instruments. This led to the coincident observation of a GeV-flare of the blazar TXS~0506+056, consistent with the direction of the neutrino event 
\citep{icecube_txs_fermi2018}. After this correlation at a level of $3\sigma$, the IceCube archival data of 9.5 years were analyzed in the direction of TXS~0506+056 \citep{icecube_txs2018}. While the
results were background compatible most of the time, there was an excess of events in the time period of March 2014 to September 2015, corresponding to a deviation from the 
background-only hypothesis of $3.5\,\sigma$. The spectrum is very flat, $\sim E^{-2.2\pm0.2}$, strengthening the scenario of an astrophysical source. Theoretical models now need to 
explain such a long-term access about 2 years before the possible gamma-ray flaring event and combine it with the coincidence detection. What all parts of the observation indicate is that a fraction of the events seem to originate from extragalactic sources, in this case potentially from a blazar. It should be noted, 
however, that the detections are at the $3-4\,\sigma$ level at this point and need to be confirmed by future measurements before final conclusions can be drawn.

 In this review, we will consider the possibility that there is some contribution from Galactic cosmic-ray sources to the detected astrophysical neutrino flux. Figure \ref{skymap_neutrinos_jakob:fig} shows the 
skymap from IceCube with those events that have  a high probability of being of astrophysical origin: High-energy cascades are shown as blue crosses, through-going tracks 
above $200$~TeV are represented by green circles, starting-tracks are displayed with orange plus-type symbols. What becomes clear from the figure is that there is neither significant 
clustering along the Galactic plane (black line), nor in the Galactic Center region (open circle with black boundary). This is a first experimental indication that the dominant 
part of the detected signal is of extragalactic nature. This fact is underlined by two arguments: (a) a dedicated combined IceCube-ANTARES search for a contribution from the 
Galactic plane result in limits that lie more than a factor of five below the detected signal.  \citep{galacticnus_icecube_antares2018}; (b) theoretical predictions that are quite general show that the expected neutrino flux from the Galaxy must be smaller than the detected signal, see e.g.\ \citet{mandelartz2015} and references therein.

\paragraph*{Third Generation: aiming for source identification}
While the first hints toward a point source have been published by IceCube now, it is clear that a full identification and interpretation of the entire diffuse flux is in need of a next generation of telescopes.

IceCube is now preparing for the new array \textbf{IceCube-Gen2} \citep{icecube_gen22014}, planned to be $\sim 10$ times larger in volume than IceCube. A larger spacing 
between strings is planned for the high-energy array to be able to perform the extension by a factor of $10$. In addition, a dense core is planned, dedicated to low-energy analysis to understand neutrino oscillations. This way, the sensitivity of IceCube can 
be enhanced by at least a factor 10, through the optimization of instruments and analysis methods in the upcoming years, most likely even more. The future array currently includes R\&D activities for a surface array detector as a combination of different techniques --- a scintillation array (IceScint) \cite{icescint_icrc2019}, an array of Cherenkov telescope stations (IceAct \cite{iceact_icrc2019}) and a radio array in combination will be able to serve as a calibration device as well as a veto detector for the in-ice events, and also as a cosmic-ray detector.

Concerning the in-ice array, a first extension, referred to as the \textbf{IceCube-Upgrade}, has 
recently been approved, implying that 7 more strings will be installed until 2023 \citep{kowalski2018}. 

In lake Baikal, the next generation telescope \textbf{Gigaton Volume Detector (Baikal-GVD)} is in preparation. Two first clusters of 8 strings have already been installed, with the plan to install $8-12$ of such cluster arrays, resulting in a km$^3$-scale array \citep{baikal_gvd2017}.

In the Mediterranean, \textbf{KM3NeT 2.0} will succeed ANTARES, located at two sites, off-shore of France and Italy. Similar to Baikal-GVD, KM3NeT will be organized in string clusters 
and the final array aims for the deployment of a super-km$^3$-sized array \citep{km3net2016}.

For the more distant future, there are more R\&D activities, for instance with the Pacific Ocean Neutrino Explorer (P-ONE) off shore of Canada \citep{holzapfel2019}  and the exploration of the possibility to detect the radio signals of neutrino-induced showers in Greenland and Antarctica \citep{nelles2017}.

With these future initiatives, the full $4\,\pi$ neutrino sky can be seen from different perspectives, minimizing systematic uncertainties and moving neutrino astronomy toward source physics.

\section{Cosmic ray fundamental properties \label{candidates:sec}}
Section \ref{data:sec} summarizes the wealth of cosmic-ray data that has emerged in the past decade(s). The different pieces of information are in need of interpretation and can be used to constrain the candidate sources. In this section, a basic overview will be given concerning the background theory needed to investigate multimessenger aspects of Galactic cosmic-ray physics and to present arguments that define which sources can be responsible for which energy range of the spectrum. This includes a thorough discussion of why Galactic sources are generally believed to produce the cosmic-ray spectrum up to the knee, why extragalactic sources should be the origin of the flux above the ankle, and why nobody knows (yet) where exactly the transition happens in the shin region of the cosmic-ray spectrum. 
In order to make the above arguments, six basic aspects derivable from cosmic-ray observations can be taken into account when suggesting a class or list of sources as the 
dominant contributer to the cosmic ray spectrum in a certain spectral range:
\begin{enumerate}
\item \textbf{All-particle spectral behavior:} The all-particle spectrum of cosmic rays follows a broken power-law which any potential dominant source (class) needs to fulfill. In 
particular, the detected spectrum reveals a spectral behavior $\Phi_{\rm CR}=A\cdot E_{\rm CR}^{-\gamma}$ in the spectral boundaries of $2.5<\alpha<3.1$, depending on the energy 
range as discussed in section \ref{allparticle:sec}. Any source and propagation model must 
reproduce such a behavior.

\item \textbf{Maximum energy:} The sources need to reproduce the relevant maximum energy: If we are looking for a source (class) that can reproduce the all-particle spectrum 
below the knee, this source or these sources need to be able to accelerate particles up to at least $10^{15}$~eV.

\item \textbf{Luminosity criterion:} Those sources that are dominantly responsible for producing the cosmic-ray energy spectrum in a certain spectral range need to be capable of 
producing the total luminosity of the cosmic-ray energy spectrum in that range.

\item \textbf{Composition:} The composition of the cosmic-ray spectrum is dominated by protons at the lowest energies, i.e.\ in the GeV-range. However, the fraction of nuclei 
contributing to the spectrum changes drastically with the energy scale as discussed in section \ref{sec:abundance}. Depending on the energy range of interest, this evolution of 
composition needs to be reproduced by the model.

\item \textbf{(An)isotropy:} The isotropy level of the cosmic-ray arrival direction is extremely high at all energies, i.e.\ anisotropies are revealed at the sub-percent-level up 
to PeV energies, only the recently discovered dipole structure at ultra-high energies is at the level of a few percent. Nevertheless, a clear dipole-anisotropy has been revealed 
during the past two decades, in particular in the TeV-range and upward. In addition, a small-scale anisotropy has been shown to be present in the TeV-PeV range at a level that is 
about one order of magnitude lower than the large-scale anisotropy. A model of cosmic-ray emitters certainly needs to explain the large-scale anisotropy and needs to be compatible 
with the small-scale features as well.

\item \textbf{Neutral secondaries:} Diffuse emission signatures from gamma-rays and neutrinos have been detected during the past decades. Every cosmic-ray emission model needs to 
be compatible with these data, i.e.\ needs to demonstrate that it does not overshoot the emission. The same is true concerning gamma-ray point or extended sources.
\end{enumerate}
In this section, the first five arguments will be presented in detail and first, basic conclusions about possible source candidates will be presented. The final argument of neutral (and also charged) secondaries will be reviewed 
in Section \ref{multimessenger_modeling:sec} on the multimessenger modeling of the non-thermal emission in the Galaxy.
\subsection{Background theory \label{spectrum_discussion:sec}}
As discussed in Section \ref{data:sec}, the cosmic-ray all-particle spectrum is quite well-described by a series of power-laws, that in total span over many orders of magnitudes. 
It is striking that the observed spectrum from $10^{10}$~eV to $10^{20}$~eV is within tight spectral boundaries, i.e.\ $2.5<\gamma<3.1$. In this section, we will focus on the 
discussion of how such a behavior can be reached if the sources of this acceleration lie within the Galaxy. Here, two effects are important: (1) the particles are accelerated, 
usually at local sources; (2) the particles need to propagate diffusively through the Galaxy. As will be shown in this section, both lead to a power-law type behavior that in 
combination should be responsible for the observed power-law spectrum at Earth.

The mathematical description of such spectra which are based on a large number of particles is usually not handled by a single-particle description, but rather by the investigation 
of the distribution of particles in phase-space --- one of the primary reasons is that even if the initial conditions of the plasma would be well-known and the coupled set of 
the equations of motion could be solved, the development is highly sensitive to small pertubations of the system. The most fundamental way of treating the particle distributions is 
via the Klimontovitch or Liouville equations, in which the plasma is assumed to consist of point-like, classical particles. While the Klimontovitch-approach assigns individual 
positions to the $N_0$ particles in one phase-space, the Liouville-approach chooses one phase space for each of the $N_0$ particles. In both approaches, the ensemble-average is 
calculated --- for the Klimontovitch-approach, this leads to the plasma-kinetic equation, equivalent to the Boltzmann equation. As the Liouville equation still contains all $N_0$ 
phase spaces, with a distribution function $f_{N_0}(\vec{x}_1\,,\vec{x}_2,\,\ldots \vec{x}_{N_0};\vec{v}_1,\,\vec{v}_2,\,\ldots \vec{v}_{N_0};t)$, it is further reduced by 
integrating out $N_0-k-1$ phase spaces ($1\leq k\leq N_0-1$), which results in a description of a distribution function 
\begin{equation}
f_k(\vec{x}_1,\,\vec{x}_2,\,\ldots \vec{x}_k;\vec{v}_1,\,\vec{v}_2,\,\ldots \vec{v}_k;t)=V^k\,\int \Pi_{i=k+1}^{N_0}d\vec{x}_i\,d\vec{v}_i f_{N_0}(\vec{x}_1,\,\vec{x}_2,\,\ldots \vec{x}_{N_0};\vec{v}_1,\,\vec{v}_2,\,\ldots \vec{v}_{N_0};t)
\end{equation} 
that delivers the exact, ensemble-averaged behavior of the first $k$ particles. 
Here, $V$ is the normalizing, finite volume, for which $f_{N_0}$ is non-zero for any $\vec{x_l}$ ($l=1,\,\ldots N_0$).
The resulting differential equation describing the behavior of $f_k$ in a $6\times k-$ dimensional phase space, is  called the \textit{BBGKY-hierarchy}\footnote{After \textbf{B}ogoliubov \citep{bogoliubov_bbgky}, \textbf{B}orn \& \textbf{G}reen \citep{borngreen_bbgky}, \textbf{K}irkwood \citep{kirkwood1_bbgky,kirkwood2_bbgky}, \textbf{Y}von \citep{yvon_bbgky}, see \citet{nicholson1983} for a discussion.}:
\begin{equation}
\frac{\partial f_{k}}{\partial t}+\sum_{i=1}^{k} \vec{v}_{i}\cdot\nabla_{\vec{x}_i}f_{k}+ \sum_{i=1}^{k} \sum_{j=1}^{k}\vec{a}_{ij}\cdot\nabla_{\vec{v}_i}f_{k}+\frac{N_0-k}{V}\sum_{i=1}^{k}\int d^3 x_{k+1}\,d^3v_{k+1}\vec{a}_{i,k+1}\cdot \nabla_{\vec{v}_i}f_{k+1}=0\,.
\end{equation}
Here, $a_{ij}$ is the acceleration produced by the force between the $i$th and the $j$th particle. The phase space $\left(\vec{x}_i,\,\vec{v}_i\right)$ is associated with the $i$th particle. Thus, the solution for $f_{k}$ is in recursive form, as the differential equation depends on the next higher order $f_{k+1}$. With this general description, arbitrary (classical) particle distributions can be described on a kinetic level. However, the level of complexity is still extremely high. Astrophysical problems are typically in a regime where particle interactions can be neglected. Thus, the case $k=1$ is investigated for identical particles, for which $f_2$ can be simplified. Finally, neglecting the correlation function that describes the interactions between the particles, the BBGKY-hierarchy can be reduced to the \textit{Vlasov equation}  for $f_1 = f$:
\begin{equation}
  \frac{\partial f}{\partial t}+\vec{v}\cdot \nabla_{\vec{x}}f+\frac{\vec{F}}{m}\cdot \nabla_{\vec{v}}f=0\,.
  \label{vlasov:equ}
\end{equation}
In the above equation, the indices have been omitted and the particle acceleration has been expressed in terms of the acting force, $\vec{F}/m=\vec{a}$. Including correlation effects\footnote{These effectively represent  wave-particle or particle-particle interactions} leads to the Boltzmann equation (equivalent to the plasma-kinetic equation received from the Klimontovitch-ansatz), where collision effects are summarized by a term $d/dt (f|_{\rm coll})$ on the right-hand side of the equation. Finding solutions for the Vlasov (or Boltzmann) equation is a typical problem in astrophysics.

It can be shown in the quasi-linear theory (QLT) of this equation that effective wave-particle scattering happens when the wavelength in Fourier space is a multiple of the 
cyclotron wavelength, i.e.\ the interaction of waves and particles happens at a resonant wave number $k_{\rm res} \sim (|\mu|\,r_{g})^{-1}$, with $\mu=\cos\theta$ and $\theta$ as the pitch angle (see \citet{berezinskii1990} for a summary). This resonance condition is used in 
particular for the interpretation of the diffusion tensor, see Section \ref{diffusion:sec}.

The Vlasov equation in general poses a coupled problem, in which the electromagnetic fields $\vec{E}$ and $\vec{B}$ responsible for the force 
$\vec{F}_L=q\,(\vec{E}+\vec{v}\times\vec{B})$, interact directly with the plasma distribution $f$ and the dynamics of the distribution $f$ in phase-space can modify the 
electromagnetic fields. Thus, there are typically two different approaches to solve the problem (see e.g.\ \citet{schlickeiser2002} for a detailed review):
\begin{itemize}
\item \textit{The test wave approach} --- here, the plasma particle distribution $f$ is predefined and the electromagnetic field with all resonant waves can be calculated.
\item \textit{The test particle approach} --- here, the fields are predefined and the particle distribution $f$ can be determined.
\end{itemize}
In this review, the focus lies on the second approach and discusses the non-linear behavior of systems where appropriate.

Assuming a stationary magnetic field, it can be shown that the plasma-kinetic equation can be expressed 
in the form of a Fokker-Planck equation. In its general form, it reads
\begin{equation}
\frac{\partial f(\vec{r},\vec{p},t)}{\partial t}=-\sum_{i=1}^{6}\frac{\partial}{\partial y_i}\left(A_i(\vec{r},\vec{p})\cdot
f(\vec{r},\vec{p},t)\right)+\frac{1}{2}\cdot\sum_{i,j=1}^{6}\frac{\partial^2}{\partial y_i\partial y_j}\left(B_{ij}(\vec{r},\vec{p})\cdot f(\vec{r},\vec{p},t)\right) \,.
\label{f:equ}
\end{equation}
Here, $y_i$ are the six components of phase space, $\vec{y}=\left(\vec{r}, \vec{p}\right)$, $A_i$ and $B_{ij}$ are one and two dimensional tensors, respectively, determined by the 
physics process (drift for $A_i$ and diffusion for $B_{ij}$). Note that the Fokker-Planck equation via It\^o's calculus can be rewritten to a stochastic differential equation 
(SDE) and thus be treated as a stochastic problem (see e.g.\ \citet{gardiner2009} for a review). This fact will be discussed in more detail in Section \ref{transportmodels:sec}.

For the specific case of the transport of Galactic cosmic rays of species $l$ (e$^-$, p, He, $\ldots$, Fe), the differential equation \ref{f:equ} is usually formulated in 
terms of their differential intensity in phase space,
\begin{equation}
  n_{l}=n_{l}(\vec{r},\vec{p},\,t)=p^2\cdot f(\vec{r},\,\vec{p},\,t)\,,
\end{equation}
with the relevant terms
\begin{eqnarray}
\frac{\partial n_l}{\partial t}&=&
\sum_{j=1}^{3}\frac{\partial}{\partial x^{j}}\left[\left(D_{jk}\cdot \frac{\partial}{\partial x^{k}}\right)\,n_l\right]
-\sum_{j=1}^{3}\frac{\partial}{\partial x^{j}}\left[u^{j}\cdot \,n_l\right]
+\frac{\partial}{\partial p}\left[p^2\,D_{pp}\,\frac{\partial}{\partial p}\left(\frac{n_l}{p^2}\right)\right]\nonumber\\
&-&\frac{\partial}{\partial p}\left[\dot{p}\,n_{l}-\frac{p}{3}\left(\nabla\cdot \vec{u}\right)\,n_l\right]+\sum_{j>l}\frac{v_l}{c}\,n_{0}\,\int dp'\,\sigma_{j\rightarrow l}(p,p')\,n_{j}(p')\nonumber\\
&-&\frac{n_l}{\tau}+Q_{l}(p)\,.
\label{transport:equ}
\end{eqnarray}
To derive this equation isotropy in the momentum-phase-space is assumed. This allows to average over the components $p_\phi$ and $p_{\mu=\cos(\theta)}$, meaning the particle 
density is only sensitive to the absolute value of momentum $|\vec{p}|$ or energy.

The terms entering the equation describe the following physical phenomena:

\begin{itemize}
\item \textit{Spatial diffusion} is represented by the first term:
\begin{equation}
\sum_{j=1}^{3}\frac{\partial}{\partial x^{j}}\left[\left(D_{jk}\cdot \frac{\partial}{\partial x^{k}}\right)n_l\right]
\end{equation}
with $D_{jk}$ as the diffusion tensor, determined by the properties of the magnetic field. More details on the connection of the effective scattering rate and the diffusion tensor 
is given in Section \ref{transport:sec}. The tensor is often approximated by a space-independent scalar\footnote{Often even with the preferred direction of being aligned with the 
magnetic field, $D=D_{\parallel}$.} ($D_{jk}\approx D\,\delta_{jk}$):
\begin{equation}
\sum_{j=1}^{3}\frac{\partial}{\partial x^{j}}\left[\left(D_{jk}\cdot \frac{\partial}{\partial x^{k}}\right)n_l\right]\approx D\cdot \Delta n_l\,.
\label{diffusion:equ}
\end{equation}
It is known from observations that the scalar diffusion coefficient of the Galaxy is energy dependent with $D\propto E^{\kappa}$, with $\kappa\approx 1/3-2/3$ \citep{ams2015a,ams2015b,ams2016}. Most 
recent measurements from PAMELA and AMS-02 point to a behavior rather close to $E^{0.3}$. The theoretical discussion of this measurement and its interpretation is more complex. For instance, a range of $0.3<\kappa<0.5$ is allowed when taking into account uncertainties related to the diffusion height, convection and diffusive reacceleration \citep{trotta2011}. Thus, the question of how these measurements are to be interpreted is far from being closed and is discussed in detail in Section 
\ref{transport:sec}.
\item \textit{Momentum diffusion} is described by the term
\begin{equation}
+\frac{\partial}{\partial p}\left[p^2\,D_{pp}\,\frac{\partial}{\partial p}\left(\frac{n_l}{p^2}\right)\right]\quad .
\end{equation}
It describes the diffusive change in momentum, i.e.\ the change in momentum $\Delta p$ within a time interval $\Delta t$ does not depend on the development of the phase-space particle trajectory, but rather on the instantaneous values of $x$ and $p$, thus describing a Markov process \citep[e.g.][]{BLANDFORD19871,OSTROWSKI1997271}. It is connected to the spatial diffusion coefficient via
\begin{equation}
    D_{pp}(p)=\frac{4\,v_A\,p^2}{3\,\alpha(4-\alpha^2)(4-\alpha)D(p)}\,.
    \label{dpp:equ}
\end{equation}
This diffusive process is equivalent to the second order Fermi process as described in Section \ref{spectrum_acceleration:sec}, with more details on the connected time scales there. 
\item \textit{Spatial advection}  is described via the divergence of the flow velocity of the background plasma $\vec{u}$
\begin{equation}
-\sum_{j=1}^{3}\frac{\partial}{\partial x^{j}}\left[u^{j}\,n_l\right]\,.
\end{equation}
In a global, galactic view, this is often a galactic wind that is oriented perpendicular to the disk, as observed in many external galaxies \citep[][e.g.]{dettmar2003,heesen2009}. Such winds can, however, even exist in more local environments, for instance driven by the local star formation activity of a starforming region. Examples here are starforming regions in the Milky Way (e.g.\ the Cygnus Cocoon or the Carina Nebula) or the Galactic Center region.
\item \emph{Momentum advection} can in detail be described by the following term:
\begin{equation}
-\frac{\partial}{\partial p }\left[p^2\,\left\{\frac{p}{3}(\nabla\cdot \vec{u})\,n_l\right\}\right]
\end{equation}
where the divergence in the flow velocity leads to adiabatic expansion effects, causing losses or gains when the plasma is compressed.
\item \textit{Continuous energy loss or gain processes} are represented by the term 
\begin{equation}
-\frac{\partial}{\partial p}(\dot{p}\,n_l) 
  \end{equation}
  in Equ.\ (\ref{transport:equ}).
Here, the sign of $\dot{p}$ determines if energy is gained (positive sign) or lost (negative sign). A typical example of an energy gain process is particle acceleration at a shock front, as discussed in Section \ref{spectrum_acceleration:sec}. Loss processes are manifold in the Galactic environment. Here, the dominant processes for electrons and hadrons are very different from each other.
Electrons lose energy via the acceleration process itself (synchrotron radiation) or via interactions with photon fields (Inverse Compton scattering) and with the ambient gas 
(bremsstrahlung). A significant part of the energy of the relativistic electrons can be lost, so that these processes are usually highly relevant to take into account in a 
transport equation. Photohadronic interactions are relevant for hadrons with energies above $\sim 280$~MeV, while ionization (hadron-hadron) and Coulomb-collisions is relevant 
below GeV-energies. 
Even adiabatic loss 
processes can be formulated via a loss term $\dot{p}$, but can also be treated more explicitly as discussed below.

\item \emph{Catastrophic energy loss (or gain)}, in which the original particle is lost (or a particle is gained) in the transport equation of consideration, is mostly relevant 
for hadrons. It can be described by terms $n/\tau$ with $\tau$ as the characteristic loss time scale $\tau$. In particular, hadron-hadron interactions above $1$~GeV lead to the 
production of neutrinos, gamma-rays, electrons and positrons.  In a galactic environment, the optical depth of such processes is usually $\tau\ll 1$. Thus, the particle distribution function is often 
(close-to) unchanged by these interactions. However, the produced neutrinos and photons are detected and their energy spectra and spatial distributions are in need of 
interpretation. They can actually help us to understand cosmic-ray acceleration and transport in more detail. The loss of particles due to interactions with the gas of density 
$n_{0}$ can be included in the transport equation, represented by a catastrophic loss term, 
\begin{equation}
-
\frac{n_l}{\tau}\,.
\label{catastrophic:equ}
\end{equation}
For interactions with the gas, $\tau^{-1}=v_l/c\cdot n_{0}\cdot \sigma(p)$,
with $\sigma$ as the relevant cross-section and $v_l$ as the cosmic-ray velocity. Other possible loss terms are for instance through radioactive decay or spallation:
Spallation of nuclei by fast, light cosmic rays leads to a change in the abundances of cosmic rays with respect to the solar element abundances and is included as a gain term in the equation. Medium heavy elements like C, N and O undergo spallation and increase the number of light elements (Li, Be, B) significantly. This process couples the transport equations of light and medium heavy elements to each other. Thus, in addition to a catastrophic loss term, a gain term is included in Equ.\ (\ref{transport:equ}):
\begin{equation}
+\sum_{j>l}\frac{v_l}{c}\,n_{0}\,\int dp'\,\sigma_{j\rightarrow l}(p,p')\,n_{j}(p')\,.
\label{spallation:equ}
\end{equation}
This term describes the increase of the particle density for the particle species $l$  with time   due to the 
spallation of a nucleus of species $j$ with momentum $p'$ and mass number $A_j>A_l$. The cross section of the processes is given as $\sigma_{j\rightarrow l}(p,p')$.

\item \emph{The sources and sinks} of the systems are described by:
\begin{align}
+ Q_l(p,\vec{r},t) \quad .
\end{align}
In the case of Galactic propagation, the concept of cosmic-ray acceleration is not 
treated explicitly. Rather, it is assumed that a source population accelerates the particles in a first step, leading to the source spectrum $Q_l(p,\vec{r},t)$, and the transport 
through the Galaxy is performed separately via the solution of the transport equation (\ref{transport:equ}).
\end{itemize}

The most simple transport model is a leaky box.
   Cosmic rays are assumed to be accelerated to a power-law spectrum in local sources within the Milky Way, $Q(p)$ and diffuse within the Galaxy with a (diffusion-driven) escape time scale $\tau_{\rm esc}$. 
Here, the diffusion term is approximated through the escape loss term:
\begin{equation}
\sum_{j=1}^{3}\frac{\partial}{\partial x^{j}}\left(D_{jk}\cdot \frac{\partial}{\partial x^{k}}\right)n_l \approx -\frac{n_l}{\tau_{\rm esc}}
\label{diffusion_leaky_box:equ} \quad .
\end{equation}
This transport model can already give a lot of insights into Galactic transport in a back-of-the-envelope calculation.


Assuming a steady-state system ($\partial n/\partial t\simeq 0$) and neglecting all other terms, Equ.\ (\ref{transport:equ}) reduces to a regular equation connecting the 
source spectrum and the propagated spectrum:
\begin{equation}
n_l(p)=\tau_{\rm esc}(p)\cdot Q(p)\,.
\end{equation}
The escape time can be estimated by approximating the diffusion tensor as a scalar (see Equ.\ (\ref{diffusion:equ})), and using the height of the halo of the Milky Way as the scale 
height $H$ of the escape, the left-hand-side term of Equ.\ (\ref{diffusion_leaky_box:equ}) can be reduced to $D \Delta n_l \approx D n_l/H^2$ gives a direct correlation between 
the diffusion coefficient and the escape time,
\begin{equation}
  \tau_{\rm esc}=H^2/D(p)\,.
  \label{escape_time:equ}
\end{equation}

The combination of observational data and theoretical arguments indicates a power-law behavior for both factors: the source spectrum is expected to be power-law like as discussed in Section \ref{acc:sec}, $Q(p)\propto p^{-\alpha}$ ($\alpha\sim 2.0 - 2.3$). The escape time is basically inversely proportional to the diffusion coefficient, which behaves as $\tau_{\rm esc}\propto D^{-1}\propto p^{-\kappa}$ ($\kappa \sim 0.3 - 0.6$), as discussed in Section \ref{transport:sec}. Thus, in this naive discussion of the complex problem, one would expect a power-law behavior of the cosmic-ray spectrum at Earth as:
\begin{equation}
n_l(p)=\tau_{\rm esc}(p)\cdot Q(p)\propto p^{-\kappa-\alpha}\propto p^{-\gamma}\,,
\end{equation}
with $2.3<\gamma<2.9$ when assuming the parameter range for $\alpha$ and $\kappa$ as discussed above. This naive calculation interestingly gives a range that fits the one for the 
observed spectral behavior at Earth well. A general discussion of these two parts of the spectral behavior follows in this section, presenting the basic physics arguments for the 
energy (rigidity) dependence of the acceleration spectrum and the escape time. The more complex description of acceleration and trasport can only be done in the framework of numerical codes taking into account the most relevant effects in realistic propagation environments as it is done in the different software codes GalProp \citep{strong_propagation_1998}, Dragon \citep{dragon2013}, CRPropa \citep{crpropa30,merten2017b}, USINE \citep{2018arXiv180702968M}, PICARD \citep{kissmann2014} and others. But in particular when considering the typically chosen approach of using the transport equation, a careful evaluation of the input parameters in form of the diffusion tensor and other plasma-related parameters is necessary. Thus, in Section  \ref{transport:sec}, the current state of the art of the observations and their 
theoretical interpretation is reviewed.

\subsection{Cosmic-ray acceleration\label{spectrum_acceleration:sec}}

The Lorentz force describes how charged particles can be accelerated in electromagnetic fields, i.e.
\begin{equation}
\dot{\vec{p}}=\vec{F}_L=q\cdot \left(\vec{E}+\vec{v}\times \vec{B}\right)\,.
\label{lorentz:equ}
\end{equation} 
The fields can be expressed as the sum of a regular ($\vec{E}_0,\,\vec{B}_0$) and inhomogeneous ($\delta\vec{E},\,\delta\vec{B}$) part of the field, $\vec{E}=\vec{E_0}+\delta \vec{E}$ and $\vec{B}=\vec{B_0}+\delta\vec{B}$ with $\langle\vec{E}\rangle=\vec{E}_0$ and $\langle\vec{B}\rangle=\vec{B}_0$ as the ensemble-averaged properties. As known from basic electrodynamics, effective acceleration of the charged particle can only be achieved with average non-vanishing large-scale regular electric fields, $\left<E\right>\neq 0$, as a change in kinetic energy requires
\begin{equation}
\frac{dE_{\rm kin}}{dt}=\frac{d}{dt}\left[E_{\rm tot}-mc^2\right]=\frac{d}{dt}\left[\sqrt{p^2+m^2\,c^4}\right]=\frac{\vec{p}\cdot\dot{\vec{p}}}{E_{\rm tot}}\stackrel{(\ref{lorentz:equ})}{=}q\cdot\frac{\vec{p}\cdot\vec{E}}{E_{\rm tot}}
\end{equation}
with the total particle energy $E_{\rm tot}=\gamma\,m\,c^2$. So, for an ensemble-averaged acceleration process, an average non-vanishing electric field is needed. This can, 
however, not be sustained in large-scale magnetic fields in most astrophysical environments: such a field immediately leads to drift currents in the plasma, which counterbalance 
the field and reduce it to zero. Individual exceptions, like large-scale electric fields in pulsars, are expected to exist, but in general, the linear acceleration of particles by 
an ordered, large-scale electric field needs rather extreme conditions. Locally, however, such fields can easily be produced by inhomogeneities in the magnetic field, according to 
Maxwell's equations\footnote{$\nabla\times (\vec{B}) = \mu_0(\vec{J}+\epsilon_0\cdot \partial (\vec{E})/\partial t)$ and $\partial (\vec{ B})/\partial t = 
-\nabla\times (\vec{ E})$}. Here, while the average $\langle\vec{E}\rangle$ is still vanishing, non-linear terms like $\left<\vec{E}^2\right>=\left<\delta \vec{E}^2\right>$, however, 
take over instead, leading to the potential of a net-acceleration effect. Defining the rigidity of a particle as its momentum $\vec{p}$ divided by its charge, $\vec{R}:= 
\vec{p}/q$, the Lorentz force can be expressed in terms of rigidity, with no explicit dependence on the charge, $\dot{\vec{R}}=\left(\vec{E}+\vec{v}\times \vec{B}\right)$. 
Multiplying the equation by $\vec{R}$ and neglecting magnetic fields for the acceleration, a differential equation of the rigidity is received, see e.g.\ \citet{sigl2017} and 
references therein:
\begin{equation}
\frac{d}{dt}(R^2)=2\vec{E}\cdot \vec{R}\,.
\end{equation}
All acceleration processes are therefore expected to result in a dependence on rigidity rather than energy. This is true in particular concerning the power-law behavior $R^{\alpha}$  and the expected maximum energy, $E_{\max}=E_{\max}(R)$. This section will start with a general discussion of the rigidity dependence of the energy spectra of electrons and hadrons and the consequences for the relative abundance of the species. It will further focus on the discussion of how particles can be accelerated stochastically from a (quasi-)thermal distribution to a power-law spectrum. The discussion of the diffusive transport through the interstellar medium after acceleration is done in the following section. The question of the maximum energy is picked up in Section \ref{hillas:sec}.

In astrophysical environments, there are four different types of acceleration processes:
\begin{enumerate}
\item \textbf{Second order Fermi acceleration} via the stochastic scattering of particles off magnetic turbulence \citep{fermi1949}. As this process is of second order in velocity, i.e.\ with an acceleration efficiency of $\xi \propto (v_{\rm CR}/c)^2$, it is usually believed to be subdominant. However, in environments that lack other acceleration processes, or where the particles have been pre-accelerated to $v_{\rm CR}\sim c$, it can play an important role. In the Galactic context, this process can be relevant for the reacceleration of particles in the ISM, where magnetic turbulence works as scattering centers for the particles.
\item \textbf{First order Fermi acceleration} at (magneto-)hydrodynamic shock fronts. As this process is of first order in velocity, $\xi\propto (v_{\rm CR}/c)$, it is more efficient as compared to Fermi second order. SNR shock fronts are good acceleration site candidates in this diffusive shock acceleration scenario that has been developed further from Fermi's first idea \citep{fermi1954} in the 1970s \citep{krymskii1977,axford1977,bell1978a,bell1978b} and in the 1980s \citep{schlickeiser1989}. The scenario that particles are accelerated at SNR shock fronts is the one that has been discussed as most promising. It is still under debate how this can lead to the acceleration of particles up to the knee and possibly beyond, but is still the one that works most efficiently for any Galactic acceleration scenario.  Together with Fermi second order acceleration, Fermi first order (today referred to as \textit{Diffusive Shock Acceleration, DSA}) will be discussed in more detail below.
\item \textbf{Linear acceleration} in large-scale electric fields is a straight-forward mechanism for particle acceleration. Such a scenario is, however, difficult to find in an astrophysical environment, as large-scale electric fields are typically not maintained for a long time. Particles will typically drift in the electric force and remove the electric potential to re-establish charge equilibrium. Only extreme environments might be able to maintain large-scale electric fields, as it is for instance discussed for pulsars.
\item \textbf{Magnetic reconnection events} have the potential to accelerate particles out of thermal equilibrium. However, the energies that are expected here are typically far below knee energies \citep{lyutikov-komissarov-sironi-porth-2018}. In particular relativistic reconnection could still be of high importance for cosmic-ray acceleration to extreme energies \citep{sironi-spitkovsky-2014}, as a pre-acceleration in reconnection events could solve the injection problem that exists in first order Fermi acceleration.
  \end{enumerate}
In this section, we will first discuss the question of charge equilibrium in the astrophysical plasma and the resulting relation between electron and hadron energy spectra (Section \ref{rigidity:sec}), followed by a short review of stochastical acceleration via second and first order Fermi acceleration (Section \ref{acc:sec}).
\subsubsection{Relation of electron and hadron spectra \label{rigidity:sec}}
The fact that the particle spectrum is expected to follow a power-law in momentum, $dN/dP =\Phi(p) \propto p^{-\alpha}$, has consequences for the observed particle spectra that are typically 
expressed in energy rather than momentum. In the high-energy limit, $E\gg mc^2$, and with $p\approx E/c$, we can express $dN/dE=\Phi(E)\approx1/c\cdot \Phi(p)\propto E^{-\alpha}$ as a power-law in energy.
However, the fact that the acceleration results in a \emph{power-law in momentum}  rather than in energy is highly important when looking at the ratio of electrons to protons: a 
plasma is a neutral fluid. When idealizing the situation to an electron-proton plasma, the total number of electrons and protons must be the same, i.e.\ $N_e = N_p$ 
with $N_i = \int dp\, \Phi_{i}(p)$ ($i=e,\,p$). As the lower integration boundary is below the rest mass of the particles, it is of high importance to take into account the mass in 
the energy equation, i.e.\ $E^2-p^2\cdot c^2=m^2\cdot c^4$. In particular, the conservation of total particle numbers should also hold for the non-thermal component of the plasma: 
the acceleration process of particles that have the same amount of charge, $|q_p|=|q_e|=e$ should result in the same total number of accelerated particles when being fed from a 
neutral plasma. In addition, the acceleration process should result in the same spectral index, independent on the sign of the charge, i.e.\ $\alpha_{\rm acc}:=\alpha_{\rm 
acc,e}=\alpha_{\rm acc,p}$. With these basic assumptions and in the relativistic limit ($E_i\gg m_i\cdot c^2$, $i=e,\,p$), the ratio of electron to proton energy spectra is 
given as (see \citet{schlickeiser2002}):
\begin{equation}
f_{ep}:=\frac{\Phi_{e}}{\Phi_{p}}\approx \left(\frac{m_e}{m_p}\right)^{-\frac{\alpha_{\rm acc}-1}{2}}\,.
\end{equation}
For a spectral index after acceleration of $\alpha_{\rm acc}=2.2$, the ratio becomes $f_{ep}\approx 0.01$, which is comparable to the observed ratio of electrons to protons at Earth. 
Subtleties do exist as discussed in detail by \citet{persic2014,merten2017a}: (1) it is not clear that electrons and protons have the same minimum energy to enter the acceleration 
process: for the acceleration at a shock front, this could be a function of the gyroradius and thus depend on the mass of the particle. In that case, $N_e\neq N_p$. (2) The 
spectral indices for the electron and proton spectra could be different. Such a difference has severe consequences for the comparison of numbers from astrophysical data: the usual 
ratio of electrons to protons is given in the differential form, i.e.\ $f_{\rm ep}=\Phi_e/\Phi_p\propto (dN_{e}/dp)/(dN_{p}/dp)$. If the spectral indices are not the same, 
however, this ratio evolves with energy and the luminosity ratio needs to be considered for a comparable measure, as it is often done in astrophysics. For \textit{equal indices}, these 
ratios \textit{are} the same. But if the spectral \textbf{indices are different}, the luminosity ratios are typically significantly higher than the differential ones. Thus, instead of expecting a value of 0.01, 
one would rather expect 0.1 --- something that is actually more compatible with what is expected from the observation of extragalactic sources. For a more detailed discussion of 
these subtleties, see \citet{merten2017a}.

\subsubsection{Stochastic acceleration\label{acc:sec}}
\begin{figure}[htbp]
\centering{
\includegraphics[trim = 0mm 0mm 0mm 0mm, clip, width=0.8\textwidth]{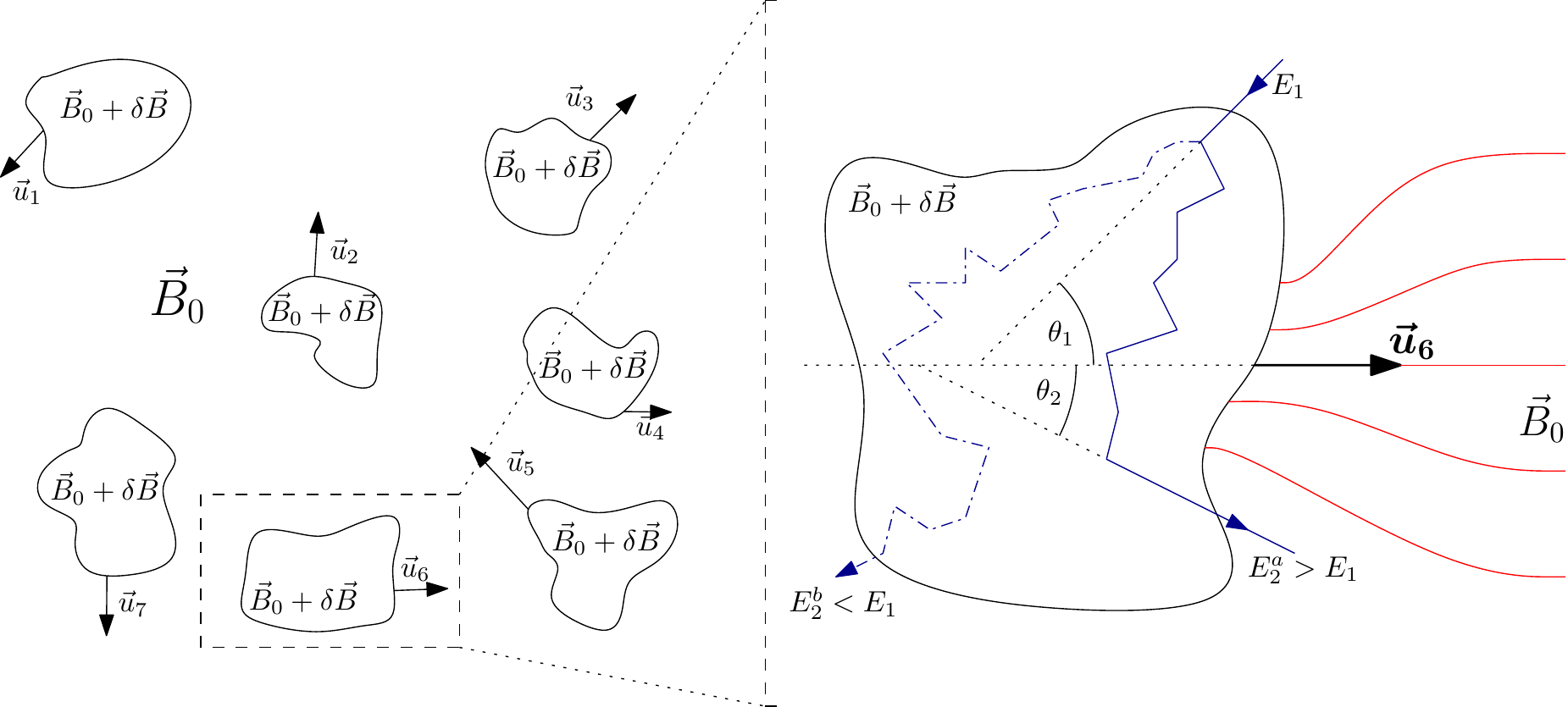}
\caption{Schematic view of a 2nd order Fermi process (sometimes referred to as \textit{Fermi-I}, as this is the original idea of stochastic acceleration)}
\label{fermi1:fig}}
\end{figure}

\paragraph{Second order Fermi acceleration}
The process of stochastic particle acceleration in astrophysical environments has first been discussed in \citet{fermi1949,darwin1949}. What these authors realized is that, on average, particles can gain energy through the encounter with magnetized plasma clouds that move randomly with respect to the interstellar medium (ISM).  A schematic view of the scenario is presented in Fig.\ \ref{fermi1:fig}. 
\begin{figure}[htbp]
\centering{
\includegraphics[trim = 0mm 0mm 0mm 0mm, clip, width=0.7\textwidth]{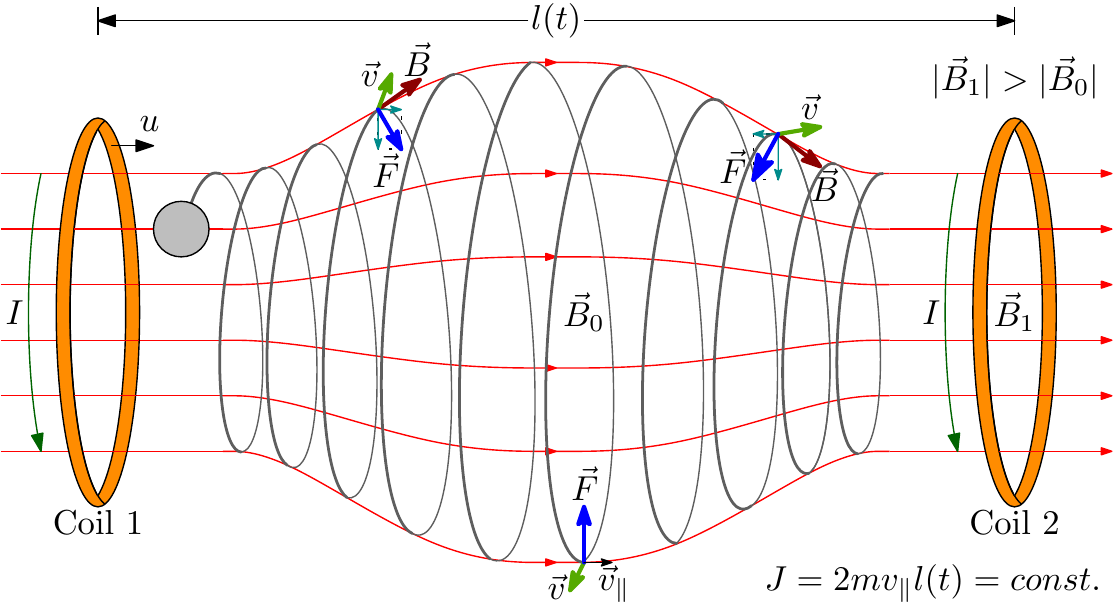}
\caption{Concept of a magnetic mirror, illustrated at the example of two coils that build up a magnetic mirror in between them via the two flowing currents $I$. The distance between the coils $l$ can be altered by moving coil 1 with a velocity $u$ toward or away from coil 2. The adiabatic invariant $J=2\,m\,v_{\parallel}\,l$ stays constant during that process, so that the parallel component of the particle trapped in the mirror needs to change in order to keep $J$ constant, i.e.\ $v_{\parallel}=v_{\parallel}(t)$.
\label{magnetic_mirror:fig}}}
\end{figure}

\begin{figure}[htbp]
\centering{
\includegraphics[trim = 0mm 0mm 0mm 0mm, clip, width=0.9\textwidth]{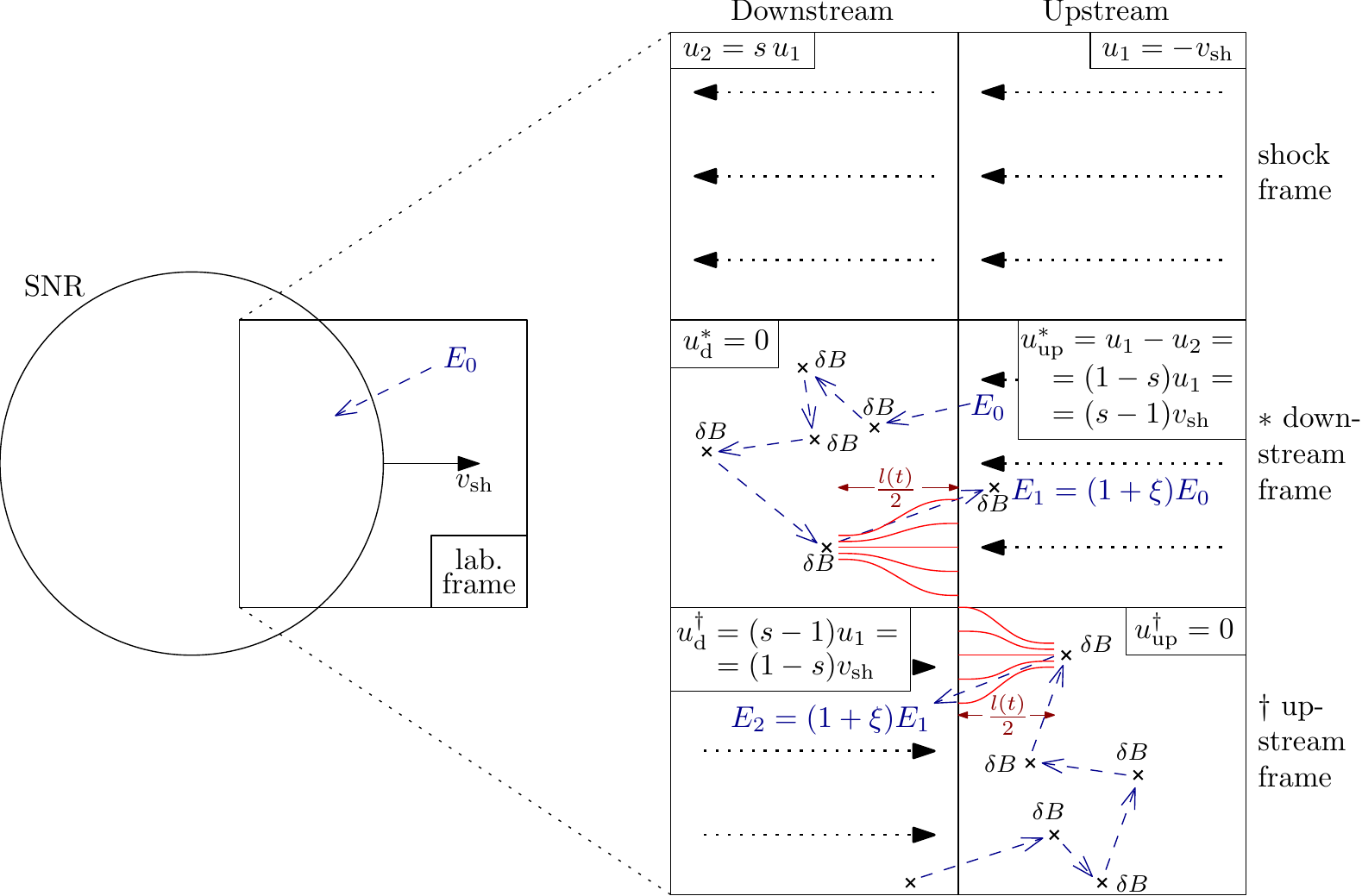}
\caption{Concept of Fermi acceleration at a shock front. The laboratory frame is displayed to the left. The uppermost right panel shows the situation in the frame of the shock front (properties labeled with $^{*}$), with $v_{\rm sh}^{*}=0$.  }
\label{fermi2:fig}}
\end{figure}

The idea is that a particle entering a magnetized cloud scatters isotropically with the magnetic inhomogeneities  in the frame of the cloud. 

The concept of this elastic scattering is explained by the magnetic mirror effect, in which a particle is reflected in a system of strong magnetic field gradients. Figure 
\ref{magnetic_mirror:fig} shows the concept of a magnetic mirror. The concept is explained in detail within plasma physics, see e.g.\ \citet{nicholson1983} for a fundamental 
discussion. In short, a magnetic mirror is contructed by a magnetic bottle, for which the homogeneous magnetic field in the central part $|\vec{B_0}|$ is smaller compared to the 
magnetic field strength at the bottle neck $|\vec{B_0}|<|\vec{B_1}|$  (see Fig.\ \ref{magnetic_mirror:fig}). In a configuration like that particles are trapped when their pitch angle $\theta_0$ at the center obeys 
$\sin^2(\theta) \geq B_0/B_1$. For any magnetic mirror, adiabatic invariants exist that remain unchanged even if the field itself is slowly 
varying, based in the adiabatic invariant of the action $J =\oint p\,dq$. Here, $p$ and $q$ are generalized momentum and coordinate, respectively. Here, the adiabatic invariant is 
applied to the periodic orbit of the particle in between the mirror characterized by the separation $l(t)$ along the magnetic field between the mirror points, $\vec{B}_0$. Hence, 
the generalized momentum and coordinate are given by the part of the momentum/coordinate parallel to $B_0$, i.e.\ $p=p_{\parallel}=m\cdot v_{\parallel}$ and $q=q_{\parallel}=2\cdot 
l$ for the entire closed orbit. Thus, the invariant action becomes $J= 2\cdot m\int_{0}^{l}v_{\parallel}\,dq_{\parallel}=2\cdot m\cdot v_{\parallel}\cdot l$.  In this context, it is 
important to note that $J$ is invariant even under slow changes of the magnetic field. This is the case here, as the cloud $i$ is moving with a velocity $\vec{u}_i$ (see Fig.\ \ref{fermi1:fig}) in the 
laboratory frame. Thus, the linear size of the mirror $l$ is changing with time. In order to keep the adiabatic invariant constant, $J=2m\,v_{\parallel}\,l(t)=const$, the velocity 
of the particle has to be adjusted accordingly.

Now, acceleration of particles in a system of moving clouds happens as follows within this concept: In the cloud frame, the magnetic field inhomogeneities move in random motion, so that the net-acceleration effect by the mirrors would be zero. The energy change happens when the particle exits the cloud: in that situation, the magnetic mirror is moving, i.e.\ the distance between the particle and the mirror (labeled $l$ in Fig.\ \ref{magnetic_mirror:fig}, corresponding to the distance of the clouds in Fig.\ \ref{fermi1:fig}) is time-dependent due to the velocity of the cloud: $l=l(t)$: it shrinks if the exit happens under an angle $\theta_2<90^{\circ}$, it is increased for $\theta_2>90^{\circ}$. Thus, in order to keep the adiabatic invariant $J= const$, the velocity of the particle increases ($\theta_2<90^{\circ}$) or decreases ($\theta_2>90^{\circ}$). 

Finally, there is a net-effect when averaging over all particles entering the cloud, as the probability that the cloud picks up particles is higher for $\theta_1>90^{\circ}$. More concretely, the distribution of particles entering the cloud is following the relative velocity between the particle and the cloud, i.e.\ $dN/d\cos\theta_1 = \alpha\cdot v_{\rm rel}=\alpha\cdot \left(v_{1}-u\cdot \cos\theta_1\right)$, generalizing $u_i=:u$, see e.g.\ \citet{gaisser1991} for a detailed discussion. Thus, for a large number of particles, the average energy change is positive, i.e.\ $\left<\Delta E\right>/E=:\xi > 0$. It can be quantified to $\xi \propto (u/c)^2$. 

In Fermi's original article \citep{fermi1949}, he points out that the typical gas clouds in the Milky Way, taking up $\sim 5\%$ of interstellar space, have gas densities $n_0\sim 
10-100$~cm$^{-3}$, extensions of $\sim 10$~pc and velocities of $u\sim 10^{-4}\cdot c$. Based on these parameters, Fermi argues that (a) the mean free path of the particles for 
collisions with clouds is too large and (b) the energy gain per encounter is too small in order to have significant acceleration. This is why he suggests the concept of stochastic 
acceleration at shock fronts as a more efficient acceleration method \citep{fermi1954}. 

\paragraph{First order Fermi acceleration}

An illustration of shock front acceleration is presented in Fig.\ \ref{fermi2:fig}. The shock front is propagating with a lab-frame velocity of $\vec{v}_{\rm sh}$ toward the
unshocked and resting upstream region. The shocked gas behind the shock is called the downstream region. In the frame of the shock front (upper panel in Fig.\ \ref{fermi2:fig}), 
the two plasma streams move with velocities $u_1=-v_{\rm sh}$ (upstream) and $u_1<s\cdot u_2$ (with $u_2$ as the downstream velocity). According to the theory of shocks, for a strong shock, these 
two velocities are related as $s:=u_1/u_2=4$. 

The concept of shock acceleration works as follows: A particle enters the shock front (lab-frame velocity $\vec{V}_{\rm sh}$) from the unshocked upstream-side. Once on the 
downstream-side of the shock, the particles are scattered at the magnetic field inhomogeneities --- the particle directions become isotropized as in the downstream-frame, as the 
scattering is isotropic. The particles have a finite probability to exit the downstream region away from the shock, $P_{\rm esc}<1$. The probability to cross the shock again into 
the upstream-region is then $(1-P_{\rm esc})$. In the latter case, the particle approaches the shock front with the same energy (on average) that it had when entering the 
downstream region. Acceleration happens at the second crossing, as the particle sees the upstream region approaching with a velocity $(1-s^{-1})v_{sh}$ (Fig.\ \ref{fermi2:fig}, middle-panel). 
Thus, the concept of an approaching magnetic mirror can be applied again. All particles exiting the process on the downstream-side away from the shock exit the acceleration process 
and represent a particle in the final particle spectrum. All particles re-entering the upstream-region and thus staying in the acceleration process, are gaining energy. There is no 
effective energy loss, as the magnetic mirror is always approaching. The particle only loses energy once when exiting the acceleration region, but gains energy in all previous crossings. This is the crucial 
difference to the cloud scenario and leads to a more effective acceleration of first order i.e.\ $\xi \propto v_{sh}/c$.

One of the most severe problems is that injection into the acceleration process in first order Fermi acceleration is not natural. There are different solutions to that problem, like pre-accelerating particles via reconnection events that can be created via the magnetoresonant instability, or via the initial \textit{surfing} on the shock front, leading to a drift-motion that can pre-accelerate the particles efficiently, today referred to as \textit{Stochastic Shock Drift Acceleration (StSDA)} \citep{katou2019}, with drift acceleration in general initially discussed by \citet{jokipii1987}. First indications that such a process does exist in nature have recently been found in the Earth's bow shock \citep{amano_prl2020}.

\paragraph{Spectral behavior}
The energy spectrum of the particles can be estimated from the escape probabilities $P_{\rm esc}$ without having to rely on a specific acceleration mechanism: For each change of frame (crossing the shock front for first order Fermi or exiting the cloud for second order Fermi), a constant energy fraction $\xi$ is gained as discussed above. This leads to an energy after $n$ encounters of $E_n=E_0\,(1+\xi)^{n}$, with $E_0$ as the initial energy. Thus, the energy at escape from the shock front region is associated to a number of $n$ crossings, i.e.\ 
\begin{equation}
n=n(E)=\ln\left(\frac{E}{E_0}\right)/\ln(1+\xi)\,.
\label{n:equ}
\end{equation}
The number of particles reaching an energy above a fixed threshold $E$ is given as
\begin{equation}
N\left(\geq E\right)\propto \sum_{m=n(E)}^{\infty}\left(1-P_{\rm esc}\right)^{m}=\frac{(1-P_{\rm esc})^{n(E)}}{P_{\rm esc}}\,.
\label{N:equ}
\end{equation}
Using Equ.\ (\ref{n:equ}) in Equ.\ (\ref{N:equ}) yields a power-law spectrum, formulated in a differential form,
\begin{equation}
\frac{dN}{dE}\propto \frac{1}{P_{\rm esc}}\left(\frac{E}{E_{0}}\right)^{-\alpha}\,,
\end{equation}
with the spectral index 
\begin{equation}
\alpha=\frac{\ln\left(1-P_{\rm esc}\right)^{-1}}{\ln(1+\xi)}+1\approx \frac{P_{esc}}{\xi}+1\,.
\end{equation}
for values of $P_{\rm esc}\ll 1$ and $\xi \ll 1$. Assuming that the escape probability corresponds to the ratio of the rate of convection downstream away from the shock and the projection of the incoming, isotropic cosmic-ray flux on a plane shock front, the probability becomes
\begin{equation}
P_{\rm esc}=\frac{4\,u_2}{c}\,.
\end{equation}
For the case of a shock front, the acceleration efficiency is $\xi\approx \frac{4}{3}\,\frac{u_1-u_2}{c}$. Thus, the spectral index becomes
\begin{equation}
\alpha\approx\frac{3}{\frac{u_1}{u_2}-1}+1\,.
\end{equation}
For the value of strong shocks, $s:=u_1/u_2=4$ and thus $\alpha \approx 2$. For a more detailed discussion of the derivation of both the values for $\xi$ and the spectral index, see e.g.\ \citet{gaisser1991,blasi2013}. It is important to note that the exact spectral index and maximum energy will depend strongly on the astrophysical environment, as it is sensitive to the shock configuration, i.e.\ the orientation of the magnetic field with respect to the shock normal \citep[][e.g.]{mbq2008}. For shock acceleration, the injection problem adds another parameter to the problem.

The considerations above neglect non-linear feedback from back-reactions of cosmic-rays on the magnetic field. These non-linear aspects lead to magnetic field amplification, which is for once observed \citep{vink2012} and secondly needed in order to make SNRs able to accelerate particles up to the knee and beyond (See Section \ref{hillas:sec}). Two different types of instabilities can govern the interplay between magnetic fields and cosmic rays:

\begin{enumerate}
    \item The streaming instability is determined by resonant modes. It moderates cosmic-ray propagation in magnetic fields via the self-generation of Alfv{\'e}n turbulence, this way enhancing the turbulent magnetic field. The latter in turn reinforces acceleration. This feedback process is finally limited by the finite size of the confinement region of the particles. These effects have been studied in detail early-on in a 2-fluid model \citep[e.g.]{drury_voelk1981a, drury_voelk1981b} and in Monte-Carlo simulations \citep[e.g.]{ellison_eichler1984,ellison_eichler1985}. Kinetic simulations of the Vlasov equation have been performed  by e.g.\ \cite[e.g.]{berezhko_voelk1997}. Analytical solutions to the problem have been presented by e.g.\ \citet{schlickeiser1993,caprioli2009}.
    \item The Bell instability \citep{bell2004} is driven by those electric fields that are created when cosmic rays propagate across magnetic fields in the acceleration region and is of non-resonant nature. As studied in kinetic simulations by e.g.\ \citep{zirakashvili2008}, it can lead to the amplification of the magnetic field by a factor of $10$. 
\end{enumerate}

Particle acceleration at shocks and their resulting spectra have been discussed in detail in the literature, see e.g.\ \citep{mbq2008,stecker2007,WINCHEN201825}. 
As shown by \citet{caprioli2010}, the different methods (Monte Carlo, numerical and semi-analytical) result in quite comparable results. A review of this concept of non-linear diffusive shock acceleration (NLDSA) is provided by \citet{malkov_drury2001}. The main consequence from non-linear effects is the expectation of a deviation from the simple power-law description. The spectra become curved, i.e.\ turn from a steeper to a flatter spectrum. It should therefore be kept in mind that the power-law description is an approximation that only holds in QLT. In the most likely relevant non-linear environment, the spectra are expected to be curved. It has been suggested that the observed upturn of the cosmic-ray spectrum for all elements at the rigidity $300$~GV could be due to NLDSA-curved cosmic-ray spectra \citep{ptuskin2013} (see Section \ref{crabundance:sec}). In general, however, it depends on the details of energy range of detection and parameter settings at the acceleration site, if the subtlety of a curved vs uncurved (power-law) spectrum can become significant in current systematic (and statistic) uncertainties of the data.

\paragraph{Acceleration terms in transport equation}
First and second order Fermi processes can be included in the transport equation (\ref{transport:equ}) as follows: Fermi second order is included via the term for momentum diffusion,
\begin{equation}
+\frac{\partial}{\partial p}\left[p^2\,D_{pp}\,\frac{\partial}{\partial p}\left(\frac{n_l}{p^2}\right)\right]\,,
\end{equation}
as it represents a diffusive process in momentum space. Under the assumption of weakly turbulent magnetic fields and a magnetic power spectrum following a 
power-law in waves-space with index $\kappa$ the momentum diffusion is connected to the spatial diffusion coefficient $D(p)$ as given in Equ.\ \ref{dpp:equ}

Fermi first order processes are included via a momentum gain term,
\begin{equation}
\frac{\partial}{\partial p}(\dot{p}\,n_l)\,.
\end{equation}

\subsection{Maximum energy \label{hillas:sec}}
There is a basic argument for each potential accelerator that delivers a sanity check if an object can be considered as a candidate source or not: the \textit{Hillas Criterion}. As first pointed out by \citet{hillas1984}, a particle escapes the accelerator as soon as the gyroradius $r_g$ of the particle exceeds the size of the object $R$,
\begin{equation}
r_g = \frac{E}{|q|\,c\,B}\leq R\,,
\end{equation}
with $E$ and $Z\,e$ as the energy and charge of the particle, $B$ as the magnetic field of the accelerator (cgs units).
 In the relativistic particle limit $v_{\perp}\rightarrow c$, the energy is thus constrained:
\begin{equation}
E^{\rm Hillas}\leq Z\,e\,c\,B\,R\,.
\end{equation}
This is the famous \textit{Hillas Formula} that gives a concrete prediction for the conditions to be present in an accelerator in order to get particles to a certain maximum energy. In typical units, the maximum energy can be expressed as
\begin{equation}
  E_{\max}^{\rm Hillas}= Z\cdot 10^{18}\,\rm{eV}\cdot \left(\frac{B}{\mu\,\rm G}\right)\cdot \left(\frac{R}{\rm kpc}\right)\,.
  \label{hillas:equ}
\end{equation}
In particular, astrophysical objects can now be studied concerning their typical magnetic fields $B$ and their extensions $R$ in order to receive a first estimate of their potential to accelerate cosmic rays to the knee ($\sim 10^{15}$~eV), to the ankle ($\sim 10^{18.5}$~eV) or to the absolute maximum of observed particle energies in the Universe ($\sim 10^{20}$~eV). A representation of the \textit{Hillas Plot} is shown in Fig.\ \ref{hillas:fig}, using this classical formula in order to derive the lines of constant maximum energies.

For the acceleration at relativistic shocks, it was pointed out by \citet{achterberg2001} that the Hillas formula needs to be modified:
\begin{equation}
E_{\max}^{\rm Hillas, rel}= Z\,e\,c\,B\,\Gamma_{\rm sh}\,\beta_{\rm sh}\,R\,.
  \label{hillas_rel:equ}
\end{equation}
Here,  $\Gamma_{\rm sh}$ is the boost factor of the shock front and $\beta_{\rm sh}=v_{\rm sh}/c$ is the shock velocity in units of the speed of light. The reason for this 
modification is that in the upstream frame, the boosting effect keeps the particles within an angular cone of $\Delta \theta \sim 1/\Gamma_{\rm sh}$ and thus, the particle only completes a fraction of a gyration period,  leading to a further enhancement of the possible maximum energy of a relativistic source. Such relativistic shocks are typically 
present in jets of active galaxies or gamma-ray bursts (GRBs), while Galactic sources usually rely on only mildly-relativistic structures. Shock fronts of 
supernova remnants, for instance, have typical shock velocities of $\beta_{\rm sh}\sim 10^{-2}$. The correction of the formula for the ultra-relativistic case is therefore only 
necessary for some of the extragalactic sources. As this review focuses on Galactic sources, the figure shows the limits for the non-relativistic case. For ultrarelativistic 
sources like shocks in AGN jets or GRBs, the lines in this plot can  be considered as a lower limit.

          It is important to note that the Hillas limit as discussed above is a \textit{necessary condition} for an astrophysical object to be considered as a cosmic accelerator up to a certain energy: any mechanism accelerating particles to energies larger than Hillas maximum energy will be cut off right at the Hillas maximum energy.

          There are other limiting factors that result in a lower maximum energy, in particular the acceleration process itself, but also loss processes that can occur via the radiation or interaction of particles in the local medium (see e.g.\ \citet{ptitsyna_troitsky2010} for a detailed discussion). The process that acts on the shortest time scale will dominate and therefore determine the actual maximum energy, i.e.\ $\tau=\min\left(\tau_{\rm Hillas},\,\tau_{\rm acc},\,\tau_{\rm loss}\right)$.

          The characteristic time scales (i = loss, acc) are defined via the energy loss or gain rates $dE_{\rm i}/dt$:
\begin{equation}
\tau_{\rm i}=\left(\frac{dE_{\rm i}}{dt}\left(E\right)\right)^{-1}\cdot E\,.
\end{equation}
Thus, the maximum energy of a source, $E_{\max}^{\rm source}$, is given by the minimum of the three maximum energies, inferred from Hillas ($E_{\max}^{\rm Hillas}$), acceleration ($E_{\max}^{\rm acc}$) and loss ($E_{\max}^{\rm loss}$):
\begin{equation}
E_{\max}^{\rm source}=\min\left\{E_{\max}^{\rm Hillas},\,E_{\max}^{\rm acc},\,E_{\max}^{\rm loss}\right\}\,.
\end{equation}
Here, the maximum energy can be determined via the integration of the energy loss rate up to the maximum possible time (or spatial extension).
In the following, a short discussion of different scenarios of acceleration and loss time scales is presented.
\subsubsection{Maximum energy from acceleration process}

\paragraph{First order Fermi acceleration}
The acceleration time scale for diffusive shock acceleration (1st order Fermi acceleration as discussed in Section \ref{acc:sec}) is given as \citep{lagage_cesarsky1983b,lagage_cesarsky1983b,drury1983}:
\begin{equation}
\tau_{\rm acc,shock}=\frac{3}{u_1-u_2}\int_{0}^{p}\frac{dp'}{p'}\,\left(\frac{D_1(p')}{u_1}+\frac{D_2(p')}{u_2}\right)
\end{equation}
with $u_1$ and $u_2$ as the upstream and downstream velocities in the shock frame as defined in Section \ref{acc:sec} and $D_1$ and $D_2$ as the upstream and downstream diffusion coefficients. This equation can be simplified to (see e.g.\ \citet{gaisser1991})
\begin{equation}
\tau_{\rm acc,shock}\approx \frac{3}{u_1-u_2}\,\left(\frac{D_1}{u_1}+\frac{D_2}{u_2}\right)\,.
\end{equation}
In the limiting case of Bohm diffusion, which represents the minimal possible value for the diffusion coefficient, i.e.\ $D_{\min}=r_g\,c/3=E/(3\,q\,E\,B)$, the acceleration time can be shown to exceed 
\begin{equation}
\tau_{\rm acc,shock} \geq 7\cdot \frac{E}{q\,B\,u_1\,c}\,.
\end{equation}
To achieve this result, a ratio between the upstream and downstream velocity of $s=u_1/u_2\approx 4$ is assumed. 
\paragraph{Second order Fermi acceleration}
The acceleration time scale for second order Fermi acceleration as discussed in Section \ref{acc:sec} is given by (see e.g.\ \citet{OSullivan2009, WINCHEN201825}):
\begin{equation}
 \tau_\mathrm{acc, cloud} = 50\left(\frac{\beta}{7\times 
10^{-4}}\right)^{-2}\left(\frac{R}{10\,\mathrm{kpc}}\right)^{2/3}\left(\frac{E}{10^{19}\,\mathrm{eV}}\right)^{1/3}\;\mathrm{Gyr} \quad . \label{eq:2OrderAcc}
\end{equation}
In a simplified picture, the magnetic scattering centers can be assumed to move with the Alv\'en speed $\beta\propto B/\sqrt{\rho_\mathrm{p}}$, with $\rho_\mathrm{p}$ as the background plasma density. If the plasma density $\rho_\mathrm{p}$ and the lifetime of the accelerating source are known,  Equ.\ (\ref{eq:2OrderAcc}) can be used to constrain possible sources in a 
Hillas like plot.  The rather long acceleration time scale rules out many source 
classes. Combining lifetime and 
plasma density only radio lobes of AGNs and superwinds of starburst galaxies are left over as possible second order Fermi accelerators \citep{WINCHEN201825}. Both classes are 
active on time scales of 
more than 10 Myrs and assume a plasma density of $\rho_\mathrm{p}\approx 10^{-4}\;\mathrm{cm}^{-3}$.
\paragraph{Linear acceleration}
For linear acceleration, it is a typical scenario that plasma instabilities lead to the generation of a particle drift, creating an electric field $E_{\rm field}=c^2\cdot 
\eta\cdot 
B$ with $\eta<1$. Thus, the energy gain by acceleration is
\begin{equation}
\left.\frac{dE}{dt}\right|_{\rm acc,lin}=q\,c\,\eta\,B\,.
\label{lin_acctime:equ}
\end{equation}
Thus, the related time scale is given as
\begin{align}
\tau_{\rm acc,lin}=\frac{E}{q\,c^2\,\eta\,B}\,.
  \end{align}

\subsubsection{Loss-dominated maximum energy}
In a \textit{diffusive acceleration environment} (Fermi first or second order), the central limiting factor is synchrotron radiation: as the acceleration does not happen in a linear, but in a stochastic way, the particle radiates continuously due to their stochastic acceleration. If the acceleration time scale is shorter than the loss time scale has to be investigated for each individual case.

The synchrotron energy loss rate is given as (see e.g.\ \citet{rybicki_lightman1979} and Section \ref{multimessenger_sources:sec} for a discussion):
\begin{equation}
  \frac{dE}{dt}=-\frac{2}{3}\,\frac{q^4}{m^4\,c^7}E^2\,B^2\,.
  \label{synch:equ}
\end{equation}
Separating the differentials and substituting $x=c\cdot t$, the maximum energy can be determined as \citep{medvedev2003}
\begin{equation}
\frac{1}{E_{\max}^{\rm loss, diff}}=-\int \frac{dE}{E^2}=\frac{2}{3}\,\frac{q^4}{m^4\,c^8}B^2\int_0^{R} dx =\frac{2}{3}\,\frac{q^4}{m^4\,c^8}B^2\,R\,. 
\end{equation}
Thus, in the Hillas plot representation, the magnetic field is given by
\begin{equation}
B(R,E_{\max})\propto R^{-1/2}\,(E_{\max}^{\rm loss,diff})^{-1/2}
  \end{equation}

In a \textit{linear acceleration environment} (also sometimes referred to as \textit{one-shot acceleration}), the synchrotron loss scale needs to be compared directly to the acceleration time scale: In the acceleration process, the particle radiates continuously and it cannot be accelerated further if the loss time scale is shorter than the acceleration time scale. 
Comparing the maximum energy of linear particle acceleration (Equ.\ (\ref{lin_acctime:equ})) to the energy loss rate for synchrotron radiation (Equ.\ (\ref{synch:equ})) yields
\begin{equation}
E_{\max}^{\rm loss, lin}=\sqrt{\frac{3}{2}}\,\frac{m^2\,c^4}{q^{3/2}}\,B^{-1/2}\,\eta^{1/2}\,.
\label{losslin:equ}
\end{equation}
For this scenario, an ordered electric field needs to be present on longer time scales, something that is rather difficult to achieve due to the tendency of the plasma to reach charge equilibrium. However, there are certain extreme conditions, like near a pulsar or in jets of active galaxies where this could be achieved, see \cite[e.g.]{klein2013,winter2014} and references therein.
\subsubsection{Maximum energy for Galactic sources}
          
Applying the above-discussed results to the representation of the Hillas plot, in which $B(R)$ is shown for constant values of $E_{\max}$, there is a critical magnetic field that 
determines if the Hillas maximum energy or the loss-dominated maximum energy is applicable. This critical energy can be determined from the criterion $E_{\max}^{\rm 
Hillas}=E_{\max}^{\rm loss,\,lin}$, even in the case of diffusive shock acceleration: the limiting case corresponds to the size of the object being the gyroradius. Even in the 
diffusive case, with the gyroradius being of the same size as the emission region, a maximum of one scattering will happen, so that in this limit, the diffusive case corresponds 
to the linear case (see \citet{ptitsyna_troitsky2010} for a detailed discussion).
          \begin{figure}[htbp]
\centering{
\includegraphics[trim = 0mm 0mm 0mm 0mm, clip, width=0.7\textwidth]{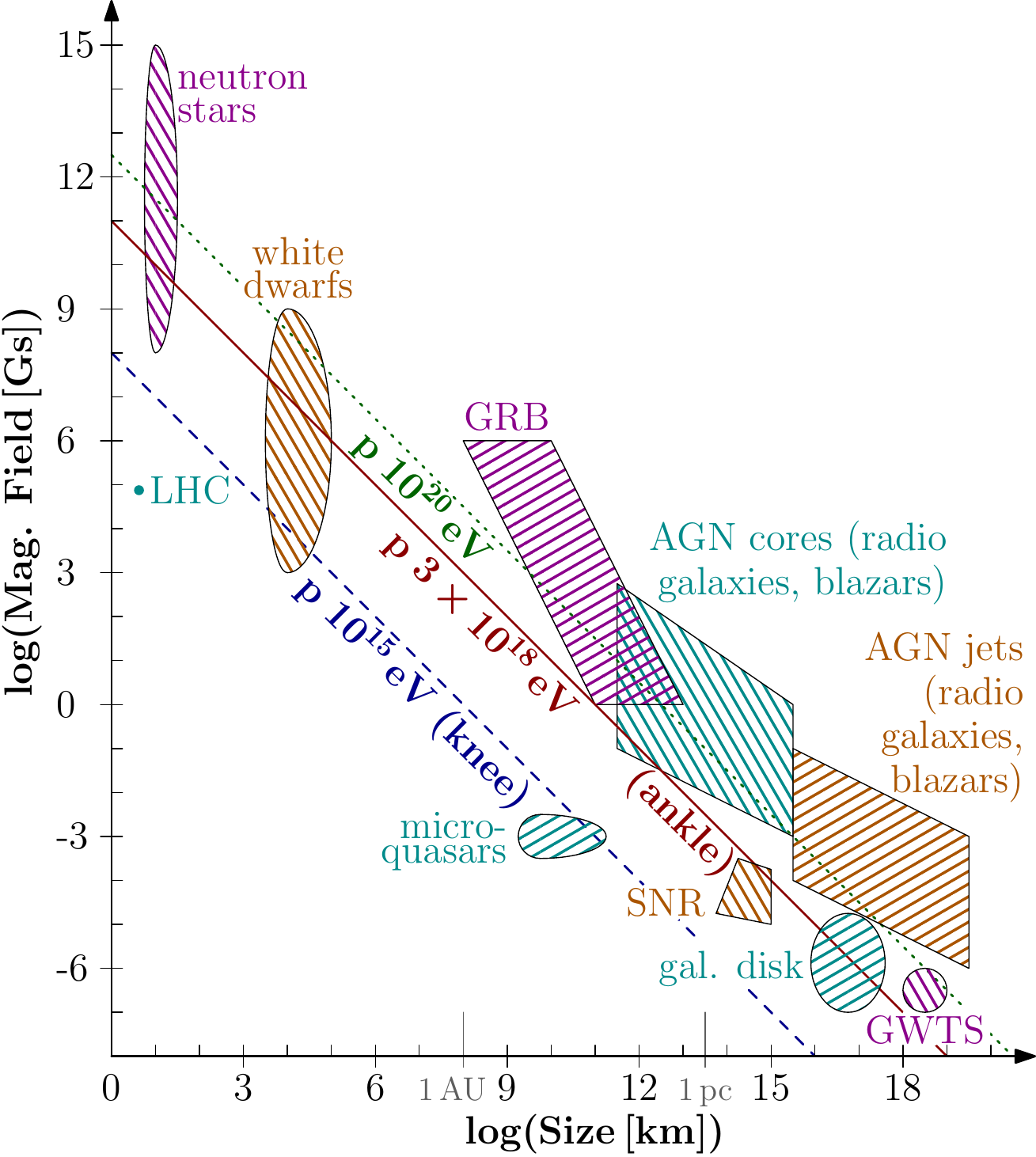}
\caption{The classical Hillas plot. The knee ($10^{15}$~eV), ankle ($3\times 10^{18}$~eV) and  maximum energy ($10^{20}$~eV) lines are shown (blue dashed, red solid and green dotted, respectively). Regions for GRBs and the different AGN components have been taken from \citet{ptitsyna_troitsky2010}, SNR from \citet{mandelartz2015}, neutron stars from \citet{bombaci1996,reisenegger2003,lattimer2001}, white dwarfs from \citet{chandrasekhar1931,kaplan2014,ferrario2015,suh2000}, LHC from \citet{rossi2012lhc}, galactic disk and halo from \citet{beck2016}, microquasars from \citet{pandey2007}.
  \label{hillas:fig}}
}
\end{figure}

\begin{figure}[htbp]
          
 \includegraphics[trim = 0mm 0mm 0mm 0mm,  width=0.48\textwidth]{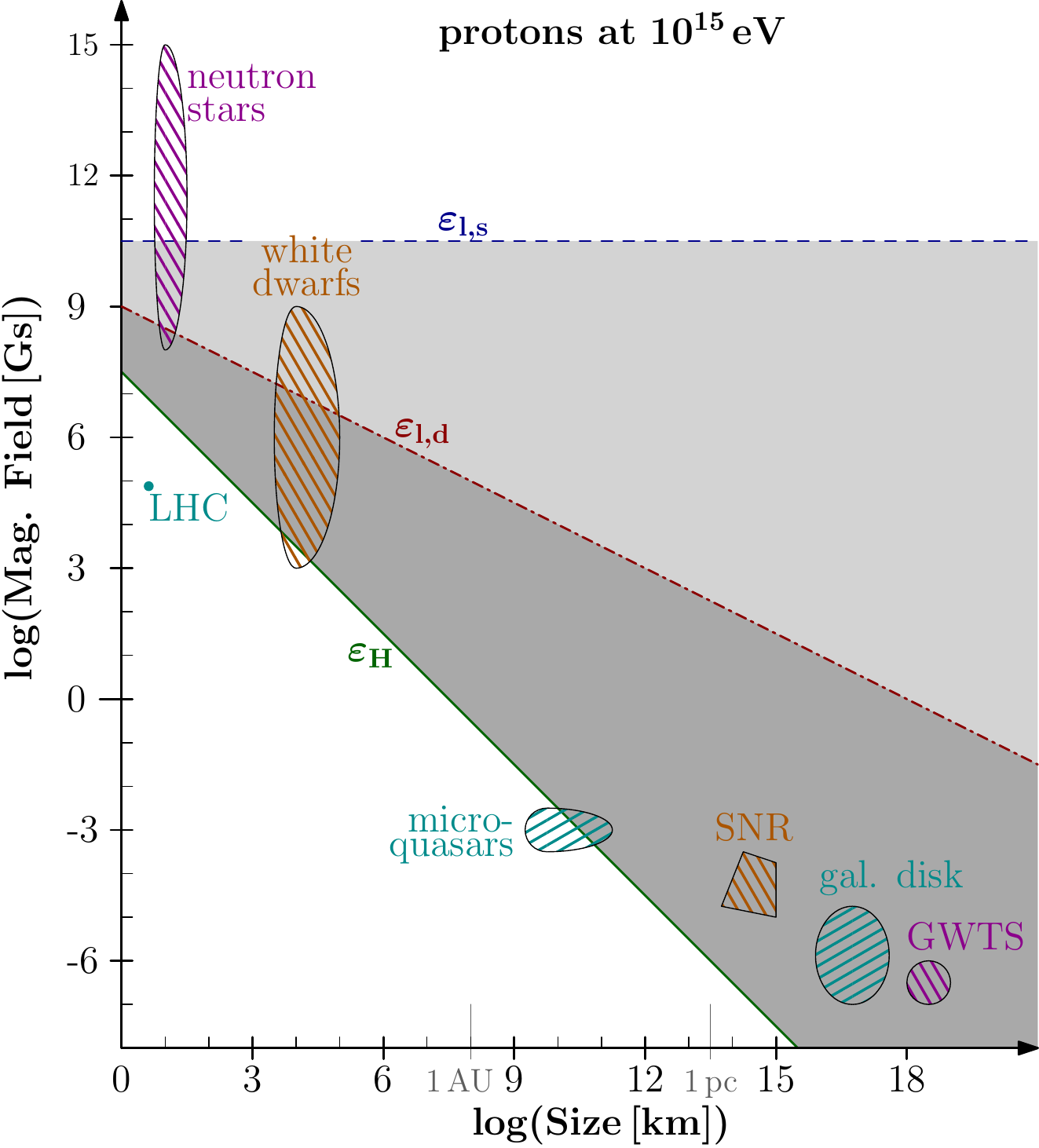}\hfill
 \includegraphics[trim = 0mm 0mm 0mm 0mm,  width=0.48\textwidth]{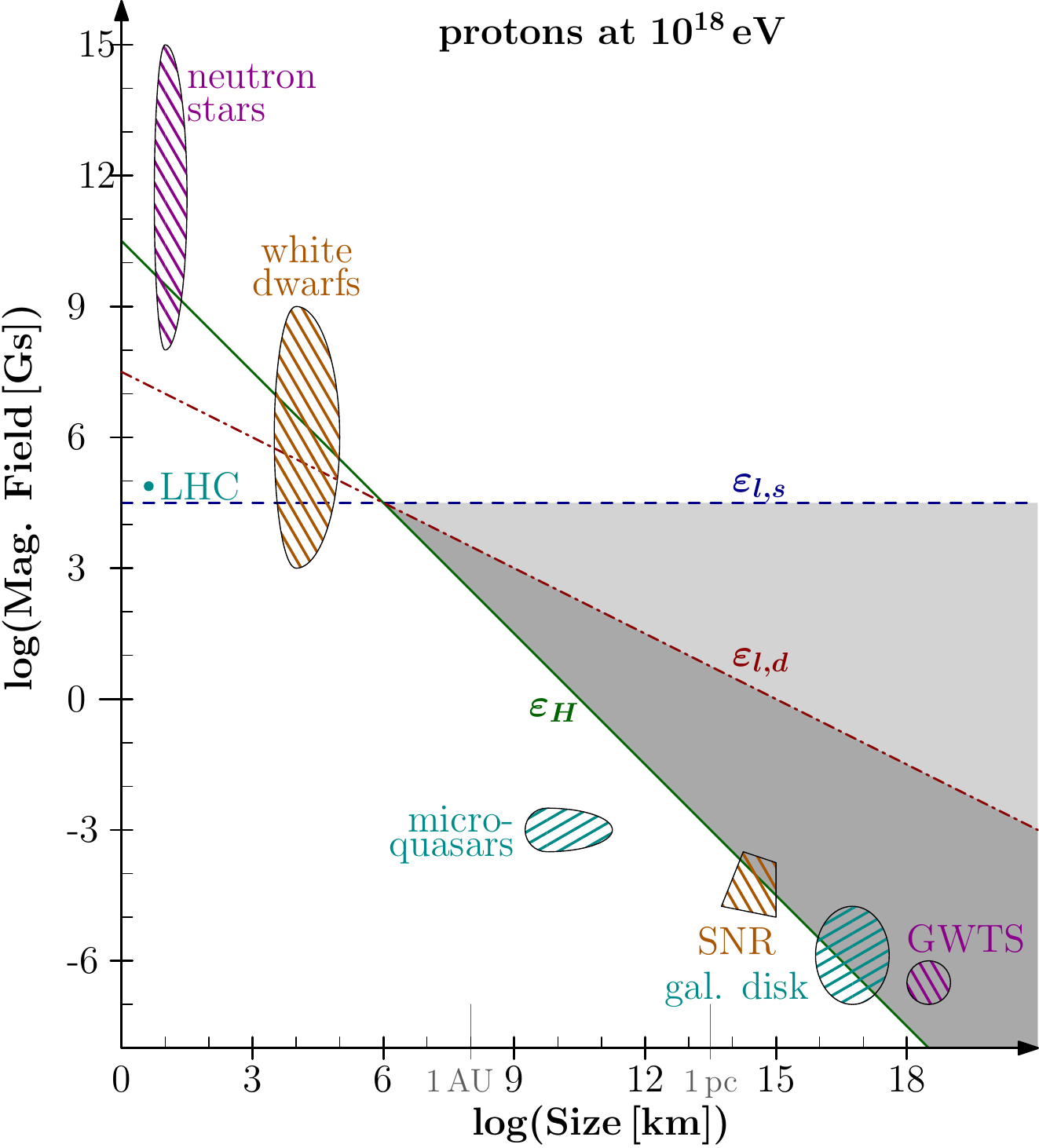}
\caption{The Hillas plot showing the region still allowed within the diffusive-loss scenario (dark-shaded region) and the one allowed within the linear-loss scenario (light-shaded and dark-shaded region) for maximum energies of $10^{15}$~eV (left) and $3\times 10^{18}$~eV (right).
References for object classes as in Fig.\ \ref{hillas:fig}. 
  \label{hillas15:fig}}
%

\end{figure}

Figure \ref{hillas:fig} shows the \textit{Hillas Plot}, in which astrophysical objects with non-thermal emission are shown with respect to their magnetic field strength $B$ and extension $R$. The lines represent the corresponding maximum energy necessary for protons to reach the cosmic-ray knee, $E_{\max}^{\rm Hillas}=10^{15}$~eV (blue, dashed), the ankle, $E_{\max}^{\rm Hillas}=3\times 10^{18}$~eV (red solid) and the highest observed energies, $E_{\max}^{\rm Hillas}=10^{20}$~eV (green dotted).
Of the Galactic objects, regular supernova remnants can reach energies above the knee, if the magnetic field is enhanced with respect to the interstellar medium's average value 
(see Section \ref{multimessenger_modeling:sec} for a more detailed discussion). In particular, if the supernova explosion is connected to a heavy progenitor star, which has ejected 
its outer shells, so that the explosion happens into a pre-existing wind (so-called wind-SNRs), then acceleration up to $10^{17}$~eV - $10^{18}$~eV is possible, see 
\citet{stanev1993,biermann95,biermann2010} for a discussion. The Galactic halo is also capable of reaching energies above the knee, this is in particular true for a possible 
Galactic wind termination shock (GWTS) with the intergalactic medium with a maximum energy of $\sim10^{19}$~eV. However, for the GWTS, the maximum energy is normally bounded by a limited acceleration time leading to lower maximum energies of $\sim 10^{17}$~eV for a Milky Way type galaxy (see e.g.\ \citet{Bustard2017}). Microquasars are a candidate for reaching $10^{15}$~eV, and so are white dwarfs 
as well as neutron stars. However, taking into account the limitations through energy loss, further constraints to the sources apply. Figure \ref{hillas15:fig} (left) shows the \textit{Hillas Plot} for protons with the knee as maximum energy. The dark-shaded region is the one allowed for sources that accelerate diffusively, the light-shaded plus dark-shaded regions are allowed for linearly accelerating sources. Figure \ref{hillas15:fig} (right) shows the same but for the ankle as maximum energy ($3\times 10^{18}$~eV). While white dwarfs and neutron stars are still reasonable candidates for the acceleration up to the knee, they can be excluded to accelerate to higher energies. The combination of their compactness and their extremely high magnetic fields lets the particles lose energy through synchrotron radiation before they can reach the classical Hillas limit.

          \clearpage

\subsection{Diffusive transport \label{transport:sec}}
In order to describe the transport of cosmic rays in the Galaxy, the transport equation as formulated in Equ.\ (\ref{transport:equ}) includes all relevant terms. In particular, diffusion and advection can be relevant, as are loss processes (continuous and catastrophic). Momentum diffusion can be present in the form of diffusive reacceleration. The spallation of elements is important to include when trying to describe the abundance of elements. The details of state-of-the-art modeling of Galactic multimessenger data are presented in Section \ref{multimessenger_modeling:sec}. In this Section, the basic arguments for the expected change in the energy spectrum within the leaky box model are discussed. In particular, it will be reviewed how diffusive transport can be governed via the Kolmogorov law and how this translates to a description via a diffusion coefficient. Finally, the role of spallation for the determination of the spectral behavior of the diffusion coefficient is discussed.

\subsubsection{Description of the diffusion coefficient within the Kolmogorov scale analysis \label{diffusion:sec}}

The diffusive behavior of the particle transport in the Galactic environment is governed by the magnetic field and in particular its turbulent nature. In this context, the wave vector spectrum of the magnetic field that results from a turbulent MHD medium determines the behavior of the diffusion coefficient. This context will be discussed in this section.

In order to derive the wave vector spectrum for a turbulent magnetic field, Kolmogorov developed a scale law that works for an isotropic and a stationary level of turbulence (kept stationary by external forces). This situation is basically the one believed to be present in the Galaxy, where isotropy and the ideal (non-dissipative) limit are granted at large scales and the turbulence level is kept up by a constant rate of supernova explosions. The (heuristic) description of the Kolmogorov scale law is shown here (paragraph \ref{kolmogorov:sec}), followed by the discussion of the diffusion coefficient and its connection to the wave vector spectrum (paragraph \ref{diffusion_coeff:sec}).

\paragraph{Kolmogorov scale analysis \label{kolmogorov:sec}}
The starting point is the Navier-Stokes equation, which describes the evolution of the flow velocity of a fluid, $\vec{U}$, where in the case of the derivation of the Kolmogorov scaling law, incompressibility is assumed ($\nabla\cdot \vec{U}=0$):
\begin{equation}
\frac{\partial \vec{U}}{\partial t}+\left(\vec{U}\cdot \nabla \right)\,\vec{U}=-\frac{1}{\rho}\nabla P + \nu\,\Delta \vec{U}\,.
\label{navier_stokes:equ}
\end{equation}
Here, $\rho$, $P$ and $\nu$ are the density, pressure and resistivity of the medium, respectively. 

The scale analysis of this equation according to Kolmogorov now works as follows: The spatial size of a vortex is defined as $\vec{\omega}:=\nabla\times\vec{U}$. By applying a rotation to Equ.\ (\ref{navier_stokes:equ}) and assuming isotropy in form of a constant density, it can be written as an equation describing the evolution of the vorticies:
\begin{equation}
\frac{\partial \vec{\omega}}{\partial t}+\nabla\times (\vec{U}\times \vec{\omega})=\nu\,\Delta \vec{\omega}\,.
\end{equation}
The temporal change of the vorticies of a fluid is determined by a convective term ($\nabla\times (\vec{U}\times \vec{\omega})$) and a viscous one  ($\nu\,\Delta \vec{\omega}$).

If energy is injected on a spatial scale $\mathcal{L}$, the Reynolds number is used as a measure for the relation of the strength of the convective and viscous terms,
\begin{equation}
\mathcal{R} = \frac{\mathcal{U}\,\mathcal{L}}{\nu}\,.
\label{Reynolds:equ}
\end{equation}
with $\mathcal{U}=|\vec{U}|$.
For $\mathcal{R}\ll 1$, the laminar flow dominates the equation, $\partial \vec{\omega}/\partial t \approx \nu\,\Delta \vec{\omega}$, for $\mathcal{R}\gg 1$, the turbulent term dominates, i.e.\ $\partial \vec{\omega}/\partial t \approx \nabla \times (\vec{U}\times \vec{\omega})$.

For arbitrary scales $l$, the flow velocity can be divided into an average, effective velocity of the fluid, $\left<\vec{U}\right>=\vec{U}_0$ with $|\vec{U}_{0}|=\mathcal{U}$ and a fluctuating part $\vec{w}_l$, acting on the scale $l$, that vanishes on averaging so that $\vec{U}=\vec{U}_0+\vec{w}_l$.

The Reynolds number can now be defined via the local change in the velocity in between two points, i.e.\ $\Delta \vec{U}=\vec{w}_l$:
\begin{equation}
\mathcal{R}_l = \frac{w_l\,l}{\nu}\,.
\label{Reynolds_l:equ}
\end{equation}
As the energy is injected on the scale $\mathcal{L}$, turbulent vorticies can only occur on scales $l\leq \mathcal{L}$. In the beginning, $\Delta U \approx \mathcal U$ and $\mathcal{R}_l \approx \mathcal{R}_L =\mathcal{R}\ll 1$: The viscous term is negligible for large vorticies and no energy dissipation happens until the vorticies are reduced to a size $l_d$ at which dissipation sets in, i.e.\ $R_{d}\approx 1$. From that point on, energy is transferred to heat and viscocity becomes important.

In order to quantify this dissipation process, the Navier-Stokes equation is solved by performing a Fourier-transform that can then be used to describe the change in energy through dissipation and transfer to different wave vectors:
\begin{equation}
\frac{\partial W(k,\,t)}{\partial t}=\dot{W}_{\rm diss}(k,\,t)+\dot{W}_{\rm trans}(k,\,t)\,.
\end{equation}
The energy transfer in $k$-space does not contribute to the total energy budget, i.e.\ $\int d^3 k\,\dot{W}_{\rm trans}=0$. It is solely responsible for an energy cascade from large scales to small scales. Thus, as energy is injected at a large scale $\mathcal{L}$, it is first transferred to small scales without causing any change in the total energy budget, and only when the dissipation scale $l_d$ is reached, the energy will be dissipated via heat with $\dot{W}_{diss}\sim \nu\,\rho\,w_{d}^{2}/l_{d}^{2}$, associating the dissipation wave number with the length scale  $k_{\max}\sim l_{d}^{-1}$.

The Kolmogorov scale analysis is based on the line of arguments presented above: With the largest (injection) scale $\mathcal{L}$ at which a typical velocity $\mathcal{U}$ is present, the injected power density is 
\begin{equation}
\dot{\epsilon}_0\sim \frac{\rho\,\mathcal{U}^{2}}{\tau}\sim \frac{\rho\,\mathcal{U}^{3}}{\mathcal{L}}\,,
\end{equation}
using the typical time scale of $\tau=\mathcal{L}/\mathcal{U}$.

At smaller length scales, $l_d<l<\mathcal{L}$, the typical velocities are $w_l$ and the power is given by the energy transfer, 
\begin{equation}
\dot{\epsilon}_l\sim \frac{\rho\,w_{l}^{3}}{l}\,,
\end{equation}
Finally, at the dissipation scale $l_d$, the dissipation velocity $w_d$ is relevant. At the same time, the local energy dissipation due to viscosity can be described as $\dot{\epsilon}_d\sim (\nu\,\rho\,w_d^2)/l_{d}^{2}$ and thus:
\begin{equation}
\dot{\epsilon}_d\sim \frac{\rho\,w_{d}^{3}}{l_d}\sim  \frac{\nu\,\rho\,w_{d}^{2}}{l_{d}^{2}}\,.
\end{equation}
In the case of stationary turbulence, which is assumed in the Kolmogorov scale analysis, these terms are the same,  $\dot{\epsilon}_0=\dot{\epsilon}_l=\dot{\epsilon}_d$, leading to 
\begin{equation}
\mathcal{U}^3/\mathcal{L}=w_{l}^{3}/l=w_{d}^{3}/l_{d}\sim \nu\,w_d^2/l_d^2\,.
\label{scale1:equ}
\end{equation}

The dissipation length scale $l_d$ and the associated time scale $\tau_d=l_d/w_d$ can now be expressed in terms of the Reynolds number, Equ.\ (\ref{Reynolds:equ}), by using the correlations above (Equ.\ (\ref{scale1:equ})) and assuming $\mathcal{R}_d\approx 1$:
\begin{equation}
\frac{l_d}{\mathcal{L}}=\mathcal{R}^{-3/4}
\end{equation}
and
\begin{equation}
\frac{\tau_d}{\tau}=\mathcal{R}^{-1/2}\,.
\end{equation}
This relation shows that small structures decay faster than large ones, as anticipated.

In order to determine the wave vector spectrum, the total energy density is used:
\begin{equation}
\rho_E=\int_{2\,\pi/\mathcal{L}}^{\infty} dk\, W(k)\sim \frac{1}{2}\rho\,w_{l}^{2}\stackrel{(\ref{scale1:equ})}{\sim}\frac{1}{2}\,\rho^{1/3}\,\epsilon_0^{2/3}\,\mathcal{L}^{2/3}\,.
\label{rhoe:equ}
\end{equation}
Here, all scales smaller than $\mathcal{L}$ are considered in the integration. Thus, there are two ways of determining $d\rho_E/d\mathcal{L}$ (2nd and 4th term in Equ.\ (\ref{rhoe:equ})). Comparing these results in a prediction of the behavior of the wave vector spectrum $W(k)$ when using $k=2\,\pi/\mathcal{L}$:
\begin{equation}
W(k)=\frac{1}{3}\,\rho^{1/3}\,\epsilon_0^{2/3}\,k^{-5/3}\,.
\end{equation}
This is the Kolmogorov scale law: an injection of energy at large scales $\mathcal{L}$ leads to a transfer in scales, toward smaller $l$, i.e.\ larger $k$, that results in a power-law energy spectrum in wave vector space with a spectral index of $-5/3$ in the one-dimensional representation. A three-dimensional calculation leads to an additional factor $k^2$, so that the three-dimensional spectrum is expected to behave as $k^{-11/3}$. Applied to the case of an energy spectrum dominated by magnetic fluctuations,
\begin{equation}
W(k) = \frac{B(k)^{2}}{2\,\mu_0}\propto k^{-5/3}\,.
\end{equation}
The spectral behavior of the magnetic field fluctuations can of course deviate from this Kolmogorov power-law, as the preconditions of an isotropic medium with stationary turbulence are not necessarily given in each astrophysical situation. A change in these preconditions typically leads to a change in the spectral index, so that the assumption of $W\propto k^{-\lambda}$ is a reasonable one \citep{berezinskii1990}. In particular, Kraichnan turbulence explicitly takes into account the magnetic field. This leads to a change in the spectral index of the power-law to $\lambda = 3/2$ in the one-dimensional case and $\lambda=7/2$, respectively, for the three-dimensional calculation.
          \begin{figure}[htbp]
\centering{
\includegraphics[trim = 0mm 0mm 0mm 0mm, clip, width=0.9\textwidth]{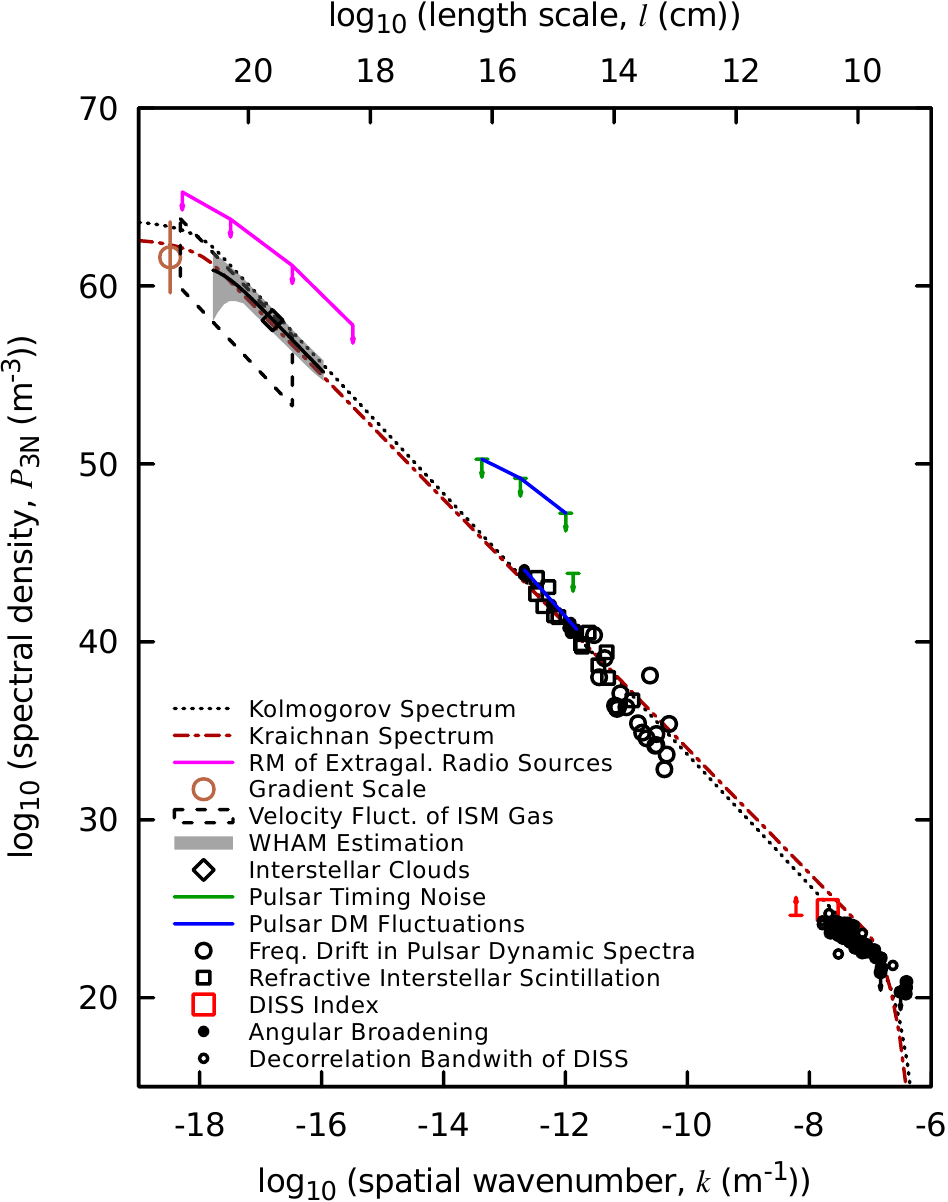}
\caption{The Big Power Law in the Sky, reproduced after \citet{chepurnov2010,armstrong1995}. Details on the data shown can be found in \citet{chepurnov2010,armstrong1995}. The lines represent a Kolmogorov spectrum (black, dotted line) and a Kraichnan spectrum (red, dot-dashed line) with cut-offs at $100$~pc and $10^{9}$~cm. Note that these are \textit{no} fits to the data, but simply lines to guide the eye. 
\label{big_power_law:fig}}}
\end{figure}

Figure \ref{big_power_law:fig} shows the observed astrophysical wave vector spectrum, often referred to as \textit{The Big Power-Law in the Sky} \citep{armstrong1995}: the data reveal a power-law behavior of the Galactic wave vector spectrum over an extremely large range of scales, covering $10^{12}$ orders of magnitude in wave numbers. The spectrum seems to be quite close to the Kolmogorov scaling law with $k^{-11/3}$ (dotted, black line), but could also be of Kraichnan-type with $k^{-7/2}$ (dot-dashed, red line). It is expected that the wave vector spectrum has a maximal scale, i.e.\ minimal wave vector $k_0$, given by the injection scale. In addition, a minimal scale, i.e.\ maximal wave vector $k_{\max}$, is expected at which energy dissipation starts. Data seem to indicate that the injection scale lies above $10^{20}$~cm, compatible with the expectation that turbulence is injected into the Galaxy by supernova explosions at $\mathcal{L} \sim 100$~pc. The dissipation scale as indicated in the figure lies below $\sim 10^{9}$~cm. Both cut-offs are predicted to exist by theory.  In the figure, we include these cut-offs. It should be noted that, while the lower wave number cut-off has a fixed scale at 100~pc as discussed above, the dissipation scale is not fixed --- data only confirm that it needs to be $10^{9}$~cm or below.

The fact that the wave vector spectrum is quite close to Kolmogorov is confirmed by the measurements of cosmic-ray spallation products as discussed in Section \ref{spallation:sec}. However, the interpretation of the spallation results relies on the results from quasi-linear theory (QLT), in which a Kolmogorov spectrum leads to an energy behavior of the diffusion coefficient of $E^{1/3}$. The full, numerical treatment of this problem indicates some tension of this interpretation as discussed below. In addition, the intermittency of turbulence may play a role for the propagation of cosmic rays, as we will discuss below as well.

          \paragraph{The diffusion tensor \label{diffusion_coeff:sec}}
          The diffusion tensor determines the directionality and amount of diffusion in a certain plasma environment. For a magnetic field $\vec{B}=\vec{B}_0+\delta\vec{B}$, composed of a regular component $\vec{B}_0$ and a turbulent one $\delta \vec{B}$ with $\left<\vec{B}\right>=\vec{B}_0$, the fraction of turbulence, defined as $\eta = |\delta\vec{B}|/|\vec{B}|$, determines the properties of the tensor.\\
          Under the assumption that the two (normal and binormal) 
perpendicular directions are degenerated\footnote{In astrophysical environments --- outside 
the solar system --- where the scattering scales are small compared to the magnetic field line curvature radius this is normally fulfilled.}, i.e.\ $D_{\perp, 
\mathrm{normal}}\approx 
D_{\perp, \mathrm{binormal}}$, components of the diffusion tensor can be expressed as (see e.g.\ \citet{snodin2016,sigl2017})
          \begin{equation}
D_{ij}=\frac{\hat{B}_{i}\,\hat{B}_{j}}{3\,r_{\parallel}}+\frac{\delta_{ij}-\hat{B}_{i}\,\hat{B}_{j}}{3\,r_{\perp}}+\frac{\epsilon_{ijk}\,\hat{B}_{k}}{3\,r_{A}}\,. 
\label{eq:DifTensor}
            \end{equation}
          Here, $r_{\alpha}$ are the effective diffusion rates for the parallel ($\alpha=\,\parallel$), perpendicular ($\alpha=\,\perp$) and anti-symmetric ($\alpha=A$) components.  In addition, 
$\vec{\hat{B}}=\vec{B}_0/|\vec{B}_0|$ is the normalized regular magnetic field vector. Choosing a coordinate system where the $x$-axis is aligned with the regular magnetic field direction therefore 
leads to a diffusion tensor of the form
\begin{align}
\hat{D} = 
\begin{pmatrix}
 D_\parallel & 0 & 0 \\
 0 & D_\perp & D_A \\
 0 & -D_A & D_\perp
\end{pmatrix} \quad .
\end{align}
The anti-symmetric diffusion $D_A$ is often neglected or absorbed into a drift term leading to $\hat{D}=$~diag$(D_{\parallel},D_{\perp},\,D_{\perp})$ (e.g.\ \citet{kopp2012}). 

In a weakly turbulent field, $\eta=\delta B/B_0\ll 1$, the particles will predominantly diffuse along the field lines, with a connection between the parallel and perpendicular diffusion coefficients as \citep{jokipii1987} 
\begin{equation}
D_{\perp}=\frac{D_{\parallel}}{1+\left(\frac{\lambda_{\parallel}}{r_g}\right)^{2}}\,.
\label{dperp:equ}
\end{equation}
Here, $\lambda_{\parallel}$ is the mean free path in the direction of the regular magnetic field. In the limit of low turbulence, $\lambda_{\parallel}\gg r_g$. Thus, $D_{\perp}\ll D_{\parallel}$, which means that the problem can be reduced to the one-dimensional case. In regions of high turbulence level, $\delta B/B_0 \geq 1$, a more complex approach including non-linear effects needs to be applied. In general, it is expected that perpendicular diffusion becomes more and more important for stronger turbulence levels.

The diffusion coefficient is connected to the diffusion mean free path as
\begin{equation}
  D_{\parallel}=\frac{\lambda_{\parallel}\,v}{3}\,,
\end{equation}
with $v$ as the particle velocity (in the case of cosmic rays, the approximation $v\approx c$ is often applicable).
The mean free path is at least the gyroradius of the particle, i.e.\ $\lambda_{\parallel}\geq r_g$ and thus, the minimal diffusion coefficient, also referred to as the \textit{Bohm limit} is given as
\begin{equation}
  D\geq D_{\min}=\frac{r_g\,c}{3}=\frac{E}{3\,q\,B}\,.
\end{equation}

The determination of the diffusion coefficient for arbitrary (power-law) wave-vector spectra within QLT limit works as follows:
The scattering rate of the particles $\nu_{\parallel}$, i.e.\ the rate on which the particles isotropize, is defined via the characteristic time scale of scattering $\tau_{\parallel}=\nu_{\parallel}^{-1}=\lambda_{\parallel}/v$. The parallel diffusion coefficient can then be expressed as (see e.g.\ \citet{berezinskii1990,schlickeiser2002} for a detailed discussion)
\begin{equation}
D_{\parallel}=\frac{v^2}{2}\int_{0}^{1}d\mu\frac{1-\mu^2}{2\,\nu_{\parallel}}\,.
\label{dparallel:equ}
\end{equation}
Here, $\mu=\cos\theta$ is the cosine of the pitch angle $\theta$, i.e.\ the angle between particle velocity and magnetic field direction. In QLT ($\eta\ll 1$), the scattering rate itself can be approximated as \citep{berezinskii1990}
\begin{equation}
  \nu_{\parallel}\approx 2\pi^2\left|\omega_B\right|\,\frac{k_{\rm res}\,W(k_{\rm res})}{B_{0}^{2}}\,.
  \label{scattering_rate:equ}
\end{equation}
Thus, the scattering rate is determined by the synchrotron frequency $\omega_B\propto k_{\rm res}$, and the one-dimensional wave vector spectrum at the wave number at resonance for particle-wave scattering, $k_{\rm res}\sim (|\mu|\,r_{g})^{-1}$. This resonant wave number is connected to the \textit{coherence length scale} as $l_c=2\pi/k_{\rm res}$.

Using the power-law behavior of the wave vector spectrum, i.e.\ $W(k)\propto k^{-\lambda}$ in combination with Equations  (\ref{dparallel:equ}) and  (\ref{scattering_rate:equ}) results  in the following correlation:
\begin{equation}
D_{\parallel}\propto E^{2-\lambda}\,\left(\frac{\delta B}{B_0}\right)^{-2}\,.
\end{equation}
A full derivation of this relation is presented in e.g.\ \citet{berezinskii1990}.  The perpendicular component can be determined using Equ.\ (\ref{dperp:equ}) and it behaves as
\begin{equation}
    D_{\perp}\propto \left(\frac{\delta B}{B_0}\right)^{2}
\end{equation}
so that the ratio of parallel to perpendicular component is highly dependent on the turbulence level, i.e.
\begin{equation}
    \frac{D_{\parallel}}{D_{\perp}}\propto \left(\frac{\delta B}{B_0}\right)^{-4}\,.
\end{equation}
For a Kolmogorov-type spectrum ($\lambda =5/3$) the energy dependence of a scalar diffusion coefficient is therefore $D_{\parallel}\propto E^{1/3}$, for a Kraichnan-type spectrum ($\lambda=3/2$), it becomes $D_{\parallel}\propto E^{1/2}$. Thus, the escape time of cosmic rays from the Milky Way (see Equ.\ (\ref{escape_time:equ})), which is inversely proportional to the diffusion coefficient, is expected to behave as
\begin{equation}
\tau_{\rm esc}\propto D_{\parallel}^{-1}\propto E^{-\kappa}\,,
\label{esc_diffcoeff:equ}
\end{equation}
with $\kappa=2-\lambda \sim 0.3 - 0.5$ by assuming that the reality lies somewhere between the Kolmogorov and Kraichnan cases. The parallel diffusion coefficient becomes stronger for larger ratios of the regular to turbulent fields, just as expected.

As a general approach to investigate whether or not the results of QLT hold, computer simulations can be performed on the basis of a method developed by  \citet{taylor1922,green1951,kubo1957}:
The so-called Taylor-Green-Kubo (TGK) formalism enables the calculation of the diffusion coefficient in the limit of long propagation times in the simulation of particle propagation in a homogeneous plus turbulent magnetic field.

As the choice of coordinate system, the homogeneous magnetic field will be aligned with the first axis, i.e.\ $\vec{e}_{x_1}\parallel \vec{B}_0$ so that $D_{x_1\,x_1}=D_{\parallel}$. Further, uniform perpendicular diffusion will be assumed, i.e.\ $D_{x_2\,x_2}=D_{x_3\,x_3}=D_{\perp}$. The diffusion equation in one direction in that case reads
\begin{equation}
\frac{\partial f(x_i,\,t)}{\partial t}=D_{x_i\,x_i}\,\frac{\partial^2 f(x_i,\,t)}{\partial x_{i}^{2}}
  \end{equation}
  with $i=1,\,2,\,3$, see \citet{shalchi_buch} for a detailed review. 
  
\begin{figure}[htbp]
\centering{
\includegraphics[trim = 0mm 0mm 0mm 0mm, clip, width=0.9\textwidth]{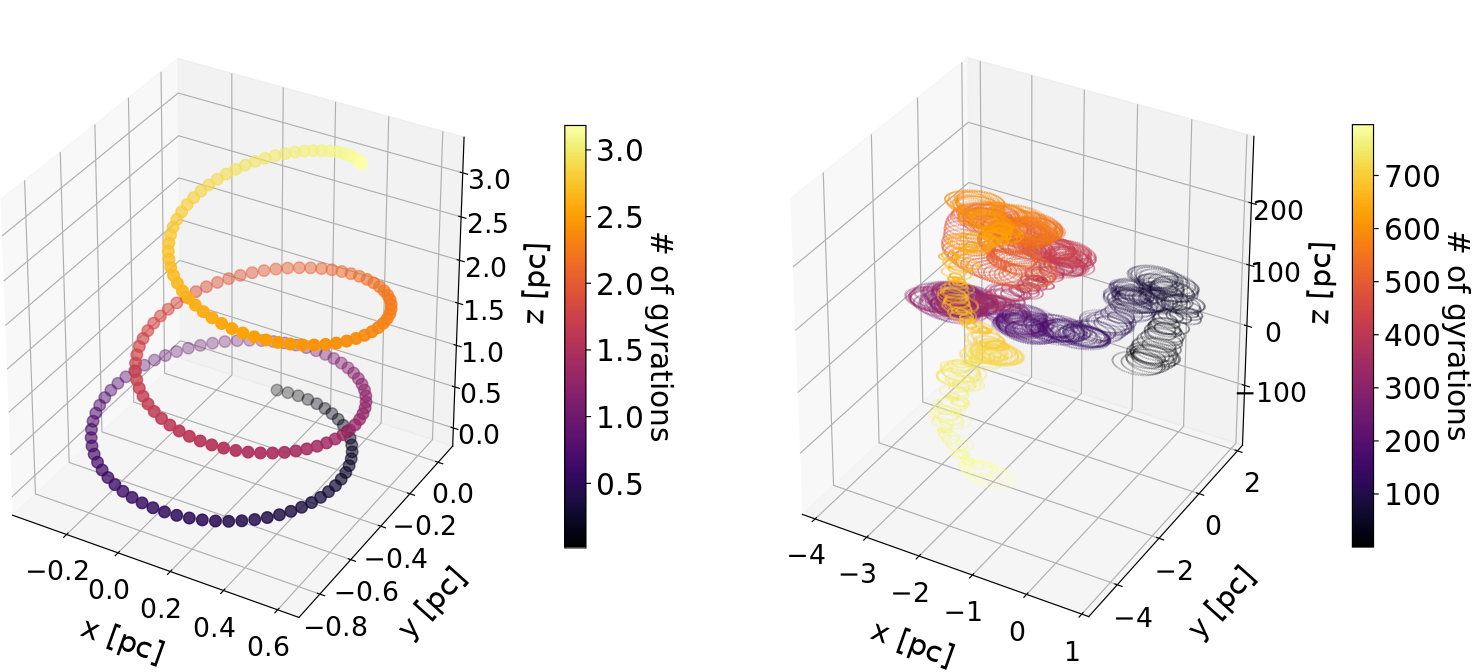}
\caption{Illustration of particle diffusion in a magnetic field $\vec{B}=\vec{B}_0+\delta \vec{B}$, with $\vec{B}_0$ as a homogeneous background component and $\delta \vec{B}$ as a turbulent component with a Kolmogorov-type power spectrum.  For this graph, the chosen parameters are:
$N_{\rm grid}=801^3$; $s=0.014$~pc; $l_d=1$~pc; $l_{\rm high}=4$~pc; $\delta B = 10^{-7}$~Gauss; $B_0=10^{-6}$~Gauss (aligned in z-direction) and $E=400$~TeV.
 Both panels show one particle trajectory, simulated via the numerical solution of the equation of motion. The particle is starting at the position $(x,\,y,\,z)=(0,\,0,\,0)$. The colors are indicating the number of gyrations (black indicating a low number of gyrations, yellow large number of gyrations, see scale). The left figure now is a zoom on the movement of the particle for the first 3 gyrations. The right figure shows the first 700 gyrations. 
  \textit{Figure courtesy: Patrick Reichherzer.}
\label{diffusion:fig}}}
          \end{figure}

The analytical solution for the injection of particles at a point source, $f(x_i,t=0)=\delta(x_i)$, is known\footnote{Here, the particle density is determined for one direction $x_i$, the density for all three directions can be constructed by building the product of the three densities and normalizing the distribution function to the total density at $\vec{x}$ and $t$.}:
\begin{equation}
    f(x_i,t)=\frac{1}{2\,\sqrt{\pi D_{x_i x_i}\,t}}\exp\left(-\frac{x_{i}^{2}}{4\,D_{x_i x_i}\,t}\right)
\end{equation}
The diffusion coefficient itself can now be determined by the following simulation setup: a large number of individual particles are emitted isotropically from a point source in a region of a homogeneous plus turbulent magnetic field.
The diffusion coefficient can then be determined by the second moment of the deviation of a particle's path from a straight line, i.e.\
\begin{equation}
  \left<\left(\Delta x_{i}\right)^2\right> =\int_{-\infty}^{+\infty}dx_i\,x_{i}^{2}\,f(x_i,\,t)=2\,t\,D_{x_i\,x_i}(t)\,.
  \label{deltax:equ}
\end{equation}
A general temporal development of the variance of the displacement can be written as $  \left<\left(\Delta x_{i}\right)^2\right>\propto t^{\sigma}$ and thus

\begin{equation}
D_{x_i\,x_i}(t)=\frac{\left<\left(\Delta x_{i}\right)^2\right>}{2\,t} \propto t^{\sigma-1}\,.
\end{equation}
The specific value of $\sigma$ depends on the regime in which the particles propagate:
\begin{enumerate}
\item In the \textbf{ballistic regime}, the change in direction is $\Delta x_i\propto t$, and thus the running diffusion coefficient also behaves as $t$ and $\sigma =2$. This case is typically realized at early times in a propagation, where the particles have not had the time to reach the diffusive limit yet, see Fig.\ \ref{diffusion:fig} and the discussion below.
\item The \textbf{super-diffusive regime} is defined for $1<\sigma<2$. Here, the propagation is anomalous. With respect to the typical Gaussian behavior of the mean square displacement, diffusion happens faster, see e.g.\ \citet{zimbardo2015}.
\item The \textbf{diffusive regime} is given for typical stochastic propagation in the Markovian diffusion regime, i.e.\ $\Delta x_{i}\propto \sqrt{t}$ and $\sigma=1$. This case is present for typical Gaussian turbulence. In simulation, this case is usually reached in the limit of $t\rightarrow \infty$, see Fig.\ \ref{diffusion:fig} and the discussion below.
\item The \textbf{Sub-diffusive regime} is defined for $\sigma <1$. This case occurs for instance if the displacement does not change over time, i.e.\ $\Delta x\sim \mathrm{const}$, as it is the case in the limit of low turbulence for the perpendicular component of the diffusion tensor.
\end{enumerate}

The different regimes of propagation have consequences for the behavior of the diffusion coefficient when simulating the propagation of single particles in a turbulent magnetic field. These will start in the ballistic limit $(\sigma = 2)$, as they have not had the time to reach a fully diffusive state. In the ballistic limit  parallel propagation yields $\Delta x_{\parallel}\propto t$, which results in a time-dependent diffusion coefficient, i.e.\ $D_{\parallel}\propto t$. The perpendicular component, on the other hand, is time-independent\footnote{apart from the gyro-motion}, i.e.\ $\Delta x_{\perp}=const$, corresponding to the sub-diffusive regime, so that is expected that the perpendicular component of the diffusion tensor decreases with time, i.e.\ $D_{\perp}\propto t^{-1}$.

Figure \ref{diffusion:fig} shows the concept of the simulation of single particle propagation in a ballistic (left panel) and diffusive (right panel) environment.
The propagation in both panels is done in a magnetic field composed of a homogeneous field going into $z-$direction ($\vec{B}_0$) and a Kolmogorov-type turbulent field $\delta \vec{B}$ with $\delta B/B=0.1$. The propagation for a particle with $E=400$~TeV is simulated via the solution of the equation of motion. The numerical propagation environment is given in the caption of Fig.\ \ref{diffusion:fig}.
In the diffusive regime ($\sigma = 1$), the stochastic motion of the particles is clearly visible, while in the ballistic regime ($\sigma=2$), the particle follows the gyro motion of an undisturbed field reasonably well.

The TGK formalism now implies that in the limit of long times, $t\rightarrow \infty$, the diffusion coefficient converges as
\begin{equation}
D_{x_i\,x_i}=\lim_{t\rightarrow \infty} D_{x_i\,x_i}(t)=\lim_{t\rightarrow \infty} \frac{\left<\Delta x_i\right>^2}{2\,t}\rightarrow constant\,.
  \end{equation}

Performing such simulations for different energies gives a quantitative prediction of the energy behavior of the diffusion coefficient in the limit of $t\rightarrow \infty$. This can be investigated for different wave vector spectra (e.g.\ Kolmogorov, Kraichnan), for different levels of turbulence, e.g.\ ranging from  $\delta B/B \ll 1$ to $\delta B/B \geq 1$. Further influence can come from the specific turbulence at work. While turbulence is chaotic by definition, it is not necessarily Gaussian in its distribution. Intermittent, non-Gaussian turbulence could change the energy dependence of the diffusion coefficient, even if the wave vector spectrum is the same, see e.g.\ \citet{shukurov2017,grauer2018}.\\

In recent years, the TGK method has been applied in order to test the prediction of QLT, in which a Kolmogorov-type turbulence spectrum, i.e.\ an isotropic turbulent field with a one-dimensional wave vector spectrum of $k^{-5/3}$ leads to the energy dependence of the parallel diffusion coefficients of $E^{1/3}$. The behavior of the perpendicular component in QLT is expected to change according to Equ.\ (\ref{dperp:equ}). 

The simulations depend on the following parameters that influence the physics of the system: first, the homogeneous magnetic field $B_0$ is chosen to be constant along one axis. The turbulent field strength is implemented according to the spectral behavior of the Kolmogorov law, i.e.\
\begin{equation}
W(k)\propto \left(\frac{k}{k_0}\right)^{-\lambda}
\end{equation}
with $\lambda =5/3$ in this one-dimensional case. Here, $k_0$ is the level of turbulence, which physically is given by the turbulence injection scale $\mathcal{L}=2\pi/k_{0}$. For the Galaxy, turbulence is believed to be injected by supernova explosions at a scale of $\mathcal{L} \sim 10-100$~pc. This scale in simulations should mirror the size of the simulation volume, $V_{\rm sim}=l_{\rm high}^{3}$, where the upper limit of the simulation should ideally be chosen to be $l_{\rm high}\geq \mathcal{L}$. The lower limit for the wave number at which the Kolmogorov spectrum cuts off in the simulation is therefore determined by the volume: $k_{\min}=2\pi/l_{\rm high}$. The upper limit is physically given by the dissipation scale, $k_{\max}=2\pi/l_{d}$ as discussed. For the numerical simulation, on the other hand, it is constrained by the grid size in the simulation, i.e.\ $2\cdot s\leq l_{d}$, with $s$ as the spacing between the grid points.  Thus, in these simulations, the choice of simulation parameters actually influences the physics of the smallest scales as well and requires interpretation: simulations are severely restricted by the number of grid points $N_{\rm grid}$. With currently typically available RAMs\footnote{At the moment, a typical local working station has around $16$~GB, larger-scale computing servers can reach more than $200-300$~GB. Only Supercomputers like JUWELS in J\"ulich, Germany have rams on the order of a few hundred TB.}, a maximum number of $N_{\rm grid}\sim 2^{10}-2^{11}$ grid points are a reasonable assumption for standard computing systems. Using a TOP500 supercomputer with hundreds of TB and thus a factor of $1000$ larger ram would improve the situation by a factor of $\sim 10$. This generates a limit for the possible largest scales, i.e.\ the injection scale: $\mathcal{L}\leq l_{\rm high}=N_{\rm grid}\cdot s/2 < N_{\rm grid}\cdot l_d/4$. As $N_{\rm grid}$ is limited in numbers, there is a direct correlation between the largest and smallest scale in the simulation. Therefore, in most simulations, the injection scale is chosen according to physical parameters, as it is believed to be better quantified as the dissipation scale, simply through the argument of injection through supernova remnants in the Galaxy. This naturally provides the simulation with an artificial dissipation scale $l_d$ and the results need to be interpreted accordingly. Most significantly, the energy range in which the diffusion coefficient is expected to behave as $E^{1/3}$ in the Kolmogorov case in simulations is typically smaller than in reality, at least according to what is suggested by the broad power-law of the wave vector spectrum (see Fig.\ \ref{big_power_law:fig}).

Simulations following the TGK method have been performed by a number of authors and a large range of values for $\delta B/B$ has been tested \citep[e.g.,][]{giacalone_jokipii1999,casse2001, demarco2007}. 
Most of the current literature is in accordance with a Kolmogorov-type spectrum leading to the behavior of the diffusion coefficient as expected in QLT, i.e.\ $\kappa=1/3$, even up to relatively large values of $\delta B/B\sim 1$ or even higher. However, several papers have pointed out that there exist tensions that need to be resolved \citep{minnie2007,lange2013,snodin2016,giacinti2018}. A first discussion of the importance of the choice of minimal and maximal length scales has been performed by \citet{snodin2016}, who reproduce the expected value of $\kappa=1/3$, but find deviations toward large energies.

The previous discussion is focused on the state-of-the-art concerning the broader context of the \textit{resonant scattering regime}. It ignores, however, the transition regions. The greatest challenge here is indeed to verify that the performed simulations are within this regime and do not reach into the transition regions in order to be able to interpret the results correctly. As one of the central aims of such studies is to investigate the energy behavior of the diffusion coefficient, it is of high importance to quantify the energy range of validity. The unvoluntary inclusion of parts of the mirror and/or quasi-ballistic regime in the spectral fit of the energy dependence of the coefficient might artifically change the spectral index.

The discussion of the Kolmogorov scale analysis is presented within the limit of QLT and the behavior of the diffusion coefficient in the limit $\delta B/B\rightarrow 1$ and larger is subject to investigation as we will discuss in more detail below. There is another restriction concerning the validity of the results, however: the calculation requires the particles to be in the \textit{resonant scattering regime}. As discussed above, the wave-vector spectrum is limited at low and high values of $k$ by the injection and dissipation scales, $l_d<l=2\pi/k<\mathcal{L}$, respectively. This defines the range in which particles can scatter resonantly on the resonance wave number
\begin{equation}
k_{\rm res}\sim (|\mu|\,r_g)^{-1}\,.
\label{kres:equ}
\end{equation}
Consider the propagation of a particle of charge $q$ in a fixed magnetic field with $B=|\vec{B}|$ at an energy $E$. The gyroradius in that problem is fixed, $r_g=r_g(E,\,B)$. This naturally gives the condition
\begin{equation}
  l_{d}<2\pi\,\mu\,r_g\,.
  \label{ld:equ}
\end{equation}

There is another limit to the problem that occurs due to the finite value of the minimal and maximal wave numbers.  Inverting Equ.\ (\ref{ld:equ}) in combination with Equ.\ (\ref{kres:equ}), there is a lower and upper limit to $|\mu|$:
\begin{eqnarray}
  |\mu|_{\rm low}&=&\frac{l_{d}}{2\pi\,r_g}=0.003\cdot \frac{l_d}{10^{9}\,{\rm cm}}\left(\frac{E}{{\rm GeV}}\right)^{-1}\cdot \left(\frac{B}{10\,\mu{\rm G}}\right)\,,\\
    |\mu|_{\rm high}&=&\min\left(1,\frac{\mathcal{L}}{2\pi\,r_g}\right)=\min\left(1,10^{4}\cdot \frac{\mathcal{L}}{100\,{\rm pc}}\left(\frac{E}{{\rm PeV}}\right)^{-1}\cdot \left(\frac{B}{10\,\mu{\rm G}}\right)\right)\,.
\end{eqnarray}
In general, pitch angle scattering is therefore limited. Inserting numbers that correspond to realistic parameters in the Galaxy reveals the following consequences for particle propagation in the Milky Way:

\begin{itemize}
\item {\bf lower limit for pitch angle scattering:}
  For values of $|\mu|\gtrsim \mu_{\rm low}$, scattering at large values of $|\mu|$ can happen more quickly, while at low values of $|\mu|\sim |\mu|_{\rm low}$, mirroring as already discussed in Section \ref{acc:sec} will occur. The mirroring criterion requires $B/B_{1}>p/p_{\perp}=|\mu|^{-1}$ for a particle starting at a total magnetic field $B$ and scattering at later position at which the total magnetic field is $B_1$.
  If we assume that the dissipation scale in the Galaxy is $l_d\sim 10^{9}$~cm, consistent with observations presented in Fig.\ \ref{big_power_law:fig}, a GeV particle has a lower limit to scattering angles of $|\mu|\sim 0.003$, corresponding to pitch angles of $89.8^{\circ}$. Thus, only a small part of the scattering regime close to $90^{\circ}$ is missing. 
  It has, however, not been investigated in detail if the mirroring effect close to $90^{\circ}$ plays a crucial role for the energy dependence of the diffusion tensor. Current investigations are ongoing \citep{patrick_master2018} and will be discussed in more detail below.
\item {\bf upper limit for pitch angle scattering:}
  If we assume that turbulence is injected by supernova remnants at $\mathcal{L}\sim 100\,$~pc, this effect would only become relevant at energies $E>10^{4}$~PeV. However, cosmic 
rays are expected to contribute to the self generation of turbulence in a significant way. If this is true, the wave vector spectrum will be influenced and possibly deviate from a 
single power-law even at lower scales. While data support that the spectrum does behave like a power-law up to a scale of $10-100$~pc, changes in the spectrum could exist. These could arise either locally in cosmic-ray birth places or even globally, as 
the wave vector spectrum has not been measured in great detail (see Fig.\ \ref{big_power_law:fig}). Such features in the spectrum would change the applicability of Equ.\ 
(\ref{esc_diffcoeff:equ}) \citep{blasi2012}. \citet{CRdrivenTurb} have shown that the break around $R\approx300$~GeV in the cosmic-ray energy spectrum can be explained by 
cosmic-ray self-induced turbulence. In that case energy is injected into the turbulent spectrum at two different wavelengths. This may lead to a broken turbulence spectrum which would lead to break in the energy spectrum, too. While a break in the turbulence spectrum is not clearly observed at this point, there are first indications that there is a change in the energy behavior of the Boron to Carbon ratio \citep{genolini2017}. It could also be an effect of local source physics \citep{ahlers2009,blasi2009,cholis2014,tomassetti_oliva2017}.
\end{itemize}

          \begin{figure}[htbp]
\centering{
\includegraphics[trim = 0mm 0mm 0mm 0mm, clip, width=0.9\textwidth]{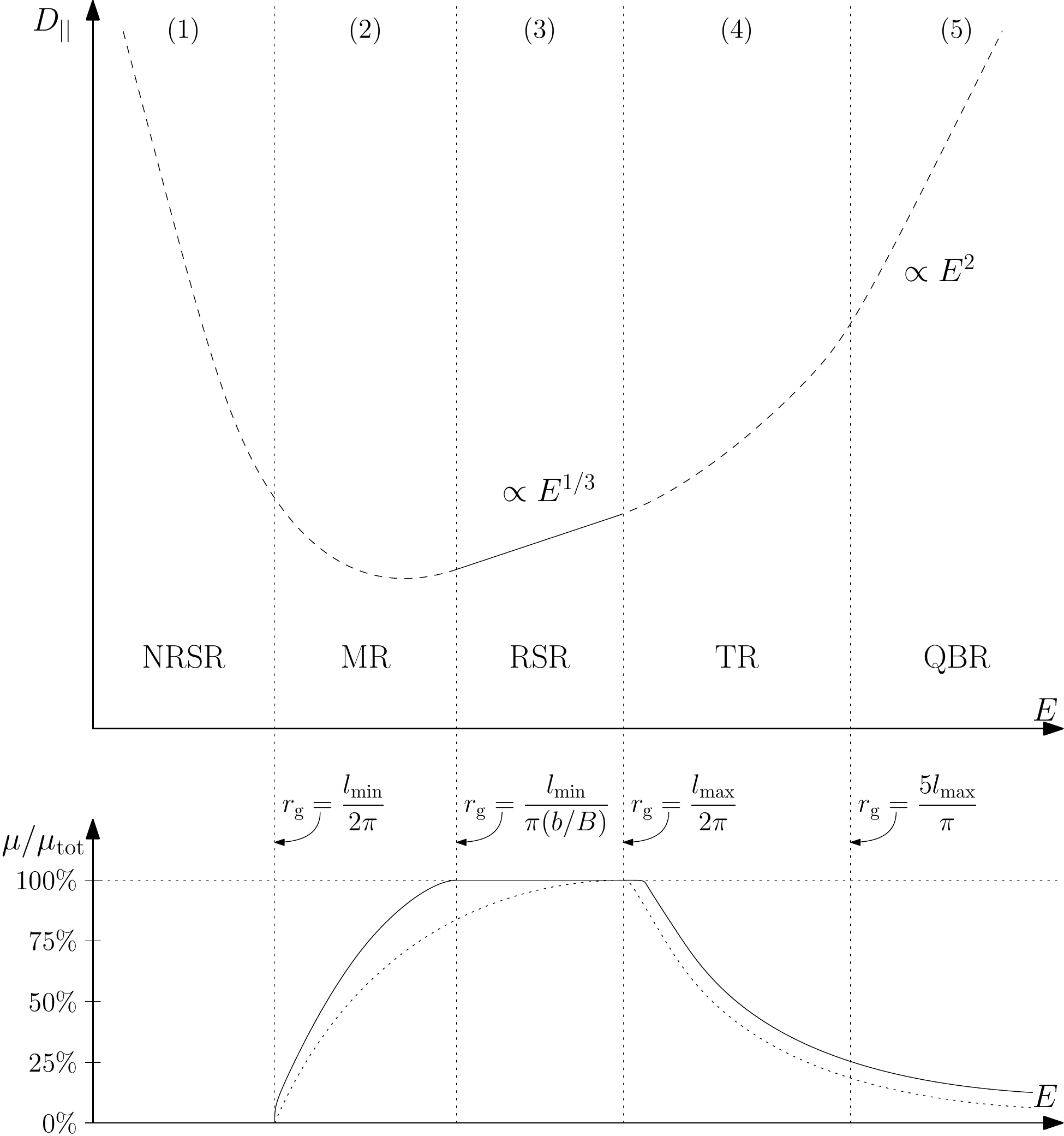}
\caption{Schematic view of the energy behavior of the diffusion coefficient for a constant fraction $\delta B/B$ and $l_{\min}$ (upper panel). The lower panel shows the fraction of angles 
$\mu=\cos{\theta}$ available for scattering for a certain energy range. While the dashed line shows the contributions from fully resonant scattering only, the solid line adds the contribution from scattering via the condition $\delta \mu = \delta B/B$ as discussed in the text\label{dparallel_schema:fig}}
}
          \end{figure}

          The behavior of the diffusion coefficient can therefore be classified in five regimes as we discuss below. The relevant length scale for the scattering process is the \textit{correlation length} $l_c=2\pi/k_{\rm res}$, on which the particles scatter resonantly. Figure \ref{dparallel_schema:fig} visualizes the concept of the change of diffusion coefficient depending on the energy of the particles. The upper part shows the expected  energy dependence of the parallel diffusion coefficient. The different regimes are described below. The lower panel shows the percentage of the range of $|\mu|$ that is available for scattering. Here, 100\% corresponds to the full range of $0<|\mu|<1$. The five different regimes are the following \citep{reichherzer2019}:\\
\begin{enumerate}
\item {\textbf{Non-resonant scattering regime, NRSR ($2\pi\,r_g/l_d < 1$)}}\\
  As discussed above, particles cannot scatter resonantly at values $l_d>2\pi\,r_g$, i.e.\ $E< l_c\cdot q \cdot B\cdot c/(2\pi)$. This is the lowest-energy regime in which no pitch angles are available for scattering (see Fig.\ \ref{dparallel_schema:fig}, lower panel). 
\item{\textbf{Mirror regime, MR ($l_d/(2\pi)<r_g < l_d/(\pi\,(\delta B/B_0))$)}}\\
At higher energies, particles
  start to find waves they can scatter with for certain angles (see lower panel of Fig.\ \ref{dparallel_schema:fig}). There are two effects acting on the particle with opposite action: As compared to the resonant scattering regime described below, the mirror regime has two effects: (a) only a fraction of pitch angles are available for scattering (see Fig.\ \ref{dparallel_schema:fig}, lower panel). This enhances diffusion and thus would increase the diffusion coefficient the stronger the effect becomes. For the mirror regime, this means that this effect increases the diffusion coefficient toward lower energies, where less and less pitch angles are available for scattering. (b) mirroring becomes more probable the smaller the range of available $|\mu|$. This in turn works to reduce diffusion and counter-works effect (a), damping the increase of the diffusion coefficient toward low energies. 
\item{\textbf{Resonant scattering regime, RSR ($l_d/(\pi\,(\delta B/B_0))<r_g<\mathcal{L}/(2\pi)$)}}\\
This energy range is what is usually referred to in the literature when the parallel diffusion coefficient is derived, see e.g.\ \cite[e.g.]{minnie2007,snodin2016,effenberger2012}. The factor $\delta B/B$ enters the boundary as a gyroresonant wave produces a change in pitch angle of the order $\delta \mu = \delta B/B_0$. This translates into a factor at the lower boundary as described in \citep{reichherzer2019} via the connection of the spatial and pitch angle diffusion coefficients.
As described above, a Kolmogorov-type wave vector spectrum leads to a behavior $\propto E^{1/3}$ in QLT, but this limit is only valid for values $\delta B/B\ll 1$ as we discuss in more detail below. 
\item {\textbf{Transition regime, TR ($\mathcal{L}/(2\pi)<r_g<5\mathcal{L}/\pi$)}}
The transition toward the quasi-ballistic regime happens when the maximum scale of turbulence is reached, $\mathcal{L}$. Here again, a certain fraction of pitch angles are available for scattering until the quasi-ballistic regime is reached. The transition region is expected to be about one oreder of magnitude until the quasi-ballistic regime is reached \citep{globus2008}.
\item {\textbf{Quasi-Ballistic regime ($r_g>5\mathcal{L}/\pi$)}}\\
Once the particles propagate quasi-ballistically\footnote{The diffusion-plateau is still reached and thus, the propagation is not fully ballistic},
   QLT predicts that the diffusion coefficient changes very quickly with energy. The diffusion coefficient within QLT is expected to behave as $D_{\parallel}^{\rm nr}\propto E^{2}$ (see e.g.\ \citet{sigl2017,subedi2017}).
\end{enumerate}
With a proper definition of the transition regions out of the resonant scattering regime, the range of the validity of a power-law can therefore be refined as done by \citet{reichherzer2019}. In doing so, it was shown that above a turbulence level of $\delta B/B_0\sim 0.07$, the QLT-index of $\kappa_{\rm QLT} = 1/3$ is \textit{not} reached. In the paper, it is discussed that the Bohm-limit of an index $\kappa \rightarrow \kappa_{\rm Bohm}$ for large values of $\delta B/B_0$ is reached at around $\delta B/B_0\sim 1$. Going toward lower values of the turbulence level decreases the diffusion index. At the lowest level that could be simulated, i.e.\ $\delta B/B_0=0.07$, the index reaches $\kappa\sim 0.6$. It is expected that at some point, for sufficiently small values, the QLT-case for the index of $\kappa=\kappa_{\rm QLT}=1/3$ will be reached. However, at this point, these cannot be reached as particularly the small-grid scales necessary in the simulations are not accessible due to interpolation problems \citep{schlegel2020}. This has direct consequences for cosmic-ray propagation in the Galaxy, but also in other astrophysical environments, as the turbulence level can be significantly higher than $\delta B/B_0\sim 0.07$ and thus, the change in the index certainly needs to be taken into account.

\subsubsection{Using spallation to determine the Galactic scalar diffusion coefficient \label{spallation:sec}}
The ratio of Boron to Carbon at Earth can be used as a measure to determine the scalar diffusion coefficient. A simlified version of the transport equation, only including the source term and the spallation loss and gain terms can demonstrate this fact: Assuming that Boron changes its abundance through the spallation of Carbon nuclei, the transport equation for Boron in the leaky box approximation can be written as
\begin{equation}
\frac{\partial n_B}{\partial t}\approx 0 = Q_B-\frac{n_B}{\tau_{esc}}-\frac{n_B}{\tau_B}+\frac{P_{C\rightarrow B}\,n_{C}}{\tau_C}\,.
\end{equation}
with $P_{C\rightarrow B}$ as the probability that a Carbon nucleus turns into Boron via spallation. The diffusion term has again been approximated as an escape term, $-n_B/\tau_{esc}$. The fundamental solution of the problem outside of the sources ($Q_B=0$) can thus easily be determined:
\begin{equation}
\frac{n_B}{n_C}\approx \frac{P_{C\rightarrow B}}{\tau_C}\cdot \left(\frac{1}{\tau_{esc,B}}-\frac{1}{\tau_{B}}\right)^{-1}
\stackrel{\tau_{esc,B}\ll\tau_B}{\approx} \frac{P_{C\rightarrow B}}{\tau_{C}}\cdot \tau_{esc,B}\,.
\end{equation}
With known (constant) spallation probabilities, the energy dependence of the ratio of Boron to Carbon elements is thus determined by the escape time, which in turn can be expressed in form of the scalar diffusion coefficient (Equations (\ref{escape_time:equ}) and (\ref{esc_diffcoeff:equ})), so that the energy dependence of the Boron to Carbon ratio reflects the one of the diffusion coefficient:
      \begin{equation}
\frac{n_B}{n_C} \propto D_{\parallel}^{-1}\propto E^{\lambda-2}\,.
      \end{equation}
      In the limit of QLT, the energy dependence of the Boron to Carbon ratio would therefore become $E^{-1/3}$ (Kolmogorov) or $E^{-1/2}$ (Kraichnan), as discussed above. State-of-the-art models of the turbulent plus homogeneous components of the Galactic magnetic field suggest that the turbulence level in parts of the Galaxy can become significantly larger than the QLT limit, i.e.\ $\delta B/B\gg 1$, see e.g.\ \citet{jansson_farrar2012,kleimann2019,shukurov2019}. Thus, a full treatment of the diffusion coefficient as reviewed in Section \ref{diffusion_coeff:sec} is needed in order to understand the diffusion behavior of cosmic rays in the Galaxy. Data are becoming more and more accurate, see Section \ref{data:sec} for a discussion of the results and point toward a scalar diffusion necessarily coefficient in a relatively narrow range, $E^{1/3 - 0.5}$ \citep{trotta2011,evoli2012,tomassetti2012,cholis2017}. These data start to reveal the average propagation properties in the Galaxy. As we review in Section \ref{multimessenger_modeling:sec}, multimessenger modeling of the Galaxy is starting to shed light on the more complex propagation behavior of cosmic rays in different parts of the Galaxy.\\

\subsection{Spectral evolution of the cosmic ray composition \label{composition:sec}}
The evolution of the composition of cosmic rays is complex: it changes significantly with energy and it can only be measured directly up to rigidities of $\sim 10^{5}$~GV. In 
addition, at the low energy end of the spectrum, below $\sim 10$~GeV, it is time-dependent due to solar modulation \citep{0004-637X-829-1-8}. In recent years, great progress was 
made to determine the composition for different energy ranges as discussed in Section \ref{data:sec}:
\begin{itemize}
\item In the MeV-GeV range, the Voyager-I spacecraft could measure the interstellar cosmic-ray spectrum for the first time, Voyager-II has now also passed beyond the 
heliosphere \citep{voyager_ii_2018}. This way, the interstellar cosmic-ray spectrum in this range without the influence of the solar magnetic field can be studied.
\item 
  In the GeV-TeV-range, CREAM, PAMELA and AMS-02 could recently shed light on the spectral evolution of the composition.
\item In the PeV range, KASCADE, KASCADE-Grande and IceTop now provide a first resolution of elements.
\item In the EeV range, Auger and TA have published first pieces of information on the composition.
\end{itemize}
The fact that the full composition -- or at least the tendency light to heavy -- is now known from below GeV to above EeV energy makes it possible to consider a fit of the broad-band cosmic-ray spectrum. Several authors have done this in recent years, most of them following the phenomenological description of a sum of source classes. The basic assumption\footnote{at least in the limit of QLT, for non-linear diffusive shock acceleration, the spectra become curved.} typically is that all components follow a rigidity-dependent power-law spectrum at Earth with an exponential cut-off, $dN/dE\propto R^{-\gamma}\,\exp{(-R/R_{\max})}$ (see Section \ref{spectrum_acceleration:sec}). The cut-off at high energies is also set to be rigidity-dependent, following the Hillas-argument (see Section \ref{hillas:sec}). For the \textit{energy} spectrum this means that protons bend at the lowest energies, iron bends last, following $E_{\max}\propto Z\cdot R_{\max}$. This concept is often referred to as \textit{Peters Cycle} after the original proposal by \citet{peters1961}.

\begin{figure}[htbp]
\centering{
\includegraphics[trim = 0mm 0mm 0mm 0mm, clip, width=\textwidth]{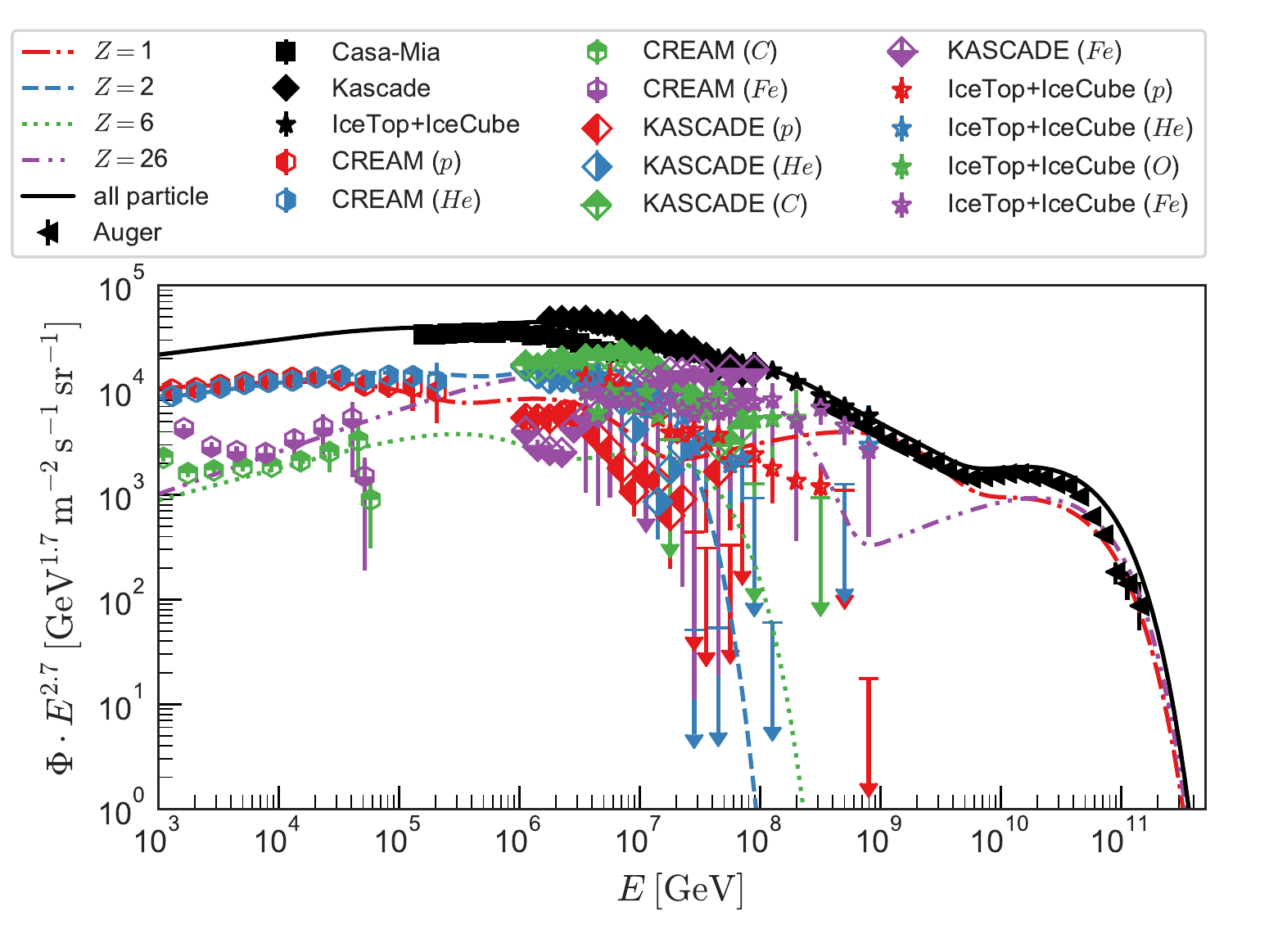}
\caption{The all-particle (black) and composition-resolved cosmic-ray spectral data are shown together with the model presented in \citet{serap2013} which is based on 4 populations. Data: CREAM \citep{0004-637X-728-2-122}, CASA-Mia \citep{CRDB_CasaMIA}, KASCADE \citep{ANTONI20051}, IceTop+IceCube \citep{AndeenPlum:2019}, Auger 
\citep{Fenu:2017hlc}.}
\label{composition_all:fig}
}
\end{figure}

\begin{figure}[htbp]
\centering{
\includegraphics[trim = 0mm 0mm 0mm 0mm, clip, width=\textwidth]{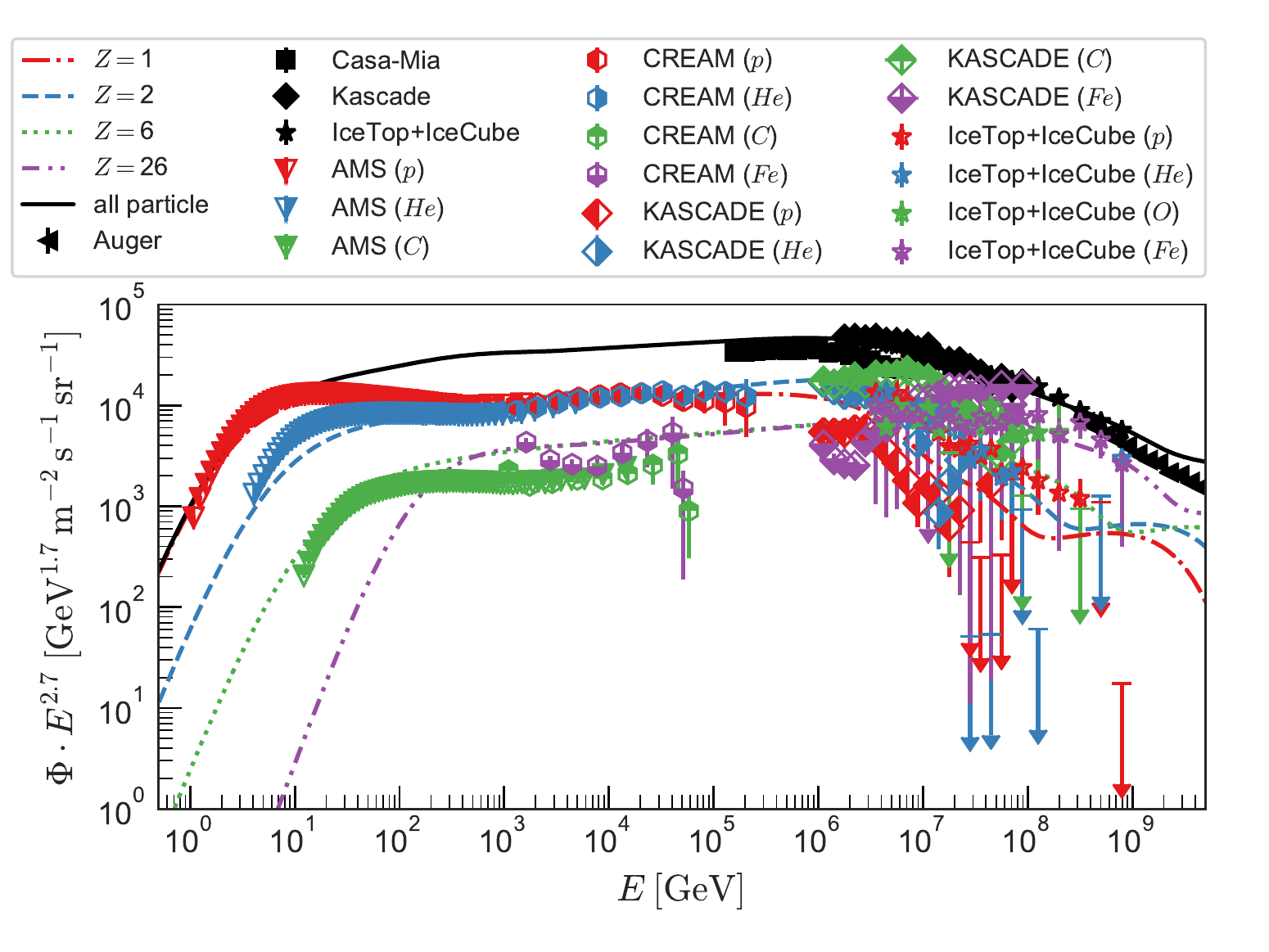}
\caption{The all-particle (black) and composition-resolved cosmic-ray spectral data are shown together with the model presented in \citet{fedynitch2012}. Here, models of \citet{gaisser2002, gaisser2012}, using a mixed composition, have been combined in order to fit a larger range of the energy spectrum. Shown data are from AMS-02 \citep{ams2015a, ams2015b, ams2017}, CREAM \citep{0004-637X-728-2-122}, CASA-Mia 
\citep{CRDB_CasaMIA}, KASCADE \citep{ANTONI20051}, ceTop+IceCube \citep{AndeenPlum:2019}, Auger 
\citep{Fenu:2017hlc}
\label{Composition_combined:fig}.}
}
\end{figure}

A first approach following the concept of the Peters cycle is the so-called \textit{Polygonato-model (short: PG)} introduced by \citet{polygonato,hoerandel2004}.
As more and more data shed light on the more complex structure of the spectra, more detailed models were presented with a focus on different parts of the energy spectrum. In 
\citet{gaisser2012}, a five-component approach was presented for the modeling of the cosmic-ray spectrum in the range of $100$~TeV to $100$~EeV in which three populations are used to fit the 
spectrum. In \citet{fedynitch2012}, the forementioned model was combined with a previous 
parametrization by \citet{gaisser2002} of the energy range below $100$~TeV in order to receive a model that fits a large energy range. This combined model (labeled \textit{comb.GH.G12} for the combined Honda-Gaisser/Gaisser2012 model) is 
presented together with the data in Fig.\ \ref{Composition_combined:fig}. Even though this model is quite recent and can reproduce the all-particle spectrum well, it still 
does not reproduce the individual components of the composition very well. The same is true for a more recent model presented by \citet{serap2013} (referred to as the \textit{GST13 model}). In addition to the five-component model of \citet{gaisser2012}, this model introduces two more components, i.e.\ around the heavy elements Xenon and Mercury.
 A fourth population of only protons is added in order to reduce the $\ln A$ values at around $10^{9}$~GV and this way fit this observable much better.
The GST-model is optimized with up to four 
populations between $100$~TV and $100$~EV and could possibly be extended to around $1$~TV, but does not describe the data below, in particular not the change in the spectrum at 
$\sim 300$~GV.
The all-particle spectrum is well-reproduced, the spectra of the individual components cannot fully be explained (see Fig.\ \ref{composition_all:fig}). In particular, none of the  
models is able to describe the element-resolved AMS-02 measurements in all detail or equivalently all components 
of the 
KASCADE observations (see Figures \ref{composition_all:fig} and \ref{Composition_combined:fig}). Another recent attempt to describe the cosmic-ray flux is presented in \citep{splineFit}, referred to as \emph{Global Spline Fit}. At the time of this work the code is not publicly available, yet. Here, the authors do not fit a specific model to the cosmic-ray flux but 
describe the available data using smooth splines. The advantage here is that such a spline fit is fully model-independent and is therefore very useful to calculate the abundance or investigate the features of the all-particle spectrum and the individual components.

\begin{figure}[htbp]
 \begin{minipage}{.5\linewidth}
 \centering
  \includegraphics[width=\linewidth]{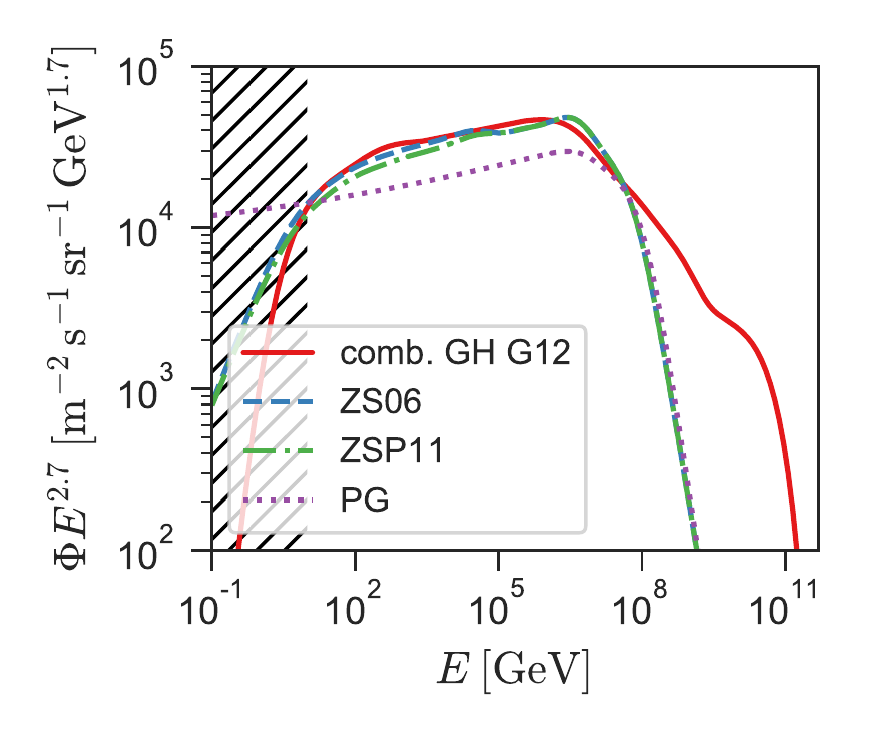}
 \end{minipage}%
  \begin{minipage}{.5\linewidth}
  \centering
    \includegraphics[width=\linewidth]{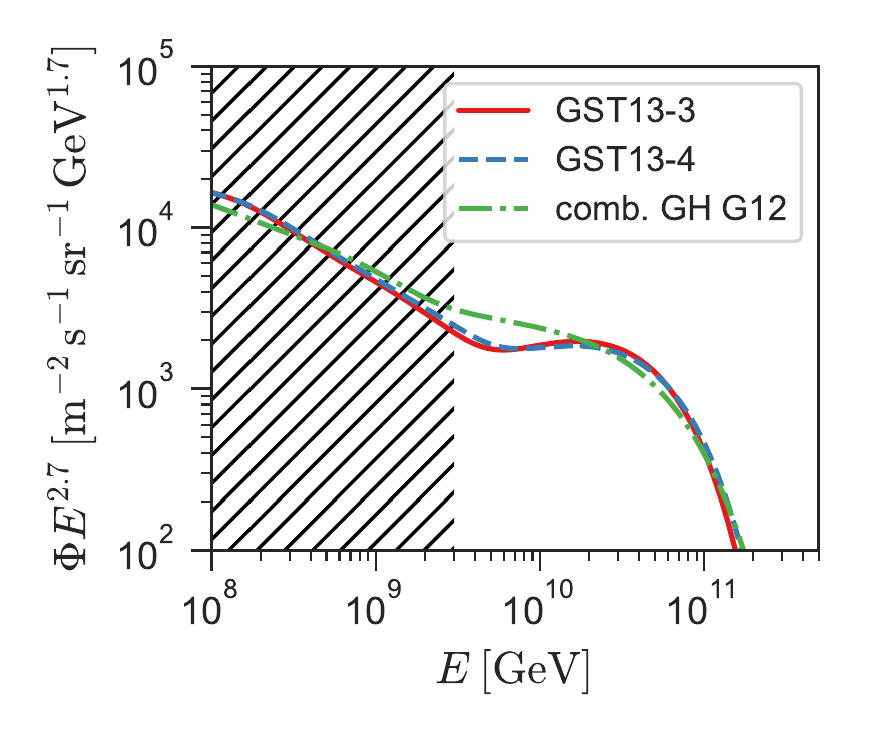}
 \end{minipage}
\caption{Models used for the calculation of the Galactic (left) and extragalactic (right) cosmic-ray power shown  in Tabs.\ \ref{tab:GalacticCRPower} and 
\ref{tab:UniverseCRPower}. Grey dashed area not included for the 
flux integration. --- Models: \emph{comb.GH.G12} \citep{fedynitch2012}, \emph{ZS06} \citep{Zatsepin2006}, 
\emph{ZSP11} \citep{PamelaFlux2011}, \emph{PG} \citep{polygonato}; \emph{SGT14} \citep{SGT2014}, \emph{GST13} \citep{serap2013}, \emph{comb.GH.G12} \citep{fedynitch2012}}.
\label{fig:FluxModels}
\end{figure}

Using generic models, the global fit models described above present first attempts to fit the all-particle spectrum and the individual components over a wide range of energies. Above, we present two models, i.e.\ \textit{comb.GH.G12} and \textit{GST13} in detail. Other models, i.e.\ \emph{ZS06} \citep{Zatsepin2006} and
\emph{ZSP11} \citep{PamelaFlux2011} provide similar descriptions. All of these models fit a very wide range of energies, but typically have a certain energy region of interest in which they works better than for other energies. In  Fig.\ \ref{fig:FluxModels}, we show those models that work well for the description of the spectrum below (and partially even above) the ankle (left panel). The right panel shows those models that have been optimized to fit the high-energy emission around the ankle. In the following, we will refer to the  models in the left panel as \textit{low-energy models} (PG, comb.G.GH12, ZS, ZSP) and to the models in the right panel as \textit{high-energy models} (comb.G.GH12, GST13 with three or four components). In particular for the calculation of the cosmic-ray luminosity, we will use the corresponding model, depending on which energy range we are looking at.

The models above present broad-band fits of the composition-resolved energy spectra of cosmic-rays. Other models have been presented that actively base their fits on a certain source model, see e.g.\ \citep{stanev1993,thoudam2016}. We do not present them in the above-context in order to stay as source-model independent as possible. Another alternative is to use the composition from the data directly. Different data bases exist from which the data can be drawn, see e.g.\ \cite{maurin2020,kascade2020}. Here, the problem is that the binning of the data typically differs for different experiments, and also that error propagation can be difficult depending on the existing public information on the errors. 

The most advanced models at this point are (1) the \textit{comb.GH-G12} --- it is the only one that connects the low-energy and high-energy range with a state-of-the-art parametrization. (2) \textit{GST13} --- while it is only valid above TeV-energies, it is the only model so far which reproduces the observed $\ln A-$behavior around and above the ankle.

The discussion above summarizes one of the difficulties of the broad-band approach: with a generic approach of a three-population power-law model, it is still difficult to describe all features. So-far undetected features cannot be predicted with this approach. In addition, there \textit{is no global theory}, but as discussed in this review, it is a viable option to use different source classes for different parts of the spectrum. And it is not entirely sure yet which ones.

Thus, in the following, we will present a short summary of physics ideas that describe the different energy ranges from MeV to TeV (Section \ref{mev:sec}), from TeV to EeV (Section \ref{tev:sec}) as well as the range above the ankle (Section \ref{eev:sec}). Details on Galactic source models will be presented in Section \ref{multimessenger_modeling:sec}.
Before going through these arguments, however, an analysis of the cosmic-ray abundance at different energies is made in the following Section \ref{crabundance:sec}.

\subsubsection{Cosmic-ray abundance \label{crabundance:sec}}
In order to receive first insights into the composition of cosmic rays, the energy integrated particle flux of different elements $N_i$ is determined, from which the abundance of 
the elements relative to the all-particle integrated flux can be calculated:
\begin{align}
N_i = \int_{E_\mathrm{min}}^\infty \Phi_i(E) \,\mathrm{d}E \quad .
\end{align}
As the composition is rigidity-dependent (see Section \ref{allparticle:sec}) and the flux decreases strongly with energy/rigidity, the result is dominated by the minimal energy 
of the integration. It also depends on the model that is chosen to 
describe the particle flux. Here, we use the two cosmic-ray  models that are described above, which cover large energy ranges, i.e.\ the one presented by \citet{fedynitch2012} which 
is a combination of the two models \citep{gaisser2002} and \citep{gaisser2012}, short \textit{comb.GH.G12}. This model can be applied from GeV-energies up to the ankle. The model presented by \citet{serap2013} (short \textit{GST13},
will be used for a comparison, it can be applied from $\sim 1$~TeV up to $10^{20}$~eV. It is important to note that the composition is not fully well-represented by these models. 
But as the binning for different elements at the same energy is not necessarily the same, it is not possible to use the data themselves at this point in order to calculate the 
composition.
 The total composition, integrated from a minimum energy $E_\mathrm{min}=1$~GeV can only be calculated in the model-approach of \citet{fedynitch2012}. According to this model the cosmic ray composition is dominated by protons ($\approx 87.1$~\%) and helium ($\approx 11.9$~\%) but medium heavy elements, e.g., carbon, nitrogen and oxygen ($\approx 1.0$~\%), and heavier elements, the so called iron group, ($\approx 4\times 10^{-4}$) are found, too. This is very similar to the composition of elements in the solar system.  
Already a minimal energy $E_\mathrm{min}=10$~GeV, above which an influence of the solar modulation is very unlikely, changes the composition significantly: The abundance of protons goes down to $\approx 69.0$~\%, while the helium abundance becomes $\approx 25.0$~\%, the CNO abundance becomes $\approx 4.3$~\%, and the iron component is around $\approx 0.4$~\%.

\begin{figure}[htbp]
\begin{minipage}{.49\linewidth}
\includegraphics[trim = 0mm 0mm 0mm 0mm, clip, width=\textwidth]{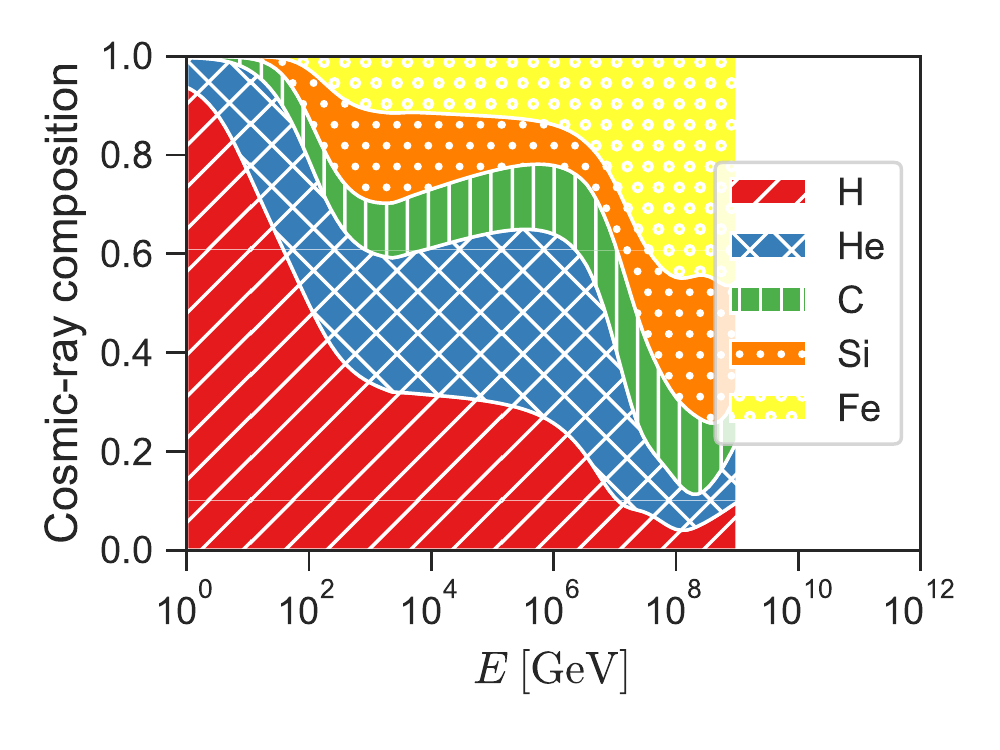}
\end{minipage}
\begin{minipage}{.49\textwidth}
\includegraphics[trim = 0mm 0mm 0mm 0mm, clip, width=\textwidth]{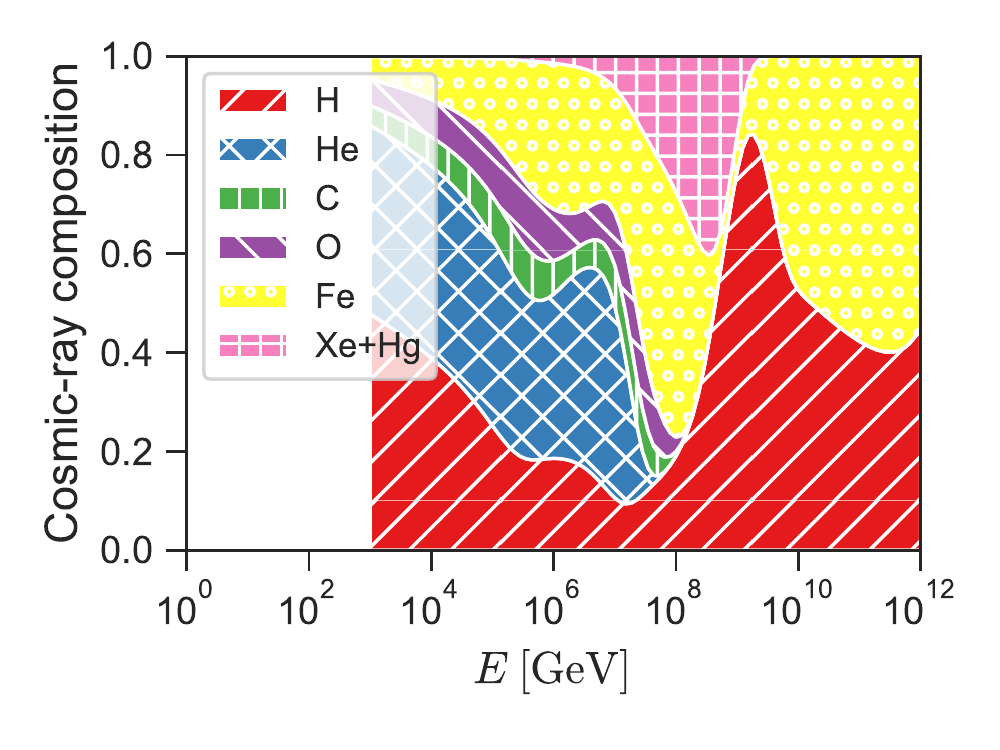}
\end{minipage}
\caption{The differential abundance of elements according to Equ.\ (\ref{ci:equ}) as a function of energy for different broad-band cosmic-ray models: (left) \citet{fedynitch2012} combining \citet{gaisser2002, gaisser2012} (mixed composition) and (right) \citet{serap2013} using the presented four population model.}
\label{Composition_stacked:fig}
\end{figure}

The energy-integrated composition is usually dominated by the lower integration boundary due to the general power-law shape of the cosmic-ray energy spectrum. The differential 
composition at $E_\mathrm{min}$ is thus a good approximation for the energy-integrated composition with $E_{\min}$ as the lower integration boundary.
Figure \ref{Composition_stacked:fig} shows the percentual differential composition, given by:
\begin{align}
 C_i(E) = \frac{\Phi_i(E)}{\sum_j \Phi_j(E)} 
\label{ci:equ}
\end{align}
for the two different models.  Starting at low energies, CREAM data reveal a turnover in the proton and helium flux at around $10^4$~GeV \citep{0004-637X-728-2-122} which is also 
present in both models.  Looking at the model presented in \citet{fedynitch2012} (left panel), the heavy component steadily becomes more important with increasing (minimum) 
energy. In particular, around $10^{7}$~GeV, there is a pronounced kink above which more than 40\% of the composition is made up by iron, corresponding to the iron knee as discussed 
in Section \ref{data:sec}. It should be kept in mind, however, that in particular the KASKADE/KASKADE-Grande data that provide the most relevant pieces of information in that 
energy range, are not well-represented by this or the GST13 model.
Thus, in general, the GST13 model (right panel) shows a somewhat different behavior 
of the composition, in particular above the knee. As already discussed above, \citet{serap2013} introduce two more components, Xenon plus Mercury (labeled \textit{Xe+Hg} here). These heavy elements contribute significantly at around $10^{8}$~GeV ($\sim 40\%$). The iron component 
contributes with up to $30\%$ at around $10^{7}$~GeV. Then, the very heavy Fe+Xe+Hg component reaches up to $80\%$ in differential flux around  
$10^{8}$~GeV. At the highest energies, i.e.\ $10^{20}$~eV, the model reproduces a mixed composition (H+Fe), consistent with the increasing mass number reported by Auger and TA above the ankle. 

In general, both models can reproduce the most important features of cosmic-ray composition: the change at around $10^{4}$~GeV and the iron knee. Still, as discussed above, those phenomenological models aiming at the description of the cosmic-ray energy spectrum from GeV to ZeV energies have difficulties to properly reproduce all aspects of the evolution of the composition. The main reason is that the models are designed to be quite generic with as little model input as possible in order to keep the number of free parameters small. Reality shows us that much more details are needed for a full parametrization of the entire energy range. The following paragraphs therefore summarize the state-of-the-art on different model ideas of the different energy ranges.

\subsubsection{MeV - TeV cosmic rays \label{mev:sec}}
The energy range for cosmic rays from MeV to several GeV is highly influenced by the solar magnetic field. The intensity of the flux changes significantly with the 22-year cycle 
of the solar magnetic field (e.g.\ \citet{0004-637X-829-1-8}). We refrain from discussing this influence here as we are rather interested in the properties of cosmic rays when they 
enter the solar system. The first human satellite that could exit the heliosphere was Voyager-I in 2012. It provides us with data of the interstellar 
cosmic-ray spectrum and composition as discussed in Section \ref{data:sec}. Most recently, Voyager-II also exited the heliosphere, soon being able to compare the interstellar 
cosmic-ray spectra with Voyager-I as well as providing us with the plasma properties at the boundary of the heliosphere \citep{voyager_ii_2018}. At energies above $10-100$~GeV, 
a power-law spectrum is present for all nuclei. At $E<10-100$~GeV the spectrum hardens. Such a turn-over is expected from cosmic-ray ionization, see e.g.\ \citet{nath_biermann1994}. As recently discussed by 
\citet{schlickeiser14}, the Voyager spectrum that provides data up to around $1$~GeV can be fitted with a cosmic-ray transport model that includs adiabatic cooling, ionization and pion production. Another possibility is a  change of propagation from advection dominated ($<10$~GeV) to diffusion-dominated ($>10$~GeV) \citep{blasi2012}.

At energies above $10$~GeV, a detailed view on the spectrum is provided by  AMS-02, PAMELA and CREAM. Most interestingly, the proton spectrum behaves differently than that of heavier nuclei. 
For heavy nuclei, the spectrum is close to $\sim E^{-2.75}$ for $R\lesssim 300$~GV (for exact fits to the data by the experiments, see Table \ref{tab:ams} in Section 
\ref{data:sec}). Toward higher rigidities, it flattens by about $\Delta\gamma \sim 0.1$ or more to $dN/dR \sim R^{-2.65} - R^{-2.5}$. This behavior seems to be the same for all 
elements between Helium and Iron, at least data are compatible with the assumption that these elements behave in the same way. Protons, on the other hand, show a steeper spectrum 
(about $\Delta \gamma \sim 0.1$  steeper as compared to the heavier nuclei), i.e.\ $\sim R^{-2.85}$. The break energy seems to be comparable to the other elements at $R_{\rm 
break}\sim 300$~GV. At higher rigidities, the spectrum continues as $\sim R^{-2.7}$, again about $\Delta \gamma \sim 0.1-0.2$ steeper.

Thus, there are two things that need to be explained:
\begin{enumerate}
  \item What process leads to a break in the spectrum at the same rigidity for all elements?
\item Why does the proton power-law index differ from the heavier elements Helium to Iron, why do the elements Helium to Iron exhibit the same indices?
  \end{enumerate}
Concerning question (1), there are two possibilities to produce a rigidity-dependent break, which could then be connected to question (2). We will discuss the general ideas below, while more details will be provided when we review the source modeling in Section \ref{multimessenger_modeling:sec}.
\begin{itemize}
\item \textbf{A source-related break}\\
  A kink in the spectrum can be produced by the mechanism in the source itself. Here, the most prominent example is that the accelerator reaches its maximum energy $E_{\max}\propto Z$. In this case, 
the largest contribution by hydrogen cuts off first and the heavier elements follow toward higher energies. This scenario does provide a change at the same rigidity. However, it does not work here, as what is needed is an \textit{upturn} of the spectrum, not a faster decrease. For a change in the spectrum that corresponds to an upturn, a new component 
is needed. For a source-related scenario, this could be a change in the acceleration mechanism in the same source. As proposed in \citet{stanev1993,biermann2010}, an upturn of the 
spectrum at the same rigidity is expected from supernova remants. Their magnetic field structure as a reminiscence of the progenitor star magnetic field, follows a Parker spiral. 
As discussed in \citet{biermann2010,biermann2018} and as summarized later in this review in Section \ref{multimessenger_modeling:sec}, this axis-symmetric structure leads to the 
emission of basically two components that cross at a fixed rigidity. The location of the break depends on the intensity ratio of the two components and is compatible with what is 
measured by CREAM, PAMELA and AMS-02 --- a break close to $300$~GV (see Section \ref{data:sec}). The difference between the hydrogen spectrum and the heavier elements can be 
explained in this scenario, as two populations of SNRs exist --- higher-mass stars explode after having lost their hydrogen envelop. While these so called wind-SNRs 
are expected to accelerate cosmic-rays to above-knee energies \citep{stanev1993}, SNRs from lower-mass stars will dominate the hydrogen spectrum as in the explosion, hydrogen is 
still present in larger parts as compared to those stars which already ejected the hydrogen shell. These SNRs that produce shocks interacting with the magnetized ISM will not reach knee energies, but can dominate the spectrum at the lowest energies and in particular the 
hydrogen part. The steep component of the wind-SNRs is dominant for the heavier elements up to $\sim 300$~GV, where the flatter component takes over. While the flat component cuts 
off at knee-energies, the steep component continues to above the knee, providing a Galactic contribution in the shin region of the cosmic-ray knee. In such a way, the source class 
of SNRs is able to reproduce the entire spectrum of (Galactic) cosmic rays. As noted above, this does not only contain a change in the acceleration mechanism of one and the same 
source, but also a change in (sub-) population by dividing the class of SNRs in ISM-SNRs and wind-SNRs as also noted in \citet{tomassetti2015a,tomassetti2015b}. More details on the 
specific source modeling are presented in Section \ref{multimessenger_modeling:sec}. An alternative scenario is presented in \citet{ptuskin2013}. Here, the 
concave nature of the NLDSA (see Section \ref{acc:sec}) is discussed to lead to the rigidity-independent break. The difference between the proton and helium spectra is explained by the 
acceleration of cosmic-rays in both the forward and the reverse shock of the SNRs \citep{ptuskin2013}. In the model of \citet{thoudam_hoerandel2013},  the break is caused by the change from the background sea of Galactic cosmic rays (low energies) to the dominance of those nearby SNRs in the nearest kpc (high energies), with e.g.\ Vela as one of the strongest contributors. \citet{erlykin2013,wolfendale2014} even discuss the Vela SNR as the cause for the cosmic-ray knee.
In general, what these source-based models have in common is that two different types of
acceleration sources are responsible for the production of the different slopes in protons and helium.
This could even be an ancient local supernova with an age of $\sim 2\times 10^{6}$~yr \citep{kachelriess2015} or the superbubbles \citep{parizot2004}, with a two component model of SN ejecta in the superbubble core and in the regular ISM \citep{ohira2016}.
\item \textbf{A transport-related break}\\
  In the Galaxy, it is expected that the transport is diffusion-dominated, with possible contributions from advection. A scenario of producing the breaks at $10$~GV and $300$~GV in the cosmic-ray spectrum is presented in   \citet{blasi2012}. The authors discuss that the turnover in the spectrum at $10$~GV can be produced by a change from advection-dominated to diffusion-dominated cosmic-ray transport: In the rigidity-range between $10$~GV and $200$~GV, the turbulence spectrum that determines cosmic-ray diffusion is assumed to be self-generated by cosmic rays via the streaming instability \citep{CRdrivenTurb}. At higher rigidities, the turbulence spectrum is then expected to be generated by an external mechanism, as usually discussed by the explosion of supernovae in this model. As discussed in  Section \ref{diffusion:sec}, a change in the $k$-dependence of the wave-vector spectrum results in a change of the energy behavior of the diffusion coefficient. As discussed in \citet{blasi2012}, the index of the diffusion coefficient would change from $\kappa\sim 0.7$ at low rigidities ($R<200$~GV) to $\kappa \sim 1/3$ according to Kolmogorov at $R>200$~GV. This in turn leads to a break in the cosmic-ray spectrum that is transport-dominated.  An alternative reason for a change in the spectrum would be a two-zone diffusion model, considering different diffusion in the Galactic halo and Galactic disk \cite{tomassetti2015c}.
  In the transport-related break scenario, the same break needs to be present in the B/C ratio (see Section \ref{data:sec}). While this is certainly true at $\sim 10$~GV, data at $300$~GV show first evidence for a break as well \cite{genolini2017}. 
  The difference between proton spectra and heavier elements could be explained by the fact that proton acceleration is not bound to the metallicity of the environment, while this is the case for the heavier elements, so this difference is expected to be due to the source physics. A break in the B/C ratio is also expected from local source physics as discussed in e.g.\ \citep{ahlers2009,blasi2009,cholis2014,tomassetti_oliva2017} and it is thus a question that remains unsolved what the exact origin of the break is.
\end{itemize}
In summary, there are reasonable explanations of the spectral break of the spectra at $\sim 300$~GV both in transport-related and source-related phenomena. More data in the rigidity range around $300$~GV will help to distinguish between the two models. The change of the proton to Helium ratio on the other hand seems to be source-related. 
 \subsubsection{TeV - EeV range \label{tev:sec}}
 In the TeV to EeV range, the most prominent break is the cosmic-ray knee at around a PeV. The most common model is that a Galactic accelerator reaches its maximum energy, the proton spectrum cuts off first (at $\sim 1-3$~PeV) and the heavier elements reach a factor of $Z$ higher, up to iron with $Z=26$. This scenario is compatible with the composition of cosmic rays as measured with KASCADE, KASCADE-Grande and IceTop (see Section \ref{data:sec}). It has been noted early-on, however, that this scenario would produce a gap in the cosmic-ray shin region, as the extragalactic spectrum kicks in too late. A \textit{Component B} is therefore introduced by \citet{hillas2006} that fills this gap in the cosmic-ray shin region. The details of the possible sources in that range are discussed in Section \ref{multimessenger_modeling:sec}. Possible sources in that range are PWN \citep{bednarek_bartosik2004}, wind-SNRs \citep{stanev1993,biermann_prl2009,biermann2010,thoudam16b} or the Galactic wind \citep{Bustard2017,merten2018}.

\subsubsection{EeV - ZeV range \label{eev:sec}}
Following the arguments presented in Section \ref{hillas:sec}, there are basically no Galactic sources that can accelerate cosmic rays up to the highest observed energies, i.e.\ $\sim 10^{20}$~eV. While the Hillas criterion does allow for the acceleration up to these high energies for objects like pulsars and magnetars, loss processes suppress the acceleration to these energies. The sources above the ankle, i.e.\ at $E>3\times 10^{18}$~eV are therefore typically explained by an extragalactic source scenario.

\subsection{Luminosity criterion}
\label{ssec:LuminosityCriterion}
           The cosmic-ray energy spectrum as measured at Earth, as displayed in Figures \ref{all_particle:fig} (all particle) and \ref{fig:compositionVoyager} (composition-resolved)  is directly connected to a luminosity or luminosity density budget that needs to be provided by the source(s) responsible for cosmic-ray emission. In this section, we go through the arguments of what is necessary and which source(s) could provide the required budget.

           The differential cosmic-ray spectrum at Earth, $\Phi_{\rm CR}$, given in units per energy, area and time, can be converted into a luminosity in the 
following way: First, the flux is converted into the local energy flux $j_{\rm CR}$ by integration:
           \begin{equation}
           j_{\rm CR}(E_{\min})=\int_{E_{\min}}^{E_{\max}}\Phi_{\rm CR}\,E\,dE\,.
           \end{equation}
           As the cosmic-ray energy spectrum at Earth drops as $E^{-\alpha}$ with $\alpha>2.5$, the dependence on the lower integration limit $E_{\min}$ is much stronger than the 
upper one, which is why only the minimum energy appears as a dependency.
           A local cosmic-ray energy density $\rho_{\rm CR}$ is received from this flux with
           \begin{equation}
           {\rho}_{\rm CR}(E_{\min})=\frac{4\,\pi}{c}\,j_{\rm CR}\,.
           \label{rhocr:equ}
           \end{equation}
           Here, the relativistic limit, $v\approx c$ has been assumed. For low-energy cosmic rays with a kinetic energy scale of $E_{\rm kin} \leq m_p\cdot c^2\approx$~GeV, the velocity $v$ becomes significantly lower than $c$. In that case, the velocity-dependence would need to be taken into account explicitly using $E=\gamma(v)\cdot m_p\cdot c^{2}$.
           
           If the local cosmic-ray density represents the average value throughout the emission volume, homogeneity and isotropy can be assumed. The corresponding luminosity is then 
           calculated from the density by multiplying with the emission volume $V_{\rm em}$ and dividing by the cosmic-ray residence time $\tau_{\rm esc}$ in the corresponding volume:
           \begin{equation}
           L_{\rm CR}(E_{\min}, \tau_{\rm em}, V_{\rm em})=\frac{\rho_{\rm CR}\cdot V_{\rm em}}{\tau_{\rm esc}}\,.
           \end{equation}
           The two generic emission regions 
are the Galaxy (Galactic emission) and the Universe (extragalactic environment).           For extagalactic sources, it is more common to use the cosmic-ray density rate, i.e.\
           \begin{equation}
             \dot{\rho}_{\rm CR} =\frac{\rho_{\rm CR}}{\tau_{\rm esc}}\,,
           \end{equation}
           as it is difficult provide an integrated volume for the Universe.
           Tables \ref{tab:GalacticCRPower} and \ref{tab:UniverseCRPower} show a summary of the 
parameters entering the calculation, and the final result for the three most relevant lower energy thresholds, i.e.\ the cosmic-ray luminosity necessary to reproduce the emission below the knee, above the knee and above the ankle.

\begin{table}[htbp]
\begin{threeparttable}

\caption{\textbf{Cosmic-ray power on Galactic Scales} It has been assumed that the emission 
volume is a cylinder of height $H=300$~pc and radius $r = 15$~kpc, with $V_{\rm Gal}=\pi\cdot r_{\rm Gal}^{2}\cdot H$.}
\label{tab:GalacticCRPower}

\begin{tabular}{l | r r c }
\toprule
$E_{\min}$~[GeV] & \multicolumn{1}{l}{$V_\mathrm{em}$~[Mpc$^3$]} & \multicolumn{1}{l}{$\tau_{\rm esc}$~[yr]} & \multicolumn{1}{l}{$L_{\rm CR}(E_{\min})$~[erg/s]}\\
\midrule

$10$ \tnote{(*)} & $2\times 10^{-7}$ & $10^{6}$ & $(5.09 - 6.88)\times 10^{40}$\\
$10^{6}$ \tnote{(*)} &  $2\times 10^{-7}$ & $10^{6}$ & $(2.72 - 4.10)\times 10^{37}$\\
$10^{6}$ \tnote{(**)} &  $2\times 10^{-7}$ & $10^{6}$ & $(3.97 - 4.41)\times 10^{37}$\\
$3\times 10^{9}$ \tnote{(**)} & $2\times 10^{-7}$ & $10^{6}$ & $(6.46 - 7.14)\times 10^{33}$ \\

\bottomrule
\end{tabular}
\begin{tablenotes}
      \small
      \item[(*)] low energy models ---  \citep{fedynitch2012, Zatsepin2006, PamelaFlux2011, polygonato}
      \item[(**)] high energy models ---  \citep{SGT2014, serap2013, fedynitch2012}
    \end{tablenotes}
\end{threeparttable}
\end{table}

\begin{table}[htbp]

\begin{threeparttable}

\caption{\textbf{Cosmic-ray power density on scales of the Universe} It is assumed that galaxies back to $z=20$ contribute. With a $\Lambda$CDM-cosmology and approximate parameters 
of $\Omega_M\sim 0.3, \Omega_\Lambda \sim 0.7, h=0.7$, this results in the active time of accelerators as given in the table.}
\label{tab:UniverseCRPower}

\begin{tabular}{l | r c }
\toprule
$E_{\min}$~[GeV] & \multicolumn{1}{l}{$\tau_{\rm esc}$~[yr]} & \multicolumn{1}{l}{$w_{\rm CR} (E_{\min})$~[erg/yr/Mpc$^3$]}\\
\midrule

10\tnote{(*)} & $1.3\times 10^{10}$ & $(6.1 - 8.2)\times 10^{50}$\\
$10^{6}$ \tnote{(*)} &  $1.3\times 10^{10}$ & $(3.2-4.9)\times 10^{47}$ \\
$10^{6}$ \tnote{(**)} &  $1.3\times 10^{10}$ & $(4.7-5.3)\times 10^{47}$ \\
$3\times 10^{9}$ \tnote{(**)} & $1.3\times 10^{10}$ & $(7.7 - 8.5)\times 10^{43}$ \\

\bottomrule
\end{tabular}
\begin{tablenotes}
      \small
      \item[(*)] low energy models ---  \citep{fedynitch2012, Zatsepin2006, PamelaFlux2011, polygonato}
      \item[(**)] high energy models ---  \citep{SGT2014, serap2013, fedynitch2012}
    \end{tablenotes}
\end{threeparttable}
\end{table}

Now, these values can be compared to the estimate of cosmic-ray luminosities from potential cosmic-ray emitters. Here, we focus on the discussion of those classes that we have identified as fulfilling the Hillas criterion for the given energy range, see Section \ref{hillas:sec}. In 
particular, we will discuss what is realistic for Galactic sources, for which the emission volume is the Galactic plane and the residence time is the escape time of cosmic 
rays from the Galaxy, see Section \ref{lumi_gal:sec}. Further, we will shortly summarize why the energy budget of extragalactic sources only suffices to explain the cosmic-ray flux 
at the highest energies, certainly not the part below the knee, see Section \ref{lumi_extragal:sec}.

\subsubsection{Galactic sources \label{lumi_gal:sec}}

The emission volume of cosmic rays in the Galaxy in our simplified calculation that leads to the numbers in Table \ref{tab:GalacticCRPower} is approximated with a cylinder of radius $r_{\rm 
GP} = 15 $~kpc and a scale height of $H=300$~pc. The residence time is given by the cosmic-ray 
escape time, $\tau_{\rm esc}\sim 10^{6}$~yrs, see e.g.\ \citet{yanasak2001} and the discussion in Section \ref{spallation:sec}. The total cosmic-ray luminosity is calculated by 
integrating from $E_{\rm min}=10$~GeV (so, starting at the energy from which an influence of both ionization and the solar wind should be negligible). Using the cosmic-ray flux 
models given in Table \ref{tab:GalacticCRPower},  the range of total cosmic-ray luminosities is determined to $L_{\rm CR} = (6.26\pm 0.72)\times 10^{40}$~erg/s. 

There is another scenario in which nearby sources (i.e.\ with a distance of less than a kpc from Earth) contribute significantly to the total cosmic-ray luminosity at Earth. Here, it is not only the ensemble-average of the cosmic-ray sea, but the assumption of a few nearby, contributing sources -- or even a single one.
The cosmic-ray density of a single point source, $\rho_{\rm CR}^{\rm source}$ is given as \citep{sigl2017}\footnote{It can be found by first determining the Green's function (with the mirror method) for a source that injects instantaneously at a given position and then folding this according to the Green's method with a proper source term $Q(t',\,p)$}:
\begin{equation}
  \rho_{\rm CR}^{\rm source}(\vec{r},\,p,\,t)\propto \frac{1}{D_{\parallel}^{1/2}\,D_{\perp}}\int dt'\,\exp\left\{-\frac{\tau_{\rm diff}(\vec{r},\,p)}{t-t'}\right\}\,Q(t',\,p)\cdot (t-t')^{-3/2}
  \label{point_source:equ}
\end{equation}
with the diffusion time as defined as
\begin{equation}
  \tau_{\rm diff}(\vec{r},p) \sim  \left(\frac{r_{\parallel}^{2}}{4\,D_{\parallel}(p)}+\frac{r_{\perp}^{2}}{4\,D_{\perp}(p)}\right)\,.
  \end{equation}
Here, $r_{\parallel}$ is the component of the distance vector parallel to the homogeneous magnetic field $\vec{B}_0$, $r_{\perp}$ is the corresponding perpendicular component.

\paragraph{Supernova Remnants}
In the case of SNRs, the total cosmic-ray luminosity can be calculated from the fraction of energy going from the total energy budget of the supernova explosion into cosmic rays (see e.g.\ \citet{drury2014}):
\begin{equation}
  L_{\rm CR}\approx 2\times 10^{41}\,{\rm erg/s}\cdot \left(\frac{\eta}{0.1}\right)\cdot \left(\frac{\dot{n}}{0.02\,{\rm yr}^{-1}}\right)\cdot \left(\frac{E_{\rm SN}}{10^{51}\,{\rm erg}}\right)\,.
  \label{cr_lumi:equ}
\end{equation}
Here, we assume that a typical SN explosion releases a total energy of $E_{\rm SN}\sim 10^{51}$~erg and that a fraction of about $\eta =10\%$ is transferred to cosmic rays. If we then assume a supernova explosion rate in the Milky Way of $\dot{n}\sim$0.02~yr$^{-1}$ \cite{diehl2006}, we end up with a cosmic-ray energy budget on the order of what is needed from our simplified calculation above. Thus, Supernova Remnants are the prime candidate for the acceleration of cosmic rays below the knee. The discussion of the maximum energy (Section \ref{hillas:sec}) shows that it is not clear yet if (regular) SNRs really can accelerate particles up to the knee. Certainly, the magnetic field in SNRs needs to be enhanced in order for SNRs to reach knee energies --- and there is observational evidence in support of such an enhancement \citep{uchiyama2007,vink2012}. Acceleration to above knee energies seems possible by so-called wind-SNRs, that occur for heavy stars, which explode into their own wind, see e.g.\ \citet{stanev1993,biermann_prl2009,biermann2010} and Section \ref{multimessenger_modeling:sec} for a detailed discussion.

As for a scenario of nearby point source contributors to the cosmic-ray spectrum, most prominent candidates are Geminga ($d = 0.15$~kpc; age $\sim 3.4\times 10^{5}$~yr), Loop-I ($d = 0.17$~kpc; age $\sim 2.0\times 10^{5}$~yr), Vela ($d = 0.3$~kpc; age $\sim 1.1\times 10^{4}$~yr), Monogem ($d = 0.3$~kpc; age $\sim 8.6\times 10^{4}$~yr), G299.2-2.9 ($d = 0.5$~kpc; age $\sim 5.0\times 10^{3}$~yr), Cygnus Loop ($d = 0.54$~kpc; age $\sim 1.0\times 10^{4}$~yr) and Vela Junior ($d = 0.75$~kpc; age $\sim 3.5\times 10^{3}$~yr). \citet{ahlers2016} discusses that Vela is an excellent candidate to account for the cosmic-ray anisotropy at TeV energies (see also Section \ref{anisotropy:sec}). In this scenario, Vela contributes to the cosmic-ray energy spectrum at the $1\%-5\%-$level in the TeV energy range. While other nearby sources contribute as well, Vela is identified as the strongest individual source. \citet{thoudam_hoerandel2013} explain the break in the individual element spectra at $\sim 300$~GeV with the contribution of nearby sources and in particular with a strong contribution (also on a $\%$ level) from Vela.

\paragraph{X-ray binaries and microquasars}
In X-ray binaries, particles can be accelerated by the accretion process, or in jet structures that are visible for those types of binaries that are called microquasars. Assuming that the energy budget of the jet is fed by the accretion process (jet-disk symbiosis model), the cosmic-ray luminosity in both cases is limited by the Eddington luminosity \citep{gaisser1991}, $L_{\rm CR/XRB} \leq \eta L_{\rm edd}\sim 10^{37}$~erg/s. Here, $\eta\sim 0.1$ is the efficiency factor, describing what part of the total energy budget is going into cosmic rays. Now, for the total luminosity budget, the number of X-ray binaries in the Milky Way needs to be quantified. This is obviously a function of the mass. However, we present a simplified calculation here to get an order-of-magnitude estimate: The number of low-mass X-ray binaries (LMXB, $M\leq M_{\odot}$) detected in the Galaxy is approximately $200$. High-Mass X-ray Binaries (HMXB), where massive O or B stars are in the system, thus $M >10\,M_{\odot}$ are present in similar numbers. These more massive systems are expected to dominate the total luminosity due to their higher individual luminosity. Thus, the total cosmic-ray luminosity is on the order of $L_{\rm CR/XRB, tot}\lesssim 10^{39}$~erg/s. This is much lower than the luminosity budget required to reproduce the cosmic-ray flux down to GeV energies. These sources could, however, contribute to the luminosity at higher energies by representing the so-called \textit{component B} (see Section \ref{composition:sec}). 

\paragraph{Pulsars and Pulsar Wind Nebulae}
In the case of pulsars, the rotational energy can be available for the acceleration of cosmic rays. Here, the generated plasma turbulence in the expanding medium could lead to the effect of linear particle acceleration in large-scale electromagnetic fields that can be maintained by the pulsars. Or, alternatively, the energy can be transferred by acceleration at a standing shock in the relativistic wind that belongs to the pulsar wind nebula (PWN). Here, it is expected that the dominant part of cosmic-ray production in pulsars can happen in their early life-time as the pulsar spins down most rapidly then. The cosmic-ray luminosity budget has been discussed in e.g.\ \citet{bednarek_bartosik2004}. The budget is expected to be on the order of $10^{37}$~erg/s and thus too low to produce the flux below the knee. \textit{Component B}, on the other hand, could have its origin in pulsars or PWN. In addition, nearby pulsars are discussed as a source of the flux of electrons and positrons.
In particular, individual, local pulsars like the Vela pulsar are discussed to contribute at knee energies, see e.g.\ \citet{erlykin_wolfendale_ecrs2014} for a discussion.

\paragraph{The Galactic Wind Termination Shock}
The basic idea is that a Galactic wind eventually forms a shock far outside the Galactic plane where cosmic rays may be accelerated similar to acceleration at SNR but on much larger scales. Here, a wind, driven e.g.\ by Supernova explosions, with a mass load $\dot{M}$ is streaming out of the Galaxy with a shock speed close to the sound speed, $v_\mathrm{shock}\approx c_{s}$ depending on the central temperature $T_0$. The shock position $R_\mathrm{shock}$ is determined by an equilibrium between shock ram pressure and the pressure of the IGM $\rho(R_\mathrm{shock})\,v_\mathrm{shock}^2=P_\mathrm{IGM}$.
When the energy flux of the shock, given by $\rho\,v_\mathrm{shock}^3$, is integrated over a spherical shell of radius $R_\mathrm{shock}$ the cosmic-ray luminosity at the shock can be written as:
\begin{align}
L_\mathrm{CR} = \eta\cdot 4\pi\, R_\mathrm{shock}^2 \, \rho \, v_\mathrm{shock}^3 \approx \eta\,\dot{M}\,v_\mathrm{shock}^2 \label{eq:GWTS_Luminosity} \quad .
\end{align}
Here, it has been assumed that only a fraction $\eta$ of the total shock energy is converted into cosmic-ray energy. 

Realistic simulations of the Galactic wind following \citep{everett2008, Bustard2017} give $R_\mathrm{shock}=260$~kpc and  $v_\mathrm{shock}=600\;\mathrm{km}\,\mathrm{s}^{-1}$ for a mass loading of $\dot{M}=2M_\odot\;\mathrm{yr}^{-1}$. Following Equ.\ (\ref{eq:GWTS_Luminosity}) the GWTS may provide in an optimistic scenario a maximum luminosity of $L_\mathrm{CR}\approx 4.5\times 10^{40}\;\mathrm{erg}\,\mathrm{s}^{-1}$. However, not all of these cosmic rays will diffuse back into the Galaxy especially when the Galactic wind is taken properly into account. Which fraction $\xi$ of the GWTS cosmic ray can contribute to the observed flux does strongly depend on the details of the transport process. \citet{merten2018} have shown that this fraction can be  between $10^{-5}\lesssim \xi \lesssim 0.1$ leading to a Galactic cosmic-ray luminosity of $L_\mathrm{GCR}\approx 4.5\times (10^{35}-10^{39})\;\mathrm{erg}\,\mathrm{s}^{-1}$.

\paragraph{Galactic Luminosity - Discussion\label{lumi_discussion:sec}}

\begin{figure}[htbp]
\centering{
\includegraphics[trim = 3mm 5mm 4mm 3mm, clip, width=0.9\textwidth]{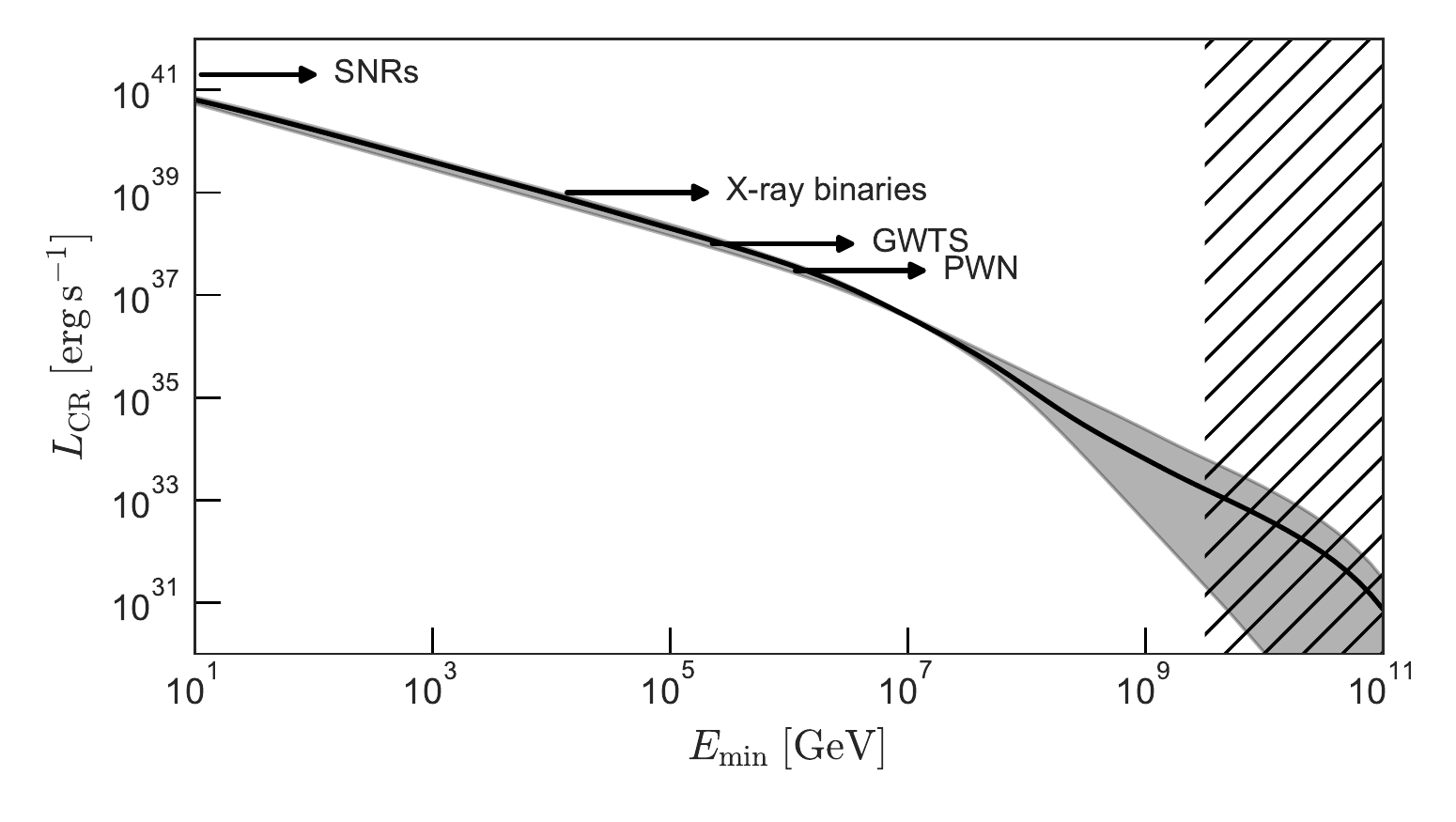}
\caption{The cosmic ray luminosity as a function of the minimum energy. They filled grey area is the uncertainty coming from differences in the models of the CR flux whereas the 
dashed area is forbidden by isotropy arguments see \ref{anisotropy:sec} for the  line of argument. The models we use to determine the uncertainty band are \citep{fedynitch2012, Zatsepin2006, PamelaFlux2011, polygonato}}
\label{lumi:fig}}
\end{figure}

Figure \ref{lumi:fig} shows the cosmic-ray luminosity as a function of the minimum energy. The range of uncertainty comes from the choice of different cosmic-ray flux models as described above.
In the figure, the possible luminosity range of SNRs, Pulsars and X-ray binaries is indicated. 
It becomes clear that SNRs are the only sources that on the natural scale of average parameters can explain the total luminosity of cosmic rays down to the lowest energies: there is 
no other object class that is able to achieve the total cosmic-ray luminosity down to GeV energies. 

Finally, it should be noted that a lot of simplifiations enter the above calculation. Having a closer look at SNRs as cosmic-ray sources, it is assumed that all SNRs end up producing the 
same cosmic-ray spectrum and that they end up showing the same energy spectrum at Earth. However, we know from gamma-ray data that the observed spectra differ significantly from 
each other, see Section \ref{data:sec} and references therein. The reason might lie in intrinsic differences, but also in the existence of a temporal evolution of the cosmic-ray 
spectra \citep{cox1972,funk2017}. 
The conclusion from this back-of-the-envelope estimate is that only SNRs can easily fulfil the energy requirements of cosmic-rays. Other sources like PWN, XRBs or the Galactic Termination Shock are candidates for \textit{Component B}, i.e.\ they can contribute to the energy range in between knee and ankle, the so-called \textit{cosmic-ray shin}.

           \subsubsection{Extragalactic sources \label{lumi_extragal:sec}}
For extragalactic sources, it is useful to compare luminosity densities rather than volume-integrated numbers, as it is unreasonable to try to quantify the emission volume in this 
case, i.e.\ the total volume of the Universe. This local luminosity  density, $\dot{\rho}_{\rm CR}=\rho_{\rm CR}/\tau_{\rm em}$ in case of extragalactic emission when integrated from the cosmic-ray ankle 
$(E_{\min}=10^{18.5}$~eV) is $\dot{\rho}_{\rm CR}\sim 5\times 10^{43}$~erg/(Mpc$^3$\, yr) \citep{vietri1995,waxman1995}. The emission time 
scale in this case corresponds to the lifetime of the Universe, so in approximation the Hubble time, $\tau_{\rm em}\sim t_{\rm Hubble}\sim 10^{10}$~yr.
The most energetic extragalactic, non-thermal sources are gamma-ray bursts, active galaxies and galaxy clusters. 
The total luminosity density can be calculated via the following integral:
\begin{equation}
\dot{\rho}_{\rm CR} = \int \frac{dn}{dL\,dV}(z=0)\,L\,dL\,.
\end{equation}
Here, $dn/(dL\,dV)$ is the luminosity function of the objects, i.e.\ the source density per unit luminosity and comoving volume. For non-thermal emission like in the radio band, a typical luminosity function follows a Schechter function,
\begin{equation}
\frac{dn}{d\log(L)\,dV}=A_L\cdot \left[ \left(\frac{L}{L_{\star}}\right)^{\gamma_1}+\left(\frac{L}{L_{\star}}\right)^{\gamma_2}\right]^{-1}\,,
\end{equation}
in which the number of sources decreases with a power-law of index $\gamma_1$ for increasing luminosities, with the spectral index steepening at a characteristic luminosity 
$L_{\star}$. Thus, the result is strongly influenced by the lower threshold of the luminosity integral. This part of the luminosity function is obviously most difficult to 
determine, as the measurement of low-luminosity sources requires high sensitivities. Despite these uncertainties, it can be concluded that all three source classes, gamma-ray 
bursts, active galaxies and galaxy clusters, are in principle able to reach the level needed to produce the flux above the ankle. For a detailed discussion of the potential of 
gamma-ray bursts, see e.g.\ \citet{vietri1995,waxman1995}, active galaxies are reviewed in e.g.\ \citet{biermann_strittmatter1987,bbr2005,becker2008}, galaxy clusters in \citet{brunetti2014}. All three source classes result in a total luminosity density of $L_{\rm CR, tot}^\mathrm{AGN/GRBs/Clusters} \sim 10^{43}-10^{44}$~erg/(Mpc\,yr). This corresponds quite exactly to what is needed above the ankle (see Table \ref{tab:UniverseCRPower}). No other source class is able to provide a higher total luminosity density. Therefore, extragalactic sources are not considered as a candidate for the production of cosmic rays that are detected below the knee --- they are simply not able to match the energy requirement.

\newpage
          
\subsection{(An)isotropy \label{anisotropy:sec}}

As discussed in Section \ref{data:sec}, both the dipole and small-scale anisotropies of the cosmic-ray flux have been studied in detail by a large number of experiments in the past decades. These measurements on the one hand provide valuable pieces of information on the properties of cosmic rays. On the other hand they make an interpretation of the multimessenger picture more complex as we will show in the following paragraphs. Here we will review the state of the art of possible origins of these large-scale (Section \ref{dipole:sec}) and small-scale (Section \ref{multipole:sec}) anisotropies.
\subsubsection{The dipole anisotropy}
The dipole anisotropy is given as the relative intensity on the largest scales on the sky, i.e.\ $l=1$, with the dipole vector defined as the local deviation from the average 
cosmic-ray flux. The
dipole approximation neglects terms with $l\geq 2$ (cf.\ Section \ref{data:sec}) and the relative intensity can be expressed as
\begin{equation}
I_{\rm rel}\approx 1+\vec{\delta}\cdot \vec{e}_{n}\,.
\label{irel_dipole:equ}
\end{equation}
with $\vec{\delta}$ as the dipole vector.

As summarized in e.g.\ \citet{berezinskii1990,sigl2017}, the dipole vector can be expressed in terms of the local cosmic-ray flux
\begin{equation}
  \vec{j}(\vec{r}, p)=2\,\pi\,\beta\,\vec{n}\int_{-1}^{1}d\beta \,\beta f(\vec{r}, p\,\vec{e}_{n})\,,
\end{equation}
given in terms of the absolute amount of momentum $p$ and the cosine of the angle between the  
direction $\vec{e}_{n}$ and the momentum $\vec{p}$, $\beta:=\vec{e}_{n}\cdot\vec{p}/p$.
In the isotropic diffusion limit the dipole vector is proportional to the particle flux $\vec{j}$,
\begin{equation}
  \vec{\delta}(p)\approx \frac{3\,\vec{j}}{v_{\rm CR}\,n_{\rm CR}}\,.
  \label{delta:equ}
\end{equation}
Here, $v_{\rm CR}$ is the cosmic-ray velocity and $n_{\rm CR}$ is the 
local cosmic-ray density.

There are two possible contributions to the local flux,
\begin{equation}
\vec{j}=\vec{j}_{\rm diff}+\vec{j}_{\rm flow}
\end{equation}
\begin{enumerate}
\item a cosmic-ray \textbf{density gradient} in the diffuse flux, which leads to a dipole-anisotropy depending on the strength of the gradient and the diffusion coefficient.
  \item a possible \textbf{relative flow} between the observer and the cosmic rays themselves, $j_{\rm flow}$, leading to a dipole anisotropy due to the Lorentz transformation between the different systems.
\end{enumerate}
The dipole vector can  expressed as a composition of these two components, 
\begin{equation}
  \vec{\delta}(p,\vec{r}))=\frac{3\,(\vec{j}_{\rm diff}+\vec{j}_{\rm flow})}{v_{\rm CR}\,n_{\rm CR}}\,,
\end{equation}
Depending on the relation of the strength of the two, one of them dominates the total flux.

Either way, the model will have to explain the broad-band energy behavior of the dipole anisotropy:
\begin{enumerate}
\item \textbf{$\mathbf{E\lesssim}\,$(10-100)~TeV:} The amplitude rises approximately with $E^{1/3}$, the phase is constant at $30^\circ \lesssim \alpha \lesssim 60^\circ$.\footnote{These values are shifted in Fig.\ \ref{anisotropy:fig} by $+360^\circ$ for a better visualization.}
\item \textbf{$\mathbf{E\sim}\,$(100-1000)~TeV:} The amplitude drops significantly, behaving as $\sim E^{-1}$. The phase begins to shift.
  \item \textbf{$\mathbf{E\gtrsim}\,$(100-1000)~TeV:} The amplitude rises faster than $E^{1/3}$ at constant phase ($ \alpha \approx 270^\circ$).
\item \textbf{$\mathbf{E\gtrsim}$~EeV:} The amplitude is at a percent level at a different phase, i.e.\ $80^\circ \lesssim \alpha \lesssim 110^\circ$.
\end{enumerate}

In the following paragraphs, we will discuss the option of a local flow dominating the cosmic-ray anisotropy (paragraph \ref{local_flow:sec}) and of a gradient coming from a local source and/or the total distribution of cosmic-ray sources in the Galaxy (paragraph \ref{gradient:sec}).
\paragraph{Local flows \label{local_flow:sec}}
A possible relative flow between the rest frame of the observer and the cosmic-ray rest frame results in a directed flux $\vec{j}_{\rm flow}$ due to the necessity of transforming the coordinate system from an (isotropic) cosmic-ray restframe to the observer's frame, which moves with a flow velocity $\vec{v}_{\rm flow}$ relative to the isotropic cosmic-ray flux. A general behavior for such a flux can be derived by using the Lorentz invariance of the particle distribution function, and thus the invariance of the intensity, i.e.\ $f(\vec{r},\,p\vec{e}_n)=f'(\vec{r'},\,p'{\vec{e}'_n})$. Here, the primed system is the reference system of the cosmic rays, the unprimed system the one of the observer.
The Taylor approximation in the limit $\vec{p}'-\vec{p}=p\cdot \vec{v}_{\rm flow}/v_{\rm CR} +\mathcal{O}((v_{\rm flow}/v_{\rm CR})^{2})\ll 1$, i.e.\ assuming flow velocities much smaller than the cosmic-ray speed, is given by
\begin{equation}
f(\vec{r},\,p\vec{e}_n)\approx f'(\vec{r},\,p\vec{e}_n)-\frac{p}{v_{\rm CR}}\,\vec{v}_{\rm flow}\cdot \frac{\partial f'(\vec{r},\,p\vec{e}_n)}{\partial \vec{p}}=f'(\vec{r},p\vec{e}_n)\cdot \left(1-\frac{\vec{v}_{\rm flow}\cdot \vec{p}}{v_{\rm CR}\,p}\cdot \frac{d(\ln f')}{d(\ln p)} \right)\,.
\label{flow1:equ}
\end{equation}
Comparing Equ.\ (\ref{flow1:equ}) and (\ref{irel_dipole:equ}), we find that:
\begin{equation}
\vec{\delta}_{\rm flow}=-\frac{\vec{v}_{\rm flow}}{v_{\rm CR}}\cdot \frac{d(\ln f')}{d(\ln p)} \,.
  \end{equation}
Assuming that the isotropic particle distribution at Earth is a power-law, $f= A\cdot p^{-2-\gamma}$ (as $n_{\rm CR} \propto f\cdot p^2\propto p^{-\gamma}$), the dipole anisotropy from a local flow is given as
\begin{equation}
\vec{\delta}_{\rm flow}=-\frac{\vec{v}_{\rm flow}}{v_{\rm CR}}\cdot \frac{d(\ln f')}{d(\ln p)}=\frac{\vec{v}_{\rm flow}}{v_{\rm CR}}\cdot(2+\gamma)\approx \frac{\vec{v}_{\rm flow}}{c}\cdot 4.7 \,.
  \end{equation}
  The latter approximation uses the relativistic limit $v_{\rm CR}\approx c$ and $\gamma\approx 2.7$, see Section \ref{data:sec} for a more detailed discussion on the exact spectral index. 
  
  Thus, an anisotropy caused by a local flow is pointing in the direction of the flow velocity and its intensity is determined by the strength of the flow relative to the cosmic-ray velocity, which is close to the speed of light in the limit we discuss here. 
  
First ideas of an anisotropy due to a local flow were already discussed by \citet{compton_getting1935}. The authors predict an anisotropy from the movement of the solar system 
through the isotropic cosmic-ray background, with the velocity determined by the circular speed of the Sun around the Galactic center, i.e.\ $v_{\odot}=220\pm 20$~km/s. The 
intensity of such a dipole signature would be $|\delta_{\rm C-G}|\sim 3\times 10^{-3}$, where the latter is close to what is observed (see Section \ref{data:sec}). However, the direction of 
this dipole is expected to lie at $l=270^{\circ},\,b=0^{\circ}$ (Galactic longitude and latitude), not compatible with the observed direction of the dipole. 
There is a simple explanation for the absence of this \textit{Compton-Getting effect}, which is named after the 
original authors: if cosmic rays are indeed accelerated at SNRs in the Galaxy, or any other type of object in the Milky Way, their frame of emission would be the one co-rotating 
with the Sun in the Milky Way. Thus, the reference frame would roughly be the same, or at least the level of anisotropy would be reduced due to the co-moving cosmic-ray emission frame.

The question is now: what \textit{is} the relevant flow velocity in order to determine the contribution from a possible local flow? As discussed in \citet{AHLERS2017184}, natural choices are either the local standard rest frame with a velocity of $v_{\rm flow}\sim 18$~km/s or the Sun's movement through the local ISM ($v_{\rm flow}\sim 20$~km/s. As the level of anisotropy would be smaller than the observed level ($\delta_{\rm local}\sim 10^{-4}$), these effects would rather contribute subdominantly to the total anisotropy and possibly to a small phase shift toward the directions of the flow.

A further possibility for a significant contribution of a local flow to the observed anisotropy is the existence of a local streaming velocity \citep{schlickeiser2019}
The flow velocity as measured in the direct vicinity of Earth is $\sim 30$~km/s. 
The relevant coherence scale for the streaming effects at TeV - PeV energies is, however, larger, as the particles have a mean free path up to $100$~pc. At these scales, the streaming velocity is not know. 
As discussed in \citet{biermann_anisotropie}, such a flow could have been induced by an ancient SNR. It could now exist on scales of $100$~pc with a speed on the order of $\sim 100$~km/s, thus resulting in an anisotropy of the level observed in the data when the isotropic flux of incoming cosmic rays gets carried along with the local flow. If the last effective scattering of the cosmic rays lies within these $\sim 100$~pc, i.e.\ if the mean free path of the particle is within the flow field ($\lambda_{\rm mfp}(E)<100$~pc), a flow-dipole on the order of $\delta_{\rm flow}\sim 10^{-3}\cdot (v_{\rm flow}/(100$~km/s)) will be present. The flow field in this case concerns the hot phase of the ISM,  since the ancient SNR is expected to deliver a gas with X-ray temperatures and densities of $\rho\sim 10^{-2.5}$~cm$^{-3}$. This hot phase, however, cannot be measured and the possible flow remains undetected so far. Thus, the direction of the dipole cannot be predicted at this point. However, it is expected that the anisotropy level increases with $E^{\kappa}$ until the mean free path reaches the maximum extension of the ancient remnant. At this maximum energy, the dipole is expected to disappear. In addition, smaller-scale structures will appear, as the boundary of the transition is not expected to be smooth: the remnant explosion is not likely to have been centered at the Sun, and the expansion must not have happened isotropically, so that anisotropies at angular scales $<2\pi$ are expected. Even at these smaller scales, the structure should generally show a symmetry, as an excess in one direction would always result in a deficit in the direct opposite direction. However, if several of these smaller-scale structures overlap, such a symmetry would be broken again. These predictions fit today's observations, that show both smaller-scale structures as well as an increase of the dipole amplitude in energy, close to $E^{1/3}$, consistent with a Kolmogorov-type spectrum. At around $10-100$~TeV, the observed dipole intensity drops and the phase shifts. This could indicate the maximum energy due to the edge of the ancient SNR, at higher energies, a directed flow at larger scales could dominate, or alternatively, a gradient in the source distribution could be responsible, as discussed in the next paragraph.

\paragraph{Cosmic-ray gradients \label{gradient:sec}}
Assuming isotropic and spatially homogeneous diffusion, the flux arising from a cosmic-ray gradient is determined by

\begin{equation}
  \vec{j}_{\rm diff}=-D(p)\,\nabla n_{\rm CR}\,.
  \label{jdiff:equ}
\end{equation}

The fundamental solution of the diffusion problem can be formulated for a point source located at distance $\vec{r}$ from the observer as in Equ. (\ref{point_source:equ}).
Using this result in Equ.\ (\ref{delta:equ}) yields an expression for the dipole anisotropy amplitude induced by the cosmic-ray gradient of a point source:
\begin{equation}
|\vec{\delta}| \sim \frac{r}{c\,\tau_{\rm diff}}\,.
\end{equation}
The local environment up to a distance of a few 100~pc provides a variety of potential cosmic ray emitters -- Loop-I, Vela, Geminga and Cygnus Loop being the most prominent ones.
Assuming a diffusion coefficient compatible with the recent results from AMS-02 \cite{ams2016}, $D(p) = 10^{28}\,$cm$^2$~s$^{-1}\,(p/p_0)^{1/3}$ results in an amplitude of
\begin{equation}
|\vec{\delta}| \sim \frac{D(p)}{c\,r}\sim 0.01 \left(\frac{r}{{\rm kpc}}\right)^{-1}\cdot \left(\frac{p}{p_0}\right)^{1/3}\,,
\end{equation}
normalizing to $p_0\,c=20$~TeV. The predicted amplitude for an anisotropy caused by a local source at a distance of a kpc from Earth at an energy of $20$~TeV is thus on the order of $\sim 1\%$.  This simplified picture of a one-dimensional transport model results in an overestimate of the dipole anisotropy, as observations show that the isotropy level is about an order of magnitude higher (see Section \ref{data:sec}). So, how could the anisotropy level be reduced? A natural scenario is a broader source distribution with a gradient caused by an enhanced cosmic-ray density toward the direction of the Galactic center. A first prediction was made by \citet{erlykin_wolfendale2006}, where sources have been distributed stochastically in time and space in the Galaxy. The anisotropy is explained by a dominating background sea of cosmic rays plus a contribution from a nearby supernova remnant at higher energies. \citet{blasi_amato2012_2} find that the size of the spiral arm structure has a significant impact on the anisotropy level when injecting sources following the pulsar distribution in space and time. It is also discussed that randomly distributed near-Earth sources could dominate the anisotropy. As the exact (time-dependent) distribution of sources is not known, it is difficult to predict the exact amplitude and phase of the dipole. The authors discuss that constraints can be set this way. For instance, they find that the spectral index of the diffusion coefficient needs to be $\kappa\ll 0.6$, supporting the scenario of a Kolmogorov-type scalar diffusion coefficient $\kappa\sim 1/3$. However, in the presence of a significant number of recent nearby cosmic-ray sources, the general problem of the overestimation of the amplitude level still exists, see also \citet{sveshnikova2013} for a discussion.

More elaborate transport models could provide an answer to this question:

\begin{itemize}
\item \citet{evoli_anisotropy2012} show that the anisotropy level as observed can be attained when assuming isotropic diffusion in a spatially inhomogeneous plasma.
\item
\citet{pohl2013} argue that intermittency plays a crucial role in order to determine the level of dipole anisotropy. However, a flat distribution of sources is needed in order to keep the amplitude as low as the level of observation suggests.
\item
  \citet{effenberger2012} and \citet{kumar_eichler2014} investigate that a reduction of the level of anisotropy as needed for the nearby-source scenario can be accomplished by assuming anisotropic diffusion.
 \item Local effects of turbulence are investigated as well. The local structure of the ISM on the scale up to $\sim 100$~pc has been discussed to exhibit inhomogeneties that should be imprinted in the magnetic field structure \citep{frisch2012}. This local magnetic turbulence influences cosmic-ray diffusion \citep{giacalone_jokipii1999,yan_lazarian2002} and thus the level and phase of anisotropy. The anisotropic structure of the heliosphere, on the other hand, has effects on the magnetic field on a scale of $\sim 100-1000$~AU. It affects cosmic rays at $\lesssim 10$~TeV \citep{desiati_lazarian2013}.
\end{itemize}
Finally, the local homogeneous magnetic field has been pointed out to play a crucial role for both amplitude and phase of the dipole \citep{mertsch_funk2015}.
\citet{ahlers2016} discusses that the dipole aligns with the local magnetic field structure. In the presented model, the anisotropy of one or several local sources can be mapped onto the field direction. This projection onto the local field  reduces the dipole amplitude. In addition, it naturally explains the direction, thus circumventing the arguments of random phase and amplitude made in the papers discussed above.

\subsubsection{The small-scale anisotropy}
As the particles diffuse in the interstellar magnetic field, all information about their origin(s) is generally washed out and only a large-scale anisotropy due to a source gradient or a relative plasma flow remains. However, there is the possibility to generate small-scale structures from local effects, where local means scales on which the diffusion process is not fully developed. There are three options to produce these smaller features:
\begin{itemize}
\item (A) local source/s;
\item local electromagnetic field structures;
 \item beyond standard model effects\,.
\end{itemize}
Due to the localized structure of the signature, small-scale anisotropies require  non-diffusive propagation, expected if the source is at distances much smaller than the gyroradius, $r_{\rm g}\sim 200/Z\,(E/$TeV$)\,(B/\mu$G$)^{-1}$~AU. This local-source model is usually combined with the effect of a local electromagnetic field structure in order to explain the small-scale features.

Here, we will present a short summary of these possibilities. A contemporary and extensive review of the interpretation of the oberved small-scale anisotropy can be found in \citet{AHLERS2017184}. A summary of the different types of models that can explain (parts of) the small-scale structures in the cosmic-ray anisotropy is presented in Table \ref{small_scale:tab}. The following paragraphs review the different options shortly.

\paragraph{The heliosphere}
As a local source, the heliosphere is discussed by several authors: while the \textit{flux level and spectral behavior} is unaffected by cosmic-ray propagation in the heliospheric field, the directionality can be influenced, given that the gyroradius for a TeV particle is $r_{\rm g}\sim 200/Z\,(E/$TeV$)\,(B/\mu$G$)^{-1}$~AU and thus on the order of the extension of the heliosphere and siginficantly smaller than the elongated shape of the heliotail. Therefore, the heliosphere was suggested early-on as a possible source to induce an anisotropy \citep{nagashima1998}. It was discussed that a larger-scale anisotropy can be induced from the structure of the heliosphere,  the deficit region would be in the range $(\delta,\,\alpha)=(180^{\circ},\,20^{\circ})$, the access region would be located around the coordinates $(\delta,\,\alpha)=(90^{\circ},\,-24^{\circ})$. The reason for this dipolar structure is the axis-rotational symmetry of the heliosphere due to the movement of the Sun in the interstellar medium, where the access region in the direction of the heliotail is labeled the \textit{tail-in} region. The broad deficit region is referred to as the  \textit{loss-cone}.

Now that data reveal the sub-structures in the cosmic-ray anisotropy, in particular with the \textit{tail-in} region pointing in the direction of the heliotail, the original idea has been widened and quantitatively extended in the recent decennium. In particular, the transport in this local field structure can be modeled by solving the equation of motion. Here, the Lorentz force is acting on the particles in the local field structure. Solving the equation numerically is being done by  backtracking the particles. In this approach, anti-particles are started at the position of Earth. Due to Liouville's theorem, the anisotropic picture received from the backtracking of the antiparticles is the same as from forward propagation of particles. In forward-tracking, far too many trajectories need to be simulated, most of them not ending at Earth. By backtracking anti-particles from Earth, a computer-time efficient approach has been found in order to produce anisotropy maps. Due to the stochastic nature of the problem, there is no in-advance filter of those trajectories ending at Earth. By backtracking the particles, only those tracks that actually end at the position at Earth are simulated, making this approach very effective.  Heliospheric anisotropy models are still mostly tailored to explain in particular regions A and B, but not necessarily the entire angular power-spectrum of the small-scale anisotropies, see e.g.\ \citet{desiati_lazarian2013,zhang2014,lopez_barquero2017}. 

An explanation for the quite hard cosmic-ray energy spectrum associated with the cosmic-ray excess region A has been provided by \citet{lazarian2010}. The authors suggest the reconnection of regions with opposed polarity in the heliotail, coming from unipolar domains originating in the polar cap region of the solar magnetic field. Such a concept would naturally be able to explain hard-spectrum small-scale features in the direction of the heliotail.

It was further suggested \citep{drury2013} that localized electric fields induced from solar modulation with potentials on the order of a few 100 GV could induce anisotropies like the ones observed in regions A and B.

\paragraph{Propagation in local magnetic field}
Propagation in the local magnetic field structure can generate small-scale anisotropies in different ways:
\begin{itemize}
\item \textbf{Non-diffusive Propagation}\\
In order to explain the excess of cosmic rays in MILAGRO region A, non-diffusive transport at scales beyond the heliosphere was suggested by several authors, see \citet{drury2008,salvati_sacco2008,drury2008,battaner2015,harding2016}. Here, the central idea is that a nearby supernova remnant like Geminga could emit cosmic rays, which would stream along a magnetic flux tube. Due to the magnetic mirror effect (see Section \ref{acc:sec}), the particles would be focused into a cone around the local magnetic field direction. This way, larger-scale structures could be explained individually. A full multipole spectrum cannot be explained systematically.
\item \textbf{Diffusive propagation: pitch angle diffusion and turbulence}\\
  Assuming that a source emits particles anisotropically, it could be shown by \citet{malkov2010} that this anisotropy can be attained if the pitch-angle diffusion coefficient is anisotropic, i.e.\ $D_{\mu\mu}/(1-\mu^2)\neq const$ \footnote{The pitch angle diffusion coefficient is connected to the spatial parallel diffusion coefficient as $D_{\parallel}=v^2/8\int_{-1}^{+1}d\mu\,\left[(1-\mu^2)^2/D_{\mu\mu}(\mu)\right]$.}. Alfv{\'e}nic Goldreich-Sridhar turbulence could generate such a behavior \citep{giacinti_kirk2017}. Now, this effect becomes visible in the relative intensity map as a narrow feature, such as the Milagro region A. A Goldreich-Sridhar-type turbulence can also explain the change of the anisotropy shape observed at large scale at around 100~TeV \citep{giacinti_kirk2019}.

 Magnetic turbulence has been discussed above as the source of diffusion in general. A large number of authors have presented and developed the idea of turbulence as the source of the multipole spectrum of anisotropies at the smallest scale. As first discussed in \citep{giacinti_sigl2012}, the phase space density is typically observed as its average $\langle I\rangle$. However, for the scattering centers closest to Earth, the fluctuations of  $I=\langle I \rangle +\delta I$ do play a crucial role. \citet{giacinti_sigl2012} show that small-scale structures comparable to the observed multipole-spectrum can arise from a turbulent magnetic field with a scale of $150$~pc and a strength of $4~\mu$G. The reproduction of these small-scale anisotropies has been confirmed by other authors \citep{ahlers2014,ahlers_mertsch2015,pohl_rettig2016,lopez_barquero2016}, with e.g.\ different initial conditions. Recently, \citet{giacinti_kirk2019} showed that, assuming that the above interpretation is correct, the observed local turbulence spectrum is consistent with anisotropic MHD turbulence. In the future, possibly with data at lower energies, the anisotropy could therefore be used as a probe for the nature of the turbulence.

  The discussion on the influence of local turbulence on the multipole spectrum above refrains from taking into account the influence of a local source. It simply quantifies what local turbulence produces as a multipole spectrum. The presence of sources can change this picture: as discussed in \citet{pohl2013}, the anisotropy level can become much larger than what is observed at PeV energies when considering non-uniform initial conditions. This effect is reduced by anisotropic pitch-angle diffusion as investigated in e.g.\ \citet{effenberger2012}. 
  \end{itemize}

\paragraph{Beyond standard-model effects}
There is some discussion if (parts of) the small-scale anisotropy can arise from non-standard model effects. In particular, a nearby dark matter source, i.e.\ matter sub-halos, could produce high-energy particles. This scenario would fit particularly well on the detection of region A, for which the particle spectrum is quite flat with a sharp cut-off --- both are features that are expected from a dark matter scenario with a dark matter mass somewhere in the range $30$~TeV$<m_{\rm DM}<200$~TeV.

An alternative scenario would be the acceleration of strangelets that are of intermediate mass ($A\sim 10^{3}$). These nuclei that include strange quarks in addition to up and down quarks are expected to be stable at high mass numbers could be created and accelerated in the rapid transition of a neutron star to a strange star \citep{angeles2014}. While the charged strangelets would diffuse to Earth just as regular cosmic rays, they could be neutralized if the neutron star is embedded in a molecular cloud by the capture of electrons or by spallation. As hotspot A lies close to the Taurus molecular cloud, which is a star-forming region around $140$~pc from Earth, it can be argued that this hot-spot could originate from such a strangelet-source.

\begin{threeparttable}
\caption{Summary of small-scale anisotmodels\label{small_scale:tab}}
\begin{tabular}{l|ll}
\toprule
Origin & Model & Reference(s)\\
\midrule

Heliosphere&anisotropic transport&\citet{desiati_lazarian2013}\\
&&\citet{zhang2014}\\
&&\citet{lopez_barquero2017}\\

&magnetic reconnection&\citet{lazarian2010}\\

&Large-scale electric field&\citet{drury2013}\\
\midrule

Transport &non-uniform pitch-adiffusion&\citet{giacinti_kirk2017}\\
&&\citet{malkov2010}\\
          
&non-diffusive, ballistic CR transport&\citet{drury2008}\\
&&\citet{salvati_sacco2008}\\
&&\citet{drury2008}\\
&&\citet{battaner2015}\\
&&\citet{harding2016}\\

&local turbulence: mapping of global dipole&\citet{giacinti_sigl2012}\\
&&\citet{ahlers_mertsch2015}\\
&&\citet{ahlers2016}\\
&&\citet{lopez_barquero2016}\\
&& \citet{giacinti_kirk2019}\\
\midrule

Beyond SM&Strangelet production in neutron stars&\citet{angeles2014} \\
&dark matter decay in local source&\citet{harding2013}\\
 \bottomrule
\end{tabular}
\end{threeparttable}

\clearpage

\subsection{Summary and Conclusions}


          \begin{figure}[htpb]
\centering{
\includegraphics[trim = 0mm 7mm 5mm 10mm, clip, width=0.9\textwidth]{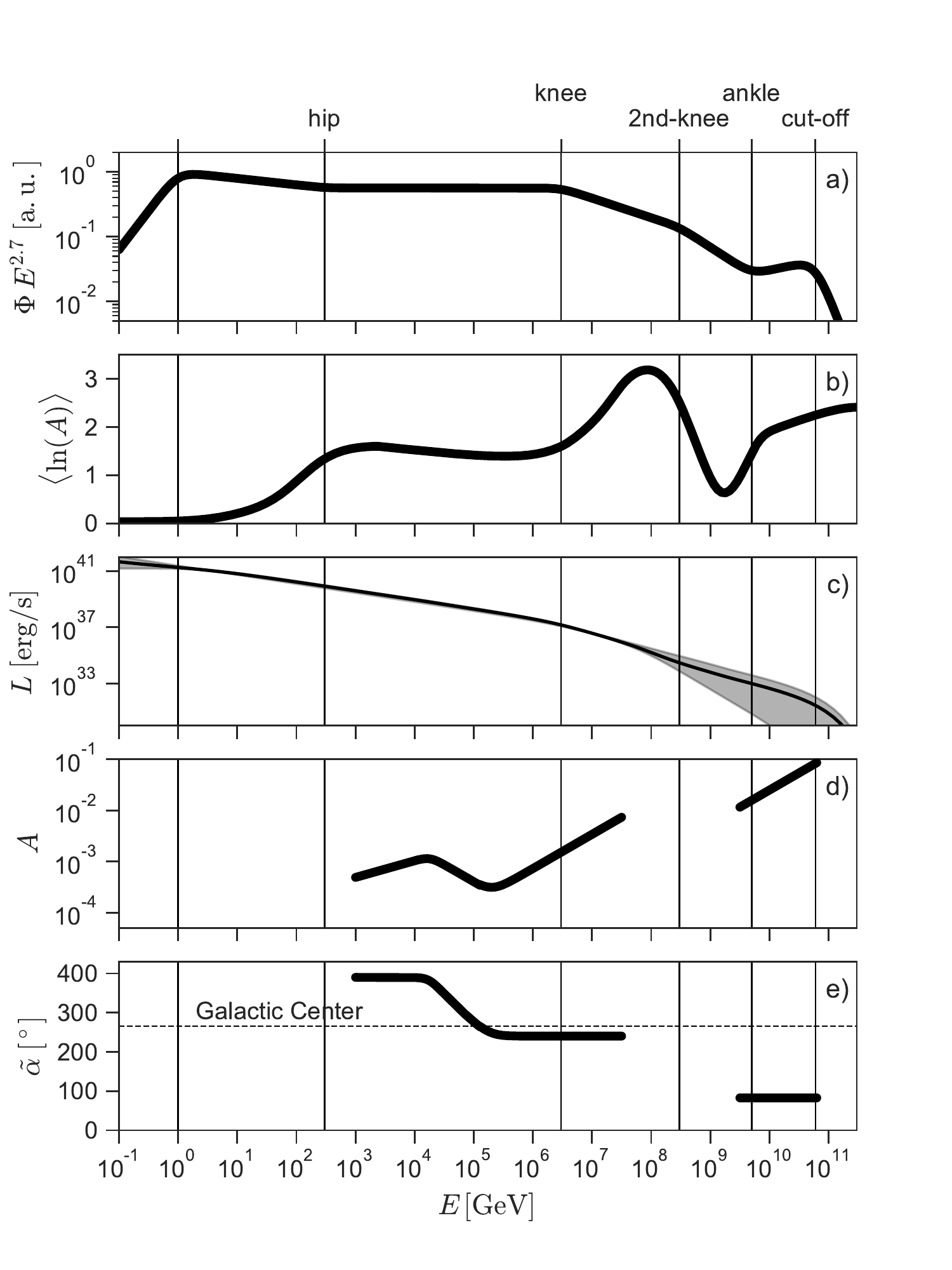}
\caption{Overview of some of the most important charged cosmic-ray observables: (a) shows the weighted all particle flux with all major breaks as listed in tables \ref{tab:ams}-\ref{tab:cutoff}; (b)  shows $\langle\ln(A)\rangle$ composed of data from the models in \citep{fedynitch2012, serap2013}. The low energy model \citep{fedynitch2012} is smoothly merged into the high energy model \citep{serap2013} using a logistic function in with a transition energy of $E_\mathrm{trans}=2.4\times10^{16}$~eV, where both models are consistent with each other. This merging is necessary as non of the models is valid over the whole energy range.  (c) shows the cosmic-ray luminosity as discussed in Section \ref{ssec:LuminosityCriterion}. (d) shows a schematic view of the projected dipole amplitude approximated by a broken power law $A\propto E^{1/3, -2/3, 2/3}$ with breaks at $18$~TeV and 200~TeV. Note that these lines are not statistically fitted to the data, but rather a rough approximation to understand the main features of the observations. (e) shows the dipole direction in the equatorial plane.}
\label{observables:fig}}
\end{figure}

          In summary, only Galactic sources are reasonable candidates to produce the full cosmic-ray energy budget. Therefore, SNRs are likely to be the sources producing the cosmic-ray luminosity budget below the knee. This is, however, no solid proof exists that this is really the case. Those three SNRs that show a turn-over in the spectrum at low energies for once have very steep spectra and do not contribute significantly at TeV-PeV energies. Secondly, it is not clear yet if this break is not rather due to the kinematic break that occurs for the prediction of power-law acceleration spectra  in momentum space. Final proof therefore remains to be found. First results show that there is a Galactic PeVatron in the center of the Milky Way \citep{hess_gc_2016}. What the exact acceleration mechanism and/or source(s) are is still unclear.

          Thus, in this section, the different pieces of information from charged cosmic rays have been  presented that all provide constraints on what a Galactic source distribution needs to look like in order to explain all features. The different observables are summarized in Fig.\ \ref{observables:fig}. The figure shows the all-particle spectrum in panel (a) and its global features (hip, knee, second knee, ankle and cutoff) are indicated. With the hypothesis that these kinks in the spectrum represents changes in the physics involved, we can now look at the other observables to develop scenarios. For the change at the hip at a few 100 GeV, a clear change in composition (panel (b)) can be seen. The cosmic-ray luminosity at this changing point is around $10^{40}$~erg/s, thus already about an order of magnitude lower than at GeV energies. This could potentially point to a change in the source population. The change at the knee of the all-particle spectrum is accompanied with a further change in composition toward a dominance of heavy elements, again consistent with a change of source population. In this energy range, i.e.\ between TeV and PeV energies, there is now detailed information of the dipole anisotropy strength (panel (d)) and direction (panel (e)). An $E^{1/3}-$ increase of the amplitude is observed as expected from a nearby source. This increase is connected to a relatively constant dipole phase of about $\tilde{\alpha}\approx 40^\circ$ and both phase and amplitude change at a few tens of TeV. This change comes somewhat earlier than the change in the spectrum and composition and is in need of explanation. After the change, the amplitude increases as $\sim E^{2/3}$ and the phase has shifted toward the Galactic center.

Looking at the highest end of the cosmic-ray spectrum, Galactic sources are not likely to 
produce the flux above the ankle. This statement is supported by the fact that a Galactic component with cosmic-ray emission above $10^{18}$~eV would lead to a highly anisotropic flux, pointing 
back to the sources, which is not observed. In addition, recent findings with Auger show a dipole anisotropy at above $E\geq 8$~EeV that does not point toward the Galactic center, 
in which direction a Galactic component would certainly point, as the majority of potential sources is located there \citep{augerDipole2018}. Extragalactic sources do match the energy budget of the cosmic-ray flux above the ankle, but they cannot accommodate for the flux below the knee.

One of the central remaining questions concerns the exact transition between the Galactic and extragalactic components: Depending on the details, extragalactic sources might contribute to energies below the ankle, see \citep{ahlers2010,berezinsky2014}. At this point, it is not clear what fraction. Global models in any case seem to indicate the necessity for a \emph{Component B} in the cosmic-ray shin region: an additional (Galactic) cosmic-ray accelerator scenario with sources that produce high maximum energies but are not dominant below the knee. Candidates are PWNs, XRBs and the GTS.

Thus, these pieces of information, combined with the neutral messengers discussed in Section \ref{multimessenger_sources:sec}, now form the basis for multimessenger modeling of Galactic cosmic rays, which will be reviewed in the following Section.

\clearpage
\section{Multimessenger modeling \label{multimessenger_modeling:sec}}
Adding information from cosmic-ray secondaries adds significant information to the problem of cosmic-ray source identification. What state-of-the-art models have to provide is a proper description of both the signatures from charged cosmic rays as well as of their secondaries. In this section, this type of multimessenger modeling is reviewed by starting with the discussion of the state-of-the-art of cosmic-ray simulation codes (Section \ref{transportmodels:sec}), followed by the review of Supernova remnant emission models (Section \ref{snr:sec}), cosmic-ray propagation signatures from the Sun (Section \ref{sun:sec}) and diffuse gamma-ray emission in the Milky Way (Section \ref{ism_multimessenger:sec}).
\subsection{Numerical concepts of multimessenger modeling \label{transportmodels:sec}}

There are two different fundamental approaches for the modeling of \cora transport:
\begin{enumerate}
\item \textbf{The single-particle picture (SPP):} For each single particle that is propagated, the Equation of Motion (EoM) is typically solved with an adaptive step length 
approach,
\begin{equation}
m\cdot \ddot{\vec{r}}=q\cdot \left(\vec{v}\times \vec{B}\right)\,.
\end{equation}
Here, $q$ is the charge of the particle with mass $m$ which is moving with a velocity $\vec{v}$ through a magnetic field $\vec{B}$.

The SPP has the advantage of being able to follow each particle individually and to use arbitrary magnetic field configurations. Further, the results can be reweighted, thus being 
able to quickly change the input parameters, e.g.\ the  composition or spectral index. While this ansatz is precise in the ballistic regime, in which the typical transport  
parameter (correlation length, mean free path etc.) size is comparable or smaller than the gyroradius of the particle, it becomes extremely CPU time consuming in the diffusive 
regime, in which these parameters are much larger than the typical particle gyroradius. For the Galactic environment 
($B_0\sim 3\mu \mathrm{G}$), given a gyroradius of $r_g\approx 3.6$~pc$\cdot\left(q/e\right)^{-1}\cdot  \left(E/\mathrm{PeV}\right)\cdot \left(B/3\mu \mathrm{G}\right)^{-1}$, the 
propagation above $\sim 10^{17}$~eV works well in the SPP. At lower energies, this method becomes highly inefficient. This implies that the simulation of the transport of Galactic 
cosmic rays, in particular below the knee, the SPP method does not work efficiently.
\item 
\textbf{The multi-particle picture (MPP):} Solving the transport equation is typically a computationally faster approach when compared to the single-particle picture described above. This particularly concerns the solution of complex problems in diffusive environments as discussed above. For simplified magnetic field and density configurations, it can be possible to 
find analytical solutions for the particle distributions. For a proper description of Galactic propagation, it is useful to include the full magnetic 
field structure as well as a detailed gas density distribution. Thus, different propagation tools exist to model Galactic propagation quantitatively. One of the larger challenges 
for the MPP is that it relies on an assumption of a diffusion tensor, often simplified to a diffusion coefficient, thus simplifying the diffusion process to a 1-dim process as performed in \citet{strong_propagation_1998}, with more recent updates of the code and other numerical tools that take into account 3-dim propagation as discussed below.
  \end{enumerate}
The properties of the two different methods described above lead to the development of the following methods: On Galactic scales --- globally in the ISM, but even on local ones for the propagation in the vicinity of the acceleration site --- the solution of a set of transport equations is 
commonly used. It describes the flux for the different particle species and the change of their energy spectrum through processes like diffusion, advection, losses from interactions 
with the field (e.g.\ synchrotron radiation) or with matter (e.g.\ spallation and inelastic scattering). It typically provides spectra for the secondary species from the interaction processes as well. Such a multi-particle solution is realized in different numerical tools 
describing Galactic transport. GALPROP \citep{strong_propagation_1998} has been pioneering this field in the late 1990s in order to properly interpret the first MeV-GeV gamma-ray maps produced with the EGRET detector on board of the Compton Gamma-Ray Observatory (CGRO). More recently, DRAGON has build on the technical and physical knowledge gained with GALPROP to develop a state-of-the-art code \citep{dragon} and extend its possibilities to a radially changing diffusion coefficient. An update of the software introduced also a diffusion tensor and more complex magnetic field structures \citet{gaggero2013}. The PICARD code was developed in order to be able to use a full description of the diffusion tensor, including perpendicular diffusion terms. Furthermore, it introduced faster and especially more precise numerical scheme including a dedicated stationary solution solver \citep{kissmann2014}. For cosmic-ray propagation of extragalactic origin, 
a single-particle approach is used by solving the equation of motion.   CRPropa \citep{kampert2013,crpropa30} is a state-of-the-art tool including all features of extragalactic propagation in a modular way, which makes it possible 
to easily extend the tool.
Most recently, CRPropa has been extended as the only tool as of today in which the user can propagate particles with the single-particle approach \emph{or} the multi-particle approach 
\citep{merten2017b, merten2018}. This way, CRPropa can now be applied to the propagation of lower energy cosmic rays below the knee. More details on the different transport 
codes are presented in the next paragraph, i.e.\ Section \ref{transport_modeling:sec}.

\subsubsection{Numerical transport modeling \label{transport_modeling:sec}}

As mentioned above the simulation of the propagation of cosmic rays through the interstellar medium, but also locally at the sources is usually done in the multi-particle picture. Beside the obvious advantage of a smaller computational effort compared 
to the SPP it is often not even reasonable to simulate individual cosmic rays as the propagate as an ensemble.

For the propagation in the Galactic ISM,  a variety of different approaches have been applied to the problem of solving the transport equation for Galactic cosmic rays, see Equ.\ (\ref{transport:equ}). Here, 
the most common ones are discussed in more detail: 1) Dicretization on a multi-dimensional grid, 2) Transformation into a set of SDEs and 3) Simplification to allow for (semi-)analytical solutions. An overview of available simulation frameworks for the propagation of cosmic rays in the Galactic ISM or in the extragalactic context is given in Tab.\ \ref{tab:proptools}. The physics results achieved with these tools are discussed in the following sections. Here, we focus on the technical development and the technical comparison of the three different approaches. Most of the tools do implement a variety of different interaction and loss processes in their simulation chain. For leptonic particles usually at least synchrotron, IC and bremsstrahlung losses are considered. Some of the simulation frameworks do also include ionization losses and Coulomb scattering. For higher energetic leptons different variations of pair-production become relevant. The fast developing electro-magnetic cascade can be treated with analytical approximations or by explicit tracking of all secondary particles. The implemented processes for hadronic particles usually depend on the main energy regime the software was designed for. This means that up to now no simulation framework is able to treat the particle transport from the very highest to the lowest energies including all interactions. For extragalactic very high energy particles, Bethe-Heitler pair production, photo-disintegration for heavier nuclei and photo-pionproduction are considered. Here, too, the exact implementation of the losses differes from tool to tool. Below GeV energies, ionization and other losses may become relevant again.

\paragraph{Discretization} The discretization on a more dimensional grid is often seen as the standard approach to solve the transport equation. Today this usually done in three 
spatial and one momentum (or energy) dimension but also simpler models including only one or two spatial dimension are still being used. The advantage of this approach in general
that it is computationally fast. The more than two decade long development history of this ansatz has led to a good understanding of all the subtleties of the programs.  Furthermore, the simulation frameworks using the approach do include a lot of interactions 
beside the pure propagation of cosmic rays. On the other hand, the necessity of the usage of a full diffusion tensor that can be spatially varying has long been overseen and only recently been taken into account.

However, when a good spatial and energetic resolution of the Galaxy is needed these programs need a huge amount of memory. In fact, a bisected resolution for the simulation 
leads to an increase in memory request by $2^{n+1}$ for $n$ spatial and one momentum dimension.
  Furthermore, most of the common integration schemes do need very small time 
integration steps to be numerically stable. New algorithms to overcome some of the numerical problems, e.g.\ an explicit stationary solver, are introduced with the PICARD code. 
But even with these improvements it is hard to include discontinuities like shocks into such grid based simulations. 

\paragraph{Transformation to a set of Stochastic Differential Equations} Stochastic differential equations can be used to describe stochastic processes like diffusion. Instead of describing the particle density $n$ of the whole ensemble they describe the trajectory of an infinitesimal phase space element $x$. In the one-dimensional version such a SDE can be written as:
\begin{align}
\mathrm{d}x = A\,\mathrm{d}t + B\,\mathrm{d}w \quad .
\end{align}
Here, $A$ is a advection scalar and $B$ is the diffusion scalar. An ordinary differential equation part ist described by first term $A\,\mathrm{d}t$ and can describe effects that evolve linearly with the time element $\mathrm{d}t$ like it is the case for terms like advection. Diffusion as a stochastic process in temporal development is proportional to $\sqrt{t}$ \footnote{For sub- or super-diffusive motion this term is $\propto t^\alpha$ with $0<\alpha<0.5$ or $0.5<\alpha<1$, respectively.}. In this set of differential equations, it has been shown that it described by a Wiener process $\mathrm{d}w$, i.e.\ a stochastic process of real values in a time continuum. The textbook by Gardiner \citep{gardiner2009} and the code description by \citet{kopp2012} \citep{kopp2012} provide an excellent review of the topic.
 Transforming the transport equation into its equivalent set of stochastic differential equations 
 does, of course, not change the underlying physics 
description: While the equation looks like the description of individual pseudo-particles a single phase space trajectory, calculated in this ansatz, the results cannot be treated as single particles. This 
important fact in mind, the transformation can still be extremely useful. The first advantage is that the problem no longer has to be discretized in space and momentum but only in time. The 
spatial resolution is no longer bound by the available amount of memory but only by computation time, which of course might lead to an effective limitation, too. Since the 
individual phase space elements --- or pseudo-particles --- are treated independently the simulation scales linearly with the number of available CPUs, making this approach very 
interesting for modern highly parallel computing architectures. Furthermore, a reweighting of results after the actual simulation, known from Monte Carlo simulations in the SPP, 
are possible allowing for an efficient sampling of parts of the configuration space. In contrast to older grid based methods, here, also dedicated treatment of discontinuities is possible. A further advantage is the possibility of directly coupling simulations within the framework of the transport equation to the SPP.  

To achieve statistically meaningful results the simulation has to be averaged over a large number of pseudo-particles, because the Monte Carlo error only slowly decreases
with the number of simulated pseudo-particles $\Delta_\mathrm{MC}\propto 1/\sqrt{N}$.\footnote{For a spatial resolution of $\delta s=10$~pc of the whole Galaxy about 100 billion 
pseudo-particles are needed to reduce the error to the percent level $\Delta_\mathrm{MC}<5$~\%.} On the other hand, most of the available implementations of this ansatz lack in 
completeness of interactions. This, although not a conceptional problems, limits the use cases of these frameworks. 

\paragraph{Simplification} Reducing the complexity of the transport problem to a certain level that shortens simulation times has always been a possibility. With the improvement 
in grid based methods over the last years, the semi-analytical approach, e.g. \citep{2018arXiv180702968M}, has been moved a little bit of the focus. However, the downside, its lack in detail, is also an advantage of this method. 
USINE \citep{2018arXiv180702968M} is a semi-analytical simulation framework that is based on analytical solutions in space combined with finite-differences calculations for the energy dependence. Due to the short computation time, several simulations including sophisticated chemical networks and ionization losses can be run on the order of seconds. This allows for a full Markov Chain Monte 
Carlo (MCMC) analysis of the complete parameter space, which is normally not feasible with more complicated and thus computationally more demanding simulation frameworks. In addition, the derivation of analytical expressions as an approximation for complex problems is a necessity in order to validate the simulation results.

\begin{landscape}
\begin{table}[htbp]
\begin{threeparttable}
\caption{Simulation Tools listed with their main propagation features. Included is also a list of available interactions (see notes below the table for explanation).}
\label{tab:proptools}

\begin{tabular}{l|llllllllllll}
\toprule
    Name  & Picture & Tech.\tnote{1} & Sim. & Lept.\ Int.\tnote{2} & Hadr.\ int.\tnote{2} & Dif. & Adv. & Re-Acc. & AC & Special features & Ref. \\

\midrule
 USINE & MPP   & sa    & 1D/2D & CS, I & CS, I, $N+p$ & iso. & part. & part. & ?     & Interface with MCMC &  \cite{2018arXiv180702968M}  \\
    GALPROP & MPP   & rg    & 3D    & I, CS, BS,  & $N+p$, RD, EC, & iso., & yes   & yes   & yes   & webrun & \cite{strong_new_1998}, \cite{GALPROP_v56} \\
    &       &       &       & IC, SYN &  ES, I, CS & var. &       &       & \\
    
    DRAGON 2  & MPP   & ig    & 2D/3D & SYN, IC,  & $N+p$, I, RD & aniso.,  & yes   & yes   & yes   &       & \cite{dragon, gaggero2013} \\
    &       &       &       & BS, I, CS & & var. &       &       & \\
    
    PICARD & MPP   & ig    & 3D    & I, CS, BS,& $N+p$, RD, EC, & aniso.,  & yes   & yes   & yes   & stat.\ solver & \cite{kissmann2014}\\
    &       &       &       & IC, SYN &  ES, I, CS & var. &       & \\
    
TransportCR & MPP   & g     & 1D    & CS, SYN, & $p+\gamma$,  & iso.,  & no    & no    & no    &  & \cite{TransportCR_I, TransportCR_II} \\
    &       &       &       &  PP, TPP, IC & $n$-decay &  const.     &       & \\
    
    CRPropa3.1 & MPP   & sde   & 3D    & SYN, IC, & IC, $N+\gamma$, RD & aniso., & yes   & no    & yes   & Backtr., Reweight.,& \cite{merten2017b, merten2018} \\
    &       &       &       &   & PP, TPP &  fixed EV &    & & &  arb.\ magn.\ field\\
    
   PIERNIK & MPP   & g     & 2D    & SYN   & --    & aniso.,  & yes   & no    & yes   & MHD, coupled to gas & \cite{Piernik_I, Piernik_II, Piernik_III} \\
       &       &       &     &   &  &   const.    &       & & & & \cite{Piernik_IV, Piernik_CRESP}\\
           
    AREPO & MPP   & mm    & 2D/3D & CS, SYN & STR, CS, $p+p$ & aniso. & yes   & yes   & yes   & MHD, coupled to gas & \cite{2017MNRAS.465.4500P} &  \\
    
    CRPropa3.0 & SPP   & bp, ck & 4D/3D/ & SYN, IC,& IC, $N+\gamma$, RD & --    & --    & --    & --   & Backtr., Reweight., & \cite{crpropa30} \\
    &       &       & 1D      &   & PP, TPP &       &       & & &\emph{real} trajectories \\
        
    SimProp & SPP   & --    & 1D    &       & $p+\gamma$, CAC & --    & --    & --    & --   &       &  \cite{SimProp2017}  \\ 
    
    HERMES & SPP   & --    & 3D    & via EleCa & $p+\gamma$, CAC & --    & --    & --    & --    &       & \cite{HERMES2013}   \\ 
    
    PriNCe & MMP & m & 1D & & IC, $N+\gamma$, RD & --    & --    & --    & --    & Fast and non-superposition    & \cite{Heinze2019} \\
    &       &       &      &   & PP, TPP &       &       & & & for $N+\gamma$ \\

\bottomrule
\end{tabular}
\begin{tablenotes}
      \small
      \item[1] g---grid based, rg---regular grid, ig---irregular grid, sde---stochastic differential equation, bp---Boris push, CK---Cash-Karp, sa---semi-analytical, mm---moving mesh, m---matrix
      
      \item[2] I---Ionization, CS---Coulomb scattering, BS---Bremsstrahlung, IC---Inverse Compton scattering, SYN---Synchrotron radiation, PP---Pairproduction, TPP---Triple pairproduction, (C)AC---(Cosmological) adiabatic cooling, RD---radioactive decay, EC---Electron capture, ES---Electron stripping
\end{tablenotes}
\label{tab:SimulationParametre}
\end{threeparttable}
\end{table}
\end{landscape}

\subsection{Supernova remnants\label{snr:sec}}
As demonstrated above, the class of SNRs is the only one that easily matches the luminosity criterion for the cosmic-ray flux above $1-10$~GeV. Maximum energies of $10^{15}$~eV or even higher can be reached with magnetic field amplification in the SNRs, predicted by non-linear theory \citep{bell1978a,bell1978b,drury1983} and observed for a larger number of SNRs \citep{uchiyama2007,vink2012}. Power-law type spectra (possibly with a curvature) are expected in the acceleration of cosmic rays at the SNR shock fronts. Those features that are more difficult to explain are the observed low level of anisotropy, which is expected to be larger given the gradient of the SNR population in the Galaxy, as well as the complex, rigidity-changing composision. But even the trivial-seeming criteria on luminosity, maximum energy and power-law spectrum have their subtleties.

\begin{figure}[htbp]
\centering{
\includegraphics[trim = 0mm 0mm 0mm 0mm, clip, width=0.9\textwidth]{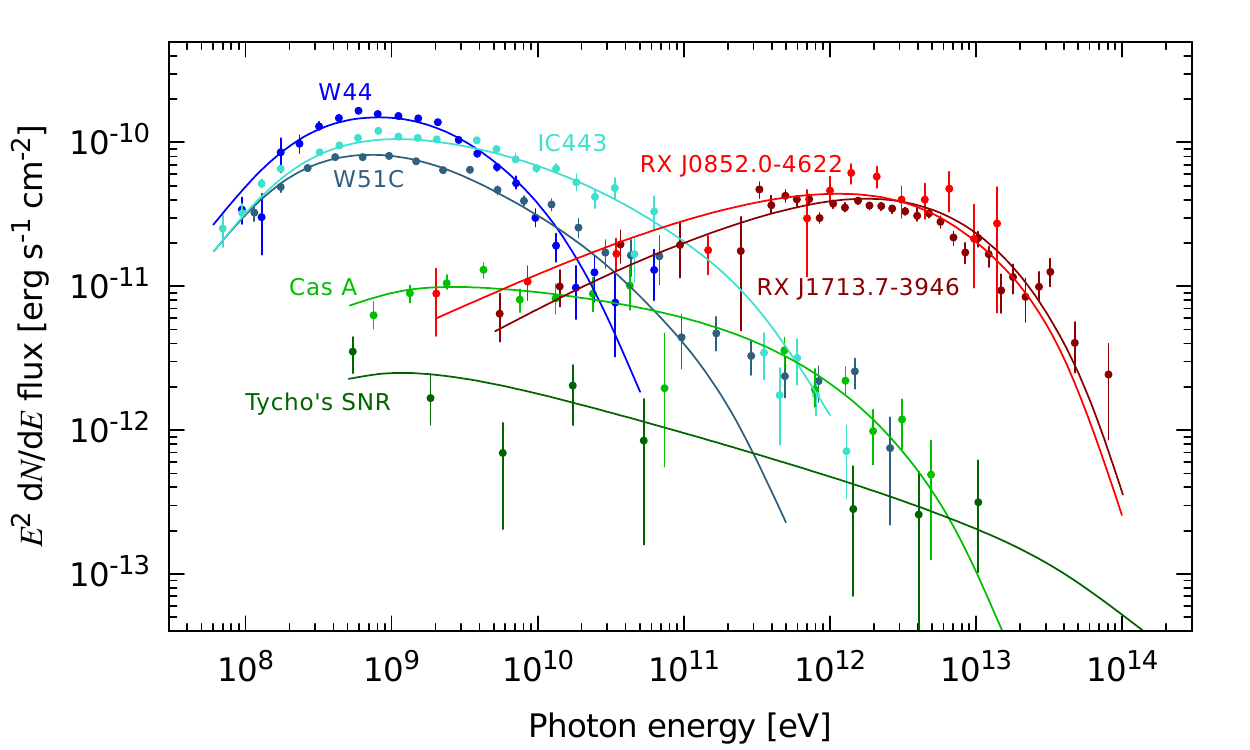}
\caption{Selection of SNRs with high-energy gamma-ray emission showing three old SNRs (order $\sim 10.000$~years, W44, IC443 and W51C) with low-energy cutoffs, young SNRs with flat spectra and high-energy cutoffs (Tycho and CasA) as well as flat sources with TeV-energy turnovers (RX J0852.0-4622 and RX J1713.7-3946). Figure reproduced after \citep{funk2017}.
\label{gamma_snrs:fig}}}
          \end{figure}

The three SNRs IC443, W44 \citep{fermi_ic443} and W51 \citep{jogler_funk2016} have been discussed to reveal the so-called pion bump and therefore representing hadronically produced gamma-ray spectra. This bump is expected at $E_{\gamma}=70$~MeV, as the low-energy limit of photon production is determined by the neutral pion mass $m_{\pi^{0}}=140$~MeV, which at lowest energy production the two photons share. To produce a sharp turnover at exactly this energy is impossible with the competing processes Inverse Compton scattering or bremsstrahlung. Figure \ref{gamma_snrs:fig} shows those shell-type SNRs that have been measured up to $>$TeV-energies. What becomes clear is that those SNRs with a confirmed pion-bump have extremely steep spectra, which at TeV energies are as steep as $E_{\gamma}^{-4}$. Assuming that the hadronic gamma-ray spectral behavior is quite close to the inducing cosmic-ray spectrum, a source population dominantly consisting of such sources cannot be responsible for the detected cosmic-ray flux up to the knee, which after the steepening diffusive transport is still more than one power flatter.  There is still a debate that is ongoing if the detected bump is not rather due to the kinematic break in the energy spectrum. Such a break is expected in the spectra at around $\sim 100$~MeV photon energy in the typical acceleration scenario, in which the particle spectra follow a power-law in momentum space. Going from momentum to energy ($E^2-p^2\,c^2=m^2\,c^4$) results in a break in the primary spectrum around the rest mass $m$ of the particle, i.e.\ $938$~MeV for protons. As the photons receive around 10\% of the proton energy on average, this results in a break at $\sim 100$~MeV for the photons \citep{strong_icrc2015,strong2018}. Thus, even this alternative explanation builds on a hadronic origin and thus supports the idea of SNRs being particle accelerators, at least up to $\sim 100$~GeV.

Figure \ref{gamma_snrs:fig} vizualizes that gamma-ray emitting SNRs occur with a large variation concerning their spectral behavior. It is interesting to note that --- although the numbers are still very small of course --- there appear to be three groups of spectra \citep{funk2017}:
\begin{itemize}
\item GeV-peaked steep-spectrum sources, $\propto E_{\gamma}^{-4}$ (examples IC443, W44, W51), all mid-aged SNRs interacting with molecular clouds, $\sim$~20,000~years old.
\item TeV-peaked flat-spectrum sources, significantly flatter than $E_{\gamma}^{-2}$ (examples RXJ1713.7-3946, RXJ0852.0-4622). These are young SNRs with an age of $\sim 2.000$~years.
  \item Flat, $E_{\gamma}^{-2.3}$-type sources (examples Tycho, CasA). These are extremely young sources with an age of $<1.000$~years. However, the spectrum of CasA has been proven to have a cutoff at $\sim 3.5$~TeV recently \citep{magic_casa_2017} and Tycho becomes as steep as $\sim E_{\gamma}^{-3}$ above 400 GeV \citep{archambault2017}, also interpretable as a cutof at $E_{\gamma}^{\rm cut}\sim 1.7$~TeV.
\end{itemize}

          Such a characterization of the source spectra according to their ages is not unexpected. As reviewed in e.g.\ \citet{cox1972}, SNRs are dynamical systems concerning shock velocity, radius, temperature, magnetic field, ect. These dynamics must be reflected in the features of cosmic-ray emission as well, i.e.\ a time-dependent maximum energy that at some point during the evolution reaches the energies of the cosmic-ray knee. It is striking, though, that no SNR PeVatron could be identified until today. The only PeVatron detection is a central source in the Galaxy, probably Sgr A$^{*}$ \citep{hess_gc_2016}. In general, the differentiation between hadronic and leptonic spectra with current data is not unambiguous as shown in \citet{mandelartz2015}. While Inverse Compton spectra in general should be flatter than $E^{-2}$, the combination of IC, bremsstrahlung and $\pi^{0}$-decay signatures can easily lead to a steep spectrum even with a leptonic dominance.

From a theoretical perspective, the general evolution scenario after a supernova explosion is the following:
The first few hundred years, the SNR is in the \textit{free expansion phase}, in which the pressure of the interstellar gas is still negligible. The shock radius $R_{\rm sh}$ is expected to increase as $R_{\rm sh}=v_{\rm ej}\cdot t$ with $v_{\rm ej}$ as the ejection velocity and $t$ as the time since explosion. As soon as mass accumulated from the ISM inside the SNR shock front equals the ejected mass, $M_{\rm ISM}\approx M_{\rm ej}$, the SNR goes into the \textit{Sedov-Taylor phase}. The self-similar solution of the evolution of the remnant in these $\sim 10.000$~years  has been presented by \citet{sedov1959} and \citet{taylor1950}. 
As presented in \citep{cox1972}, the shock radius is expected to expand as $R_{\rm sh}^{\rm ST}\propto t^{m}$ with $m$ depending on the nature of the supernova explosion \citep{chevalier1982}. The upstream magnetic field at the shock front is expected to decrease as $B(t)\propto R^{-1}-R^{-2}$, see e.g.\ \citep{biermann2018}. Thus, in the best case, the maximum energy in this simplified scenario is expected to decrease as $E_{\max}(t)\propto t^{-m+\epsilon}$, with $0<\epsilon <m$. The acceleration time itself is limited by the age of the remnant \citep{lagage_cesarsky1983b}.

These further factors, however, constraint the actual evolution of the maximum energy of the SNR, as the true maximum energy will be determined by the minimum function
\begin{equation}
E_{\max}(t)=\min\left(E_{\max}^{\rm Hillas/age}(t),\,E_{\max}^{\rm loss}(t),\,E_{\max}^{\rm acc}(t)\right)\,.
\end{equation}

\citet{pohl2012} discuss that even in the early free-expansion phase, maximum energies above knee-energies can be reached when assuming a realistic velocity profile in the MHD modeling of the evolution of the SNR shell. In particular, the authors discuss that for an explosion into the ISM, the maximum energy increases before decreasing before the Sedov-Taylor phase. For wind-SNRs, it is highest at the earliest times and is monotoneously decreasing afterwards. Taking into account magnetic field amplification, \citet{zirakashvili_ptuskin2012}  find that the highest maximum energies for cosmic rays can be reached when the transition between the free expansion phase and the Sedov-Taylor phase is happening. Based on these findings,  \citet{gaggero_time2018} discuss the evolution of maximum energy for the case of SN-Ia and SN-II events assuming  that $3.5\%$ of the ram pressure is converted into magnetic field energy. They find that both progenitor types have their maximum ($E_{\max}>$~PeV) in the maximum energy at early times (within the first hundred years), dominated by the acceleration time scale in the case of hadrons. Electrons, on the other hand, are basically loss-dominated in their maximum energy. the SN-II type sources have a second local maximum at $\sim 1000$~years that is somewhat below $1$~PeV, at the time where the transition from the thick wind to the cavity is happening. This would explain why young sources like Tycho and CasA do not reach PeVatron energies: being a few hundred years old, they have already passed the absolute maximum energy and in the case of the SN-type-II CasA has not reached the second peak yet. This scenario is supported by findings in M82, where a sample of 23 recently exploded SNe of red and blue supergiant progenitor stars reveal a combination of radius and magnetic field that interestingly always result in a constant value of their product, i.e.\ $R\,B = 10^{16 \pm 0.08}$~G~cm \citep{biermann2018,biermann2019}. Such a combination of radius and magnetic field corresponds to a Hillas maximum energy of $3\times 10^{18}$~eV, thus providing the possibility to early-on in the evolution contribute to energies well beyond the knee. Overall, as already pointed out in \citep{pohl2012}, the time evolution of the maximum cosmic-ray energies in SNRs is still largely unknown, since it strongly depends on the evolution of the magnetic field, which is far from being well-constrained. What still becomes clear from the results discussed above is that it is rather reasonable to assume that SNRs can accelerate particles up to the knee early-on in their life-time and that this energy is maintained for about  hundred years, but most likely not longer than that. In order to explain the observed cosmic-ray spectrum from GeV to PeV energies, this phase must be the one to dominate the cosmic-ray luminosity, otherwise, the spectrum would not be so smooth, at least if it is a large number of sources responsible for the flux in that energy range. Alternatively, the flux could be explained by assuming that one single source dominates the cosmic-ray energy spectrum.

As summarized by \citet{gabici2019}, the measurement of the energy spectra of electrons and positrons can help in the future to shed more light on the question of how smooth the energy spectra are. While there is an indication of a break in the electron plus positron spectrum in Fermi data at around $\sim 53\pm 8$~GeV, the change in the spectrum from steep ($\gamma \sim 3.2$) to flat ($\gamma \sim 3.1$) is not significant yet when considering systematic uncertainties of Fermi. The measurement of the electron spectrum by PAMELA does not reveal a break at this point and it is well-described by a powerlaw with $\gamma\sim 3.2$ in the energy range $30-625$~GeV.  The Dark Matter Particle Explorer (DAMPE) provides direct measurements in the range $25- 4,600$~GeV and detects a break at around $900$~GeV, which is compatible with H.E.S.S.\ measurements \citep{aharonian2008}. The Calorimetric Electron Telescope (CALET)  measurement the spectrum in the energy range  $11 - 4800$~GeV and is in agreement with the other detections below energies of $300$~GeV. At higher energies, however, the spectrum is significantly softer \citep{calet2018}.

If one or more SNRs in the Galaxy are responsible for the cosmic-ray flux, the different features in the spectrum, in particular what we call the \textit{cosmic-ray hip} and the \textit{cosmic-ray knee} in Section \ref{candidates:sec} need to be explained.

\begin{figure}[htbp]
\centering{
\includegraphics[trim = 0mm 0mm 0mm 0mm, clip, width=0.9\textwidth]{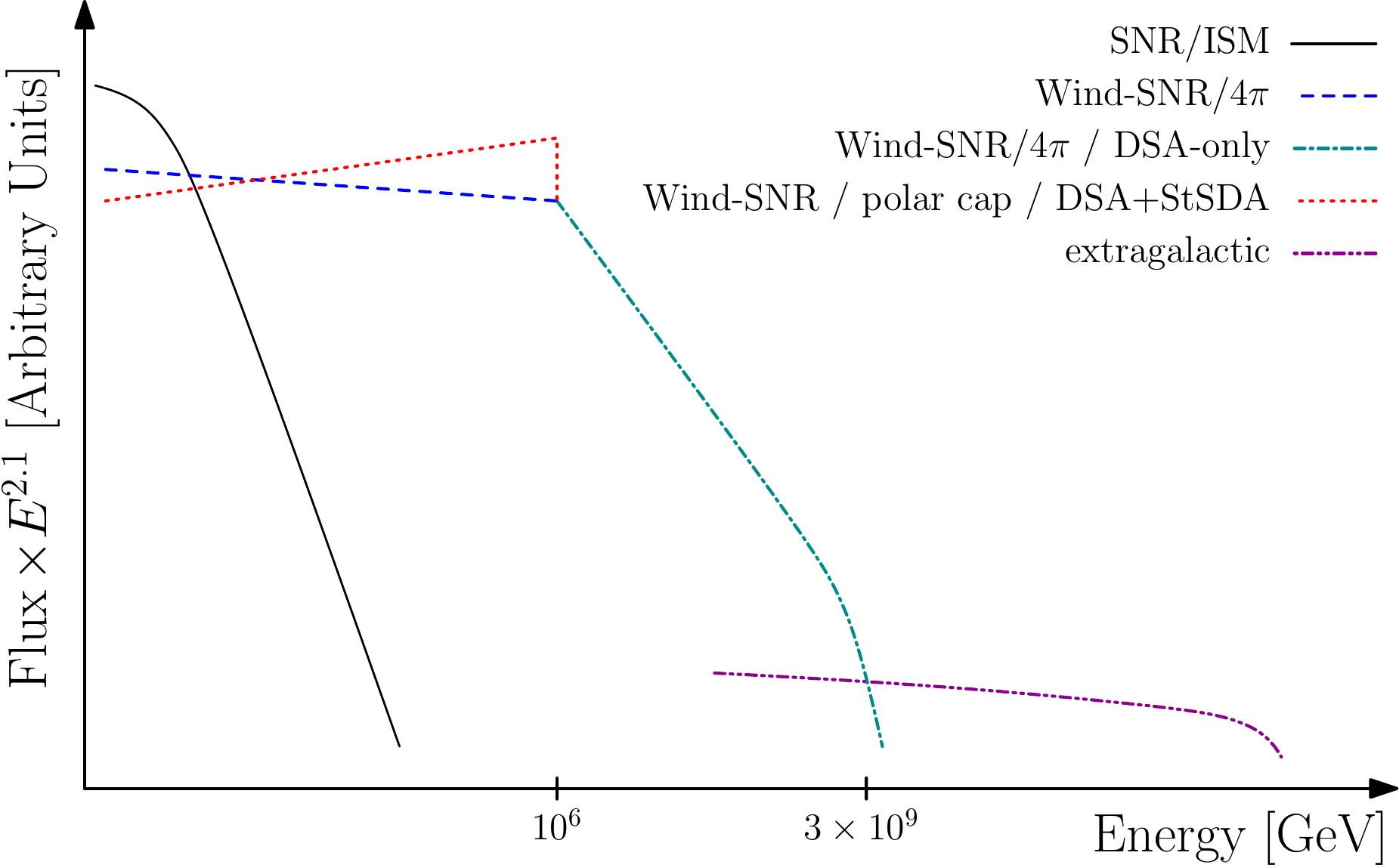}
\caption{Sketch of the different components expected from supernova remnants. black-solid: Contribution from interaction with the ISM; Blue-dashed: Wind-SNR contribution from the $4\pi$-region, where both DSA and StSDA are at work; green-dot-dashed: Wind-SNR contribution from $4\pi$-region with only DSA at work. Red-dotted: Wind-SNR contribution from the polar cap region. The dot-dot-dashed pink line shows the extragalactic contribution, not connected to emission from the Milky Way. Figure reproduced after \citep{biermann2018}.
\label{snr_spectrum:fig}}}
          \end{figure}

In doing so, SNRs can be divided into different sub-classes: SNRs that explode into the ISM (referred to as \textit{ISM-SNRs} in the following) and those that explode into their winds (labeled \textit{wind-SNRs} from now on). In the following, it will be shown that using the different types of SNRs discussed below, it is possible to explain the entire cosmic-ray energy spectrum from GeV-energies up to the ankle. As reviewed below, the explanation is able to explain all spectral kinks (see also Fig.\ \ref{snr_spectrum:fig}), as well as the primary and secondary composition.
\begin{enumerate}
\item \textbf{ISM-SNRs:} Massive stars with zero age main sequence (ZAMS) masses on the order of $10\,M_{\odot}<M<25\,M_{\odot}$ explode into the ISM and create a shock front as discussed above. The upstream density at these shocks is teneous which limits the acceleration process for these ISM-SNRs \citep{stanev1993}. This way, ISM-SNRs therefore are expected to have a relatively steep spectrum, i.e.\ $\left.dN_{\rm ISM-SNR}/dE\right|_{\rm inj}\propto E^{-2.4}$ and the spectrum will be dominated by hydrogen \citep{biermann_apj2010}. In Fig.\ \ref{snr_spectrum:fig}, the black-solid line represents this contribution at Earth. The ISM-SNRs are not suited to reach knee-energies.
\item \textbf{Wind-SNRs:}
  It has been argued for a long time that the acceleration of cosmic rays toward the highest energies if far more efficient in SNRs that explode into their own, dense wind \citep{biermann1993,stanev1993,thoudam16b}. In this scenario, the powerful magnetized wind that is emitted from stars with ZAMS masses $M>25\,M_{\odot}$ is in favor of cosmic-ray acceleration via both diffusive shock acceleration in highly amplified fields, with additional contribution from stochastic shock drift acceleration. An elaborate review on the properties of wind-SNRs and their cosmic-ray spectra is given in \citet{biermann2018}. 
  For the heavier stars, there are two sub-classes: Red supergiants (RSGs) with a high-density wind and Blue supergiants (BSGs). The latter have a thin, high-velocity wind and barely any hydrogen left, as they have ejected the hydrogen envelop before the explosion. Depending on the initial mass, these stars have a high fraction of helium (low-mass BSGs) or tend toward a higher ratio of Carbon and Oxygen, moving toward the highest mass stars, Helium is negligible for these stars as even the helium envelop is ripped off. SNRs from RSG stars can accelerate particles up to the knee, while BSG star progenitor SNRs can reach to even higher energies
  As for the BSGs, in the lack of these lightest elements, cosmic-ray spectra are therefore expected to be dominated by heavier elements, starting at helium. The proton spectrum is dominated at the lowest energies by the abundant ISM-SNRs. With their steep spectrum, these are overtaken by the RSGs at some point.
  
  In the cosmic-ray acceleration at the shock front, the orientation of the magnetic field becomes important. When assuming a Parker-spiral as the underlying structure of the large-scale magnetic field, two general components can be identified, with different orientations of the magnetic field and the shock front normal toward each other:
  \begin{enumerate}
  \item \textit{$4\pi$-component:} For most of $4\pi$, the magnetic field orientation is basically perpendicular to the shock front normal. In such a scenario, cosmic rays are accelerated by a combination of DSA and StSDA up to knee-energies. At the acceleration site, the spectra are $dN_{\rm 4\pi}/dE|_{\rm inj}\propto E^{-2}$ as they are injected into the local medium. There, they are further steepened via propagation in the local medium by diffusion in a highly turbulent medium, thus leading to a local spectrum of $\left.dN_{\rm 4\pi}/dE\right|_{\rm inj}\propto E^{-7/3}$. At knee energies, StSDA becomes less and less important and the spectrum steepens by about  $E^{-0.5}$ \citep{biermann2018}. This scenario can reproduce the steep component contributing between the knee and the ankle, $dN_{\rm 4\pi,high}/dE\propto E^{-2.8}$. Figure \ref{snr_spectrum:fig} shows the contribution before steepening as the blue dashed line and after steepening as the green dot-dashed line.
    \item \textit{Polar-cap component} For this geometrically small component at the poles, the shock front normal is parallel to the magnetic field. In such shock configurations, the acceleration process itself is slower and the effect of turbulence lower. This way, the spectrum is not locally steepened, but obtains its spectral features directly from the acceleration, $dN_{\rm polar}/dE\propto E^{-2}$, while due to the slower acceleration process, time for the production of secondaries at lower energies is larger than in the $4\pi$-component. With no StSDA at work for these parallel shocks, the spectrum is expected to have a sharp cutoff at knee-energies. In relation to the $4\pi$-component, the polar cap component is (a) significantly flatter and (b) of lower luminosity due to the smaller angle that is covered. Thus, a change in the spectrum is expected at the energy for which the $4\pi$-component becomes weaker than the polar-cap component. This is shown as the red dotted line in Fig.\ \ref{snr_spectrum:fig}. This model is in accordance with newest simulation results, where \citet{pais2018} can show that the dominance of a large-scale field results in shock acceleration at parallel shocks, while for purely turbulent fields, emission is seen over $4\pi$.
  \end{enumerate}
\end{enumerate}
Such a concave turn-over of the spectrum is expected at rigidities $100$~GV$<R_{\rm turn}<$~TV for all elements \citep{biermann2010}. This could explain the change in the spectrum at $\sim 300$~GV. In this composition-resolved view that became available with CREAM, PAMELA and AMS-02-data, the proton spectrum is much steeper than the heavier elements. Within the model presented above, this is well-explained by the dominance of ISM-SNRs in protons, as Wind-SNRs have ejected their hydrogen-envelope before the explosion and leave little hydrogen for acceleration, but rather dominate the heavier nuclei spectrum.

The scenario presented above is well-suited to explain the multimessenger picture of Galactic cosmic rays, including the observation of composition and all-particle spectrum \citep{biermann2010,thoudam16b}, ratios of anti-protons to protons as well as the measured positron fraction \citep{biermann_prl2009}. It is compatible with gamma-ray observations and expects neutrino emission that is currently too weak to be detectable with IceCube \citep{mandelartz2015}.

\begin{figure}[htbp]
\centering
\includegraphics[width=0.8\textwidth]{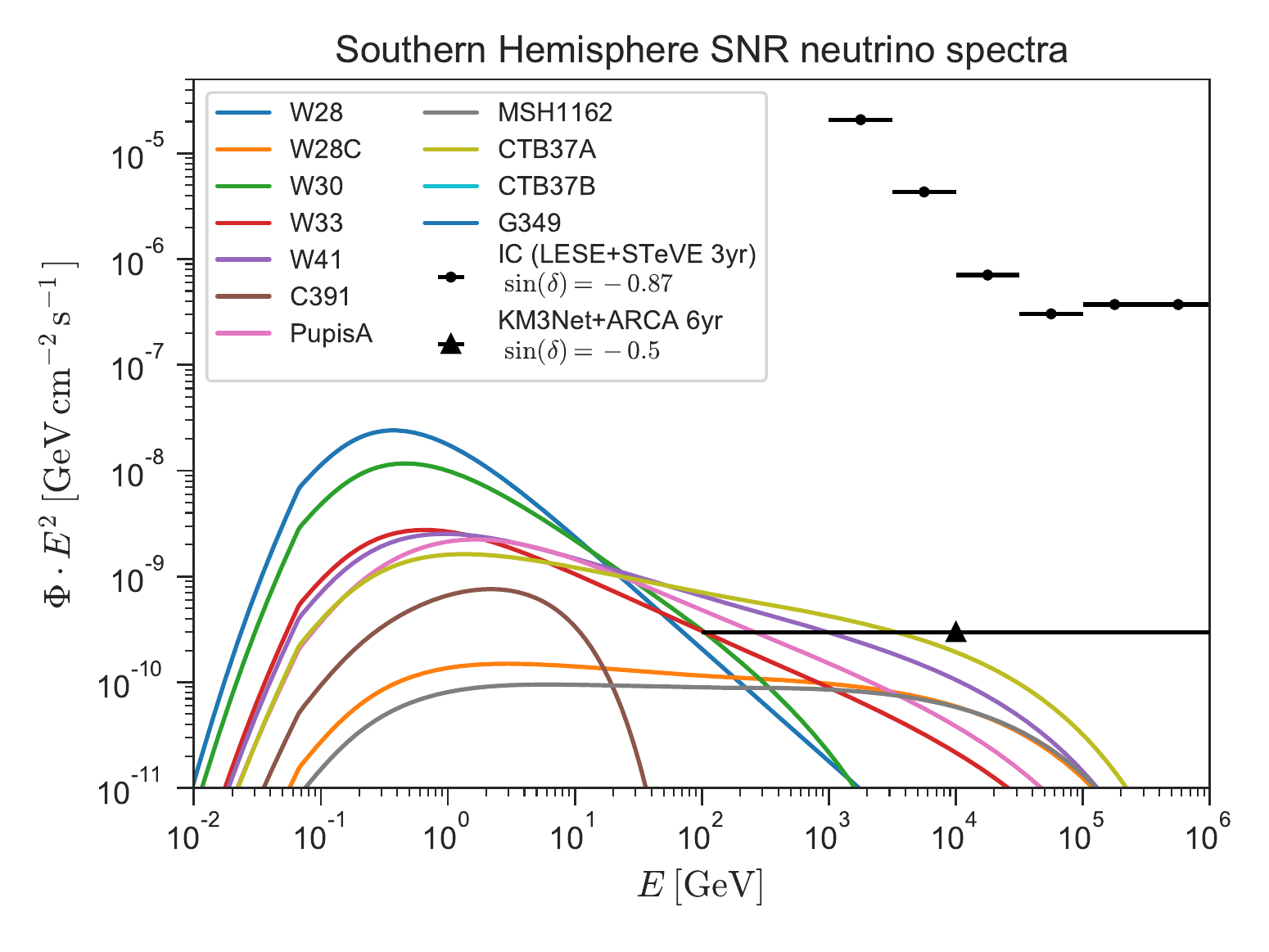}
\caption{Neutrino spectra from SNRs in the southern hemisphere in comparison to IceCube \citep{icecube_steve} and KM3NeT \citep{km3net_snrs2019} sensitivities. Figure reproduced after \citep{mandelartz2015}.  
\label{neutrinos_snrs_south:fig}}
\end{figure}

\begin{figure}[htbp]
\centering
\includegraphics[width=0.8\textwidth]{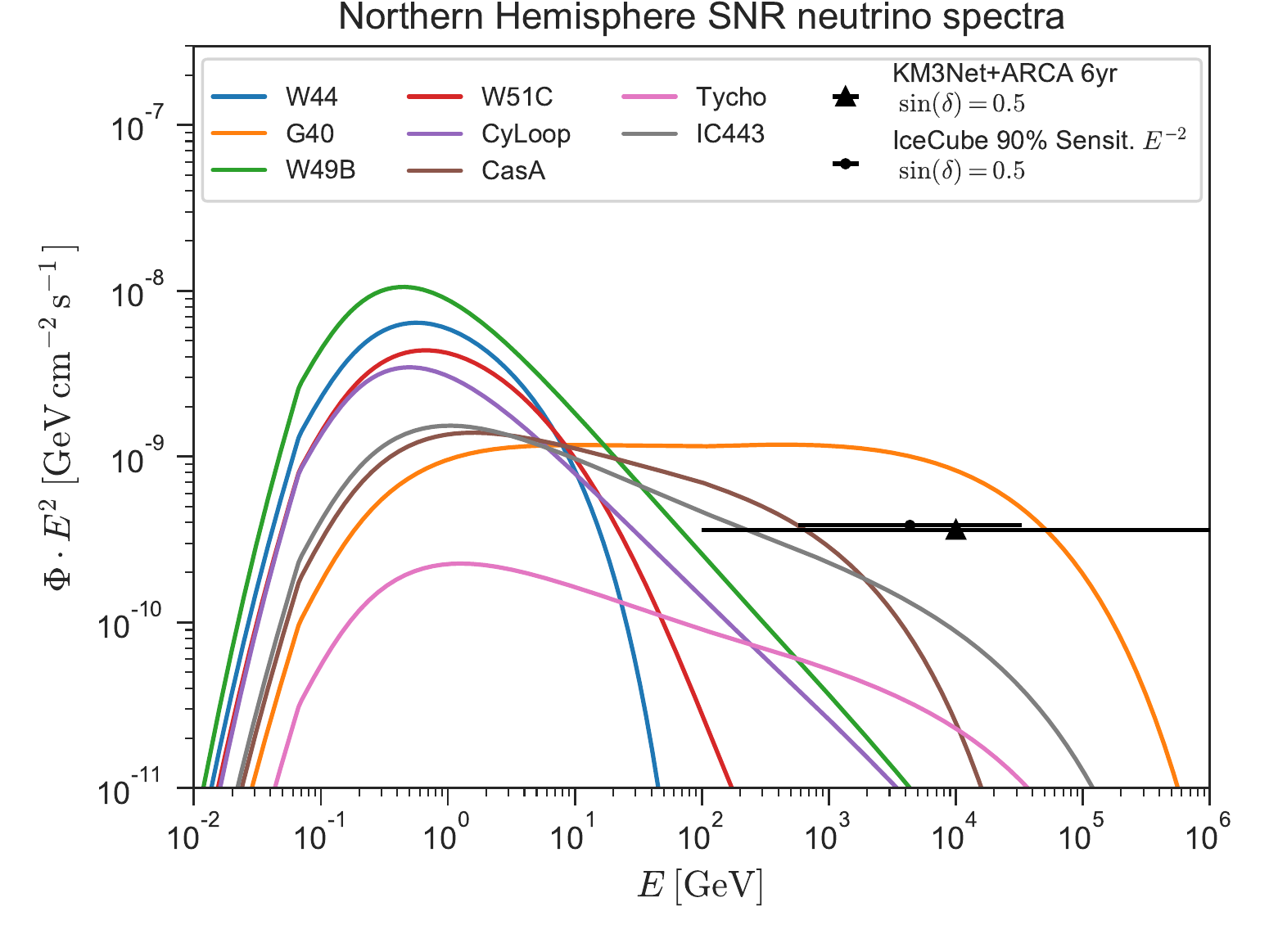}
\caption{Neutrino spectra from SNRs in the northern hemisphere in comparison to IceCube \citep{icecube_ps2020} and KM3NeT \citep{km3net_snrs2019} sensitivities. Figure reproduced after \citep{mandelartz2015}.}
\label{neutrinos_snrs_north:fig}
\end{figure}

The general problem that still exists more than 80 years after Baade \& Zwicky's prediction that SNRs dominate the cosmic-ray flux \cite{baade_zwicky1934}: there is no positive proof yet for this statement. Also the features discussed above could come from the alternative scenario to explain the turn-over of the spectrum at $\sim 300$~GV is due to the streaming instability  \citep{CRdrivenTurb}. The following measures will help in the future to come to solid conclusions about SNRs as the origin of cosmic rays up to the knee and possibly above:
\begin{itemize}
\item \textbf{Origin of the break at $300$~GV} While the transport-related break predicts a break at $300$~GV in the B/C ratio, it could also arise in the source-related scenario \citep{ahlers2009,blasi2009,cholis2014,tomassetti_oliva2017}. Recent results presented by \citet{ams2016} indicate the possibility for such a break in the rigidity dependence of B/C. However, for solid conclusions, the ratio needs to be measured at high precision to values $\gg 300$~GV in the future, in particular to be able to detect potential subtle differences in the transport- and source-related break scenario.  Gamma-ray spectra should reveal a concave structure in the energy spectrum at around $E_{\gamma}^{\rm break}\sim E_{\rm CR}^{\rm break}/10\sim 30$~GeV. In order to properly observe this change by only $1/3$ in the spectral index, higher-precision measurements over a broad range of energies are necessary, thus relying on detections in the GeV-range by a mission that would need to exceed the Fermi precision in energy resolution and sensitivity significantly. Such a level of precision is unfortunately not foreseeable in the near future. In general, synchrotron observations also yield the possibility to detect a break in the local cosmic-ray electron spectrum of an SNR. The $300$~GeV electrons dominantly radiate at a critical energy $\nu_c=3\gamma^2\,q\,B/(4\pi\,mc)\sim 1,5$~THz\,(B/$\mu$G)\,$(E/(300$~GeV$))^2$. At theses super-microwave frequencies, it is not possible to detect synchrotron radiation unfortunately and this option therefore does not seem viable.
\item \textbf{Acceleration up to knee and beyond energies} Only a study of a broad (complete) sample of GeV-emitting SNRs in combination with TeV measurements will make it possible to identify those sources that can accelerate up to the knee and model the entire population to fit the cosmic-ray spectrum. At this point, using the existing sample of gamma-ray sources to explain the all-particle cosmic-ray spectrum with a QLT Kolmogorov diffusion index of $1/3$ results in a flux that is still compatible with the all-particle spectrum up to the knee  \citep{galprop_sources2016}. However, the median expectation is in principle too low and the compatibility is mainly due to the high uncertainty in the proton spectra due to uncertainties in the gamma-ray measurement and disentangling hadrons from electrons. It is also not possible to identify a possible break at $\sim 30$~GeV in gamma-rays. In the future, the Cherenkov Telescope Array with increased gamma-ray sensitivities at $100$~TeV energies and beyond, corresponding to cosmic-ray energies of a PeV and above, is of great importance to pin-point the sources of cosmic rays and to investigate if the variety of
  spectra is possible to explain the relatively smooth cosmic-ray distribution up to knee energies.
\item \textbf{Distinguishing between hadronic and leptonic signatures}
            to go for higher spatial and energy resolution in order to understand the gamma-ray evolution of the SNR and to have a proper view of the spectral behavior. The CTA Observatory will be valuable in the future, aiming at a spatial resolution of $\theta \sim 0.03^{\circ}$ and the improvement of sensitivity by a factor of $\sim 10$ as compared to this generation's IACTs \citep{cta}. The latter is connected to an energy extension of the measurements toward the highest energies, helping to identify more PeVantrons in the Galaxy via a systematic plane scan that is expected to detect several hundreds of individual sources, among these on the order of 100 SNRs \citep{zanin2017}. In addition, young SNRs will be studied in more detail \citep{cta_snrs2015}.  
            Adding information from CTA, studying the morphology of the multimessenger emission will help to distinguish electron-dominated from hadron-dominated scenarios. In particular, a sharply peaked gamma-ray emission near the  shock front of the remnant is something that is expected in a hadronic scenario, as interactions would dominantly happen in the dense region at the shock front. IC emission, sensitive to the electron density and the magnetic field, on the other hand, would not decay that quickly, providing a clear prediction for differences in the morphology. 
            As an example, the morphology of one of the brightest SNRs at TeV energies,  RX~J1713.7-3946, can be investigated at high resolution and hadronic and leptonic scenarios are expected to be clearly separable \citep{rxj_cta2017}.
                        Adding synchrotron information to this multimessenger picture will help further, as IC and synchrotron are expected to be correlated. The 
                        scenario becomes more complicated for bremsstrahlung-dominance, which also scales with the density. Adding information on spatially-resolved ionization measurements as i.e.\ done for the case of W28 in \citep{vaupre2014}. Theoretical investigations show that depending on the decay of the ionization signature with the distance from the shock front strongly depends on the local environment and can either be dominated by photohadronic ionization in the case of strong X-ray emission --- in these cases, the photohadronic component can be maintained up to large distances via the subsequent production of secondary electrons \citep{schuppan2014}. For regions with lower X-ray emission, hadrons will dominate the ionization level \citep{becker2011,schuppan2012,schuppan2014}. Finally, the detection of neutrinos from SNRs would make the association of hadronic cosmic-ray acceleration bullet proof, see \citep{becker2008} for a review. Figures \ref{neutrinos_snrs_south:fig} and \ref{neutrinos_snrs_north:fig} show the neutrino flux from those SNRs that are measured at TeV energies with IACTs for southern and northern sources, respectively. Using existing multiwavelength information from radio to TeV energies, the two energy bumps for the sources in these plots have been fitted by first fitting the synchrotron spectrum in the radio to X-ray range. This fixes the product of the electron density and the square of the magnetic field, $n_e\cdot B^2$. To fit the high-energy bump, the magnetic field is varied from the lowest value in which typically the leptonic processes (IC, brems) dominate toward higher B-fields, where the hadronic component becomes more important. The best-fit results have then been used to derive the neutrino spectra from the hadronic gamma-ray part with the results shown in the respective figures. While the fluxes are too low to be detected with current instruments (IceCube, ANTARES), future observatories IceCube-Gen2 and KM3NeT/ARCA have point-source sensitivities on the order of $\left. E_{\nu}^{2} dN_{\nu}/dE_{\nu}\right|_{\rm sens} \sim 10^{-10}$~GeV/(s\,cm$^2$) as shown in the figures. KM3NeT has the sensitivity to detect the sources VelaX and RX J1713.7-3946 within $6-15$ years \citep{km3net_snrs2019} and in a recent IceCube cascade analysis, RX J1713.7-3946 was the warmest spot in the analysis, however not significant for evidence of detection \citep{icecube_cascades2019}. The exact detection potential for detection in the end depends on the subtleties of the sources, i.e.\ how steep/flat they are, at what maximum energies they cut off, at what declination they are located and how point-like or extended they are. A detection of neutrinos will be the ultimate proof for the acceleration of cosmic-rays up to TeV-PeV energies. In particular for the case of RXJ 1713.7-3946 , it is not clear if the emission is of hadronic nature, as the spectral shape follows the one of an Inverse Compton spectrum, having a very flat spectrum ($\sim E^{-1}$) up to a maximum energy at $\sim$TeV above which the flux becomes very steep, compatible with the KN limit. There is the possibility to get such a behavior from a hadronic component as well when taking into account a clumpy medium as discussed in \citet{gabici_aharonian2014}. Thus, in order to resolve the question of the origin of cosmic rays for this bright gamma-ray emitter, it is of high importance to search for high-energy neutrinos from RX J1713.7-3946.     \end{itemize}

\newpage
\subsection{The Galactic sink: cosmic-ray propagation around the Sun\label{sun:sec}}
Cosmic-ray propagation is influenced by the Sun, the solar atmosphere and its magnetic field in various ways:
\begin{enumerate}
    \item The Sun iteself is opaque to cosmic-rays and their secondaries. It therefore represents a cosmic-ray sink, referred to as the \textit{cosmic-ray shadow of the Sun}. Gamma-rays and even neutrinos are absorbed by the Sun itself. In the solar atmosphere, cosmic rays interact as well and the secondaries produced in these interactions are observed (gamma-rays) or expected to be observed in the future (neutrinos).
    \item Cosmic-ray propagation is highly influenced by the solar magnetic field. 
\end{enumerate}

First predictions of the expected multimessenger signatures from cosmic-ray propagation and interaction were presented by \citet{Seckel-etal-1991,moskalenko1991,ingelman_thunman1996}.

The expectations from these early works were the following:
\begin{enumerate}
    \item \textit{Charged cosmic rays:} the cosmic-ray Sun shadow was expected to change in shape and shadow depth over time. In particular, the 11-year cycle and the connected changes in the magnetic field influence all properties of the shadow.
    \item \textit{Neutrinos:} The solar interior was shown to be opaque to neutrinos because of the large optical depth $\tau_{N\nu,\odot}=N_{\rm column}\,\sigma_{N\,\nu}\gg 1$. Here, $N_{\rm column,\odot}\sim 10^{38}$~cm$^{-2}$ is the slant depth and $\sigma_{N,\nu}\sim 0.8\times 10^{-38}$~cm$^2\,(E_{\nu}/$GeV$)$ is an approximation for the nucleon-neutrino cross section valid up to $\sim 10-100$~TeV. The solar atmosphere, however, provides an environment in which a large number of neutrinos is produced \textit{and} the optical depth is significantly smaller so that the neutral particles reach Earth. 
    \item  \textit{Gamma-rays:} While the Sun itself can produce gamma-rays up to GeV energies, this only happens in flare events. Gamma-rays induced from the impact of Galactic cosmic rays are expected to deliver the following signatures: (1) \textit{The disk component} --- when cosmic rays travel toward the Sun, they can be mirrored. If this happens in the solar atmosphere, the hadrons can interact and produce gamma-rays (and neutrinos). Thus, particles entering the Sun from the direction of Earth produce a signature toward Earth if they are mirrored and interact and the solar disk is (seemingly) illuminated \citep{Seckel-etal-1991}. (2) \textit{The halo component} is produced when relativistic cosmic-ray electrons Inverse Compton scatter with the photons in the outer solar atmosphere. 
\end{enumerate}
The above-expected signatures have been tested observationally and through detailed theoretical signal simulations during the past decade. In particular, the following conclusions could be drawn:

\textbf{Cosmic rays:} 
The cosmic-ray Sun shadow was first detected by \citet{Amenomori-etal-2013} at $\sim 10$~TeV median energy. The observations were compared to cosmic-ray propagation simulations in two different solar magnetic field models. The originally developed \textit{potential field source surface (PFSS)} model \citep{Altschuler-Newkirk-1969,Schatten-etal-1969} uses magnetograms of the solar surface as inner boundary conditions and derive a three-dimensional magnetic field model for a current-free corona. By assuming $\nabla\times \vec{B}=\vec{0}$, a magnetic potential $\phi_B$ can be introduced as $\vec{B}=:-\nabla\phi_B$, which then fulfills the Laplace equation via $\nabla\cdot \vec{B}=\Delta \phi = 0$. At the so-called \textit{solar base} at $2.5\,R_{\odot}$, the outer boundary condition of a purely radial field is applied in the PFSS models. The advantage of this type of model is that a three-dimensional magnetic field is provided without having to take the (computation) time to solve a full set of MHD equations. However, sheet currents are neglected here. These are in turn taken into account in the \textit{current sheet source surface (CSSS)} model \citep{Zhao-Hoeksema-1995,Arge-Pizzo-2000}. This  way, MHD solutions do not need to be derived and a computationally quick and relatively exact solution is available.

It was shown by \citet{Amenomori-etal-2013} that these current sheets are of importance for the three-dimensional modeling of the magnetic field: the CSSS model fits data taken with the Tibet-III Air Shower Array during the period 1996 to 2009  well. The PFSS model, on the other hand, cannot fully explain the detected temporal variations. It could further be shown with the same data that the shadow projection detected at Earth was shifted by the Interplanetary Magnetic Field. It is interesting to note that the detected shift would require a field strength about a factor of $1.5$ larger than indicated by measurements \citep{Amenomori-etal-2018prl}. This mismatch is actually in accordance with the modeling and measurements of the solar wind, which are in need for an even higher field strength \citep{Linker-etal-2017, Yeates-etal-2018}.

Finally, \citet{Amenomori-etal-2018apj} investigate the influence of Coronal Mass Ejections (CMEs) on the depth of the shadow at $3$~TeV energy. The authors conclude that the detected shadow depth at these lower energies in the period 2000 to 2009 clearly varies in time. The absolute depth of the shadow, however, is significantly smaller than expected from simulations. \citet{Amenomori-etal-2018apj} attribute this mismatch of data and simulation to an influence of CMEs on the shadow.

At higher energies, i.e.\ a median cosmic-ray energy of $\sim 40$~TeV, the IceCube Neutrino Observatory could detect the shadow and its temporal variation for the first time  during the years 2010 - 2015 \citep{bos2019}. In further investigations of an enlarged data set (2010 - 2017) and combining simulations with data, it could be shown that the CSSS model fits these higher energies well, while even here, the PFSS model has deficits \citep{tenholt_icrc2019}. In the future, HAWC will be able
to add information to the high-energy shadow and all observatories in combination can provide us with an energy dependence of the shadow, which is highly interesting with respect to the interpretation of the results as discussed below.

\begin{figure}[htbp]
 \begin{minipage}{.5\linewidth}
 \centering
  \includegraphics[width=\linewidth]{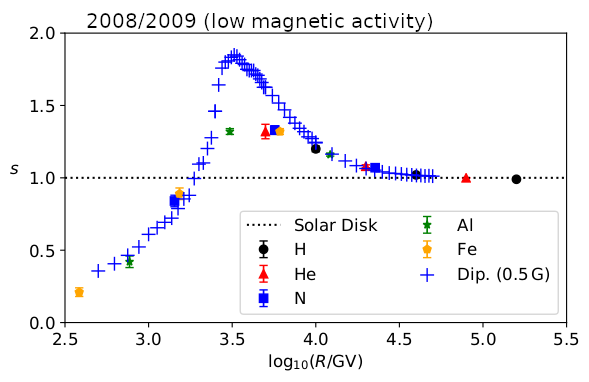}
 \end{minipage}%
  \begin{minipage}{.5\linewidth}
  \centering
    \includegraphics[width=\linewidth]{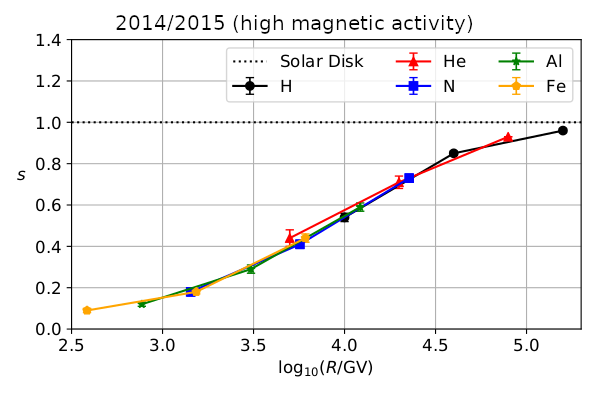}
 \end{minipage}
\caption{The simulated depth of the cosmic-ray shadow as a function of the rigidity for five different elements (see legend of the plots) -- left panel: Season 2008/2009 in the solar minimum (colored marker symbols). Also shown is the theoretical expectation of the shadow depth for a pure dipole field with strength $0.5$~Gauss at the solar equator. Right panel: Season 2014/2015 in the solar maximum. The dashed, black line represents the expectation for the shadow of the geometrical solar disk. Figures adapted  following \citep{beckertjus2020}.
\label{sunshadow:fig}}
\end{figure}

 Measuring the energy dependence of the shadow can provide valuable information on the magnetic field in the future: It has been shown in \cite{beckertjus2020} that for years of low magnetic activities, the shadow basically follows the expectation for a dipole-type field (see Fig.\ \ref{sunshadow:fig}, left panel). Here, the shadow rises with rigidity up to energies of around a few TeV to a level larger than the geometrical shadow and decreases toward higher energies to finally converge to the geometrical shadow depth at around $100$~TeV. For years of high magnetic activity, the shadow is expected to increase linearly with energy instead toward the geometrical depth (see Fig.\ \ref{sunshadow:fig}, right-hand side).

\textbf{Gamma-rays:} The halo component could be quantified and validated by \citet{Orlando-Strong-2008} using EGRET data. Measurements with Fermi could confirm these early results \citep{Abdo-etal-2011} so that this emission due to IC scattering is believed to be understood today.  The same work announced the detection of the disk component in the $10-100$~GeV energy range. In addition, strong limits on the TeV $\gamma$ emission from the Sun's disk component have been presented as well \citep{hawc2018} that start to exclude the largest possible predictions already. Toward lower energies, a signal has been detected as presented by \citet{Ng-etal-2016} ($E_{\gamma}=1-10$~GeV). In this case, however, the detected intensity level is about one order of magnitude higher than expected by \citet{Seckel-etal-1991}, newer estimates quantify this overproduction to a factor $6$ \citep{Nisa-etal-2019}. A time dependence of the gamma-ray signal in the $1-10$~GeV energy range has been detected as well \citep{Ng-etal-2016}, in particular as anti-correlated to the magnetic field strength. It has been discussed that the change in particle propagation through the presence of the solar magnetic field could account for a change in $\sim 15\%$, while observations show a much higher variation, i.e.\ by a factor of $2-3$. Thus, while the influence by the solar magnetic field cycle itself appears to be clear, it is not entirely sure where the strong variation comes from. A further mismatch toward the early prediction is found in the power-law shape of the gamma-ray energy spectrum. While the expectation is a spectrum following the one of cosmic rays, i.e.\ $E_{\gamma}^{-2.7}$, the observed spectrum behaves as $E_{\gamma}^{-2.2}$. Finally, the observed spatial distribution of gamma-rays shows a specific structure that needs to be explained: as found by \citet{Linden-etal-2018}, the gamma-ray component close to the polar-cap regions is constant, while the equatorial gamma-ray flux shows a time-variability which appears to be energy-dependent.

\textbf{Neutrinos:}

Early-on, \citet{hettlage_mannheim2002} calculate the yearly rate that was expected in a --- at that time still to be built --- IceCube detector to $3-30$. \citet{fogli2006} could show that this number is reduced to about $1-2$ neutrinos per year when taking into account three-flavor-interactions and updates on the diffuse cosmic-ray flux. In fact, the most common background in IceCube from neutrinos produced in the Earth's atmosphere basically can be removed completely, making the detection of such a low flux feasible. However, in addition to the expected signal from the solar atmosphere, dark matter signatures are predicted as well and have been investigated in IceCube extensively during the past decade. In particular, the sensitivity works well for an energy range of supersymmetric WIMP masses between $10-1000$~GeV. Most recent stringent limits for the spin-dependent dark matter cross sections are presented in  \citet{icecube_dark_matter2017}. These searches test the dark matter theory by using the Sun as a place of enhanced local gravity, in which even dark matter particles should cluster. Neutrinos could be detected via the annihilation of those WIMPs in the Sun. Cosmic-ray interactions in the solar atmosphere represent a background for these searches that also define the so-called \textit{neutrino floor} for high-energy neutrinos \citep{Argueelles-etal-2017,edsjoe2017,Ng-etal-2017}. These most recent estimates of the neutrino flux from solar atmospheric neutrinos predict $\sim 1$~event per year \citep{Argueelles-etal-2017}, $\sim 2-3 $~events per year \citep{edsjoe2017} and $\sim 5-6$~events per year \citep{Ng-etal-2017} within IceCube, respectively. Current limits reached with IceCube lie about an order of magnitude above the \textit{neutrino floor} that, once it is hit, will not allow dark matter limits to improve anymore, but rather lead to a possible detection of the solar atmospheric neutrino flux. Above the dark matter scale, i.e.\ at energies of $E_{\nu}>1$~TeV, IceCube now starts to reach sensitivies for which the solar atmospheric flux could become detectable. Thus, the reasoning for a need of precise and time-dependent predictions of the solar atmospheric neutrino flux is twofold, for once as a potential high-energy neutrino source ($E_{\nu}>$~TeV) and as a guaranteed background for dark matter searches ($E_{\nu}<$~TeV).

These multimessenger signatures are highly valuable, firstly to constrain the solar magnetic field models and secondly to enhance the understanding of the nature of dark matter, possibly even detecting a signal in the future. The full potential of multimessenger searches of the Sun at the highest energies has been summarized in the white-paper of \citet{Nisa-etal-2019}. Research in this area is only beginning to start. Future time- and energy-dependent observations of charged and neutral messengers with a proper modeling of the solar magnetic field will help to reveal the basic physics of the Sun.

\newpage

\subsection{Diffuse gamma-ray emission in the Milky Way\label{ism_multimessenger:sec}}
\begin{figure}[htbp]
\centering
\includegraphics[width=0.87\textwidth]{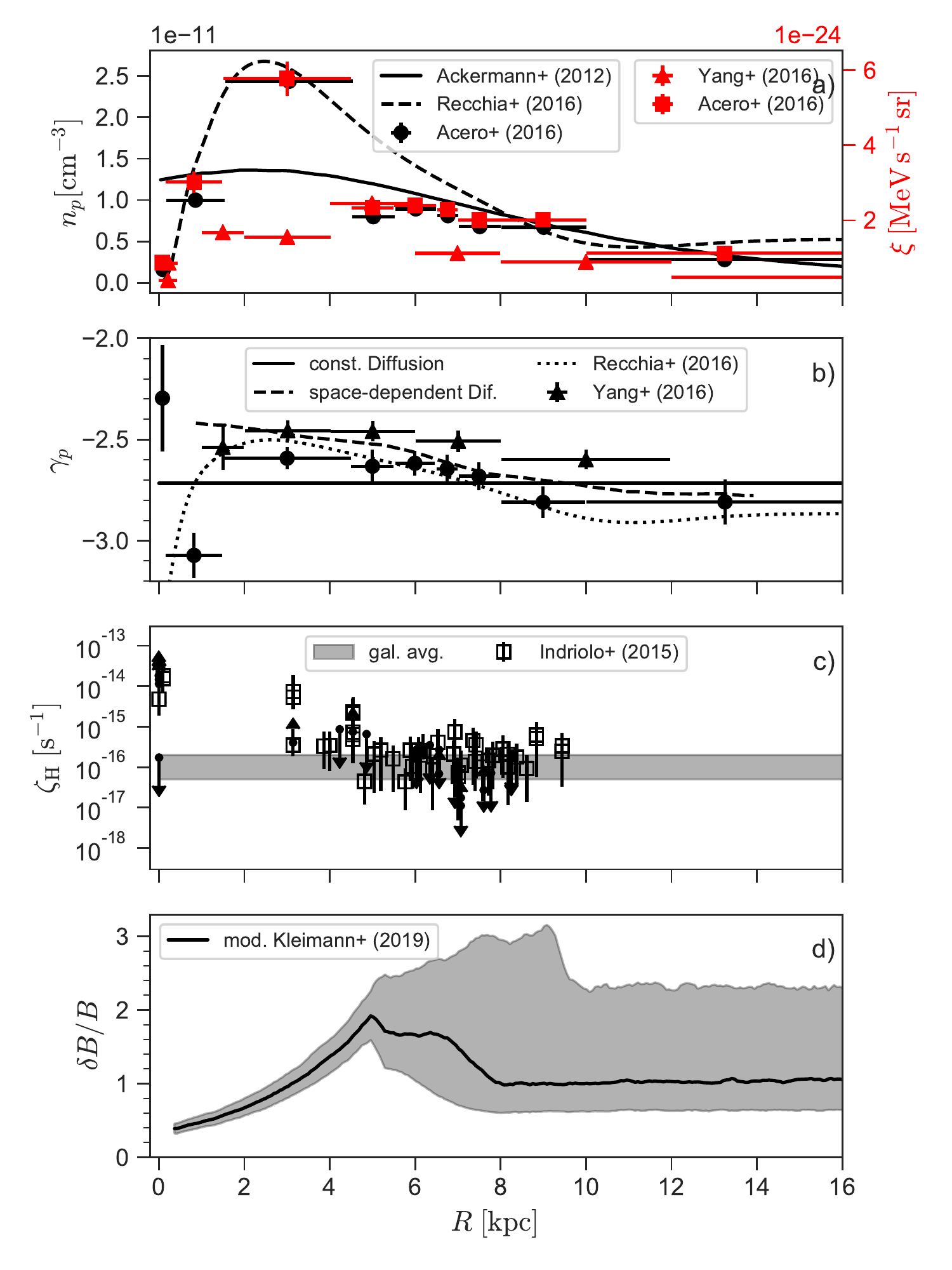}
\caption{The gamma-ray gradient in the light of different observables. Panel (a) shows a combination of the cosmic ray density $n_p$ \cite{acero2016} and the emissivity $\xi$ \cite{acero2016, yang2016} together with fits for the density \cite{2012ApJ...750....3A, recchia2016} (solid and dashed lines, respectively). The second panel (b) gives the observed and modelled proton spectral index $\gamma_p$ while in the third panel (c) the ionization rate $\zeta$ is plotted \cite{indriolo2015}. The last panel (d) shows the turbulence level $\delta B/B$ as implemented in \citet{kleimann2019} reduced by a factor 5. This reduction is due to a likely overestimation of the turbulent component because of the underlying WMAP data (see also  \citep{Planck_magField, Unger2019}). The grey band gives the 0.5 quantile with respect to all angles $\phi$ in galacto-centric polar coordinates. The data are smoothed by five-datapoint-wide wolling mean.}
\label{gamma_diffuse:fig}
\end{figure}
The diffuse component of gamma-rays in the range $\sim 100$~MeV to tens of GeV has been measured in detail by Fermi in the past decade \cite{2012ApJ...750....3A,acero2016}.

Two distinct features are in need of explanation: (a) an excess of cosmic rays in the inner Galaxy; (b) a gradient in the spectral index of the protons, going from flat spectra in the innermost part to steep spectra in the outer Galaxy. These two features are summarized in Fig.\ \ref{gamma_diffuse:fig}(a) \& (b). The Fermi data (black, filled squares) are reproduced from \citet{acero2016}, data from \citep{yang2016} (red triangles) are added. In panel (a), the black scale (left) on the y-axis shows the cosmic-ray number density, while the red scale (right) represents the gamma-ray emissivity $\xi$. The two graphs  summarize what is known as the \textit{cosmic-ray gradient problem}: The direct measurement of Fermi is the gamma-ray emissivity as a function of the galactocentric radius 
(panel (a), right scale). From this, by modeling cosmic-ray electron and hadron propagation in the Galaxy, the local proton density (panel (a), left scale) and spectral index (panel (b)) 
can be deduced. Investigations of the inner Galaxy have been of high interest already since first EGRET measurements showed a discrepancy between cosmic-ray modeling and the measured gamma-ray flux. This excess of gamma-rays in the inner part of the Galaxy as detected by EGRET, however, seemed to have arosen by an instrumental effect \citep{grenier2015}. Still, even with Fermi data, there is a mismatch with a peak at around $\sim 3$~GeV when compared to state-of-the-art modeling \cite{2012ApJ...750....3A}. Recent investigations \citep{acero2016} have been done by systematically scanning the parameter space of cosmic-ray modeling in the context of Galactic diffuse background estimation with the GALPROP code \citep{strong_new_1998}. Here, the cosmic-ray source distribution, the scale height and the radius of the Galaxy enter as parameters for cosmic-ray propagation. Further and equally important, the target for these interactions is interstellar hydrogen. Thus, the temperature of spin excitation $T_S$ together with the $E(B-V)$ magnitude determine the equations via the derivation of the local gas densities in the Galaxy. It was shown that this best-fit scenario still deviates significantly from the measured distribution (see solid line in panel (a)). Looking at the energy behavior of the flux, it turns out that this excess comes from the higher-energy part of the measured spectrum, i.e.\ the spectra in the inner Galaxy are harder than expected from the simulations as seen from Fig.\ \ref{gamma_diffuse:fig} (b). This mis-match between data and simulation gave rise to the interpretation of a contribution from dark matter annihilations in the inner Galaxy, see e.g.\ \citet{fermi_dm2017} for a thorough analysis of the Fermi collaboration and \citet{bertone2010, conrad_reimer2017} for reviews. It is known today, however, that the difference between model and data could be of standard-model nature. These deviations can rely on a number of effects that concern the modeling of cosmic-ray interaction and propagation. Those parameters that govern the results of the simulations are the gas density as a target for the interaction (Section \ref{gas:sec}), the spectral behavior of the sources (Section \ref{spect:sec}), the turbulent and large-scale magnetic field structure in which the particles propagate diffusively (Section \ref{bfield:sec}) and non-diffusive effects that can influence the transport (Section \ref{deviation:sec}).

\subsubsection{Gas densities \label{gas:sec}}
The gas density in the inner part of the Galaxy has large systematic uncertainties due to possible self-absorption effects together with optical depth corrections  in the measurements. These effects are believed to be largest in the central kpc \citep{acero2016}. The gas distribution in the Central Molecular Zone (CMZ) has been discussed in detail in \citet{guenduez_bfield2019} in order to properly model the propagation of high-energy cosmic rays in the inner $200$~pc. Still, in order to fully understand the GeV signatures, this region needs to be treated with great care and it is not clear if the derived flat proton spectra in the GeV-range could be due to a misinterpretation of the gas densities in this central region \citep{acero2016}. But even when excluding the central part of the Galaxy, a gradient in the cosmic-ray density and spectral index from a few kpc distance up to 20~kpc remains and is in need of explanation.

\subsubsection{Spectral behavior and source distribution \label{spect:sec}}
The number of unresolved point sources and extended ones is expected to be particularly large in the Galactic Center region as the number of cosmic-ray sources is large here as well. This could artificially flatten the spectrum of the diffuse emission \citep{grenier2015}. While \citet{pothast2018} argue that these effects are rather small, \citet{carlson_prl2016,carlson_prd2016} discuss that modeling of the GeV excess is possible when emitting $20\%-25\%$ of the cosmic rays directly fom the star forming region within the CMZ. Intrinsic differences in the acceleration spectra of the cosmic-ray sources have been discussed as the reason for the gradient in \citet{yang2016}. Such an adjustment reduces the proton density peak at $3-4$~kpc and flattens the gradient in the spectrum (see Fig.\ \ref{gamma_diffuse:fig} (a) and (b), triangles). \citet{berezhko_voelk2013} discuss that even emission from Supernova Remnants can contribute to the flux.
A further hint toward the existence of an additional cosmic-ray source component in the CMZ comes from the ionization rate, shown in Fig. \ref{gamma_diffuse:fig}(c)  as a function of Galactocentric radius \citep{indriolo2015}. For the central molecular zone, the cosmic-ray induced ionization rate becomes as large as $2\times 10^{-14}$~s$^{-1}$ \citep{oka2019}, thus more than 2 orders of magnitude as compared to the Galactic average, see Fig.\ \ref{gamma_diffuse:fig}(c). Such high ionization rates are expected at the sources of cosmic-ray acceleration \citep{becker2011,schuppan2012,schuppan2014} and indicate the existance of point sources in the CMZ. Thus, the cosmic-ray flux is expected to be higher in the CMZ as compared to the outer regions, at least up to $\sim$GeV proton energies, expecting a flat component in the central region this way. It is interesting to note that the gamma-ray spectrum at TeV energies in the CMZ has also been shown to be extremely flat, reaching energies to $E_{\gamma}>100$~TeV and thus indicating for the first time the existence of a PeVatron in the Galactic Center region \citep{hess_gc_2016}. \citet{hess_gc_2016} argue that this spectrum and radial dependence ($\propto 1/r$) of the flux can be explained with a central source compatible with the position of Sgr A$^*$ using a $1-$dimensional propagation model and is neither compatible with an advection-dominated scenario, which would be $1/r^2$, nor with a burst-like emission, resulting in a constant profile. \citet{guenduez_gamma2020} confirm that a central source describes the TeV gamma-ray data best and that an additional contribution of pulsars or other point sources in the CMZ does not describe TeV data well. To summarize, the question of a contribution from point sources in the GC region is not fully resolved and needs more investigations into the future, in particular by connecting multimessenger information and developing a consistent picture from radio wavelengths up to TeV energies.
\subsubsection{Magnetic field modeling \label{bfield:sec}}
\begin{figure}[htbp]
\centering
\includegraphics[width=0.8\textwidth]{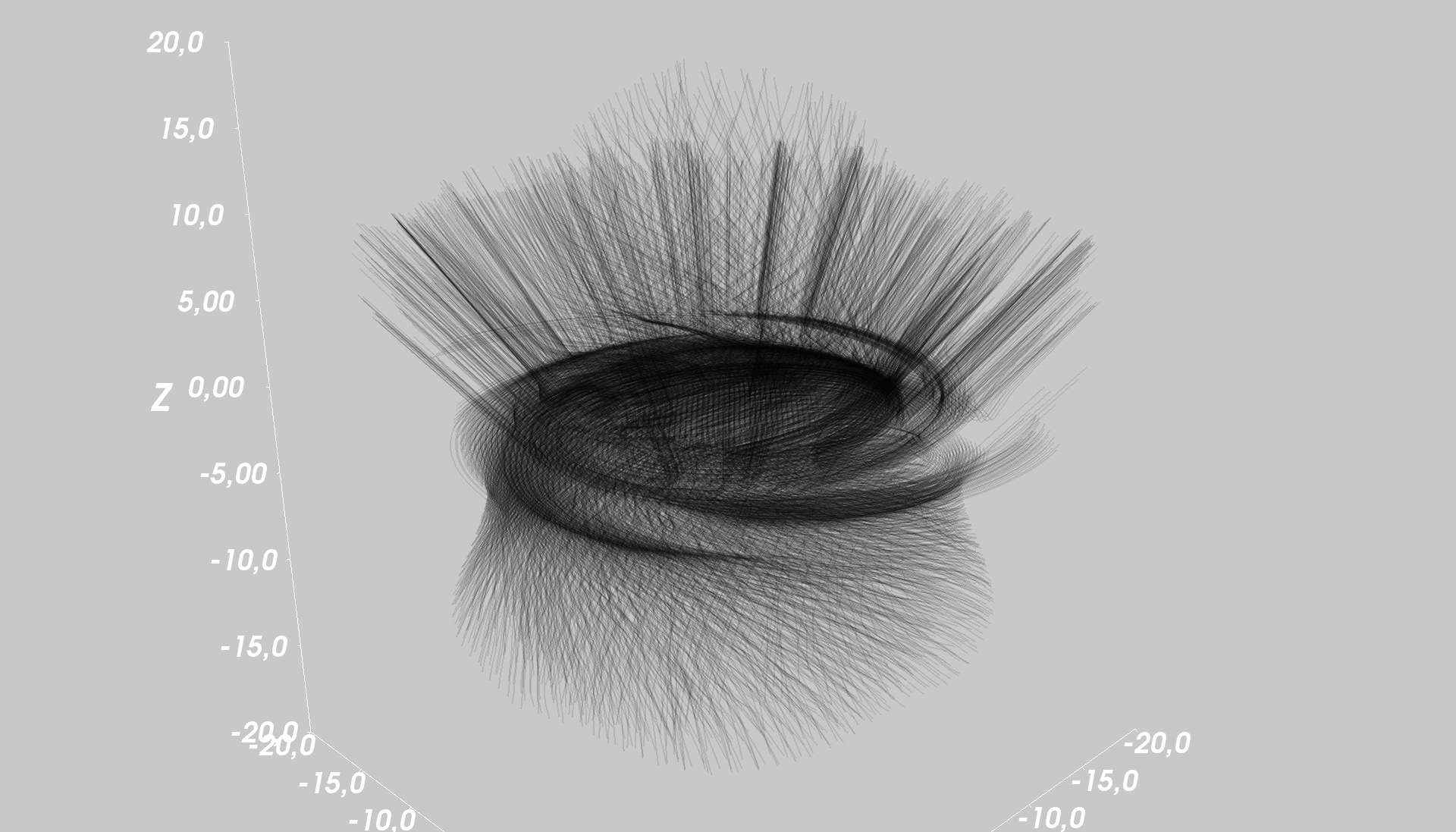}
\caption{Vizualization of the large-scale magnetic field according to \citep{jansson_farrar2012}.} 
\label{jf12:fig}
\end{figure}

\paragraph{Large-scale field effects}
Both the orientation of the large-scale field and the turbulence level could play a role in explaining the cosmic-ray gradient problem with its different features detected at GeV-TeV gamma-rays. \citet{cerri2017}
discuss that the orientation of the field is of high importance when assuming a transport scenario that is mostly parallel to the magnetic field direction. In particular, concerning the escape from the Galaxy, the $z-$direction in Galactic coordinates is most relevant as the scale height $H$ is much smaller than the $x-y$-extension of the Galaxy. As the field configuration changes significantly with the galactocentric radius (see Fig.\ \ref{jf12:fig}), even the escape-behavior does change. 

The magnetic field lines in the innermost kiloparsec of the disk are oriented perpendicular to the Galactic plane, see Fig.\ \ref{jf12:fig}. Here, only the halo component is taking into account as the magnetic field in the central kpc of our Galaxy is not well-constraint. 
Moving outward along the Galactic Plane, the field lines become parallel to the Galactic disk. Assuming that escape happens in the z-direction, that means that the escape must be dominated by parallel propagation in the inner part of the Galaxy, along the field lines, but perpendicular in the outer Galaxy, i.e.\ perpendicular to the field lines.

This shows us that taking the field geometry into account is certainly of high importance. However, there are different issues with this global field modeling, which can be deduced from comparisons with external galaxies, see \cite{beck2019} for a review: spiral arms, the transition to the inter-arm regions and the inter-arm regions themselves are typically less well-defined as indicated in the model, the pitch angle is usually not constant, but varies with the radius, regions in between the spiral arms. On average, the inter-arm regions do have less turbulence as the spiral arms, but there typically do exist arms with low turbulence level as well.

Further, the JF12 model shown in Fig.\ \ref{jf12:fig}  only includes the halo component, which is basically defined via field lines that are perpendicular to the disk itself. As a possible contribution from the disk component itself is missing, a proper interpretation of the innermost kpc is therefore difficult at this point. A parameteric description of the magnetic field in the CMZ has recently been derived in \citep{guenduez_bfield2019} and might be able to shed light on this question in the future. In particular, \citep{guenduez_bfield2019} show that including a diffuse component in the Central Molecular Zone (CMZ) together with structures derived from the observed molecular clouds and filaments in this region will result in propagation that has a larger component along the disk itself.  It could be shown \citep{guenduez_gamma2020} that such a field configuration indeed is better suited to describe the PeVatron emission in the Galactic Center region detected with H.E.S.S.\ \citep{hess_gc_2016}. Such a component should therefore be included for global cosmic-ray propagation modeling as well. This statement is supported by observation of external galaxies. For instance, NGC891 is quite similar in the general field structure as the Milky Way and does not show a vertical field in the center, but rather a horizontal one \citep{beck2019}. In general, \textit{if} a galaxy is dominated by a halo-type magnetic field in the central part, this is typically true even for the outer parts \citep{krause2020}.

Thus, these differences discussed above might not be as strong as they appear when only considering the global magnetic field model.

\begin{figure}[htbp]
\centering
\includegraphics[width=0.8\textwidth, trim=0 0mm 0 0mm, clip]
{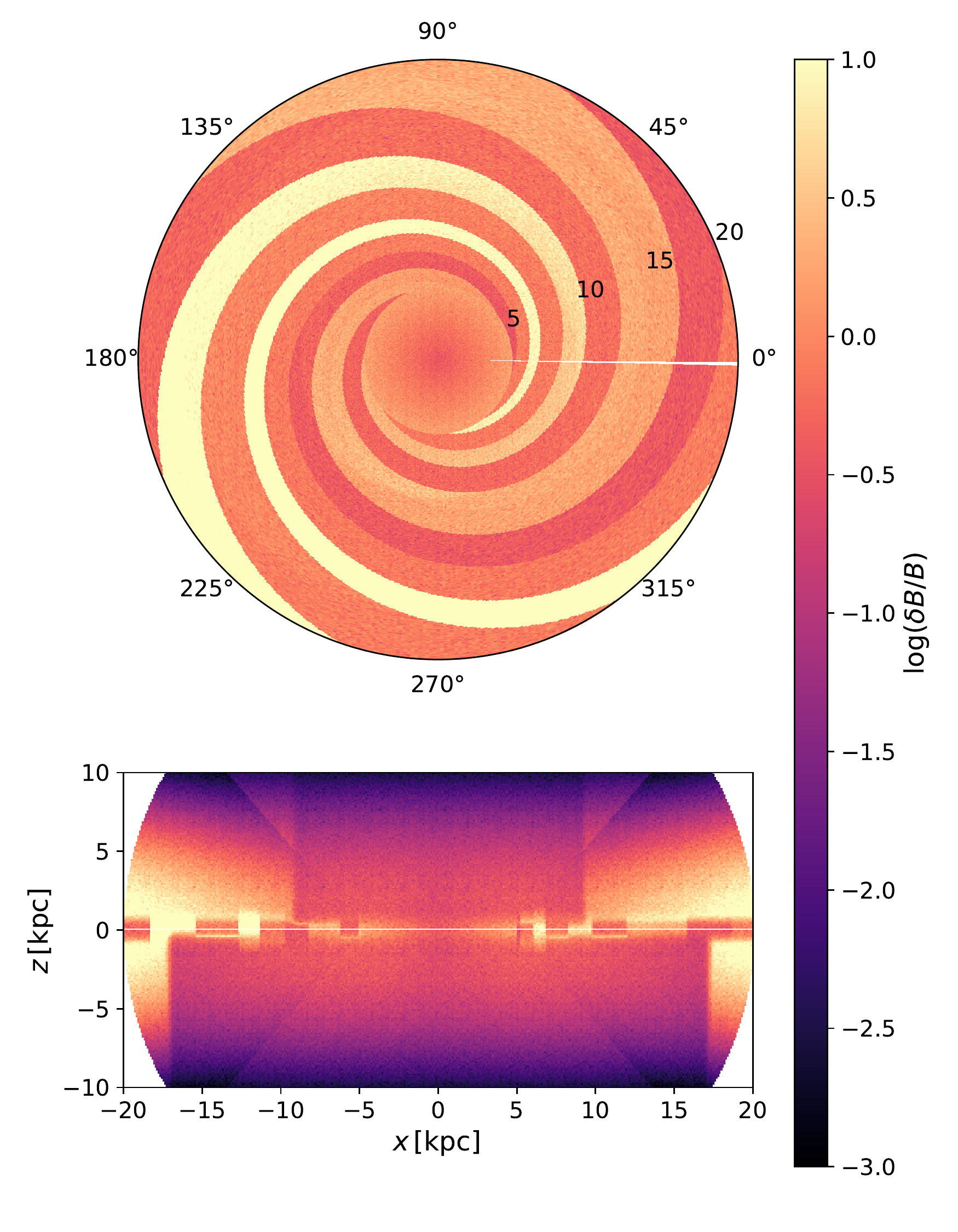}
\caption{Modified Jansson \& Farrar global Galactic field \citep{jansson_farrar2012, kleimann2019}. Shown is the ratio $(\delta B/B)/5$ in a face-on (top) and edge-on (bottom) view of the Galaxy. The decrease by a factor of five is due to an overestimation of the turbulent component in the GMF models , see e.g., \citet{Unger2019, Planck_magField}. The turbulence ratio is largest in the inter-arm region. This is due to the fact that the turbulent component in the GMF model is less structured than the regular one, leading to a possible overestimation of the ratio between the spiral arms. Each pixel shown, is averaged over 11 data points in the direction perpendicular to the plane with a maximum distance of $\Delta =250$~pc.}
\label{turbulence_galactic:fig}
\end{figure}

\paragraph{Influence of the turbulent component}
It has been discussed by \citet{evoli_anisotropy2012,gaggero2015} that a radial change in the diffusion coefficient could explain the galactocentric gradient in the spectral index of the gamma-ray spectra. The phenomenological description can actually base its reasoning on a change in the tubulence level $\delta B/B$. Typical global magnetic field models of the Milky way show a gradual rise in the turbulence level $\delta B/B$. Figure \ref{turbulence_galactic:fig} shows a face-on (top) and edge-on (bottom) view of $\delta B/B$ according to the model of \citep{kleimann2019}, which represents a modification of the analytical model by \citep{jansson_farrar2012} taking into account divergence-freeness and ensuring continuity. In addition, following the line of arguments presented in \citet{farrar_unger2019}, who discuss that the turbulent component is overestimated in \citet{jansson_farrar2012}, we have reduced $\delta B$ in the figure by a factor of $5$ as compared to the original publications \citep{jansson_farrar2012,kleimann2019}. It can be seen that in this model, the turbulent component is smaller with respect to the regular magnetic field in the spiral arms as compared to the inter-arm regions (top panel).

\begin{figure}[htbp]
\centering
\includegraphics[width=0.8\textwidth]{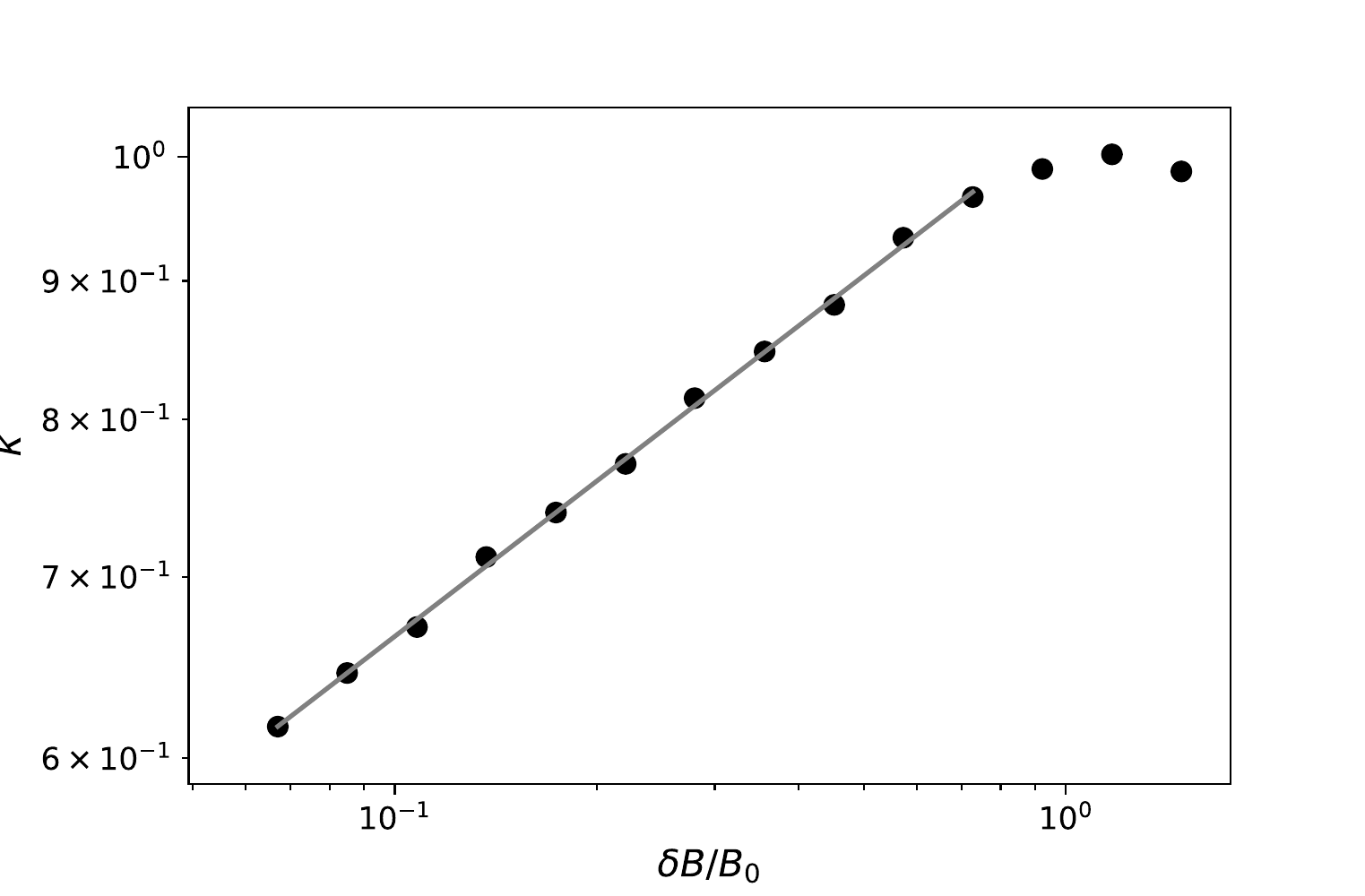}
\caption{Spectral index of the parallel diffusion coefficient, $D\propto E^{\kappa}$ as a function of the turbulence level $\delta B/B_0$. \textit{Figure courtesy:} Patrick Reichherzer, data from \citep{reichherzer2019}. \label{deltab_b:fig}}
\end{figure}

Further, the edge-on view reveals that the turbulence level is highest in the disk ($z=0$) in the central $\pm 10$~kpc of the Galaxy, but increases significantly even above the disk at larger galactocentric radii. In order to analyze the change of turbulence level with galactocentric radius, we derive the median value, 
shown as the black line in Fig.\ \ref{gamma_diffuse:fig} (d), the grey area covers 50\% of all data. Here, the values are averaged over eleven equidistant positions in $|z|<250$~pc. The underlying distribution for the median derivation is derived for 500 positions in $\phi$. 
This way, we can quantify that the turbulence level is low in the inner part of the Galaxy and rises up to a level of $\delta B/B\sim 1$ and larger at galactocentric radii above $\sim 3$~kpc. This high turbulence level is maintained up to galactocentric radii up to $\sim 20$~kpc. Recently, using the TGK method (Section \ref{candidates:sec}), a correlation between the turbulence level and the diffusion index has been shown to exist in \citep{reichherzer2019}.  The results are shown in Fig.\ \ref{deltab_b:fig}. Above values $\delta  B/B\gtrsim 1$, the Bohm limit of diffusion is reached and $D(E, \delta B/B\gtrsim 1)\sim E^{1}$. Moving toward lower values of $\delta B/B$, the index of the diffusion tensor gradually  decreases to a value of $D(E, \delta B/B\sim 0.05)\sim E^{0.6}$. A fit to the region $0.067<\delta B/B<0.727$ reveals a correlation between the diffusion spectral index $\kappa$ and the turbulence level as $\kappa \propto (\delta B/B)^{0.1904\pm0.0021}$.
Values below $\delta B/B\sim 0.067$ are currently not accessible in simulations due to interpolation effects studied in \citet{schlegel2020}, but it is expected that moving to smaller $\delta B/B$ results in the limit of QLT with $D(E)\sim E^{1/3}$. Thus, with a leaky box approximation $n(E) \propto E^{-\gamma}$ and  $\gamma=\kappa+\alpha$ as a function of $\delta B/B_0$ due to $\kappa \propto (\delta B/B_0)^{\delta}$, a steepening of the cosmic-ray flux with increasing galactocentric radius is certainly expected.
The considerations above are so far only indications that the turbulence level matters for a proper description of the spatially resolved cosmic-ray component in the Milky Way. The results discussed above show that two things are of high importance for future modeling of cosmic-ray transport: 
\begin{itemize}
    \item[(a)] precise knowledge on a spatially-resolved turbulent and homogeneous component in the Galaxy is necessary; 
    \item[(b)] a fundamental understanding of the diffusion coefficient. The latter is making progress with an improved understanding on the diffusion coefficient calculations as in \citep{snodin2016,reichherzer2019}. 
    \end{itemize}
    The point of improved measurements of the Galactic field is certainly very difficult, as high-resolution results in the Galactic plane are difficult to achieve and the field is largely unknown. Learning from the observation of distant galaxies, there are indications that  the turbulence level in galaxies like M51, NGC342 and M83 rather \textit{decreases} with galactocentric radius \citep{beck2019}. However, the interpretation of these measurements is difficult, as the measurements rely on large structures between $\sim 0.5^{\circ}-3^{\circ}$. This means that the scales the observations are sensitive to are quite large and do not necessarily correspond to those relevant for cosmic-ray propagation as the magnetic field waves scatter with particles of energy $E$ in a magnetic field $B$  according to the resonance criterion $k_{\rm res}\sim r_{\rm g}(E,\,B)^{-1}$. This determines a length-scale $l$ for the scattering via $k_{\rm res}=2\pi/l$. Further, the definition of the ordered and turbulent field are different in the observations as compared to the theoretical ones. Thus, in the future, a quantitative comparison of the theoretical to observational measures will enable the usage of the astronomical data for the modeling of galactic magnetic fields and cosmic-ray propagation in these structures. Another source of mis-interpretation can be the  anisotropic turbulence as discussed in \citep{beck2019}. For instance, if the degree of anisotropy increases toward the outer parts of the galaxy due to differential rotation and the connected sheering effects, the polarization level does as well without changing the degree of turbulence. Thus, future studies of the field rely on a proper understanding of the data together with a detailed consideration of the underlying plasma physics that leads to the production of the turbulence.

        \subsubsection{Deviation from diffusive behavior \label{deviation:sec}}
        There are different situations in which propagation deviates from the typically assumed diffusive behavior. In this context, in particular the dominance of a Galactic wind in the Galactic Center region can play an important role as advective transport does not change the energy spectrum of the particles and could explain a flat component in the inner region of the Galaxy. In addition, the streaming instability can lead to a change in the spectrum at the point where it loses importance. Both effects are described below.
  \paragraph{Outflows from the Galactic Center region}
  From the discussion above, it becomes clear that the Galactic Center region is of special interest in order to understand the gradient problem in the Galaxy: On the one hand, a large fraction of the sources of cosmic rays is expected to be located here. On the other hand, uncertainties for modeling cosmic-ray transport and interacion are large in the GC as knowledge on gas distributions and magnetic fields is quite poor for this region of the Galaxy. Most large-scale magnetic field models actually actively exclude the description of the field in the Galactic Center region, e.g.\ \citep{jansson_farrar2012,Unger2019,kleimann2019}. Only recently, a first parameter-based model of the magnetic field in the Central Molecular Zone (CMZ) has been presented \citep{guenduez_bfield2019}, so that a first more detailed transport model in this region can be developed \citep{guenduez_gammarays2019}. While this model is able to describe the PeVatron signatures detected by H.E.S.S.\ in the CMZ both concerning the energy spectrum, radial extension and intensity profile, the specific outflow signatures are not reproduced by this field component that is dominantly present in the disk. It is argued in \cite{hess_gc_2016} that the $1/r$ profile that is observed for the PeVatron is characteristic for diffusive behavior of a point-source in the Galactic center and cannot be reproduced by an advective model. Thus, features that have been called \textit{WMAP haze}, \textit{Fermi bubbles} etc.\ cannot be reproduced in the same description. These largely extended structures have been observed at radio- \citep{carretti2013,Heywood2019} and micro-wavelengths \citep{dobler2012} and up to X-ray \citep{finkbeiner2004} and gamma-ray \citep{finkbeiner_galactic_wind2010}  energies. 
    The presence of such outflow signatures suggests that a specific halo component  of the B-field needs to be present. A structure that drives out cosmic-ray electrons and/or protons in order to create these bubbles has to be the cause. The origin of this emission perpendicular to the Galactic plane can be associated with an ancient AGN jet- or wind-structure or an outflow induced by nuclear star formation. Cosmic-ray electrons and/or protons need to be accelerated somewhere, either at the base of the outflow via a jet or wind structure or in-situ, at a possible shock front that terminates the bubbles. These accelerated particles then propagate, either diffusively or advectively. Depending on the type of particles that dominate the emission, low- and high energy emission can be connected (leptonic radiation from synchrotron emission at low frequencies and Inverse Compton scattering off the interstellar radiation field) or not (hadronic scenario with the high-energy emission explained via pion production). A detailed, contemporary review on this topic concerning the observations and theoretical modeling of the bubbles can be found in \citet{yang2018}, where it is summarized that three types of models are compatible with the different observables of the bubbles: wind-driven models dominated by hadrons,  the jet-type models dominated by leptons and in-situ acceleration models. Future observations by TeV instruments like HAWC \citep{hawc_bubbles2017} and CTA \citep{yang_razzaque2019} will help to distinguish between these scenarios as emission going beyond TeV particle energy is expected to come from hadronic processes. The detection of neutrinos associated to the Fermi bubbles could help as well. However, as the signal is relatively faint and highly extended, it is hard to detect \citep{unger_phd2019}.
    
    In summary, if the PeVatron-results by H.E.S.S.\ are directly related to the Fermi GeV data, the latter cannot be explained via advection and the flat spectrum that is observed at GeV energies must be explained differently. Still, outflows do exist, so that at some level, particles must be able to propagate out of the Galaxy in perpendicular direction.

        \paragraph{Self-generation of magnetic waves by cosmic rays} The streaming instability predicts a growth of the energy density in Alfv{\'e}n waves due to the interaction of cosmic-rays with the background magnetic field. This interaction between particles and waves leads to the effective modification of the cosmic-ray diffusion coefficient as discussed in \citet{recchia2016}. As  discussed in \citet{blasi2012}, the streaming instability can be  important up to rigidities of $R\lesssim 300$~GV, this way being able to explain the break in the cosmic-ray spectrum at these rigidities. In this context, a break in the B/C ratio is predicted as well and the detection of such a break could confirm the scenario of a diffusion coefficient that is modified by the streaming instability. For the gamma-ray spectra, this means that the results are influenced up to energies of $E_{\gamma}\lesssim 30 $~GeV. As the GeV-peak in the inner Galaxy is at $\sim 3$~GeV, the results are expected to be highly influenced by streaming. A dependence on the galactocentric radius, however, can only be achieved when including the dominance of the Galactic wind in the Galactic Center region and assuming a background magnetic field that has an exponential cutoff at $\sim 10$~kpc as discussed in \citet{recchia2016}. Their results are shown as the dashed line in Fig.\ \ref{gamma_diffuse:fig}. With this combination of effects, the data can be fitted quite well. This scenario can be verified in the future with CTA data --- at higher energies, a change in the behavior is expected as the influence of the streaming instability will disappear. In addition, the model needs to be tested in a more elaborate Galactic magnetic field configuration in order to cross-check if the change in turbulence level and global magnetic field structure can change the effect of the streaming instability locally.
\subsubsection{Neutrino flux predictions}
The first detection of an astrophysical neutrino flux by IceCube opens another window toward the identification of Galactic cosmic rays. While gamma-rays are basically restricted to test the local cosmic-ray spectra up to knee energies, neutrino measurements are sensitive to the region from the knee on and above \citep{icecube2013,icecube_globalfit2015,numu_signal2016,icecube_cascades2017,ic79_spectrum2017}. The diffuse flux that has been detected in recent years has, however, been shown to be dominated by extragalactic sources: firstly, there is no significant clustering of the detected neutrino events in the Galactic plane. Secondly, the flux that is expected from the Galactic plane when being related to the hadronic gamma-ray emission of the Galaxy is about $1-2$ orders of magnitude lower than the measured flux, both concerning the contribution from point sources \citep{mandelartz2015}.
A dedicated search for neutrinos from the Galactic plane with IceCuba data has shown that the component from cosmic-ray interactions in the interstellar medium makes up less than $14$ percent of the total diffuse astrophysical neutrino flux, compatible with theoretical predictions of the diffuse emission  \citep{neronov2014,winter_galactic2014,ahlers_galacticnus2016,gaggero_neutrinos2015}.

  The possibility to identify Galactic sources in neutrinos in the future does exist, though: Searching for SNRs as point sources/extended sources with IceCube-Gen2 and KM3NeT will result in increased sensitivities, mostly because the background can be reduced by specifically focussing on the regions of interest and by adding sensitivity through better pointing due to less scattering in the mediteranean (KM3NeT) \citep{km3net2016}, a significantly larger detection array for improved pointing and energy resolution and connected to the large size a generally enhanced sensitivity (IceCube-Gen2 \citep{icecube_gen22014,kowalski2018}). This way, the brightest gamma-ray emitting SNRs are within reach of detection in neutrinos within the next decades \citep{km3net_snrs2019}.

  \begin{figure}
\centering{
\includegraphics[trim = 0mm 0mm 0mm 0mm, clip, width=0.9\textwidth]{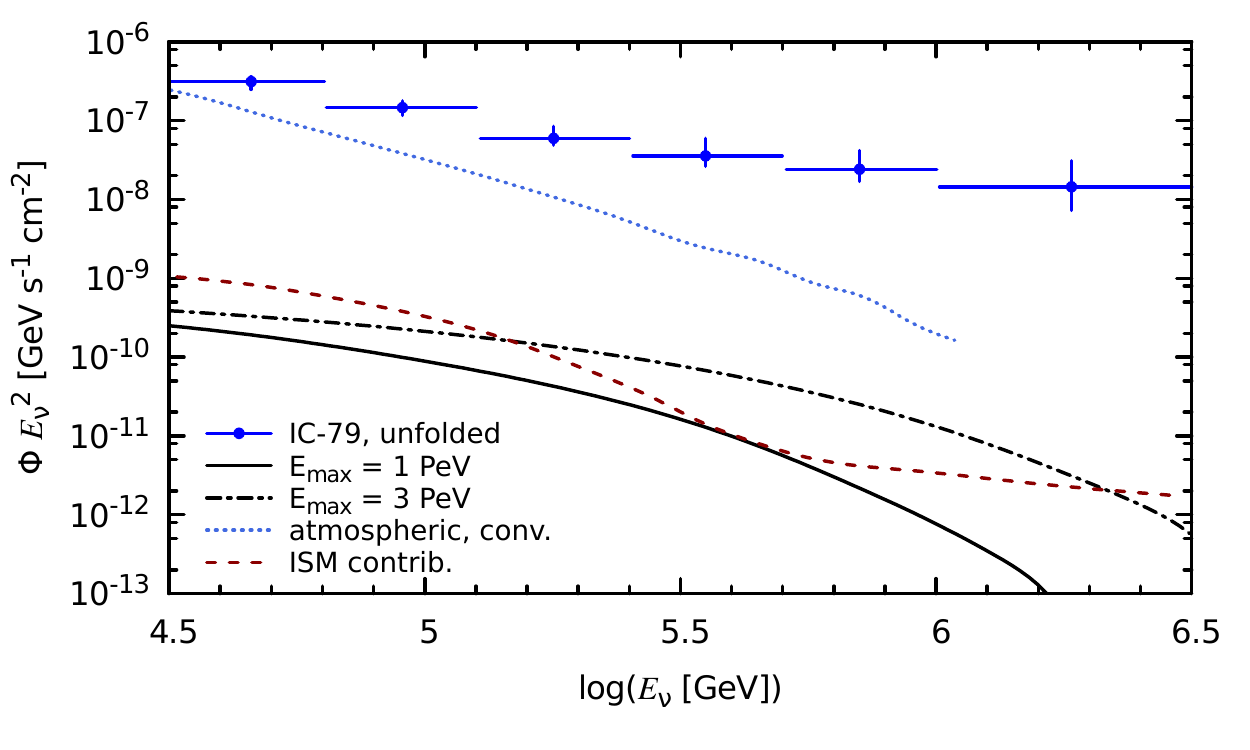}
\caption{Expected galactic diffuse neutrino emission. Data (blue circles): IC-79 unfolded \citep{ic79_spectrum2017}. Theoretical prediction of the conventional atmospheric neutrino spectrum (blue dotted line) from \citet{fedynitch2012}. Red dashed line: prediction of the contribution of the neutrino flux from cosmic-ray interactions with the ISM \citep{winter_galactic2014}. The diffuse contribution from cosmic-ray interactions in the vicinity of SNRs with a maximum cosmic-ray energy of 1 PeV and 3 PeV are shown as the black solid and black dot-dashed lines, respectively \citep{mandelartz2015}.
\label{galactic_diffuse_nus:fig}}
}
\end{figure}

\subsubsection{Signatures from the Galactic wind \label{winds:sec}}
Galactic winds are believed to play an important role in various stages of the evolution of the Universe; they are, e.g., discussed as sources for the metal enrichment of the IGM \citep{loewenstein2001}. Furthermore, in the last decade galactic winds have been observed in many evolutionary stages of the Universe, e.g., in star-forming galaxies with redshifts $2<z<3$ in \citet{steidel2010}. Therefore, many different effects have been proposed that may launch an advective flow, streaming out of the the Galactic plane, e.g., \citet{spitzer56}.

The formation of a stable wind solely due to conversion of the thermal energy into kinetic energy is only possible in objects with low gravitational potential, such as dwarf galaxies, e.g., \citet{dorfi_2012}. For a recent study of the wind formed in the LMC see, e.g., \citet{bustard2018}. In late-type spiral galaxies, like our Milky-Way, the thermal pressure alone is not sufficient to reach the escape velocity. However, in localized regions with a very high star-forming and therefore also increased supernova-rate, hit gas may reach significant heights above the Galactic plane \citep{deavillez2004,deavillez2005}.

In connection with cosmic-ray transport Parker's approach of a cosmic-ray driven wind is especially interesting \citep{parker1965}: A magnetic cloud, like the ISM of the Galaxy, 
is not stable when cosmic rays are generated within this cloud. No matter how small the cosmic-ray pressure $p_\mathrm{CR}$ is, it will dominate over the magnetic pressure after a 
some distance $l_0$: $p_\mathrm{CR}\geq B^2/8\pi$, leading to an inflation of the cosmic ray occupied volume beyond that distance $l_0$. This inflation would grow further and 
further in the case of a quasi-equilibrium treatment. In reality the growth is most likely limited by, e.g., instabilities and cooling processes releasing the cosmic rays eventually.
The general importantce of cosmic rays on the evolution of galaxies especially theier influence on outflows is also confirmed in a number of modern simulations: 1) The influence of cosmic ray diffusion is discussed, in e.g.\ \citet{Hanasz_2013, Pfrommer_2017} and 2) the influence of cosmic-ray streaming in e.g.\ \citet{Uhlig_2012}. They use different implementation of coupled MHD codes which include the influence of CRs on the background medium (see also Section \ref{transport_modeling:sec} and Table \ref{tab:proptools}). 

However, cosmic rays do not only influence the evolution of a galactic wind, but a such a wind can also influence the cosmic-ray population. The idea of cosmic-ray re-acceleration in a Galactic wind was developed by, e.g., Jokipii and others \citep{Jokipii1987b}. Here, a model of a termination shock, forming when the boundary of the wind expansion is reached, was described. At such a shock, of course, cosmic rays might be re-accelerated by DSA or adiabatic heating. Where \citet{Jokipii1987b} proposed a maximum energy of $E_\mathrm{max}=10^{20}$~eV for iron more recent simulation show a much lower maximum energy of $\sim 10^{17}$~eV \citep{zirakashvili_2006} or even $E_\mathrm{max}=10^{16}$~eV \citep{Bustard2015, Bustard2017} taking the finite lifetime of the shock into account. In addition to acceleration at the termination shock other concepts of re-acceleration exist. In \citep{zirakashvili_1996, voelk_2004} the concept of cosmic-ray acceleration at propagating waves connected to the rotation of the galactic spiral arms are discussed and in \citep{dorfi_2012, dorfi_2019} the influence of time dependent and close-by shocks on cosmic ray acceleration is elaborated.

The need for a \emph{second} Galactic component of the cosmic-ray flux is long standing idea to explain the detailed features in the region between knee and ankle. As mentioned above the maximum energy of re-accelerated cosmic rays of most of the models falls directly into that interesting regime. Still, the possibility to accelerate cosmic rays to energies above $>10^{15}$~eV alone is not yet a sufficient criterion to conclude that the cosmic-ray flux in the shin region comes from the termination shock. The question whether these cosmic rays can be transported back into the Galaxy against the advective flow of the wind is not trivially answered. In the following three different models, which aim to describe the cosmic-ray flux between knee and ankle by cosmic rays accelerated at the Galactic termination shock and include transport effects, are compared with each other: 

\begin{enumerate}
\item \citet{zirakashvili_2006} assume azimuthal symmetry and neclect cross-field or perpendicular diffusion \cite{zirakashvili_2006}. Allowing the termination shock efficiency to depend on the Galactic latitude they get a rather smooth steady-state solution for the flux in the shin region. They conclude that only a non-spherical symmetric termination shock can provide a continues transition between the SNR- and the GW-component of the flux. The termination in their model is latitude depend $R_\mathrm{S}(\theta)=150\sqrt{1+3\cos^2(\theta)}\;\mathrm{kpc}$ and has a wind speed of $u=500\;\mathrm{km}\,\mathrm{s}^{-1}$.
\item \citet{thoudam2016} use a steady state transport equation including, 1-dimensional diffusion, advection and adiabatic cooling. They assume a linearly increasing wind velocity $u(r)=15\times (r/\mathrm{kpc})\;\mathrm{km}\,\mathrm{s}^{-1}$ and the shock to terminate at $R_\mathrm{S}=96\;\mathrm{kpc}$. In addition to the former study, they include also heavier elements than proton in their simulation. Their results show that the termination shock has the potential to explain the observed \emph{total} cosmic-ray flux. However, their fits for the expected mass composition shows tension with observed PAO data at the high-energy end of the shin region.  

\item  In \citet{merten2018}, a full three dimensional study of the cosmic-ray transport including all relevant processes for high-energy cosmic rays was done. Furthermore, the finite source duration of the shock was taken into account, for which a time dependent transport equation had to be solved. Perpendicular diffusion was allowed with a diffusion ratio of $\kappa_\parallel / \kappa_\perp =10$. The shock was assumed to be spherical symmetric at $R_\mathrm{S}=250\;\mathrm{kpc}$ and the wind speed was $u=600\;\mathrm{km}\,\mathrm{s}^{-1}$. The results show, that under these assumptions, cosmic rays in the shin region can not solely originate at the GWTS. This is mainly caused by a too large anisotropy of the arrival directions when realistic non-spherically symmetric models of the Galactic magnetic fields are included. 
\end{enumerate}

\begin{figure}[htbp]
\centering
\includegraphics[width=0.8\textwidth]{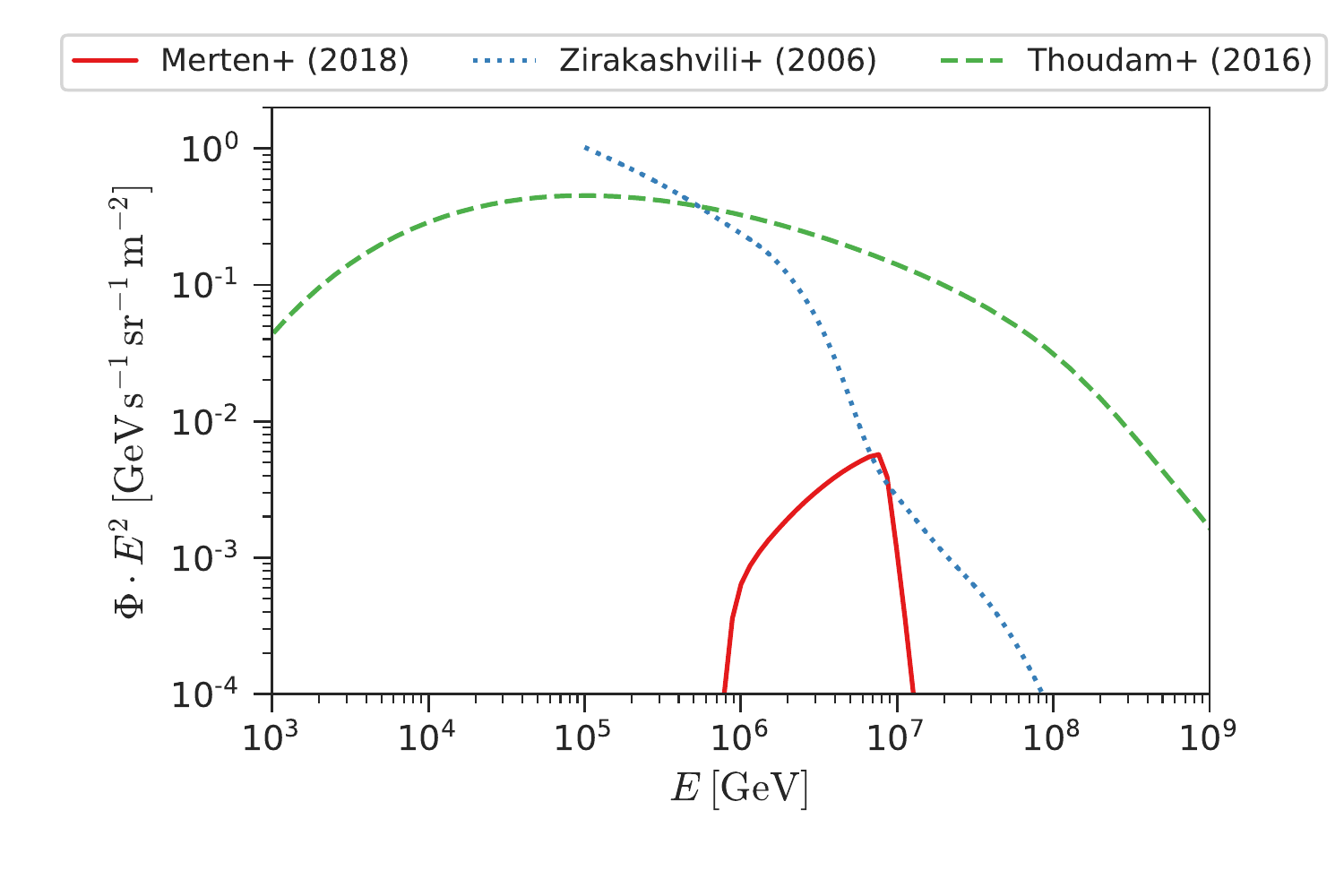}
\caption{Comparison of three models for cosmic rays re-accelerated at the galactic termination shock. For \citet{zirakashvili_2006} the data of the non-spherical symmetric model is normalized to a cosmic ray energy density of $w_\mathrm{CR}=10^{16}$~eV for the shown energy range. For \citet{merten2018} the simulation parameters of the shown flux are: diffusion index $\delta=0.6$, diffusion ratio $\epsilon=0.1$ and observation time $T_\mathrm{obs}=210$~Myr, which is one of highest simulated fluxes. For the \citet{thoudam2016} the all particle best fit scenario for a galactic wind is shown.}
\label{fig:GWTS_Overview}
\end{figure}

Figure \ref{fig:GWTS_Overview} shows the best scenarios for the cosmic ray flux that can be expected from the three different termination shock models. The small energy range of the model by \citet{merten2018} is due to a limited injection range ($10^{15}-10^{16}$~eV). Therefore, one may expect the flux to be extended to lower energies in a larger simulation. 
 
\paragraph{Neutral messengers}

\begin{figure}
\begin{minipage}{.49\textwidth}
\includegraphics[width=\textwidth]{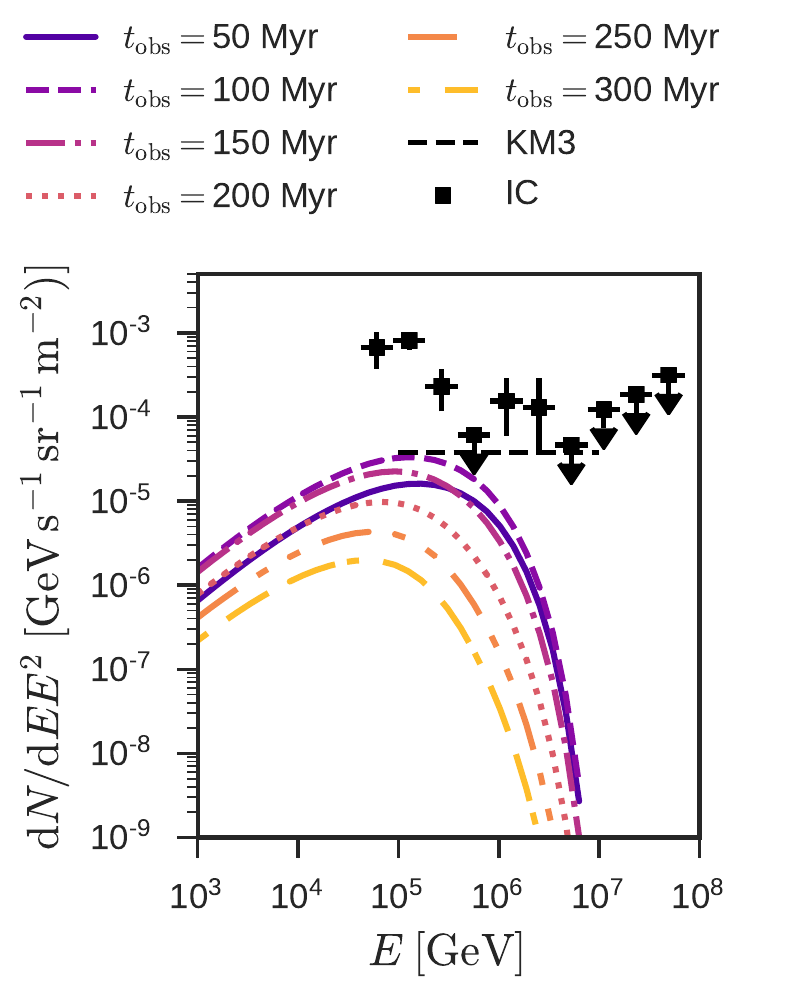}
\end{minipage}
\begin{minipage}{.49\textwidth}
\includegraphics[width=\textwidth]{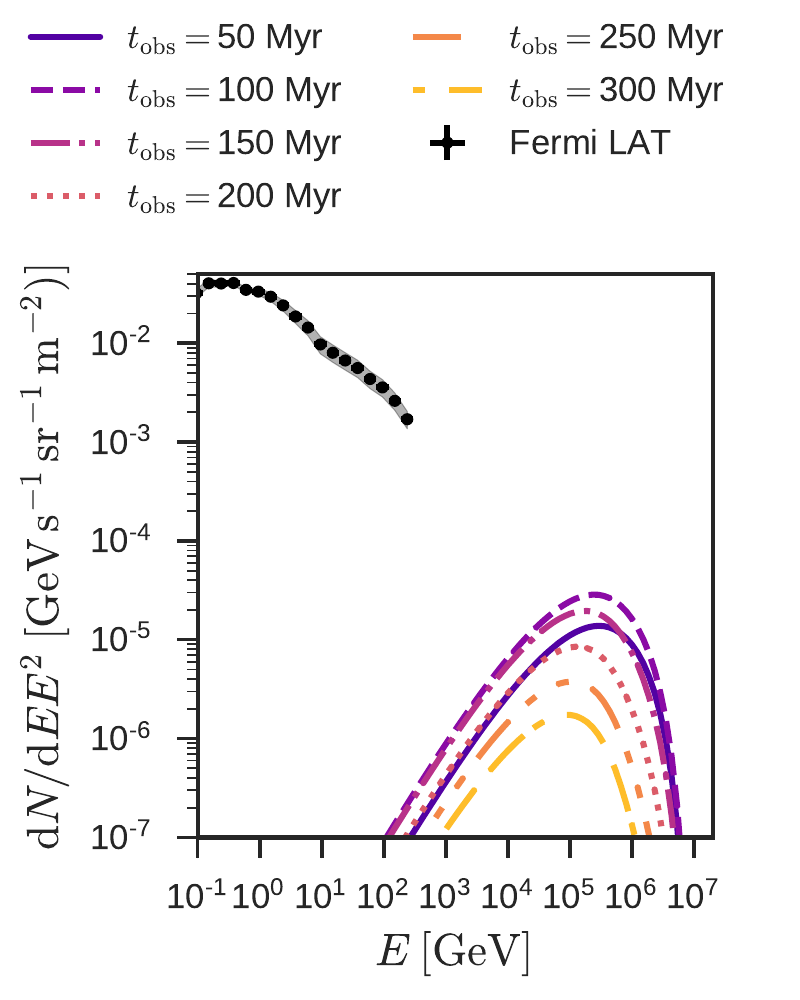}
\end{minipage}
\caption{Estimated neutrino flux produced by galactic termination shock protons (left). For comparison the diffuse neutrino flux measured by IceCube \cite{kopper2017observation} and the expected Km3Net sensitivity are shown \citep{Merten2019}. In addition, the right panel shows the estimates gamma-ray flux. For comparison the total non-Galactic-plane flux observed by Fermi-LAT is shown \cite[Fig.\ 12t]{2012ApJ...750....3A}. Both models used a diffusion index of $\delta=0.5$.}
\label{fig:GWTS_Neutrinos}
\end{figure}

In the upcoming era of multi-messenger astroparticle physics not in addition to the cosmic-ray flux also neutral messenger, like neutrinos and gamma.rays may provide valuable information. Unfortunately, only \citet{merten2018} provides estimates for the produced neutrino flux and non of the studies derives the expected gamma-ray flux. Exemplary results for the neutrino flux are shown in Fig.\ \ref{fig:GWTS_Neutrinos} in addition with estimates of the gamma-ray flux (this work). It can be seen the (upper limit) of the neutrino flux is below the observed diffuse IceCube-neutrino flux observations.

The presented neutral messenger fluxes have to be carefully interpreted as they are not modeled in a fully self-consistent way. The neutral messenger fluxes are calculated, based on the column density that was accumulated by the protons during the propagation from the termination to the galaxy. This column density was then used as to apply the semi-analytic formulas from \citet{kelner_pp2006} to calculate the corresponding fluxes, assuming a instantaneous production of neutrinos and gamma-rays at the galactic boundary of $r_\mathrm{obs}=10$~kpc (see \citet{merten2018} for details of the calculation).

For the future it would be desirable to include a self-consistent description of the production of neutrinos and gamma-rays in models of the termination shocks. This will then allow to draw better conclusions on possible parameter restriction of the model by comparing, e.g., additionally to the observed diffuse gamma-ray flux.
\clearpage
\section{Summary, Conclusions and Outlook \label{summary:sec}}
The field of astroparticle physics has its origins in understanding the signatures of charged cosmic-rays. While their air showers were detected for the first time in 1912, the rapid development of the field only started in the late 1980s, when detection techniques started to reach a level that allowed for measurements that now have reached the precision to determine the following cosmic-ray properties concerning the luminosity, the spectral behavior and composition from GeV to ZeV energies, with several breaks in the spectrum and significant changes in the average mass number of the particles. The level and direction of the dipole in arrival directions could be quantified as a function of energy as well as the properties of the higher multipole spectrum. Since the late 1990s, TeV gamma-ray astronomy was added to the picture, with the most recent success of the first identification of a PeVatron in the Galaxy, more specifically in the Galactic Center region, possibly associated with the central black hole Sgr A$^*$. Another great success was the detection of hadronic emission from Supernova Remnants at GeV energies with Fermi. The latest additions to the multimessenger view on the Galaxy was the detection of astrophysical high-energy neutrinos (although the dominant part of the diffuse flux does not originate from the Milky Way) and gravitational waves from binary mergers. This vast progress at the frontier of multimessenger astronomy has lead to the development of a detailed theory of cosmic-ray acceleration, propagation, interaction and radiation as reviewed in this paper. The many details of the measured cosmic-ray properties is remindful of the parable of blind people investigating different parts of an elephant, seeing very different local pictures that can only be interpreted correctly when they are put together in the context of the entire elephant. The elephant in the room has been the class of supernova remnants since the first suggestion in the 1930s. It is quite intriguing that there is still no smoking gun for this hypothesis, despite the tremendous progress that has been made.

The following central conclusions can be drawn from the combination of theory and observations with open questions attached to them:
\begin{itemize}
    \item The \textbf{luminosity criterion} is the most established hint pointing toward the class of supernova remnants as the origin of cosmic-rays. Given the relative smoothness of the spectrum, the sources that dominate the spectrum at $\sim$GeV energies are expected to be the same that govern the energy range around the knee as well. GeV emission from SNRs is confirmed, PeVatron-emission has only been detected from the direction of Sgr~A$^*$. The contribution of this accelerator is energetically too weak to explain the entire spectrum (proton luminosity at the source of $\sim 10^{37}-10^{38}$~erg/s), but could contribute in the knee-to-ankle region. 
    \item The \textbf{spectral evolution of the all particle spectrum and the composition} is still one of the largest challenge in modern astroparticle physics. The break in the cosmic-ray spectrum at a rigidity around $\sim 300$~GV for basically all observed elements indicates that there is a change either in the transport properties via the self-generation of magnetic waves below the break, or a change in the acceleration scenario at the SNR. Future measurements of the B/C ratio and other tracers of interstellar spallation can distinguish between these two scenarios, measurements of local gamma-ray spectra that correspond to the energy range in which the break is occurring will further help to deduce a consistent model of cosmic-ray acceleration and transport. The anomalies in the antiparticle spectra provide further constraints to the spectral evolution of the sources and one of the challenges in the future is to clarify if these are due to astrophysical scenarios like via the contribution of different magnetic field configurations at the SNR shock front, or from dark matter annihilations.
    \item The \textbf{cosmic-ray anisotropy} adds important information about possible sources. In particular, the level of the dipole is significantly smaller (order $\delta I/I \sim 10^{-3}$) as compared to simple transport models from a cosmic-ray gradient (expected level $\sim 0.02$). The measured dipole intensity increases with energy until it drops at around $1$~PeV, where it also changes direction by around $180^{\circ}$. At higher energies, the amplitude rises again. Local streaming effects or the mapping of a gradient onto the local magnetic field structure have been suggested. The latter model naturally adds a multipole component via correlations in the arrival direction. But other options like the influence of the heliosphere could also explain the high-order multipole struture. The anisotropy problem might be connected to the cosmic-ray gradient problem that is derived from gamma-ray data as both problems arise due to a simplification of the diffusion model to a one-dimensional, space-independent problem. Introducting anisotropic diffusion with a tensor that changes with the turbulence level in the Galaxy might deliver the solution for both problems. There are two major challenges: (1) the turbulent component of the magnetic field in the Galaxy is not well-known. If it is available, it is only given on those specific scales that the telescopes are sensitive to, limited by their resolution. In addition, even the large-scale magnetic field is not completely known. While global models do exist, they rely on inter- and extrapolations especially for very distant regions. The Galactic Center region poses the largest problem here due to line-of-sight effects. (2) Simulations of the diffusion tensor are limited in performance and rely on the underlying turbulence model. These questions need to be solved before a solid answer to the anisotropy problem can be delivered.
\end{itemize}

In the future, 
the many facets of modern multimessenger-modeling will continue to shed more light on the diffuse component of the non-thermal component of the interstellar medium.
Data from radio to X-rays are needed in order to understand the electron population in combination with the magnetic field. Hence, we are in need of high-resolution synchrotron data in our Galaxy, in combination with an advanced understanding of line-of-sight effects and the coupling between magnetic fields and the electron population. Studies that help to understand the transport properties of cosmic-ray electrons in external galaxies are ongoing in e.g.\ the CHANG-ES poject \citep{Wiegert-etal-2015,Stein2018,Miskolczi2019}. The edge-on galaxies that are studied here at radio wavelengths provide valuable information on the outflow properties, i.e.\ if these are diffusion or advection dominated. It is interesting that the outcome is not the same for all galaxies and that there seem to exist both advection \citep{mora-partiarroyo2019,dettmar2019} and diffusion dominated \citep{stein2019} galaxies. For our own Galaxy, making such a case is much more difficult, since we are sitting in the middle of the system and cannot judge the process systematically from the outside. In particular, the spiral arm that is on the other side of the galaxy is almost impossible to detect properly.
At the highest energies, the new generation of gamma-ray telescopes will be dominated by the Cherenkov Telescope Array. The increased intensity in a large energy range up to 100 TeV gamma-ray energy, together with improved pointing will yield the possibility to study SNRs and other Galactic sources at high resolution. The morphology can be studied at multiwavelengths and leptonic and hadronic scenarios can be discriminated this way. Ionization signatures play a crucial role here for the identification of the cosmic-ray signal at GeV energies. But also synchrotron radiation is important to constrain the magnetic field and the electron population. By using all spatially resolved pieces of information, it might be possible to distinguish the $4\pi$ component from the polar cap component. The possibility of investigating a large field of view with CTA will provide a systematic view of the Galaxy at TeV gamma-ray energies. Furthermore, population studies will be possible. If at the same time, the morphology studies of nearby objects can help to identify hadronic accelerators, the detected population can be used to cross-check if their luminosity and spectrum is able to reproduce the cosmic-ray energy spectrum up to the knee and beyond. The smoking gun for the existence of cosmic ray acceleration up to the knee can be delivered by neutrino telescopes. Today's most sensitive neutrino telescope IceCube already starts to reach the sensitivity of the predictions for the diffuse neutrino flux produced by cosmic-ray interactions in the diffuse interstellar medium. The brightest gamma-ray SNRs will be detectable within the lifetimes of the next generation telescopes IceCube-Gen2 and KM3NeT.

Finally, putting together all details of the pieces of information of cosmic rays requires the development of state-of-the-art propagation codes, which are able to provide predictions for all observables with high-precision knowledge that enters from astro-, particle- and plasma-physics. 

\section*{Acknowledgments}
First of all, we would like to thank the anonymous referee to provide valuable feedback that largel improved the manuscript. Further, we would like to thank Steve Barwick, Rainer Beck, Peter L.\ Biermann, John Black, Ralf-J\"urgen Dettmar, Horst Fichtner, Stefan Funk, Rainer Grauer, Francis Halzen, Andreas Haungs, Marijke Haverkorn, Elisabeth A. Hays, Jörg Hörandel, Gudlaugur J{\`o}hannesson, Marita Krause, Philipp Mertsch, MJ P\"uschel, Mike Richman, Frank Schr\"oder, Anvar Shukurov, Andrew Strong, Satyendra Thoudam and Ellen Zweibel for detailed and valuable discussions that significantly improved this review article. JBT would further like to thank the entire IceCube collaboration for continuous and inspiring interactions. JBT and LM warmly thank Julien D\"orner, Bj\"orn Eichmann, Frederik Tenholt, Johan Wulff and Michael Zacharias for reading and commenting on the manuscript. We also thank Patrick Reichherzer for providing us with several figures as indicated in their captions. We would further like to give a specially large thank to  Lennart R.\ Baalmann, who during the (long) time period of writing this review continuously helped us with figures and technical details of the manuscript, always under efficient consideration of the contents of the review. Finally, we would like to thank the RAPP Center and its PIs (\url{www.rapp-center.de}) as well as the Department of Plasmas with Complex Interactions for the helpful discussions and other support.

\bibliographystyle{elsarticle-harv}
\bibliography{lib,lib2}
\end{document}